\documentclass[
aip,
jcp,
reprint,
superscriptaddress,
 amsmath,amssymb,
floatfix,
]{revtex4-2}

\usepackage{graphicx}% Include figure files
\usepackage{dcolumn}% Align table columns on decimal point
\usepackage{bm}% bold math
\usepackage{listings}
\usepackage{dsfont}
\usepackage{color}
\usepackage{silence}
\usepackage[normalem]{ulem}
\usepackage{braket}
\usepackage[version=4]{mhchem}
\usepackage[margin=2cm]{geometry}
\usepackage{tikz}
\usetikzlibrary{shapes.geometric, arrows, arrows.meta}
\usepackage[ colorlinks,
    linkcolor={blue},
    citecolor={blue},
    urlcolor={blue}]{hyperref}

\DeclareFontShape{OT1}{cmss}{bx}{it}{<->ssub*cmss/bx/n}{}
\DeclareFontShape{OT1}{cmss}{bx}{sc}{<->ssub*cmss/bx/n}{}
\DeclareFontShape{OT1}{cmss}{bx}{scit}{<->ssub*cmss/bx/n}{}

\makeatletter
\def\@bibdataout@aip{%
 \immediate\write\@bibdataout{%
  @CONTROL{%
   aip41Control%
   \longbibliography@sw{\true@sw}{\aip@jtitx@sw{\false@sw}{\true@sw}}%
   {%
    ,author="48",pages="1",title="0"%
   }{%
    ,author="48",pages="0",title=""%
   }%
  }%
 }%
 \if@filesw
  \immediate\write\@auxout{\string\citation{aip41Control}}%
 \fi
}%

\let\SIsection\section
\let\SIsubsection\subsection
\let\SIsubsubsection\subsubsection

\def\section{%
  \@startsection{section}{1}{\z@}%
    {0.5cm \@plus.4ex \@minus .2ex}{0.3cm}%
    {\normalfont\small\sffamily\bfseries\raggedright}%
}
\def\subsection{%
  \@startsection{subsection}{2}{\z@}%
    {0.5cm \@plus.4ex \@minus .2ex}{0.3cm}%
    {\normalfont\small\sffamily\bfseries\raggedright}%
}
\def\subsubsection{%
  \@startsection{subsubsection}{3}{\z@}%
    {0.5cm \@plus.4ex \@minus .2ex}{0.3cm}%
    {\normalfont\small\sffamily\bfseries\itshape\raggedright}%
}
\def\bibfont{%
  \fontsize{8}{9}\selectfont
  \@clubpenalty\clubpenalty
  \labelsep\z@
}
\makeatother

\newcommand{\br}{\mathbf{r}}

\newcommand{\bG}{\mathbf{G}}
\newcommand{\bk}{\mathbf{k}}
\newcommand{\bR}{\mathbf{R}}

\newcommand\eqt{\hspace{0.17em}{=}\hspace{0.17em}}

\newcommand\mt{\hspace{0.17em}{-}\hspace{0.17em}}
\newcommand\eigvalksdft{\varepsilon_n^\text{KS}}

\newcommand{\vecr}{\ensuremath{\mathbf{r}}}

\newcommand{\bre}{\br_e}
\newcommand{\brh}{\br_h}

\newcommand{\Xian}{{X}_{ia}^{(n)}}
\newcommand{\Yian}{{Y}_{ia}^{(n)}}

\newcommand{\Psiexc}{\Psi_\mathrm{exc}^{(n)}}
\newcommand{\Omexc}{\Omega^{(n)}}
\newcommand{\bX}{\mathbf{X}}
\newcommand{\bY}{\mathbf{Y}}

\definecolor{maxviolet}{rgb}{0.52,0,0.56}

\newcommand{\Tr}{\mathrm{Tr}}
\newcommand{\dd}{\mathrm{d}}

\DeclareRobustCommand{\TDKout}[1]{{\color{red}\ifmmode\text{\sout{\ensuremath{#1}}}\else\sout{#1}\fi}}

\definecolor{tdkbrown}{rgb}{0.50,0.25,0.05}

\begin{document}

\title{\textbf{CP2K: An electronic structure and molecular dynamics software package -- Dynamics, Transport, and Spectroscopic Response
%Atomistic Dynamics, Transport, and Response Properties
%First-Principles Dynamics, Response, and Spectroscopy
%First-Principles Dynamics and Spectroscopic Response
%From First-Principles Dynamics to Spectroscopic Observables
}}%

\author{Jan Wilhelm}
\affiliation{\mbox{Institute of Theoretical Physics and Regensburg Center for Ultrafast Nanoscopy (RUN),} \mbox{University of Regensburg, 93053 Regensburg, Germany}}
\author{Anna-Sophia Hehn}
\affiliation{\mbox{Institute of Physical Chemistry, Christian-Albrechts-University Kiel, 24118 Kiel, Germany}}
\author{Hossam Elgabarty}
\affiliation{\mbox{Department of Chemistry, Paderborn University, 33098 Paderborn, Germany}}
\author{Beliz Sertcan G\"okmen}
\affiliation{\mbox{Department of Chemistry, University of Zurich, 8057 Zurich, Switzerland}}
\author{Maximilian Graml}
\affiliation{\mbox{Institute of Theoretical Physics and Regensburg Center for Ultrafast Nanoscopy (RUN),} \mbox{University of Regensburg, 93053 Regensburg, Germany}}
\author{Štěpán Marek}
\affiliation{\mbox{Institute of Theoretical Physics and Regensburg Center for Ultrafast Nanoscopy (RUN),} \mbox{University of Regensburg, 93053 Regensburg, Germany}}
\author{Ritaj Tyagi}
\affiliation{\mbox{Institute of Theoretical Physics and Regensburg Center for Ultrafast Nanoscopy (RUN),} \mbox{University of Regensburg, 93053 Regensburg, Germany}}
\affiliation{\mbox{Department of Chemical Sciences, Tata Institute of Fundamental Research, Mumbai 400005, India}}
\author{Frederick Stein}
\affiliation{\mbox{Center for Advanced Systems Understanding (CASUS), 02826 Görlitz, Germany}}
\affiliation{\mbox{Helmholtz-Zentrum Dresden-Rossendorf, 01328 Dresden, Germany}}
\author{Johann V. Pototschnig}
\affiliation{\mbox{Center for Advanced Systems Understanding (CASUS), 02826 Görlitz, Germany}}
\affiliation{\mbox{Helmholtz-Zentrum Dresden-Rossendorf, 01328 Dresden, Germany}}
\author{Christian S. Ahart}
\affiliation{\mbox{Department of Physics, School of Science and Research Center for} \mbox{Industries of the Future, Westlake University, Hangzhou, 310030, China}}
\author{Zehua Chen}
\affiliation{\mbox{Theoretical Chemistry Institute, Department of Chemistry, University of Wisconsin--Madison,} \mbox{Madison, Wisconsin 53706, United States}}
\author{Linnéa Andersson}
\affiliation{\mbox{Department of Chemistry-Ångström Laboratory, Uppsala University, Uppsala, Sweden, 75121}}
\author{Zdenek Futera}
\affiliation{\mbox{Faculty of Science, University of South Bohemia, 370 05 Ceske Budejovice, Czech Republic}}
\author{Filip Ivanovic}
\affiliation{\mbox{Department of Physics and Astronomy and Thomas Young Centre,}
\mbox{University College London, London, WC1E 6BT, United Kingdom}}
\author{Margherita Buraschi}
\affiliation{\mbox{Department of Chemistry, Imperial College London, London, W12 0BZ, United Kingdom}}
\author{Christoph Schran}
\affiliation{\mbox{Cavendish Laboratory, Department of Physics, University of Cambridge,} \mbox{CB3 0HE Cambridge, UK}}
\author{Rémi Pasquier}
\affiliation{\mbox{Institute of Theoretical Physics and Regensburg Center for Ultrafast Nanoscopy (RUN),} \mbox{University of Regensburg, 93053 Regensburg, Germany}}
\author{Leonard Prokisch}
\affiliation{\mbox{Institute of Theoretical Physics and Regensburg Center for Ultrafast Nanoscopy (RUN),} \mbox{University of Regensburg, 93053 Regensburg, Germany}}
\author{Bibek Samal}
\affiliation{\mbox{Institute of Theoretical Physics and Regensburg Center for Ultrafast Nanoscopy (RUN),} \mbox{University of Regensburg, 93053 Regensburg, Germany}}
\author{Jelena Schmitz}
\affiliation{\mbox{Institute of Theoretical Physics and Regensburg Center for Ultrafast Nanoscopy (RUN),} \mbox{University of Regensburg, 93053 Regensburg, Germany}}
\author{Shridhar Sanjay Shanbhag}
\affiliation{\mbox{Institute of Theoretical Physics and Regensburg Center for Ultrafast Nanoscopy (RUN),} \mbox{University of Regensburg, 93053 Regensburg, Germany}}
\author{Harald Forbert}
\affiliation{\mbox{Center for Solvation Science ZEMOS, Ruhr-Universität Bochum, 44801 Bochum, Germany}}
\author{Ole Schütt}
\affiliation{\mbox{CP2K Lab, 8049 Zürich, Switzerland}}
\author{Augustin Bussy}
\affiliation{\mbox{Swiss National Supercomputing Centre (CSCS), ETH Zurich, CH-6900 Lugano, Switzerland}}
%\author{Hossein Mirhosseini}
%\affiliation{\mbox{Center for Advanced Systems Understanding (CASUS), 02826 Görlitz, Germany}}
%\affiliation{\mbox{Helmholtz-Zentrum Dresden-Rossendorf, 01328 Dresden, Germany}}
\author{Franz P\"oschl}
\affiliation{\mbox{Center for Advanced Systems Understanding (CASUS), 02826 Görlitz, Germany}}
\affiliation{\mbox{Helmholtz-Zentrum Dresden-Rossendorf, 01328 Dresden, Germany}}
\author{Sebastian Ehlert}
\affiliation{\mbox{Microsoft Research AI for Science, 10178 Berlin, Germany}}
\author{Stefano Battaglia}
\affiliation{\mbox{Microsoft Research AI for Science, 1118 CZ Schiphol, Netherlands}}
\author{Michele Nottoli}
\affiliation{\mbox{Institute of Applied Analysis and Numerical Simulation, University of Stuttgart,} \mbox{70569 Stuttgart, Germany}}
\author{Benjamin Stamm}
\affiliation{\mbox{Institute of Applied Analysis and Numerical Simulation, University of Stuttgart,} \mbox{70569 Stuttgart, Germany}}
\author{Vamsee Voora}
\affiliation{\mbox{Department of Chemical Sciences, Tata Institute of Fundamental Research, Mumbai 400005, India}}
\author{Rustam Z. Khaliullin}
\affiliation{\mbox{Department of Chemistry, McGill University, Montreal QC H3A 0B8, Canada}}
\author{Sergey K. Chulkov}
\affiliation{\mbox{School of Mathematics and Physics, University of Lincoln, Lincoln LN6 7TS, United Kingdom}}
\author{Matt B. Watkins}
\affiliation{\mbox{School of Mathematics and Physics, University of Lincoln, Lincoln LN6 7TS, United Kingdom}}
\author{Clotilde S. Cucinotta}
\affiliation{\mbox{Department of Chemistry, Imperial College London, London, W12 0BZ, United Kingdom}}
\author{Jochen Blumberger}
\affiliation{\mbox{Department of Physics and Astronomy and Thomas Young Centre,}
\mbox{University College London, London, WC1E 6BT, United Kingdom}}
\author{Chao Zhang}
\affiliation{\mbox{Department of Chemistry-Ångström Laboratory, Uppsala University, Uppsala, Sweden, 75121}}
\affiliation{\mbox{Wallenberg Initiative Materials Science for Sustainability, Uppsala, Sweden, 75121}}
\author{Yang Yang}
\affiliation{\mbox{Theoretical Chemistry Institute, Department of Chemistry, University of Wisconsin--Madison,} \mbox{Madison, Wisconsin 53706, United States}}
\author{Dominik Marx}
\affiliation{\mbox{Lehrstuhl für Theoretische Chemie, Ruhr-Universität Bochum, 44780 Bochum, Germany}}
\author{Matthias Krack}
\affiliation{\mbox{PSI Center for Scientific Computing, Theory and Data, Paul Scherrer Institute,} \mbox{5232 Villigen PSI, Switzerland}}
\author{Jürg Hutter}
\affiliation{\mbox{Department of Chemistry, University of Zurich, 8057 Zurich, Switzerland}}
\author{Marcella Iannuzzi}
\affiliation{\mbox{Department of Chemistry, University of Zurich, 8057 Zurich, Switzerland}}
\author{Thomas D. Kühne}
\email{tkuehne@cp2k.org}
\affiliation{\mbox{Center for Advanced Systems Understanding (CASUS), 02826 Görlitz, Germany}}
\affiliation{\mbox{Helmholtz-Zentrum Dresden-Rossendorf, 01328 Dresden, Germany}}
\affiliation{\mbox{Institute of Artificial Intelligence, Technische Universit\"at Dresden, 01069 Dresden, Germany}}

\date{\today}% It is always \today, today,

\begin{abstract}
%One of the unique aspects of CP2K is its ability to combine various structural and transition state optimization methods, as well as (rare event) sampling methods such as Monte Carlo, meta- and molecular dynamics (MD), with a large variety of energy and force methods reaching from classical simulations, over mixed quantum-classical and semi-empirical approaches to sophisticated quantum mechanical electronic structure methods. At the core of the latter is the Gaussian and plane wave approach and its augmented all-electron generalization, which has been extensively discussed in our previous code review paper [T. D. Kühne et al., J. Chem. Phys. 152, 194103 (2020)]. Therefore, the present focus is on dynamics and spectroscopy as a manifestation of CP2K to unify quantum chemistry with quantum and statistical mechanics.

One of the distinguishing aspects of CP2K is its seamless integration of diverse structural and transition-state optimization techniques with advanced sampling approaches including Monte Carlo, molecular dynamics, and metadynamics, enabling the efficient exploration of complex potential- and free-energy landscapes, including rare events. These capabilities are combined with a broad hierarchy of energy and force evaluation methods, ranging from classical and machine-learned interaction potentials and mixed quantum--classical multiscale and semiempirical schemes, to highly accurate quantum-mechanical electronic-structure approaches. At the heart of the latter lies the Gaussian and plane-wave framework, along with its augmented all-electron generalization, which have been described in detail in our previous code review [T. D. Kühne et al., J. Chem. Phys. 152, 194103 (2020)]. Building on this foundation, the present work revisits the methods within CP2K that turn electronic structure into dynamics, transport, and spectroscopic response. Particular emphasis is placed on the coupling between static response calculations and nuclear motion: spectra may be evaluated at optimized structures, averaged over thermally sampled configurations, obtained from time-correlation functions along \textit{ab-initio} or path-
%dm -
integral molecular trajectories, or followed in real time together with electronic and nuclear dynamics. The same modular structure also enables equilibrium and biased transport simulations, from Kubo-type linear response to open-boundary approaches under external potentials, highlighting CP2K's unique capability to unify electronic-structure theory with statistical mechanics as well as classical and quantum nuclear dynamics within a versatile, holistic simulation environment.
\end{abstract}

%Use showkeys class option if keyword

\maketitle

%\tableofcontents

\section{Introduction}

Over the past decades, electronic-structure theory has reached a level of maturity that enables the reliable prediction of structural, energetic, and spectroscopic properties across a wide range of chemical systems.\cite{Helgaker2000,martin2004} In this context, static, zero-temperature approaches, based on optimized geometries and harmonic approximations, have become the workhorse for interpreting and predicting experimental observables.\cite{Barone2021ComputationalMolecularSpectroscopy} Although remarkably successful, such approaches neglect finite-temperature nuclear motion, anharmonicity, and dynamical disorder, which are often essential for quantitatively accurate and sometimes even qualitatively correct spectra.\cite{Jansen2021ComputationalSpectroscopyComplexSystems}

Spectroscopy, by its very nature, probes not only electronic structure but also the coupled dynamics of electrons and nuclei.\cite{Mukamel1995} Vibrational, infrared, Raman, and terahertz spectra,\cite{Perakis2016} as well as core- and valence-level spectroscopies,\cite{deGroot2008CoreLevel} are strongly influenced by thermal fluctuations, solvent environments, and anharmonic nuclear motion.\cite{HammZanni2011} In condensed-phase systems, heterogeneous environments and fluctuating hydrogen-bond networks further broaden and shift spectral features, rendering static pictures insufficient.\cite{Bakker2010VibrationalWater} As a consequence, a consistent theoretical description of spectroscopy increasingly requires going beyond the potential energy surface at a single minimum and instead sampling the underlying free energy landscape at finite temperature.\cite{vib_martin,Ditler2022}

This shift from static to dynamical descriptions naturally calls for the integration of electronic-structure methods with statistical mechanics.\cite{CarParrinello1985,Kuehne2007} Molecular dynamics (MD),\cite{FrenkelSmit2002} whether classical, \textit{ab-initio} MD (AIMD),\cite{Marx2009,HutterIannuzziKuehne2024} or enhanced by advanced sampling techniques,\cite{PLUMEDConsortium2019Transparency} provides a rigorous framework to account for thermal fluctuations and time-dependent correlations.\cite{Tuckerman2010} In this setting, spectroscopic observables can be computed from time-correlation functions,\cite{Kubo1957/10.1143/JPSJ.12.570} enabling a direct connection between microscopic dynamics and experimentally measurable quantities.\cite{Ceperley1999Microscopic} Importantly, this framework also allows for the inclusion of nuclear quantum effects (NQEs),\cite{Cao1994a,Craig2004} which can play a decisive role for light atoms,\cite{Ceperley1995} particularly in hydrogen-bonded networks and proton-transfer processes.\cite{Marx1999HydratedProton}

Within this broader context, CP2K occupies a unique position. By combining efficient and accurate density-functional theory (DFT) methods, most notably the Gaussian and Plane-Wave (GPW) approach and its Gaussian Augmented Plane-Wave (GAPW) all-electron extension, with a wide range of MD, enhanced sampling, and response methods such as variational density-functional perturbation theory (DFPT),\cite{Putrino2000,Baroni2001} CP2K provides a unified platform for exploring both potential- and free-energy landscapes.\cite{Kuehne2020,Iannuzzi2025} This enables the seamless transition from static structure determination to fully dynamical simulations under realistic, finite-temperature conditions.\cite{Hutter2012CPMD,Kuehne2014SecondGenerationCPMD} Table~S1 in the Supplementary Information (SI) lists further boundary cases.

\providecommand{\ReviewTableCell}[2]{\begin{minipage}[t]{#1}\raggedright #2\strut\end{minipage}}

% --- adjustable table column widths: table fills the FULL text width ---------
% Available width = \textwidth minus the column padding (5 columns x 2 x \tabcolsep(3pt) = 30pt).
% The five fractions below must sum to 1.000; adjust them to redistribute space.
\newlength{\ReviewTableAvail}
\setlength{\ReviewTableAvail}{\dimexpr\textwidth-30pt\relax}
\newlength{\ReviewColA}\setlength{\ReviewColA}{0.126\ReviewTableAvail}% Spectrum / response
\newlength{\ReviewColB}\setlength{\ReviewColB}{0.253\ReviewTableAvail}% CP2K method / keyword
\newlength{\ReviewColC}\setlength{\ReviewColC}{0.165\ReviewTableAvail}% Applied probe; varied parameters
\newlength{\ReviewColD}\setlength{\ReviewColD}{0.133\ReviewTableAvail}% Detected signal
\newlength{\ReviewColE}\setlength{\ReviewColE}{0.323\ReviewTableAvail}% Theoretical description
% --- adjustable table spacing ------------------------------------------------
\newlength{\ReviewRowSep}        % vertical space after every table row
\setlength{\ReviewRowSep}{0.20em}
\newlength{\ReviewGroupSkipAbove}% strut height above a bold group heading
\setlength{\ReviewGroupSkipAbove}{1.0em}
\newlength{\ReviewGroupSkipBelow}% vertical space below a bold group heading
\setlength{\ReviewGroupSkipBelow}{0.5em}
\newlength{\ReviewRowGapTopMinus}
\setlength{\ReviewRowGapTopMinus}{-0.15em}
% -----------------------------------------------------------------------------
% REVTeX ignores the optional argument of \\ inside tabular, so all vertical
% spacing is inserted via \noalign, driven by the lengths above.
\providecommand{\ReviewRowGap}{\noalign{\vskip\ReviewRowSep}}
\providecommand{\ReviewTableGroup}[1]{\multicolumn{5}{l}{\rule{0pt}{\ReviewGroupSkipAbove}\textbf{#1}} \\ \noalign{\vskip\ReviewGroupSkipBelow}}
\definecolor{SpecCoveredGreen}{rgb}{0.00,0.45,0.20}
\definecolor{SpecMentioned}{rgb}{0.70,0.45,0.10}
\DeclareRobustCommand{\SpecCovered}[1]{#1}
\DeclareRobustCommand{\SpecPartial}[1]{#1}
\DeclareRobustCommand{\SpecMissing}[1]{#1}

\begin{table*}[p]
  \caption{CP2K spectroscopy and response capabilities. Each row lists a spectroscopic or response observable together with the corresponding CP2K methods and keywords, the applied probe and its experimentally varied parameters, the detected signal, and a brief theoretical description with pointers to the relevant manuscript sections.} \label{t1}
\vspace{0.3em}
  \linespread{0.868}\scriptsize
  \setlength{\tabcolsep}{3pt}
  \renewcommand{\arraystretch}{0.92}

  \begin{tabular}{lllll}

    \hline\hline
    \rule{0pt}{1.1em}%
    \ReviewTableCell{\ReviewColA}{\textbf{Spectrum / response}} &
    \ReviewTableCell{\ReviewColB}{\textbf{CP2K method / keyword}} &
    \ReviewTableCell{\ReviewColC}{\textbf{Applied probe and varied parameters}} &
    \ReviewTableCell{\ReviewColD}{\textbf{Detected signal}} &
    \ReviewTableCell{\ReviewColE}{\textbf{Theoretical description of the perturbation / spectroscopy}} \\ \noalign{\vskip 0.25em}
    \hline
    \ReviewTableGroup{Magnetic and nuclear-spin response}
    \ReviewTableCell{\ReviewColA}{\SpecCovered{Nuclear magnetic resonance chemical shifts and nucleus-independent chemical shifts}} &
    \ReviewTableCell{\ReviewColB}{\SpecCovered{\texttt{NMR} in \texttt{\&PROPERTIES\%LINRES}; \texttt{LOCALIZE} in \texttt{\&DFT}; Section~\ref{sec:magn-shield-tens}}} &
    \ReviewTableCell{\ReviewColC}{\SpecCovered{Static magnetic field and weak radio-frequency field. \textit{Varied:} field direction, chemical environment}} &
    \ReviewTableCell{\ReviewColD}{\SpecCovered{Shielding tensor, chemical shift, nucleus-independent chemical-shift map}} &
    \ReviewTableCell{\ReviewColE}{\SpecCovered{Magnetic-field DFPT yields shielding tensors. Finite-temperature chemical shifts can be obtained by configurational averaging over MD snapshots.}} \\[\ReviewRowGapTopMinus] \ReviewRowGap
    \ReviewTableCell{\ReviewColA}{\SpecPartial{Nuclear magnetic resonance spin--spin couplings}} &
    \ReviewTableCell{\ReviewColB}{\SpecPartial{\texttt{\&PROPERTIES\%LINRES\%SPINSPIN}; Section~\ref{sec:magn-shield-tens}}} &
    \ReviewTableCell{\ReviewColC}{\SpecPartial{Magnetic perturbations at nuclei. \textit{Varied:} nuclear pair}} &
    \ReviewTableCell{\ReviewColD}{\SpecPartial{Indirect spin--spin coupling constants}} &
    \ReviewTableCell{\ReviewColE}{\SpecPartial{Linear-response spin--spin perturbations yield reduced indirect coupling tensors or constants. Complete multiplet line shapes require external line-shape construction, as discussed in Table~S1 in the SI.}} \\[\ReviewRowGapTopMinus] \ReviewRowGap
    \ReviewTableCell{\ReviewColA}{\SpecMissing{Nuclear quadrupole interaction parameters}} &
    \ReviewTableCell{\ReviewColB}{\SpecMissing{\texttt{ELECTRIC\_FIELD\_}\linebreak\texttt{GRADIENT} and \texttt{AO\_MATRICES} \linebreak with \texttt{EFG} in \texttt{\&DFT\%PRINT}}} &
    \ReviewTableCell{\ReviewColC}{\SpecMissing{Static electric-field gradient at nuclei. \textit{Varied:} nuclear site, isotope quadrupole moment, geometry}} &
    \ReviewTableCell{\ReviewColD}{\SpecMissing{Electric-field-gradient tensor and quadrupole-coupling parameters}} &
    \ReviewTableCell{\ReviewColE}{\SpecMissing{CP2K can print electric-field gradients at atomic positions and the corresponding one-electron matrices. Conversion to nuclear-quadrupole-resonance or solid-state nuclear magnetic resonance quadrupolar splittings is not covered in the manuscript.}} \\[\ReviewRowGapTopMinus] \ReviewRowGap
    \ReviewTableCell{\ReviewColA}{\SpecCovered{Electron paramagnetic resonance}} &
    \ReviewTableCell{\ReviewColB}{\SpecCovered{\texttt{EPR} in \texttt{\&PROPERTIES\%LINRES}; \texttt{HYPERFINE\_COUPLING\_TENSOR} \linebreak in \texttt{\&DFT\%PRINT}; Sections~\ref{sec:epr-g-tensor} and~\ref{sec:hyperfine-couplings}}} &
    \ReviewTableCell{\ReviewColC}{\SpecCovered{Static magnetic field acting on an open-shell system. \textit{Varied:} field orientation, spin state}} &
    \ReviewTableCell{\ReviewColD}{\SpecCovered{$g$ tensor and hyperfine couplings}} &
    \ReviewTableCell{\ReviewColE}{\SpecCovered{Magnetic-response and hyperfine calculations provide the tensors entering an electron-paramagnetic-resonance spin Hamiltonian.}} \\[\ReviewRowGapTopMinus] \ReviewRowGap
    \ReviewTableGroup{Vibrational spectroscopy}
    \ReviewTableCell{\ReviewColA}{\SpecCovered{Infrared and terahertz absorption}} &
    \ReviewTableCell{\ReviewColB}{\SpecCovered{\texttt{VIBRATIONAL\_ANALYSIS}; \texttt{DCDR} in \texttt{\&PROPERTIES\%LINRES}; \texttt{MOMENTS} \linebreak in \texttt{\&DFT\%PRINT}; Berry-phase polarization; Sections~\ref{sec:dfpt-raman}, \ref{sec:aimd-finite-temperature-spectra}, and~\ref{sec:path-integral-dynamics-spectra}}} &
    \ReviewTableCell{\ReviewColC}{\SpecCovered{Infrared or terahertz electric field. \textit{Varied:} photon frequency, temperature}} &
    \ReviewTableCell{\ReviewColD}{\SpecCovered{Absorption intensity from dipole or polarization response}} &
    \ReviewTableCell{\ReviewColE}{\SpecCovered{Harmonic spectra from normal modes and dipole derivatives. Anharmonic finite-temperature spectra from dipole or polarization time-correlation functions along \textit{ab-initio} or path-
%dm -
integral MD trajectories.\cite{vib_martin,voronoi_vib_martin,schienbein2025,schienbein2026mimyriamachinelearnedvibrational}}} \\[\ReviewRowGapTopMinus] \ReviewRowGap
    \ReviewTableCell{\ReviewColA}{\SpecCovered{Raman scattering}} &
    \ReviewTableCell{\ReviewColB}{\SpecCovered{\texttt{\&PROPERTIES\%LINRES\%POLAR}, \texttt{DO\_RAMAN}; polarizability time-correlation functions; Sections~\ref{sec:dfpt-raman} and~\ref{sec:aimd-finite-temperature-spectra}}} &
    \ReviewTableCell{\ReviewColC}{\SpecCovered{Visible or near-infrared light. \textit{Varied:} frequency shift, polarization, temperature}} &
    \ReviewTableCell{\ReviewColD}{\SpecCovered{Polarizability derivatives or polarizability correlations}} &
    \ReviewTableCell{\ReviewColE}{\SpecCovered{Harmonic Raman tensors follow from polarizability derivatives, while finite-temperature liquid or condensed-phase Raman spectra are obtained from polarizability time-correlation functions along AIMD trajectories.\cite{raman_luber_marcella}}} \\[\ReviewRowGapTopMinus] \ReviewRowGap
    \ReviewTableCell{\ReviewColA}{\SpecPartial{Sum-frequency generation and related nonlinear vibrational spectra}} &
    \ReviewTableCell{\ReviewColB}{\SpecPartial{Finite-field response and trajectory correlation functions; Sections~\ref{sec:aimd-finite-temperature-spectra} and~\ref{sec:finite-field-aimd}}} &
    \ReviewTableCell{\ReviewColC}{\SpecPartial{Two incident optical or infrared fields. \textit{Varied:} frequencies, polarization, interface orientation}} &
    \ReviewTableCell{\ReviewColD}{\SpecPartial{Nonlinear polarization signal}} &
    \ReviewTableCell{\ReviewColE}{\SpecPartial{Nonlinear vibrational response is described through higher-order polarization or response-tensor correlation functions, evaluated along interfacial MD trajectories.}} \\[\ReviewRowGapTopMinus] \ReviewRowGap
    \ReviewTableGroup{Valence optical and excited-state spectroscopy}
    \ReviewTableCell{\ReviewColA}{\SpecCovered{Ultraviolet--visible valence absorption}} &
    \ReviewTableCell{\ReviewColB}{\SpecCovered{\texttt{\&PROPERTIES\%TDDFPT}; \linebreak \texttt{REAL\_TIME\_PROPAGATION} in \texttt{\&DFT}; \texttt{RI\_RPA\%GW\%BSE} in \texttt{\&XC\%WF\_CORRELATION}; Sections~\ref{sec-lr_tddft}, \ref{sec-rt_tddft}, and~\ref{sec-lr_bse}}} &
    \ReviewTableCell{\ReviewColC}{\SpecCovered{Ultraviolet--visible electric field. \textit{Varied:} photon energy, polarization}} &
    \ReviewTableCell{\ReviewColD}{\SpecCovered{Excitation energies and oscillator strengths}} &
    \ReviewTableCell{\ReviewColE}{\SpecCovered{Linear-response and real-time time-dependent density-functional theory calculations describe valence excitations and oscillator strengths. $GW$ plus Bethe--Salpeter equation methods provide many-body excitation spectra for selected systems.\cite{Hehn2022,Sertcan2024,vogt2025,Marek2025,Iannuzzi2025}}} \\[\ReviewRowGapTopMinus] \ReviewRowGap
    \ReviewTableCell{\ReviewColA}{\SpecCovered{Spin--orbit-corrected electronic excitation spectra}} &
    \ReviewTableCell{\ReviewColB}{\SpecCovered{Spin--orbit coupling in \texttt{\&PROPERTIES\%TDDFPT}; Section~\ref{spin_orbit_coupling_section}}} &
    \ReviewTableCell{\ReviewColC}{\SpecCovered{Optical excitation in systems with spin--orbit coupling. \textit{Varied:} spin--orbit coupling, photon energy}} &
    \ReviewTableCell{\ReviewColD}{\SpecCovered{Spin--orbit-corrected excitation energies and intensities}} &
    \ReviewTableCell{\ReviewColE}{\SpecCovered{Perturbative spin--orbit coupling extends electronic excitation calculations to spin--orbit-corrected energies and intensities.}} \\[\ReviewRowGapTopMinus] \ReviewRowGap
    \ReviewTableCell{\ReviewColA}{\SpecPartial{Semiempirical broadband optical spectra}} &
    \ReviewTableCell{\ReviewColB}{\SpecPartial{\texttt{STDA} in \texttt{\&PROPERTIES\%TDDFPT}, xTB and semiempirical kernels; Sections~\ref{sec-lr_tddft}, \ref{sec:periodic-xtb}, and~\ref{sec:nonadiabatic-md-surface-hopping}}} &
    \ReviewTableCell{\ReviewColC}{\SpecPartial{Ultraviolet--visible field. \textit{Varied:} photon energy, ensemble geometry}} &
    \ReviewTableCell{\ReviewColD}{\SpecPartial{Broad-band absorption profile}} &
    \ReviewTableCell{\ReviewColE}{\SpecPartial{Simplified time-dependent density-functional theory and semiempirical kernels approximate excitation spectra for large configurational ensembles and nonadiabatic dynamics simulations.}} \\[\ReviewRowGapTopMinus] \ReviewRowGap
    \ReviewTableCell{\ReviewColA}{\SpecCovered{Bethe--Salpeter optical absorption}} &
    \ReviewTableCell{\ReviewColB}{\SpecCovered{\texttt{RI\_RPA\%GW\%BSE} in \texttt{\&XC\%WF\_}\linebreak\texttt{CORRELATION}; \texttt{DO\_BSE} in \texttt{\&PROPERTIES\%TDDFPT}; \linebreak Section~\ref{sec-lr_bse}}} &
    \ReviewTableCell{\ReviewColC}{\SpecCovered{Optical electric field. \textit{Varied:} photon energy, broadening}} &
    \ReviewTableCell{\ReviewColD}{\SpecCovered{Excitonic absorption spectrum}} &
    \ReviewTableCell{\ReviewColE}{\SpecCovered{$GW$ quasiparticle energies combined with the Bethe--Salpeter equation describe screened electron--hole interactions and excitonic optical absorption.}} \\[\ReviewRowGapTopMinus] \ReviewRowGap
    \ReviewTableCell{\ReviewColA}{\SpecPartial{Vibronic optical spectra and Franck--Condon progressions}} &
    \ReviewTableCell{\ReviewColB}{\SpecPartial{\texttt{VIBRATIONAL\_ANALYSIS}; \texttt{\&PROPERTIES\%TDDFPT}; excited-state forces; \texttt{tools/vibronic\_spec}; \linebreak Sections~\ref{sec-lr_tddft}, \ref{sec:aimd-finite-temperature-spectra}, and~\ref{sec:path-integral-dynamics-spectra}}} &
    \ReviewTableCell{\ReviewColC}{\SpecPartial{Optical electronic excitation coupled to vibrational modes. \textit{Varied:} photon energy, vibrational quantum number, temperature}} &
    \ReviewTableCell{\ReviewColD}{\SpecPartial{Vibronic absorption or emission envelope}} &
    \ReviewTableCell{\ReviewColE}{\SpecPartial{Normal modes, time-dependent density-functional theory excitation energies, and excited-state forces feed the bundled VibronicSpec post-processing tool for Franck--Condon-type spectra. More general vibronic line-shape approaches remain boundary cases discussed in Table~S1 in the SI.}} \\[\ReviewRowGapTopMinus] \ReviewRowGap
    \hline\hline

  \end{tabular}
\end{table*}
\begin{table*}[p]
  \addtocounter{table}{-1}
  \renewcommand{\theHtable}{\thetable.cont1}
  \caption{(Continued.)}
\vspace{0.3em}
  \linespread{0.868}\scriptsize
  \setlength{\tabcolsep}{3pt}
  \renewcommand{\arraystretch}{0.92}

  \begin{tabular}{lllll}

    \hline\hline
    \rule{0pt}{1.1em}%
    \ReviewTableCell{\ReviewColA}{\textbf{Spectrum / response}} &
    \ReviewTableCell{\ReviewColB}{\textbf{CP2K method / keyword}} &
    \ReviewTableCell{\ReviewColC}{\textbf{Applied probe and varied parameters}} &
    \ReviewTableCell{\ReviewColD}{\textbf{Detected signal}} &
    \ReviewTableCell{\ReviewColE}{\textbf{Theoretical description of the perturbation / spectroscopy}} \\ \noalign{\vskip 0.25em}
    \hline
    \ReviewTableGroup{Real-time nonlinear and ultrafast response}
    \ReviewTableCell{\ReviewColA}{\SpecCovered{Nonlinear optical response from real-time propagation}} &
    \ReviewTableCell{\ReviewColB}{\SpecCovered{\texttt{REAL\_TIME\_PROPAGATION} in \texttt{\&DFT}; \texttt{RTBSE} in \texttt{\&REAL\_TIME\_PROPAGATION}; \texttt{MOMENTS\_FT}, \texttt{POLARIZABILITY}; Sections~\ref{sec-rt_tddft}, \ref{sec:rt-tddft-delta-kick}, and~\ref{sec-rt_bse}}} &
    \ReviewTableCell{\ReviewColC}{\SpecCovered{Time-dependent electric field or delta kick. \textit{Varied:} pulse shape, field strength, delay, frequency}} &
    \ReviewTableCell{\ReviewColD}{\SpecCovered{Time-dependent dipole, polarizability, harmonic or nonlinear response}} &
    \ReviewTableCell{\ReviewColE}{\SpecCovered{The pulse is treated as an external time-dependent field $\mathbf{E}(t)$. Either real-time time-dependent density-functional theory or the real-time Bethe--Salpeter equation yields the nonlinear polarization response at fixed nuclei or with Ehrenfest nuclear motion.\cite{Marek2025,Iannuzzi2025}}} \\[\ReviewRowGapTopMinus] \ReviewRowGap
    \ReviewTableCell{\ReviewColA}{\SpecCovered{Ultrafast pump--probe spectra}} &
    \ReviewTableCell{\ReviewColB}{\SpecCovered{\texttt{REAL\_TIME\_PROPAGATION} in \texttt{\&DFT}; delta-kick probe, Ehrenfest dynamics, surface-hopping interfaces; Sections~\ref{sec:rt-tddft-pump-probe} and~\ref{sec:nonadiabatic-md-surface-hopping}}} &
    \ReviewTableCell{\ReviewColC}{\SpecCovered{Pump and delayed probe pulses. \textit{Varied:} delay time, photon energy, pulse intensity}} &
    \ReviewTableCell{\ReviewColD}{\SpecCovered{Transient absorption or photoemission-like signal}} &
    \ReviewTableCell{\ReviewColE}{\SpecCovered{Real-time propagation describes nonequilibrium electron dynamics driven by pump and probe fields. Ehrenfest dynamics or surface hopping adds nuclear motion on longer time scales.\cite{Siday2024,Marek2025,cp2k_newtonx,Iannuzzi2025}}} \\[\ReviewRowGapTopMinus] \ReviewRowGap
    \ReviewTableGroup{Core-level and X-ray response}
    \ReviewTableCell{\ReviewColA}{\SpecCovered{X-ray absorption spectroscopy}} &
    \ReviewTableCell{\ReviewColB}{\SpecCovered{\texttt{\&DFT\%XAS}, \texttt{\&DFT\%XAS\_TDP}, real-time X-ray delta-kick; Sections~\ref{sec:xray-tp}, \ref{sec:xray-lr-tddft}, and~\ref{sec:xray-cysteine}}} &
    \ReviewTableCell{\ReviewColC}{\SpecCovered{X-ray photons. \textit{Varied:} photon energy, edge, donor core orbital}} &
    \ReviewTableCell{\ReviewColD}{\SpecCovered{Core-excitation energy and oscillator strength}} &
    \ReviewTableCell{\ReviewColE}{\SpecCovered{Core excitations are described by transition-potential or delta self-consistent-field calculations with enforced core holes, by core-level linear-response time-dependent density-functional theory, or by real-time X-ray delta-kick response.\cite{Iannuzzi2008,Bussy2021,Iannuzzi2025}}} \\[\ReviewRowGapTopMinus] \ReviewRowGap
    \ReviewTableCell{\ReviewColA}{\SpecPartial{X-ray photoelectron spectroscopy and core-level binding energies}} &
    \ReviewTableCell{\ReviewColB}{\SpecPartial{\texttt{\&DFT\%XAS}; \texttt{GW2X} and \texttt{XPS\_ONLY} in \texttt{\&DFT\%XAS\_TDP}; Sections~\ref{sec:xray-tp} and~\ref{sec:xray-lr-tddft}}} &
    \ReviewTableCell{\ReviewColC}{\SpecPartial{X-ray photons. \textit{Varied:} photon energy, core orbital, charge state}} &
    \ReviewTableCell{\ReviewColD}{\SpecPartial{Core binding energy and chemical shift}} &
    \ReviewTableCell{\ReviewColE}{\SpecPartial{Delta self-consistent-field calculations compare neutral and core-ionized states to obtain core-level binding energies and chemical shifts.\cite{Iannuzzi2008} GW2X-corrected linear-response time-dependent density-functional theory aligns core excitations and core binding energies through core-ionization-potential corrections.\cite{Bussy2021Correction}}} \\[\ReviewRowGapTopMinus] \ReviewRowGap
    \ReviewTableCell{\ReviewColA}{\SpecPartial{X-ray emission spectroscopy}} &
    \ReviewTableCell{\ReviewColB}{\SpecPartial{\texttt{XES\_CORE} and \texttt{XES\_EMPTY\_}\linebreak\texttt{HOMO} in \texttt{\&DFT\%XAS}; Section~\ref{sec:xray-tp}}} &
    \ReviewTableCell{\ReviewColC}{\SpecPartial{Core-hole preparation followed by photon emission. \textit{Varied:} emitted photon energy, core occupancy}} &
    \ReviewTableCell{\ReviewColD}{\SpecPartial{Emission intensity versus photon energy}} &
    \ReviewTableCell{\ReviewColE}{\SpecPartial{Core-hole and transition-potential configurations provide the initial state, while valence-to-core transitions define the emitted-photon spectrum.}} \\[\ReviewRowGapTopMinus] \ReviewRowGap
    \ReviewTableCell{\ReviewColA}{\SpecMissing{Resonant inelastic X-ray scattering}} &
    \ReviewTableCell{\ReviewColB}{\SpecMissing{\texttt{\&PROPERTIES\%RIXS} combining \texttt{XAS\_TDP} and \texttt{TDDFPT}; Section~\ref{sec:rixs}}} &
    \ReviewTableCell{\ReviewColC}{\SpecMissing{Resonant X-ray excitation and emitted X-ray photon. \textit{Varied:} incident and emitted photon energies}} &
    \ReviewTableCell{\ReviewColD}{\SpecMissing{Energy-loss spectrum}} &
    \ReviewTableCell{\ReviewColE}{\SpecMissing{Core-excited intermediate states from the X-ray absorption machinery are combined with valence excitations to build a resonant inelastic X-ray scattering response.\cite{SertcanGoekmen2026RIXS}}} \\[\ReviewRowGapTopMinus] \ReviewRowGap
    \ReviewTableGroup{Electronic-structure and scattering probe spectra}
    \ReviewTableCell{\ReviewColA}{\SpecCovered{Valence photoelectron quasiparticle energies and band dispersions}} &
    \ReviewTableCell{\ReviewColB}{\SpecCovered{\texttt{\&PROPERTIES\%BANDSTRUCTURE\%GW}; band structures, DOS and LDOS; Sections~\ref{8A}, \ref{GWfullkp}, and~\ref{GammaonlyGW}}} &
    \ReviewTableCell{\ReviewColC}{\SpecCovered{Ultraviolet or soft X-ray photoemission probe. \textit{Varied:} photon energy, crystal momentum}} &
    \ReviewTableCell{\ReviewColD}{\SpecCovered{Quasiparticle energies, ionization potentials, band dispersions}} &
    \ReviewTableCell{\ReviewColE}{\SpecCovered{$GW$ quasiparticle energies provide ionization potentials, electron affinities, and band dispersions for photoemission-relevant observables. Full angle-resolved intensities require external matrix-element and surface modeling, as summarized in Table~S1 in the SI.\cite{Graml2024,Pasquier2025}}} \\[\ReviewRowGapTopMinus] \ReviewRowGap
    \ReviewTableCell{\ReviewColA}{\SpecPartial{Density of states, local density of states, and scanning tunneling spectra}} &
    \ReviewTableCell{\ReviewColB}{\SpecPartial{\texttt{DOS}, \texttt{PDOS/LDOS}, and \texttt{STM} in \texttt{\&DFT\%PRINT}; \texttt{DOS/LDOS} in \texttt{\&PROPERTIES\%BANDSTRUCTURE}; \texttt{TIP\_SCAN} in \texttt{\&PROPERTIES}; Section~\ref{8A}}} &
    \ReviewTableCell{\ReviewColC}{\SpecPartial{Tunneling tip or energy-resolved electronic probe. \textit{Varied:} energy, bias, tip position}} &
    \ReviewTableCell{\ReviewColD}{\SpecPartial{Density of states, local density of states, tunneling-current image}} &
    \ReviewTableCell{\ReviewColE}{\SpecPartial{Kohn--Sham or quasiparticle eigenvalue spectra yield density-of-states observables. Local-density and tip-scan outputs approximate tunneling contrast from spatially resolved electronic states.}} \\[\ReviewRowGapTopMinus] \ReviewRowGap
    \ReviewTableCell{\ReviewColA}{\SpecMissing{Coherent X-ray diffraction spectrum}} &
    \ReviewTableCell{\ReviewColB}{\SpecMissing{\texttt{XRAY\_DIFFRACTION\_SPECTRUM} \linebreak in \texttt{\&DFT\%PRINT}}} &
    \ReviewTableCell{\ReviewColC}{\SpecMissing{Coherent X-ray scattering. \textit{Varied:} scattering vector}} &
    \ReviewTableCell{\ReviewColD}{\SpecMissing{Diffraction intensity versus momentum transfer}} &
    \ReviewTableCell{\ReviewColE}{\SpecMissing{Diffraction intensities are computed from coherent scattering by the nuclear and electronic density, but this spectrum-like output is beyond this review.}} \\[\ReviewRowGapTopMinus] \ReviewRowGap
    \ReviewTableGroup{Charge-transfer and transport response}
    \ReviewTableCell{\ReviewColA}{\SpecCovered{Absolutely localized molecular orbital energy decomposition and localized charge-transfer response}} &
    \ReviewTableCell{\ReviewColB}{\SpecCovered{\texttt{ALMO\_SCF} in \texttt{\&DFT}; \texttt{DELOCALIZE\_METHOD}; extended absolutely localized molecular orbital options; Section~\ref{sec:almo-eda-md}}} &
    \ReviewTableCell{\ReviewColC}{\SpecCovered{Fragment localization constraint. \textit{Varied:} fragment definition, delocalization domain, geometry}} &
    \ReviewTableCell{\ReviewColD}{\SpecCovered{Frozen, polarization, and charge-transfer energy terms}} &
    \ReviewTableCell{\ReviewColE}{\SpecCovered{A block-localized occupied space defines an absolutely localized molecular orbital reference. Controlled relaxation and extended absolutely localized molecular orbital delocalization separate polarization from local charge-transfer contributions in closed electronic systems.}} \\[\ReviewRowGapTopMinus] \ReviewRowGap
    \ReviewTableCell{\ReviewColA}{\SpecCovered{Electronic-transfer diabatic couplings}} &
    \ReviewTableCell{\ReviewColB}{\SpecCovered{\texttt{ET\_COUPLING} in \texttt{\&PROPERTIES}; constrained density-functional theory and projection-operator-based diabatization electronic couplings; Sections~\ref{sec:cdf-charge-transfer} and~\ref{sec:electronic-couplings-cdft-pod}}} &
    \ReviewTableCell{\ReviewColC}{\SpecCovered{Donor--acceptor charge constraint or Hilbert-space projection. \textit{Varied:} geometry, diabatic partition, donor--acceptor state}} &
    \ReviewTableCell{\ReviewColD}{\SpecCovered{Electronic coupling elements and diabatic energies}} &
    \ReviewTableCell{\ReviewColE}{\SpecCovered{Constrained density-functional theory constructs charge- or spin-localized many-electron states, while projection-operator-based diabatization block-diagonalizes the projected one-electron Hamiltonian to obtain diabatic energies and couplings for charge-transfer models.}} \\[\ReviewRowGapTopMinus] \ReviewRowGap
    \ReviewTableCell{\ReviewColA}{\SpecCovered{Linear-response charge-transport coefficients}} &
    \ReviewTableCell{\ReviewColB}{\SpecCovered{\texttt{KUBO\_TRANSPORT} in \texttt{\&PROPERTIES}; Section~\ref{sec:kubo-transport}}} &
    \ReviewTableCell{\ReviewColC}{\SpecCovered{Weak electric field in linear response. \textit{Varied:} temperature, chemical potential, dissipation, transport direction}} &
    \ReviewTableCell{\ReviewColD}{\SpecCovered{Conductivity tensor or isotropic conductivity}} &
    \ReviewTableCell{\ReviewColE}{\SpecCovered{The finite-volume Kubo--Greenwood formula evaluates equilibrium transport coefficients from the converged \textsc{Quickstep} Hamiltonian, overlap matrix, and atomic geometry, including lower-dimensional normalization.}} \\[\ReviewRowGapTopMinus] \ReviewRowGap
    \hline\hline

  \end{tabular}
\end{table*}

\renewcommand{\theHtable}{\thetable}

\begin{table*}[t!]
  \caption{When finite-temperature sampling and trajectory broadening matter for spectroscopy. The table summarizes how thermal configurational ensembles, nuclear motion, anharmonicity, and NQEs enter spectra in gas-phase, solid-state, liquid, interfacial, and soft condensed-phase systems, and it indicates the CP2K sampling strategies and manuscript Sections~\ref{sec:statistical-spectroscopy-framework}--\ref{sec:path-integral-dynamics-spectra} that connect static response calculations to finite-temperature observables.} \label{t4}
\vspace{0.5em}
  \setlength{\tabcolsep}{4pt}

  \begin{tabular}{llll}

    \hline\hline \\[-0.5em]
    \ReviewTableCell{2.2cm}{\textbf{System class}} &
    \ReviewTableCell{3.0cm}{\textbf{Origin of finite-temperature effects}} &
    \ReviewTableCell{3.0cm}{\textbf{Impact on spectroscopy}} &
    \ReviewTableCell{7.8cm}{\textbf{Theoretical treatment of finite-temperature effects}} \\
    \\[-0.5em]
    \hline
    \\[-0.5em]

    \ReviewTableCell{2.2cm}{Gas-phase molecule} &
    \ReviewTableCell{3.0cm}{Thermal population of different molecular geometries (vibrational distortions and conformers)} &
    \ReviewTableCell{3.0cm}{Vibronic structure and temperature-dependent peak shifts, with additional line broadening from averaging over geometry distributions} &
    \ReviewTableCell{7.8cm}{CP2K: Classical conformational and vibrational sampling from AIMD or path-
%dm -
integral MD when NQEs are important. Spectra (for example ultraviolet--visible absorption from linear-response time-dependent density-functional theory) are evaluated on every snapshot of an MD trajectory. See Sections~\ref{sec:aimd-finite-temperature-spectra} and~\ref{sec:path-integral-dynamics-spectra}.
    \linebreak\linebreak Alternative strategy: discrete conformer sampling combined with explicit electron-vibration coupling models} \\

    \\[-0.5em]

    \ReviewTableCell{2.2cm}{Solid} &
    \ReviewTableCell{3.0cm}{Thermal distribution of nuclear displacements around equilibrium positions} &
    \ReviewTableCell{3.0cm}{Peak shifts (e.g., temperature-dependent band gaps) and line widths} &
    \ReviewTableCell{7.8cm}{CP2K: AIMD (required for anharmonic or soft materials, e.g., metal--organic frameworks), ensemble averaging of single-snapshot spectra. See Sections~\ref{sec:aimd-finite-temperature-spectra} and~\ref{sec:finite-field-aimd}
    \linebreak\linebreak
    Alternative strategy: perturbative electron-phonon coupling for temperature-dependent band structure renormalization and linewidths without explicit MD sampling} \\

    \\[-0.5em]

    \ReviewTableCell{2.2cm}{Liquid, interface, and soft condensed phase} &
    \ReviewTableCell{3.0cm}{Thermal ensemble of molecular arrangements, solvent configurations, interfacial orientations, and soft structural disorder} &
    \ReviewTableCell{3.0cm}{Large ensemble-dependent peak shifts and line broadenings. Time-correlation spectra shaped by hydrogen-bond, solvation, and interfacial fluctuations} &
    \ReviewTableCell{7.8cm}{AIMD generates trajectory snapshots. Ensemble averaging of single-snapshot spectra. Time-correlation functions are used in particular for infrared, Raman, and sum-frequency generation spectroscopy. Path-
%dm -
integral MD is used when NQEs are relevant. See Sections~\ref{sec:aimd-finite-temperature-spectra}, \ref{sec:finite-field-aimd}, and~\ref{sec:path-integral-dynamics-spectra}} \\

    \\[-0.5em]

    \ReviewTableCell{2.2cm}{All systems (general strategy)} &
    \ReviewTableCell{3.0cm}{Thermal nuclear motion, configurational disorder, trajectory correlations, and, where needed, NQEs or nonadiabatic dynamics} &
    \ReviewTableCell{3.0cm}{Broadened vibrational, electronic, magnetic, or core-level spectra from snapshot distributions or time-correlation functions} &
    \ReviewTableCell{7.8cm}{\texttt{\&MOTION\%MD}, AIMD and path-
%dm -
integral MD, as well as per-spectrum engines generate thermal snapshots or time-correlation functions. The appropriate static or dynamical spectroscopy engine is then applied to the ensemble or trajectory. See Sections~\ref{sec:aimd-finite-temperature-spectra}, \ref{sec:path-integral-dynamics-spectra}, and~\ref{sec:nonadiabatic-md-surface-hopping}} \\

    \\[-0.5em]
    \hline\hline

  \end{tabular}
\end{table*}

Building on these capabilities, the present review focuses on the methods in CP2K that turn electronic structure into dynamics, transport, and spectroscopic response. The emphasis is not on spectroscopy as an isolated post-processing step,\cite{BrehmKirchner2011TRAVIS,Brehm2020TRAVIS} but on the path from an electronic-structure calculation to a measured response. Along this way, CP2K can keep the nuclei fixed, sample a thermodynamic ensemble, propagate nuclei and electrons in time, or impose electronic reservoirs and external potentials.

The organizing idea is therefore the perturbation--response relation. Magnetic and electric fields, photon pulses, core-hole constraints, charged reservoirs, or bias potentials define the probe, whereas tensors, transition strengths, time-correlation functions, excitation energies, currents, or broadened spectra define the signal. At fixed nuclei, CP2K provides normal-mode, finite-field, DFPT, time-dependent density-functional theory (TDDFT), $GW$/Bethe--Salpeter equation (BSE),\cite{Onida2002} constrained DFT (CDFT),\cite{Kaduk2012CDFT} and real-time methods. In liquids, interfaces, soft matter, and disordered solids, the same quantities become ensemble observables, so the decisive issue is not only nuclear motion itself but sampling the thermodynamic distribution from which spectra and transport coefficients emerge.\cite{AllenTildesley2017}

This is where CP2K's architecture becomes central. Energies, forces, electronic states, response tensors, trajectories, path integrals, enhanced sampling, Ehrenfest dynamics,\cite{LiTullySchlegelFrisch2005} surface hopping,\cite{tully} and property analysis are available within one simulation environment. A single electronic-structure representation can therefore support both a static response calculation and the trajectory, ensemble average, or nonequilibrium propagation needed for comparison with experiment.

The same response viewpoint also explains why charge transfer,\cite{KhaliullinKuehne2013WaterEDA} open boundaries,\cite{PanahianJandKuehneDelleSite2024} and transport are part of this review. CDFT and projection-based diabatization turn electronic localization into diabatic states and couplings. Kubo transport extracts equilibrium conductivities from thermally disordered structures.\cite{KuhneProdan2018,KuhneHeskeProdan2020,EfremkinHeskeKuehneProdan2026} Hairy Probes and nonequilibrium Green's functions (NEGF), combined with DFT as DFT+NEGF,\cite{Zauchner2018} introduce reservoirs, imposed potentials, and biased dynamics.\cite{Luisier2014TransportNanoelectronicDevices} These topics extend spectroscopy from closed, static systems to driven or electronically open situations in which currents, polarization, excitations, and nuclear motion are coupled.

Tables~\ref{t1} and~\ref{t4} turn this outline into a map of the manuscript. Table~\ref{t1} asks what CP2K can compute as a native or near-native spectroscopy, response, coupling, or transport quantity. Its rows follow the physical probe and detected signal rather than simply listing input keywords. Table~\ref{t4} then answers when a static response calculation must be embedded in finite-temperature sampling, snapshot averaging, time-correlation functions, path integrals, or explicit dynamics. Table~S1 in the SI collects complementary boundary cases where CP2K supplies first-principles ingredients but not a complete experimental spectrum-generation procedure and may require external line-shape, scattering, spin-dynamics, or photoemission modeling.

The remainder of the review follows the same order. It begins with electronic-structure engines that supply energies, forces, states, and response matrices. It proceeds through magnetic, vibrational, optical, X-ray, quasiparticle, and embedding methods. It then treats charge transfer, open boundaries, and transport. Finally, it turns to nuclear motion, finite-temperature spectroscopy, NQEs, and excited-state dynamics. The article thus moves from static electronic structure to dynamical response, with the tables serving as entry points into the methods rather than as a catalogue.

\section{Efficient Density-Functional Electronic-Structure Methods}
\label{sec:TotalEnergy}
\label{sec:gapw}

The methods collected in this section define the electronic-structure layer on which later response, transport, spectroscopic, and finite-temperature applications rely. The emphasis is therefore not on a general review of DFT, but on the CP2K machinery that makes repeated energy, force, stress, and property evaluations practical for large molecular, interfacial, liquid, and periodic systems. A common theme is the controlled reduction of computational cost: GPW representations provide an efficient first-principles baseline, modern exchange-correlation (XC) integration and Brillouin-zone sampling extend the range of accessible materials and observables, and approximate density-functional or tight-binding methods supply lower-cost force engines when sampling rather than a single electronic-structure calculation is the limiting factor.

\subsection{Gaussian and Augmented Plane Waves for Density-Functional Theory}

For the purposes of dynamics and spectroscopy, the central virtue of CP2K's density-functional layer is that it combines chemical flexibility with smooth, repeatable energy and force evaluations. This capability rests primarily on the GPW method,\cite{Lippert1999} and on its augmented all-electron generalization, the GAPW method,\cite{Krack2000} both of which were described in detail in our previous CP2K code review paper.\cite{Kuehne2020} The discussion here recalls only the ingredients that are most relevant for response calculations, AIMD, and finite-temperature spectroscopy.

The central idea of GPW is the use of a \emph{dual} representation of the electronic
structure: the Kohn--Sham (KS) molecular orbitals (MOs) $\psi_i(\mathbf{r})$ are expanded in an
atomic-orbital (AO) basis of contracted Gaussian-type orbitals (GTOs)
\begin{equation}
    \psi_i(\mathbf{r}) = \sum_{\mu} c_{\mu i}\, \phi_{\mu}(\mathbf{r}),
    \label{eq:mo_expansion}
\end{equation}
where \(c_{\mu i}\) is the coefficient of contracted GTO \(\phi_\mu\) in orbital \(i\), while \(\mu\) and \(i\) label basis functions and KS orbitals, respectively.
The corresponding electron density $n(\mathbf{r})$ is simultaneously represented on a
regular plane-wave (PW) auxiliary grid
\begin{equation}
    n(\mathbf{r}) = \sum_{\mathbf{G}} \tilde{n}(\mathbf{G})\,
    e^{i\mathbf{G}\cdot\mathbf{r}}\ensuremath{,}
\end{equation}
where $\mathbf{G}$ are reciprocal lattice vectors, \(\tilde n(\mathbf G)\) are the Fourier coefficients of the density, and \(i=\sqrt{-1}\).  The KS total energy functional in GPW
reads
\begin{align}
    E_{\mathrm{KS}}[n] &= E_{\mathrm{kin}} + E_{\mathrm{ext}} + E_{\mathrm{H}}[n]
                          + E_{\mathrm{XC}}[n] + E_{\mathrm{II}},
    \label{eq:gpw_energy}
\end{align}
where $E_{\mathrm{kin}}$ is the kinetic energy, $E_{\mathrm{ext}}$ the interaction with
external (ionic) potentials, $E_{\mathrm{H}}$ the Hartree (classical Coulomb) energy,
$E_{\mathrm{XC}}$ the XC energy, and $E_{\mathrm{II}}$ the classical core--core
repulsion.

A key advantage of GPW is the natural \emph{separation} of density-dependent and
density-independent contributions to the energy and the KS matrix.  The density-independent
terms, i.e., the kinetic energy matrix elements $\langle \phi_\mu |
{-\tfrac{1}{2}\nabla^2} | \phi_\nu \rangle$ and the core overlap integrals, are evaluated
analytically in the GTO basis and computed only once at the start of the self-consistent
field (SCF) procedure.  The density-dependent terms ($E_{\mathrm{H}}$ and
$E_{\mathrm{XC}}$), which dominate the computational cost, are evaluated on the PW grid
by mapping $n(\mathbf{r})$ via a collocation step and exploiting the $\mathcal{O}(N_G \log
N_G)$ scaling of fast Fourier transforms (FFTs) for the solution of the Poisson equation.
This decomposition avoids the $\mathcal{O}(N^4)$ four-center electron-repulsion integrals
inherent to conventional GTO codes while retaining the chemical flexibility and
transferability of localized basis sets.\cite{Lippert1999,Kuehne2020}
In GPW, ionic cores are described by pseudopotentials (PPs), such as Goedecker--Teter--Hutter (GTH) pseudopotentials,\cite{GoedeckerTeterHutter1996} which remove chemically inert core
electrons from the calculation and ensure that the resulting smooth valence density is
accurately representable on a computationally tractable PW grid without requiring prohibitively
high PW cutoffs.  The use of GTH PPs thus renders GPW particularly efficient
for systems dominated by valence-electron chemistry, including most organic and biological
molecules, transition-metal complexes, and condensed-phase materials.
A recent CP2K validation protocol disentangles Gaussian-basis incompleteness from PP error through matched calculations with systematic PW and all-electron references, and uses this diagnosis to construct improved molecularly optimized (MOLOPT) basis sets and GTH PPs.\cite{Mirhosseini2026UZH}
More broadly, reproducible code-verification workflows have benchmarked CP2K alongside other periodic DFT implementations over a chemically comprehensive test set.\cite{Bosoni2024Verification}

In GAPW, the electron density is decomposed as
\begin{equation}
\label{eq:gapw_density}
 \begin{array}{c@{}c@{}c@{}c@{}c@{}c@{}c}
  \raisebox{0.35ex}{\includegraphics[width=0.1\textwidth]{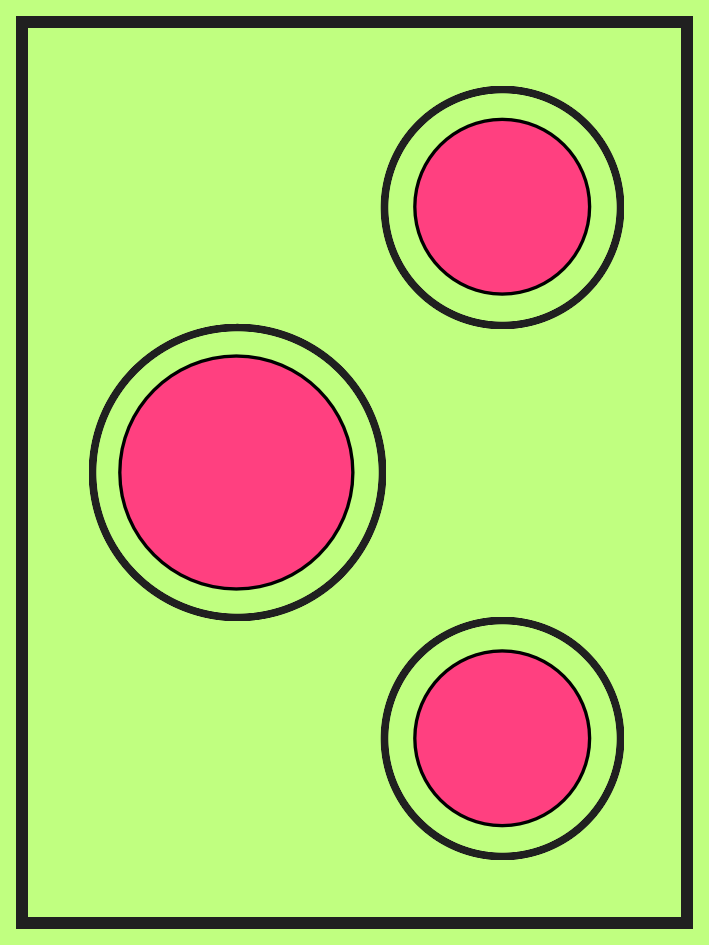}}     &&
  \raisebox{0.35ex}{\includegraphics[width=0.1\textwidth]{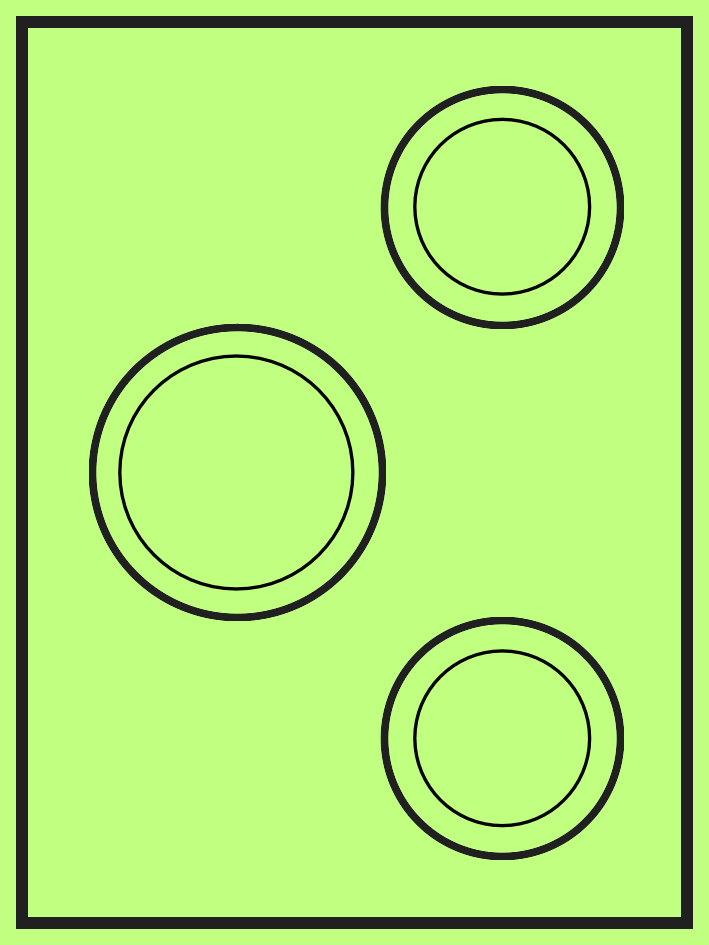}}  &&
  \raisebox{0.35ex}{\includegraphics[width=0.1\textwidth]{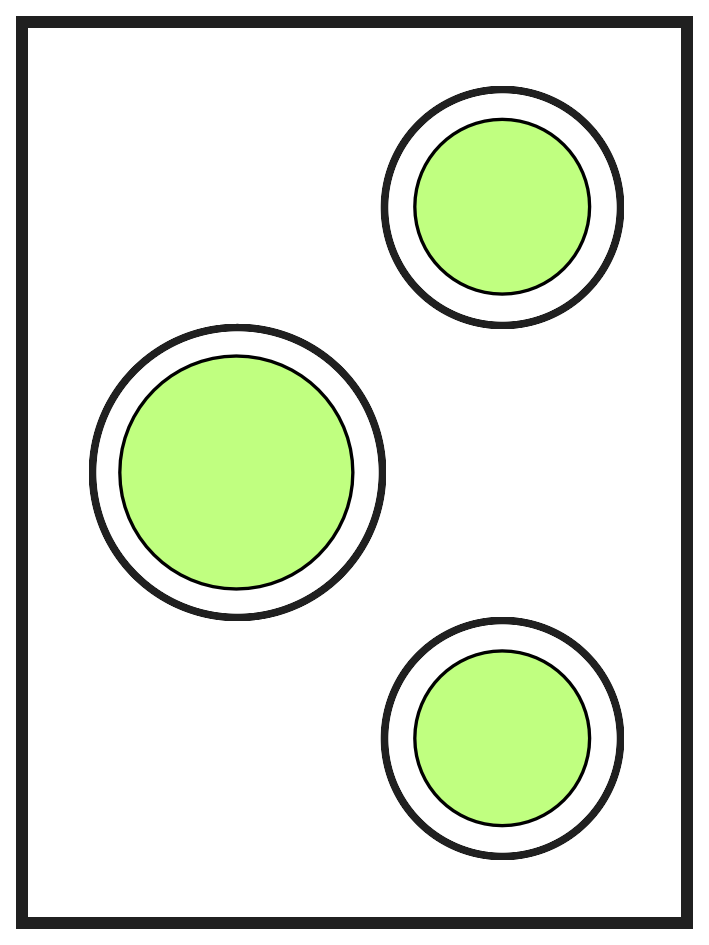}} &&
  \raisebox{0.35ex}{\includegraphics[width=0.1\textwidth]{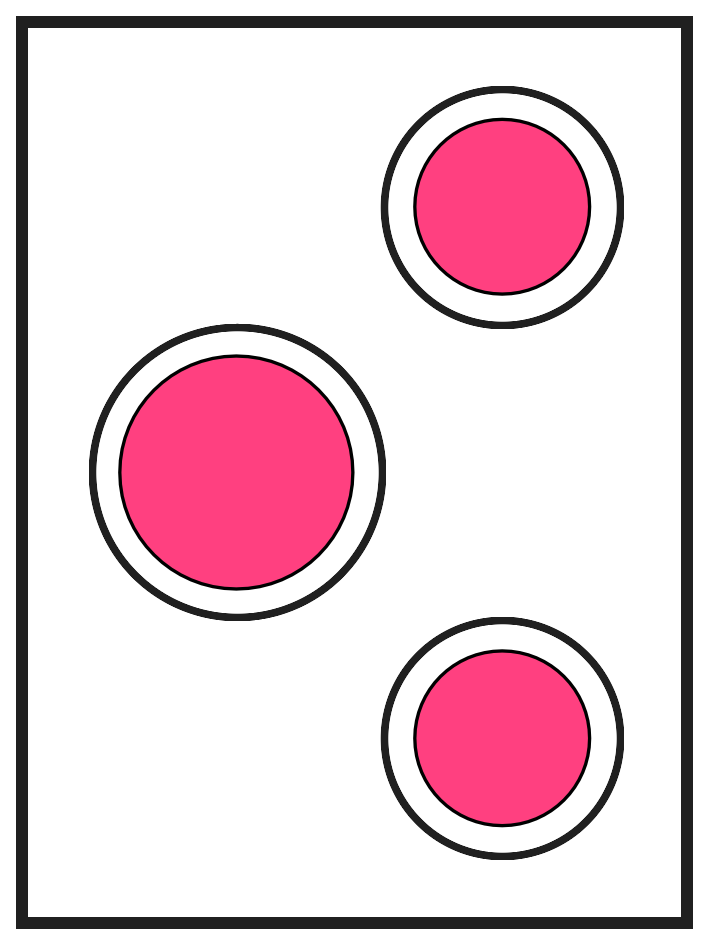}} \\[-0.15ex]
  \displaystyle n(\mathbf{r}) &{=}&
  \displaystyle \tilde{n}(\mathbf{r}) & {-}&
  \displaystyle\sum_A \tilde{n}^A(\mathbf{r}) & {+}&
  \displaystyle\sum_A n^A(\mathbf{r}),\\
 \end{array}
\end{equation}
where $\tilde{n}(\mathbf{r})$ is a smooth \emph{soft} density that is well represented on the PW grid, and $n^{A}(\mathbf{r})$ and $\tilde{n}^{A}(\mathbf{r})$ are the \emph{hard} all-electron and the corresponding soft density, respectively, localized within an atomic augmentation sphere centered on atom $A$. Both $n^A$ and $\tilde{n}^A$ are expanded in atom-centered GTO primitives, so their difference is confined to the augmentation region and never needs to be resolved on
the PW grid.  This separation removes the hard core density from the PW expansion,
eliminating the need for very high kinetic-energy cutoffs while retaining full all-electron
accuracy.  The total energy in GAPW is partitioned consistently with
Eq.~\eqref{eq:gapw_density}, and the Hartree and XC contributions are evaluated separately
for the smooth and local components.\cite{Krack2000,Kuehne2020}
GAPW is distinct from the projector augmented-wave formalism, although both use atom-centered augmentation to recover all-electron information from smooth auxiliary quantities.\cite{paw1,Bloechl2026CPPAW}

Importantly, GAPW is not restricted to all-electron calculations and can equally be
employed in conjunction with PPs or effective core potentials (ECPs). Even when core electrons are absent, the
valence density in the vicinity of atomic nuclei retains significant curvature that is
difficult to represent faithfully on a finite PW grid.  Within the GAPW decomposition of
Eq.~\eqref{eq:gapw_density}, the hard local component $n^{A} - \tilde{n}^{A}$ accounts
analytically for this curvature inside each augmentation sphere, so that the XC energy and
potential are evaluated with the full local density rather than its grid-aliased
approximation. Integration of the XC functional over a coarse grid introduces numerical
noise that, upon taking analytic derivatives of the energy, manifests as spurious
oscillations in the atomic forces, an artifact commonly referred to as the
\emph{ripple problem}.  By keeping the cusped part of the
density off the PW grid entirely, the GAPW scheme systematically suppresses this error,
yielding smoother potential energy surfaces and more accurate forces even at modest
PW cutoffs. This makes GAPW with PPs or ECPs an attractive choice when
 the accuracy of the forces directly controls the quality of the
sampled dynamics and the reliability of the resulting spectra.

Beyond KS DFT, GPW supports post-SCF random-phase approximation (RPA) and \(\sigma\)-functional correlation energies benchmarked for molecular thermochemistry and crystalline structures, with the Perdew--Burke--Ernzerhof (PBE) functional serving as a semilocal reference.\cite{Mandalia2025}

\subsubsection{\texorpdfstring{GauXC and Neural Exchange-Correlation Functionals}{GauXC and Neural Exchange-Correlation Functionals}}

% Franz und Sebastian ...

Recent developments also connect the GPW/GAPW method to XC backends that go beyond traditional semilocal quadrature loops. This is particularly useful for machine-learned or otherwise nonlocal functionals, where the functional evaluation may require external tensor libraries, automatic differentiation, and a more flexible representation of grid and basis information.

The Skala-1.1 neural XC functional,\cite{luise2025} is defined by the usual integral for the XC energy
%
% \begin{equation}
% \begin{aligned}
%     E^\theta_\text{XC} [ n ]
%     &= -\frac34\left(\frac6\pi\right)^{\frac13}
%     \int \left( n ^{(\uparrow)}(\mathbf r)^{4/3}
%     +  n ^{(\downarrow)}(\mathbf r)^{4/3}\right) \\
%     &\quad \times f_\theta[\mathbf x[ n ]](\mathbf r)\,
%     d\mathbf r,
% \end{aligned}
% \label{eq:neural-xc}
% \end{equation}
\begin{align}
    E^\theta_\text{XC} [ n ]
    &=\alpha
    {\int}   \sum_{\sigma=\uparrow,\downarrow} n ^{(\sigma)}(\mathbf r)^{4/3} \, f_\theta[\mathbf x[ n ]](\mathbf r)\,
	    \ensuremath{\mathrm{d}\mathbf r}\,,
    \label{eq:neural-xc}
\end{align}
where $\alpha=-3\sqrt[3]{6/\pi}/4$, \( n ^{(\uparrow/\downarrow)}\) are the spin-channel densities and \(f_\theta\) is a neural-network enhancement factor with parameters \(\theta\), evaluated from density-derived features \(\mathbf x[ n ](\mathbf r)\).
The input layer uses semilocal quantities
\begin{align}
	    \mathbf x_\text{input}[ n ](\mathbf r)
	    &= \big(
     n ^{(\uparrow)}(\mathbf r),
     n ^{(\downarrow)}(\mathbf r),
    \|\nabla n ^{(\uparrow)}(\mathbf r)\|,\|\nabla n ^{(\downarrow)}(\mathbf r)\|,\nonumber\\[0.3em]
	    &\quad
    \tau^{(\uparrow)}(\mathbf r),
    \tau^{(\downarrow)}(\mathbf r), \|\nabla n ^{(\uparrow)}(\mathbf r)
    + \nabla n ^{(\downarrow)}(\mathbf r)\|
    \big),
\end{align}
where \(\tau\) denotes the kinetic-energy density and \(\nabla n \) the density gradient.

These semilocal input features are regularized by applying the logarithmic squashing transformation
\begin{equation}
    \mathbf x = \log(\mathbf x_\text{input} + \epsilon).
\end{equation}
The positive regularizer \(\epsilon\) prevents singular logarithms for vanishing input features.

The Skala architecture consists of three main parts. First, the two spin channels are merged using two linear layers according to
\begin{equation}
    f_\text{repr}(\mathbf x) = \sigma(W_2\cdot\sigma(W_1\cdot\mathbf x + b_1)+b_2),
\end{equation}
where \(W_1\), \(W_2\), \(b_1\), and \(b_2\) are the corresponding weights and biases. The output is processed by the Swish activation function \(\sigma\).\cite{ramachandran2017} Spin-invariant features are obtained by evaluating the same representation with swapped spin ordering and averaging the two results as
\begin{equation}
    \mathbf h^{(0)} = \frac12\left(f_\text{repr}\bigl(\mathbf x^{(\uparrow,\downarrow)}\bigr) + f_\text{repr}\bigl(\mathbf x^{(\downarrow,\uparrow)}\bigr)\right).
\end{equation}

Second, the processed features \(\mathbf h^{(0)}\) are passed through a stack of \(L\) nonlocal layers according to
\begin{equation}
    \mathbf h^{(l+1)} = f^{(l)}_\text{nonl}\bigl(\mathbf h^{(l)}, \{\mathbf R\}\bigr),
\end{equation}
where \(f^{(l)}_\text{nonl}\) denotes a nonlocal layer that first maps the grid features \(\mathbf h^{(l)}\) to a set of coarse points \(\{\mathbf R\}\), processes the resulting latent representation, and then interpolates the updated information back to the integration grid to obtain \(\mathbf h^{(l+1)}\). In the Skala-1.1 architecture, these coarse points are associated with the nuclear positions.
Skala-1.1 uses three such layers; further details are given in Section~A of the Supporting Information of Ref.~\citenum{luise2025}.

Third, the enhancement factor is computed with two additional linear layers
\begin{equation}
    \mathbf h_\text{enh} = \sigma_\text{out}\bigl(W_4\cdot\sigma\bigl(W_3\cdot\mathbf h^{(L)} + b_3\bigr) + b_4\bigr),
\end{equation}
where \(\sigma_\text{out}(x)=2/(1+\exp[-x/2])\) is a scaled sigmoid function and \(\sigma\) again denotes the Swish activation. The final XC energy is obtained from Eq.~\eqref{eq:neural-xc}, with \(\mathbf h_\text{enh}\) defining the output of \(f_\theta\).

The evaluation of Eq.~\eqref{eq:neural-xc} is delegated to PyTorch,\cite{Paszke2019PyTorch} and derivative quantities such as the XC potential, nuclear-gradient contributions, and electronic Hessian contributions are obtained by automatic differentiation.

Although the Skala input features can be represented on any integration grid used for conventional DFT calculations, the nonlocal layers require careful assignment of grid points to coarse points for efficient evaluation. The CP2K implementation therefore uses the GauXC library.\cite{petrone2018,williams-young2020}
GauXC provides a higher-level abstraction of XC integration and manages both the numerical grid and the evaluation of density features from the Gaussian-basis density matrix. 
The molecular GauXC interface also supports an all-electron route through GAPW. For this route, \texttt{GPW\_TYPE TRUE} marks each relevant atomic kind for GPW treatment even when \texttt{METHOD GAPW} is active, irrespective of the hardness of its Gaussian primitives. The corresponding density therefore enters through the smooth global representation without a hard-minus-soft one-center term, while GauXC evaluates Skala directly from the complete AO density matrix. For conventional-functional validation, native GAPW with accurate augmented XC integration provides the corresponding all-electron reference. This density is distinct from GAPW calculations with GTH pseudopotentials or ECPs, which remain valence-density calculations and must not be interpreted as all-electron Skala. A fully periodic version of Skala based on the GPW/GAPW formalism using FTorch is also available.\cite{Atkinson2025FTorch}

\begin{figure*}[t!]
    \centering
    \includegraphics[width=1.0\linewidth]{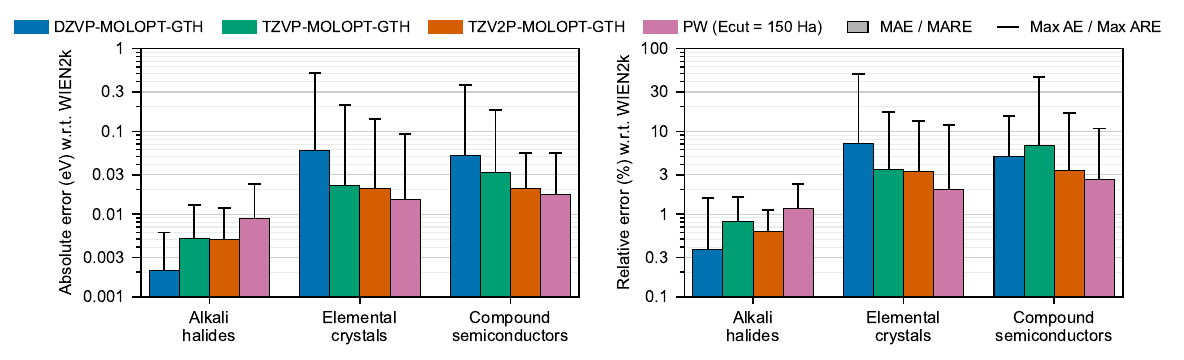}
    \vspace{-2em}\caption{Accuracy of spin--orbit-coupling splittings in the electronic band structure computed using PBE with HGH spin--orbit coupling,\cite{hartwigsen1998,Samal2026}
    evaluated against all-electron reference from WIEN2k.\cite{blaha2001augmented,PhysRevMaterials.1.033803}
The benchmark set contains spin splittings of various crystals: alkali halides, elements, and semiconductors.\cite{PhysRevMaterials.1.033803}
    We employ various Gaussian basis sets with CP2K and a PW basis set with ABINIT.\cite{Verstraete2008,Verstraete2025,Samal2026}
    %For every crystal the error is taken over the SOC-induced band splittings.
    \emph{Left panel:}
absolute errors (eV). The bar and whisker give the mean and maximum absolute errors, respectively, over the spin splittings in each material
class. \emph{Right panel:} corresponding relative errors (\%).
}
    \label{soc_error}
\end{figure*}

\subsubsection{\texorpdfstring{\textit{k}-Point Sampling and Brillouin-Zone Symmetry}{k-Point Sampling and Brillouin-Zone Symmetry}}
\label{sec:kpoint-sampling-bz-symmetry}

For periodic systems with small primitive cells, $\Gamma$-point sampling alone is often insufficient to converge total energies, forces, response properties, and spectroscopic observables. CP2K therefore supports Brillouin-zone sampling in the GPW and GAPW frameworks. In this formulation the MOs are represented as Bloch functions expanded in atom-centered Gaussian basis functions. The KS equations become a generalized Hermitian eigenvalue problem at each sampled crystal momentum:
\begin{equation}
    \mathbf{H}(\bk)\mathbf{C}(\bk)
    =\mathbf{S}(\bk)\mathbf{C}(\bk)\boldsymbol{\varepsilon}(\bk)
    \ensuremath{,}
    \label{eq:qs_kpoint_evp}
\end{equation}
where \(\mathbf H(\bk)\), \(\mathbf S(\bk)\), \(\mathbf C(\bk)\), and \(\boldsymbol\varepsilon(\bk)\) are the KS Hamiltonian, AO overlap, MO-coefficient, and diagonal orbital-energy matrices, respectively.
Brillouin-zone integrals are evaluated as weighted sums:
\begin{equation}
    \langle A \rangle_\mathrm{BZ} \simeq
    \sum_{\ell=1}^{N_k} w_\ell A(\bk_\ell)\,,
    \qquad
    \sum_{\ell=1}^{N_k} w_\ell = 1\,
    \ensuremath{,}
    \label{eq:qs_kpoint_weights}
\end{equation}
where \(A(\bk_\ell)\) is the value of an observable \(A\) at sampled point \(\bk_\ell\), \(w_\ell\) is its normalized integration weight, and \(N_k\) is the number of sampled points.
The Hamiltonian and overlap matrices are first assembled in real space for a translation vector~$\bR$ and subsequently transformed to reciprocal space:
\begin{equation}
    H_{\mu\nu}(\bk) = \sum_{\bR} e^{i\bk\cdot\bR} H_{\mu\nu}^{\bR}\,,
    \hspace{1em}
    S_{\mu\nu}(\bk) = \sum_{\bR} e^{i\bk\cdot\bR} S_{\mu\nu}^{\bR}\,,
    \label{eq:qs_kpoint_fourier}
\end{equation}
where $\mu$ and $\nu$ label Gaussian basis functions in the reference and translated unit cells, respectively. This real-space construction is the same locality principle that makes GPW and GAPW efficient at the $\Gamma$ point. The additional cost of a $\bk$-point calculation enters through an increased number of complex-valued diagonalizations~\eqref{eq:qs_kpoint_evp} and lattice sums~\eqref{eq:qs_kpoint_fourier}.

Regular or user-defined $\bk$-point meshes, including Monkhorst--Pack and shifted grids,\cite{Monkhorst1976,MacDonald1978} replace Brillouin-zone integrals by weighted sums, and independent $\bk$ points provide a natural level of parallelism in addition to the real-space grid, sparse-matrix, and diagonalization parallelism used within each $\bk$ point. Dense meshes can be accelerated by reducing the full grid to irreducible representatives under space-group and time-reversal symmetries.\cite{Worlton1972,spglibv1} In a Gaussian basis this reduction is not merely a reciprocal-space operation: accepted symmetry operations must also map atoms, local basis functions, and Bloch phases consistently with the real-space construction of Eq.~\eqref{eq:qs_kpoint_fourier}.

For high-symmetry crystals, such a reduction can lower the number of explicit KS problems by more than an order of magnitude, although the realized speedup remains method and system dependent. This acceleration is particularly useful for spectroscopic and many-body calculations requiring dense meshes, including band structures, photoemission-like observables, TDDFT for crystals, and periodic $GW$ calculations.\cite{Pasquier2025} The technical details of the symmetry-orbit construction, atom mapping, Gaussian-basis phase factors, and backend separation are summarized in Section~II~A in the SI.

\subsubsection{\texorpdfstring{Perturbative Spin--Orbit Coupling}{Perturbative Spin-Orbit Coupling}}
\label{spin_orbit_coupling_section}

Spin--orbit coupling (SOC) is essential for a reliable description of the electronic structure of molecules and materials containing heavy elements, and it can also influence spectroscopic splittings, band structures, and spin-dependent response properties.
Electronic-structure codes usually include SOC either through relativistic all-electron approaches, including four-component Dirac-based methods and two-component approximations derived from them,\cite{10.1063/1.1413510,10.1063/1.4940140,PhysRevB.99.205103,PhysRevB.103.245144} or through PP-based formulations.\cite{hartwigsen1998,doi:10.1021/cr2001383}
The latter option is comparatively inexpensive and has therefore become a workhorse of computational materials science.
PW codes such as ABINIT implement PP SOC,\cite{Verstraete2008} but their computational cost can increase steeply with the number of atoms in the unit cell, so SOC calculations for very large systems remain demanding.
Gaussian-basis codes such as CP2K are attractive in this regime because they retain locality and sparse-matrix structure for large-scale simulations. The current CP2K implementation incorporates SOC perturbatively through Hartwigsen--Goedecker--Hutter (HGH) GTH-type dual-space Gaussian PPs.\cite{hartwigsen1998,Samal2026}

The accuracy of this approach has been assessed by computing SOC splittings in the band structures of alkali halide crystals, elemental crystals, and compound semiconductors (Fig.~\ref{soc_error}).
Reference values were obtained from Ref.~\citenum{PhysRevMaterials.1.033803}, where self-consistent treatment of SOC is carried out with the inclusion of Dirac $p^{1/2}$ orbitals,\cite{PhysRevB.64.153102,PhysRevB.70.035212} using the WIEN2k code.\cite{blaha2001augmented}
The mean absolute errors (MAEs) of HGH SOC splittings computed with ABINIT and a PW basis are below 20 meV for alkali halides, elemental crystals, and compound semiconductors.
Similar MAEs are observed for CP2K with the TZV2P-MOLOPT basis,\cite{vande2007} and errors increase only slightly when using smaller basis sets (DZVP-MOLOPT and TZVP-MOLOPT, see Fig.~\ref{soc_error}).
%In contrast, the MAEs computed using molecularly optimized (MOLOPT) Gaussian basis sets range from 1\,--\,4 meV for alkali halides, 19\,--\,57 meV for elemental crystals, and 19\,--\,50 meV for compound semiconductors.\cite{vande2007}
%
%The comparatively large MAEs for compound semiconductors and elemental crystals are due to SOC splittings of higher unoccupied states.
%

These benchmarks show that CP2K's perturbative SOC implementation computes band-structure splittings cost-effectively, with competitive accuracy across diverse and large systems.
%
%The remaining errors point primarily to basis-set incompleteness, motivating tailored and larger Gaussian basis sets for SOC-sensitive applications.
%
%An extension to total energies and forces would make AIMD with SOC available in the same simulation environment.

\subsection{Harris Functional}
\label{Harris_Functional}
The Harris functional provides a non-self-consistent approximation to the KS DFT energy in which a variationally determined input density \(n^{\mathrm{in}}\) is used to build the effective potential, while the energy is evaluated for a corrected density \(n^{\mathrm{out}} = n^{\mathrm{in}} + \Delta n\).\cite{harrisfunctional1,harrisfunctional2,
harrisfunctional3,Belleflamme2023}
Expanding the energy around \(n^{\mathrm{in}}\) and neglecting terms quadratic and higher in the density perturbation \(\Delta n\) gives
\begin{align}
&E_{\mathrm{Harris}} [\mathbf{D}] = \sum_{\bar{\mu} \bar{\nu} \sigma} D_{\bar{\mu}\bar{\nu} \sigma}
h_{\bar{\mu}\bar{\nu} \sigma}
+
\sum_{\bar{\mu} \bar{\nu} \sigma} D_{\bar{\mu} \bar{\nu} \sigma} V_{\bar{\mu}\bar{\nu} \sigma}^{\mathrm{XC}}
[n^{\mathrm{in}}]
\nonumber \\
&\hspace{1em}+
\sum_{\bar{\mu}\bar{\nu}\kappa \lambda \sigma \sigma'} D_{\bar{\mu}\bar{\nu} \sigma} D_{\kappa \lambda \sigma'} ( \bar{\mu} \bar{\nu} | \kappa
\lambda)- \sum_{\mu \nu \sigma} D_{\mu \nu \sigma} V_{\mu \nu
\sigma}^{\mathrm{XC}} [n^{\mathrm{in}}] \nonumber \\
&\hspace{1em}- \frac{1}{2} \sum_{\mu \nu \kappa \lambda \sigma \sigma'} D_{\mu \nu \sigma} D_{\kappa \lambda \sigma'} ( \mu \nu | \kappa \lambda ) +
E^{\mathrm{XC}} [n^{\mathrm{in}}] \,,
\end{align}
where unbarred indices refer to the input density \(n^{\mathrm{in}}\), whereas barred indices refer to the corrected density \(n^{\mathrm{out}}\). Moreover, \(\mathbf D\) is the spin-resolved AO density matrix, \(\mathbf h\) is the one-electron core-Hamiltonian matrix, \(\mathbf V^{\mathrm{XC}}\) is the XC-potential matrix, \((\mu\nu|\kappa\lambda)\) denotes an electron-repulsion integral, and \(\sigma\) and \(\sigma'\) label spin channels. The associated variational Lagrangian imposes the KS constraint for the input orbitals through the Z-vector multiplier \(\bar{\mathbf Z}\) and enforces orthonormality of the corrected and input MO coefficients through the corresponding Lagrange multipliers \(\bar{\mathbf W}^{\mathrm{out}}\) and \(\bar{\mathbf W}^{\mathrm{in}}\), i.e.
\begin{align}
L_{\mathrm{Harris}}
&=
E_{\mathrm{Harris}}
+ \sum_{\mu \nu k \sigma} \bar{Z}_{\mu k \sigma}
\left( F_{\mu \nu \sigma} [n^{\mathrm{in}}] - \varepsilon_{k \sigma} S_{\mu \nu} \right)  C_{\nu
k \sigma} \nonumber \\
&- \sum_{\bar{k}\bar{l} \sigma} \bar{W}_{\bar{k}\bar{l}\sigma}^{\mathrm{out}} \left( \sum_{\bar{\mu}\bar{\nu} }
C_{\bar{\mu}\bar{k}\sigma}S_{\bar{\mu}\bar{\nu}} C_{\bar{\nu}\bar{l}\sigma} - \delta_{\bar{k}\bar{l}} \right) \nonumber \\
&- \sum_{kl \sigma} \bar{W}_{kl \sigma}^{\mathrm{in}} \left( \sum_{\mu \nu}
C_{\mu k\sigma}S_{\mu \nu} C_{\nu l\sigma} - \delta_{kl}  \right)
\ensuremath{.}
\end{align}
Here, \(\mathbf F\), \(\mathbf S\), \(\mathbf C\), and \(\varepsilon_{k\sigma}\) denote the KS matrix, AO overlap matrix, MO-coefficient matrix, and KS orbital energies, respectively. The indices \(k\) and \(l\) label MOs, whereas \(\mu\) and \(\nu\) label AOs.
In analogy to the Lagrangian construction used for linear-response TDDFT (LR-TDDFT) in Section~\ref{sec-lr_tddft}, the multipliers are obtained by differentiating with respect to the input and corrected MO-coefficient matrices, \(\mathbf C\) and \(\bar{\mathbf C}\), respectively, and by projecting onto their corresponding occupied subspaces or onto the virtual subspace defined by the spin-resolved projector \(\mathbf Q_\sigma\), i.e.
\begin{equation}
Q_{\mu\nu\sigma}
=
\delta_{\mu\nu}
-\sum_{\kappa i}
C_{\mu i\sigma}C_{\kappa i\sigma}^{*}S_{\kappa\nu}
\ensuremath{.}
\end{equation}
The associated stationarity conditions are
\begin{align}
\frac{\partial L}{\partial \mathbf{C}} \bar{\mathbf{C}} &\rightarrow
\bar{\mathbf{W}}^{\mathrm{in}}
\ensuremath{,} \hspace{1em}
\frac{\partial L}{\partial \bar{\mathbf{C}}} \bar{\mathbf{C}} \rightarrow
\bar{\mathbf{W}}^{\mathrm{out}}
\ensuremath{,} \hspace{1em}
\frac{\partial L}{\partial \mathbf{C}} \mathbf{Q}_\sigma \rightarrow \bar{\mathbf{Z}}
\ensuremath{.}
\end{align}
The resulting working equations for all three Lagrange multipliers are conveniently written in terms of the difference density matrix \(\Delta \mathbf D = \mathbf D^{\mathrm{out}} - \mathbf D^{\mathrm{in}}\) as
\begin{subequations}
\begin{align}
R_{\mu i \sigma} &= - 2 \sum_{\tau \nu} Q_{\tau \mu \sigma} H_{\tau \nu \sigma}[\Delta \mathbf{D}] C_{\nu i \sigma}
\ensuremath{,}
\\
W_{ij \sigma}^{\mathrm{in}} &= \sum_{\mu \nu } C_{\nu i \sigma} H_{\nu \mu \sigma}[\Delta \mathbf{D} +
\mathbf{D}^{\mathrm{Z}}] C_{\mu j \sigma}
\ensuremath{,}\\
W_{\ensuremath{\bar{\imath}}
\ensuremath{\bar{\jmath}}\sigma}^{\mathrm{out}}
&= 2 \sum_{\bar{\mu} \bar{\nu}}
C_{\bar{\mu}\ensuremath{\bar{\imath}}\sigma}
F_{\bar{\mu}\bar{\nu}\sigma}[\mathbf{D}^{\mathrm{in}}]
C_{\bar{\nu}\ensuremath{\bar{\jmath}}\sigma}
\ensuremath{.}
\end{align}
\end{subequations}

In these equations, \(R_{\mu i\sigma}\) is the right-hand side of the Z-vector equations, \(H_{\mu\nu\sigma}[\mathbf M]\) is the response kernel acting on a trial density matrix \(\mathbf M\), and \(\mathbf D^{\mathrm Z}\) is the Z-vector response density. Their implementation-level definitions are given in Section~IV~A in the SI.
The nuclear gradient contains the explicit derivative of the Harris functional and overlap-matrix derivatives from both the input and corrected orbital spaces and takes the form
\begin{align}
\frac{\partial L_{\mathrm{Harris}} }{\partial \mathbf{R}} &=
\frac{\partial E_{\mathrm{Harris}}}{\partial \mathbf{R}} + \sum_{\mu \nu \sigma} D^{\mathrm{Z}}_{\mu \nu \sigma} \frac{\partial F_{\mu \nu }
[\mathbf{D}^{\mathrm{in}}]}{\partial \mathbf{R}} \nonumber \\
&- \sum_{\mu \nu \sigma} \left( \bar{W}_{\mu \nu \sigma}^{\mathrm{in}} + \sum_{k}\varepsilon_{k\sigma} \bar{Z}_{\mu
   k \sigma} C_{\nu k \sigma} \right) \frac{\partial S_{\mu \nu}}{\partial \mathbf{R}} \nonumber \\
&- \sum_{\bar{\mu} \bar{\nu} \sigma} W_{\bar{\mu} \bar{\nu}\sigma}^{\mathrm{out}} \frac{\partial
S_{\bar{\mu}\bar{\nu}}}{\partial \mathbf{R}}
\ensuremath{.}
\end{align}
In this gradient, \(\mathbf R\) denotes the complete set of nuclear coordinates rather than the Z-vector right-hand side \(R_{\mu i\sigma}\).
Density-corrected DFT follows the same variational logic, but starts from an already converged reference density, for example a Hartree--Fock density for SCAN or r$^2$SCAN. The corresponding r$^2$SCAN working equations are obtained by adjusting the reference KS matrix and the associated density-functional contributions.\cite{Belleflamme2023} In this case, the right-hand side of the Z-vector equations contains the difference between the XC-potential matrix and the exact-exchange matrix \(\mathbf K^{\mathrm{EX}}\) of the reference Fock operator
\begin{align}
   R_{\mu i \sigma} = 2 \sum_{\kappa} C_{\kappa i \sigma} ( V_{\kappa \mu \sigma}^{\mathrm{XC}} - K^{\mathrm{EX}}_{\kappa \mu
   \sigma}) \ensuremath{.}
\end{align}

\subsection{Periodic Extended Tight-Binding}
\label{sec:periodic-xtb}

Semiempirical tight-binding methods are attractive whenever the statistical
sampling required for dynamics and spectroscopy would make repeated
first-principles calculations too expensive. In CP2K this class of methods
includes density-functional tight binding (DFTB),\cite{Seifert2007} self-consistent-charge DFTB (SCC-DFTB),\cite{Elstner1998} its third-order extension
DFTB3,\cite{Gaus2011} and the geometry, frequency, and non-covalent interaction extended
tight-binding family GFN$n$-xTB ($n=0,1,2$).\cite{Grimme2017,Bannwarth2019,Pracht2023,Alizadeh2026PeriodicGFN2XTB} All of them can be viewed as
localized-basis approximations to the KS total energy, expanded around a
superposition of neutral atomic reference densities, but they differ in the
order of the density fluctuation terms retained and in the empirical
short-range and dispersion terms used to restore transferability.
%\cite{Elstner1998,Grimme2017,Bannwarth2019,Pracht2023}
For periodic condensed-phase simulations the essential additional ingredients are lattice-summed
Hamiltonian and overlap matrices, Brillouin-zone integration, and a consistent
periodic treatment of long-ranged monopolar and, for GFN2-xTB, multipolar
electrostatics.\cite{Alizadeh2026PeriodicGFN2XTB}
These methods are particularly useful for high-throughput structure
screening, long MD trajectories, and large systems for which
fully first-principles sampling would be prohibitive. Compared with earlier
pairwise tight-binding parametrizations, the GFN-xTB family uses
element-specific fits targeted at geometries, vibrational frequencies, and
non-covalent interactions. It retains a minimal valence basis and an extended
H\"uckel-like leading term,\cite{Hoffmann1963ExtendedHuckel} while empirical repulsion, dispersion,
short-range-bond, halogen-bond, and electrostatic terms supply the additional
transferability required for diverse molecular and condensed-phase systems. The
current GFN$n$-xTB variants in CP2K are spin-restricted. Open-shell systems and
bond-dissociation regions are therefore typically treated with finite electronic
temperature through the Fermi entropy term $G_\mathrm{Fermi}$.

The periodic formulation keeps the local-orbital character of molecular tight
binding, but replaces isolated-molecule matrix elements by lattice-summed
Hamiltonian and overlap matrices. These matrices are obtained by the same
lattice Fourier transformation as in Eq.~\eqref{eq:qs_kpoint_fourier}, with
method-specific real-space matrix elements. The atomic-pair form and the corresponding density-matrix back transformation are
given in Section~II~B in the SI. This construction is shared by periodic
(SCC-)DFTB, DFTB3, and GFN$n$-xTB. The methods differ mainly in the
parametrized matrix elements, the charge variables that are optimized
self-consistently, and the electrostatic kernels used under periodic boundary
conditions (PBC).

At the level of the common periodic implementation, the total energy can be viewed schematically as
\begin{equation}
 E_\mathrm{TB}^{\mathrm{per}} =
 \mathrm{Tr}[\mathbf{D}\mathbf{H}^0]
 + E_\mathrm{rep}
 + E_\mathrm{charge}
 + E_\mathrm{disp}
 + E_\mathrm{corr}
 + G_\mathrm{Fermi}
\ensuremath{,}
\label{eq:periodic-xtb-main-energy}
\end{equation}
where \(\mathbf D\) is the periodic AO density matrix and \(\mathbf H^0\) is the zeroth-order tight-binding Hamiltonian.
This compact expression is an umbrella form: the separate (SCC-)DFTB, DFTB3, and GFN$n$-xTB energy expressions are given explicitly in Section~II~C in the SI.
The first term is the band-structure contribution from the periodic density matrix, \(E_\mathrm{rep}\) is the short-range pair repulsion, \(E_\mathrm{charge}\) contains the DFTB SCC response or the corresponding GFN-xTB isotropic electrostatics, \(E_\mathrm{disp}\) contains the D3 or D4 dispersion contribution, \(E_\mathrm{corr}\) collects method-specific terms such as short-range-bond, halogen-bond, or anisotropic XC corrections, and \(G_\mathrm{Fermi}\) is the finite-electronic-temperature entropy contribution. DFTB3 adds the usual third-order charge response, whereas GFN2-xTB adds the anisotropic electrostatic term \(E_\mathrm{AES}^{(2)}\), which depends on atomic charges, cumulative dipoles, and quadrupoles.

The long-range electrostatic part is the main additional ingredient that turns these molecular models into robust periodic methods. (SCC-)DFTB, DFTB3, GFN0-xTB, and GFN1-xTB require a periodic monopole or shell-charge Ewald treatment for the asymptotic Coulomb part of the charge-response kernel. Periodic GFN2-xTB is more demanding because its anisotropic electrostatics couples charges, dipoles, and quadrupoles. A compact real-space notation introduces Cartesian multipoles \(M_A^{(l)}\) and Coulomb interaction tensors \(T^{(n)}=\nabla^{(n)}(1/r)\):
\begin{equation}
 E_\mathrm{mp} =
 \frac{1}{2}\sum_{A,B,\mathbf{L}}^{\prime}
 \sum_{l,m=0}^{2}
 \frac{(-1)^m}{l!\,m!}
 M_A^{(l)} M_B^{(m)}
 T^{(l+m)}(\mathbf{R}_{AB}^{\mathbf{L}}),
\label{eq:periodic-xtb-main-multipole}
\end{equation}
where tensor contractions over Cartesian indices are implied. The atom labels are \(A\) and \(B\), \(\mathbf L\) is a lattice vector, \(l\) and \(m\) are the multipole ranks, \(\mathbf R_{AB}^{\mathbf L}\) is the separation from atom \(A\) to the periodic image of atom \(B\), and the prime excludes the singular self-interaction. The reciprocal-space Ewald contribution is obtained by differentiating the scalar Ewald kernel.\cite{Aguado2003,Laino2008} The recently reported periodic GFN2-xTB implementation in CP2K includes this multipolar Ewald machinery together with analytic forces, stress contributions, \(k\)-point sampling, and molecular-solid benchmarks.\cite{Alizadeh2026PeriodicGFN2XTB} The upcoming g-xTB method has also been implemented in CP2K for molecular and periodic systems.\cite{Froitzheim2026}

The practical consequence is that CP2K provides periodic (SCC-)DFTB, DFTB3, and GFN\(n\)-xTB force engines that can be used for structure screening, finite-temperature dynamics, and spectrum-oriented ensemble generation when a fully first-principles treatment would be prohibitive. Liquid-water benchmarks emphasize that SCC-DFTB and GFN-xTB require system-specific parametrization.\cite{Wu2026SemiempiricalWater} Analytic gradients and stresses require differentiating the lattice sums, charge or multipole response, dispersion and coordination-number terms, and, for GFN2-xTB, the strain dependence of the multipolar electrostatics. Sections~II~A--II~E in the SI give the full density-matrix and energy expressions, monopolar and multipolar Ewald formulas, forces, stress, \(k\)-point sampling, and Brillouin-zone symmetry.

%\section{Ground State Electronic Structure for Spectroscopy: Kohn--Sham DFT}

%Jan (Should we put this section at all??)

%\subsection{Molecules, large supercell with $\Gamma$-point-only approach and small cells with $\bk$-point sampling}
%\subsection{Periodic boundary conditions and \textit{k}-point sampling}

%Benchmarking by TDK, Matthias Krack

%\subsection{Pseudopotentials versus all-electron approach}
%

%\section{Variational Density functional Perturbation Theory}

%Density functional perturbation theory (DFPT) provides the linear-response framework used here for field- and displacement-induced response properties, including nuclear magnetic resonance (NMR) and electron paramagnetic resonance (EPR).

%\subsection{Formalism}
%Coupled-perturbed SCF

%Discussion of atomic-orbital (AO) versus molecular-orbital (MO) response solver~\cite{Belleflamme2023}

%Perturbation: electric-field (E-field) (polarizabilities) or magnetic-field (B-field) (static, frequency = 0)

%

%\subsection{Application to IR and Raman spectra}

%Modern Theory of Polarization

%IR and Raman spectra of MOFs~\cite{Bas2024}

%\subsubsection{Born effective charges}

% \subsection{Application of density functional perturbation theory to nuclear magnetic resonance}

% incl. AIMD?

% perturbation: B-field

% \subsection{Application of density functional perturbation theory to electron paramagnetic resonance}

% incl. AIMD?

% perturbation: B-field (?)

% BEGIN INLINED FROM DFPT.tex
    % -*- TeX-master: "../main.tex" -*-
\section{Density-Functional Perturbation Theory}
\label{sec:dens-funct-pert}

%\TDKtodo{TODO: harmonize the notation with the rest of the article.}
DFPT describes the change of the energy, electron density, and KS orbitals when a reference Hamiltonian is perturbed by a small parameter $\lambda$. In CP2K, variational DFPT can be used to compute responses to perturbations due to electric fields, magnetic fields, nuclear displacements, and nuclear velocities.\cite{Sebastiani2001} The derived response properties include magnetically induced current densities, magnetic-resonance parameters, polarizabilities, and dipole derivatives (infrared amplitudes, the atomic polarization tensor and Born effective charges), vibrational circular dichroism spectra, and response quantities used in Raman optical activity calculations.\cite{Gonze1995,Putrino2000,Weber_JCP_2009,raman_luber_marcella,roa_martin_2,DitlerZimmermannKumarLuber2022VCD}
Published DFPT- and trajectory-based Raman optical activity studies use CP2K response tensors and trajectories as ingredients for subsequent spectral construction.\cite{Luber2017ROA}
Liquid-phase vibrational circular dichroism has also been formulated through nuclear-velocity perturbations along AIMD trajectories.\cite{ScherrerVuilleumierSebastiani2016} An alternative trajectory formulation obtains vibrational circular dichroism from electric--magnetic dipole cross-correlation functions.\cite{ThomasKirchner2016VCD}

The starting point in DFPT is to expand the perturbed energy and KS orbitals in a power series in the perturbation parameter
\begin{subequations}
  \label{eq:dfpt_series}
\begin{align}
  \label{eq:dfpt_e_series}
  E(\lambda) &= E^{(0)} + \lambda E^{(1)} + \lambda^2 E^{(2)} + \cdots,
\\
  \label{eq:dfpt_psi_series}
  \psi_i(\lambda) &= \psi_i^{(0)} + \lambda \psi_i^{(1)} + \lambda^2 \psi_i^{(2)} + \cdots,
\end{align}
\end{subequations}
where the zeroth-order terms are already known from the reference system. The successive terms in the series are related to the derivatives with respect to $\lambda$ at the corresponding order. Its first-order truncation and derivative convention are given in Section~III~A in the SI.
The first-order orbitals determine energy corrections up to third order according to Wigner's \((2n+1)\) rule.\cite{GonzeVigneron1989,Gonze1995}

A perturbation can be externally imposed on the system, as in spectroscopy, or it can represent a small term omitted from the approximate Hamiltonian that defines the reference system. Typical examples for the latter are small changes in the nuclear positions, \(n\)th-order M{\o}ller--Plesset perturbation theory (MP\(n\)) used to describe dynamic electron correlation, including double-hybrid density functionals,\cite{Moller1934,Grimme2006B2PLYP} and various relativistic corrections, such as SOC. In every case, the response of the system is given by the derivatives of the energy or orbitals in Eqs.~\eqref{eq:dfpt_e_series} and~\eqref{eq:dfpt_psi_series}, from which the density variation follows.

The first-order orbitals are obtained by linearizing the KS equations around the reference state. In canonical-orbital notation, this response problem can be written schematically as an inhomogeneous Sternheimer equation
\begin{equation}
\left(H^{(0)}-\epsilon_j^{(0)}\right)\left|\psi_j^{(1)}\right\rangle
=
-\hat{Q}\,H^{(1)}\left|\psi_j^{(0)}\right\rangle ,
\label{eq:sternheimer}
\end{equation}

where \(H^{(0)}\) is the reference KS Hamiltonian, \(\epsilon_j^{(0)}\) and \(\psi_j^{(0)}\) are its orbital energy and orbital \(j\), \(H^{(1)}\) is the first-order Hamiltonian, and \(\psi_j^{(1)}\) is the corresponding orbital response. The projector \(\hat Q\) removes the occupied zeroth-order subspace, thereby avoiding an explicit sum over unoccupied states.
For noncanonical orbitals, the eigenvalues $\epsilon_j^{(0)}$ are to be replaced by the matrix elements of the zeroth-order Hamiltonian. CP2K solves the corresponding AO-basis equations self-consistently with a preconditioned conjugate-gradient minimizer. The detailed Hellmann--Feynman,\cite{Feynman1939Forces} density-response, AO-projection, and overlap-derivative forms are given in Section~III~A in the SI.

%For magnetic responses, the first-order orbitals are imaginary, and one explicitly includes \(\mathrm{i}=\sqrt{-1}\) so that the expansion coefficients can be treated as real. With atom-centered basis functions and vibrational perturbations, the Sternheimer equation also acquires right-hand-side contributions from overlap-matrix derivatives, as detailed in Section~III~A in the SI.

%For a perturbing magnetic field $B$, the orbital expansion can be written as
%\begin{equation}
%  \label{eq:nmr_orb_response}
%  \psi_k = \psi^{(0)}_k + B\,\psi_k^{(1)} + \dots,
%\end{equation}
%where $\psi_k^{(1)}$ is the first-order orbital correction. This explicit magnetic-field expansion provides the link to the magnetic-resonance response properties discussed below.

In general, \(H^{(1)}\) combines the explicit derivative of the external perturbation with the induced Hartree and XC response through the second-order energy kernel. Nuclear displacements and homogeneous electric fields therefore enter through perturbation-specific external-potential terms, whereas magnetic fields require the corresponding vector-potential coupling. The explicit kernel, nuclear-displacement, and electric-field expressions are given in Section~III~A in the SI. The same nuclear-displacement response is also relevant beyond conventional electronic DFPT: periodic constrained nuclear-electronic orbital DFT, recently developed in CP2K, treats selected light nuclei quantum mechanically in a multicomponent KS framework. The method is discussed with NQEs in Section~\ref{sec:nuclear-quantum-effects}.\cite{Chen2025} The case of a perturbation by an external magnetic field, as well as internal (spin) magnetic fields, is discussed with the magnetic-resonance parameters in Section~\ref{sec:NMR+EPR}.

The same response can also be formulated variationally. Whenever the perturbed quantity obeys a variational principle, the even orders of the perturbation obey a stationary principle as well. Minimizing the second-order functional under the parallel-transport condition \(\langle\psi_i^{(0)}|\psi_j^{(1)}\rangle=0\) gives the same Sternheimer response problem. This variational form is particularly convenient when the perturbation is itself specified as an energy functional, as in Berry-phase electric-field response. The explicit Gonze and Putrino functionals and the constrained minimum condition are presented in Section~III~A in the SI.

Sections~\ref{sec:dfpt-raman}, \ref{sec:born-charges}, and~\ref{sec:NMR+EPR} cover in more detail the calculation of dipole polarizabilities, dipole derivatives, and magnetic resonance parameters. Readers interested in input-level details and the interplay of the relevant CP2K input sections within the linear-response modules are referred to our user-oriented CP2K Made Simple article.\cite{Iannuzzi2025}

\subsection{Dipole Polarizability and Raman Spectroscopy}
\label{sec:dfpt-raman}

For periodic systems, CP2K follows the modern theory of polarization.\cite{Berry1984,KingSmith:1993hp,resta} In the \(\Gamma\)-point limit used for the DFPT polarizability implementation,\cite{Yaschenko1998} the electronic polarization is evaluated from the phase of a Resta single-point overlap operator. This construction replaces the ill-defined position operator of an extended system by a unitary ``twist'' operator and therefore supplies the polarization entering electric-field response, dipole derivatives, and finite-field dynamics. The explicit Berry connection, single-point overlap, electric-field perturbation, induced-polarization, and phase-unwrapping expressions are given in Section~III~B in the SI.

Solving the electric-field Sternheimer problem gives the derivative of the Berry-phase polarization with respect to the applied field. The static electronic polarizability \(\boldsymbol{\alpha}\) follows from differentiating the polarization \(\mathbf P\) with respect to the applied electric field \(\boldsymbol{\mathcal E}\)
\begin{equation}
  \alpha_{\alpha\beta}
  =
  \left.
  \frac{\partial P_\alpha}{\partial \mathcal{E}_\beta}
  \right|_{\boldsymbol{\mathcal{E}}=0}
  \ensuremath{,}
  \label{eq:dfpt_polarizability_derivative}
\end{equation}
where \(\alpha_{\alpha\beta}\) is the polarizability component relating polarization \(P_\alpha\) to a field \(\mathcal E_\beta\), and \(\alpha\) and \(\beta\) are Cartesian directions.
This expression is understood up to the volume convention used to distinguish a susceptibility from a cell polarizability. Raman intensities are then obtained from derivatives of this polarizability tensor with respect to nuclear displacements or normal-mode coordinates, whereas finite-temperature Raman spectra can also be formulated through polarizability time-correlation functions, as discussed in Section~\ref{sec:aimd-finite-temperature-spectra}.
%\TDKtodo{TODO: Add the normal-mode derivative form of the Raman activity and check consistency with the finite-temperature Raman discussion in Section~\ref{sec:aimd-finite-temperature-spectra}.}

The same polarization also enters finite-temperature infrared (IR) absorption through polarization time-correlation functions. Because that treatment is dynamical rather than a fixed-geometry DFPT response, the explicit expression appears with the polarization working equations in Section~III~B in the SI and is connected conceptually to the finite-temperature spectroscopy discussion in Section~\ref{sec:aimd-finite-temperature-spectra}.
%\TDKtodo{TODO: Decide whether to revive a short analytic dipole-derivatives/DCDR paragraph; before doing so, verify the current CP2K implementation status and any multigrid limitations.}

%\subsection{Analytic dipole derivatives}

%TODO: should we include this section?

%\begin{figure}[h]
%  \includegraphics[width=0.5\textwidth]{figures/dcdr.png}
%\end{figure}

%TODO: Can someone else confirm that there is something wrong with the DCDR implementation?

%NOTE: Doesn't work with multigrid (?).

\subsection{\texorpdfstring{Atomic Polarization Tensors (Born Effective Charges) from Finite Differences}{Atomic Polarization Tensors (Born Effective Charges) from Finite Differences}}
\label{sec:born-charges}

The atomic polarization tensor, also known as the Born effective charge tensor, \(\mathbf Z_a^\ast\), in a periodic system is a Cartesian tensor whose components \(Z_{a,\alpha\beta}^{\ast}\) measure the polarization \(\mathbf P\) linearly induced in the \(\beta\) direction by a displacement \(u_{a,\alpha}\) of atom \(a\) in the \(\alpha\) direction under conditions of vanishing electric field \(\boldsymbol{\mathcal E}\)
\begin{equation}
  \label{eq:dfpt_born}
  Z_{a,\alpha\beta}^{\ast}
  =\Omega \frac{\partial P_{\beta}}{\partial u_{a,\alpha}} \Bigr|_{\boldsymbol{\mathcal E}=0},
\end{equation}
where $\Omega$ is the volume of the simulation cell. This expression is the basis for computing the Born charge as a dipole derivative using perturbation theory.

Alternatively, one can express the Born charge of an atom as the electric field derivative of the force on that atom in the following form
\begin{equation}
  \label{eq:dfpt_born_force}
  Z_{a,\alpha\beta}^{\ast}
  = \frac{\partial F_{a,\alpha}}{\partial \mathcal E_{\beta}},
\end{equation}
where $F_{a,\alpha}$ is the force induced on atom $a$ in the direction $\alpha$ at zero displacement, and \(\mathcal E_{\beta}\) is the $\beta$ Cartesian component of a macroscopic (Maxwell) electric field. The latter expression is particularly amenable to a finite-difference implementation. For two-point finite differences, one obtains the Born charges of all atoms from six force calculations in the presence of an external field, which is more efficient for this finite-difference strategy than separate displacement-response calculations.

% \subsection{Nuclear Magnetic and Electron Paramagnetic Resonance Spectroscopy} \label{sec:NMR+EPR}
% The energy levels probed in magnetic spectroscopy correspond to transitions among nuclear and electronic spin eigenstates in the presence of an external magnetic field.
% %The calculation of magnetic response properties in CP2K is based on variational density functional perturbation theory (DFPT)~\cite{Putrino2000, Sebastiani2001, Sebastiani2003}.

\subsection{The Magnetic Shielding Tensor}
\label{sec:magn-shield-tens}
\label{sec:NMR+EPR}

An external magnetic field induces a current density and an associated magnetic field. Their Biot--Savart relation and the spin--spin branch of the magnetic-response expansion are given in Section~III~C in the SI. The tensor field %$\boldsymbol{\sigma}(\mathbf{r})$
\begin{equation}
  \label{eq:nmr_sigma_def}
  \boldsymbol{\sigma}(\mathbf{r}) = -\frac{\partial\mathbf{B}^{\textrm{ind}}(\mathbf{r})}{\partial\mathbf{B}^{\textrm{ext}}}
\end{equation}
is known as the magnetic shielding tensor. Experimentally, the induced magnetic field can only be probed at the position of a nuclear spin $\mathbf{r}^N$. Being a ratio, $\boldsymbol{\sigma}(\mathbf{r})$ is conventionally reported in units of parts per million. The experimental value is almost always referenced to the isotropic average $\frac{1}{3}\mathrm{Tr}[\boldsymbol{\sigma}(\mathbf{r})]$ of the magnetic shielding tensor of some chosen standard for the respective nucleus according to
\begin{equation}
  \label{eq:nmr_shift}
  \boldsymbol\delta(\mathbf{r}^N_i) = \sigma^{\textrm{iso}}_{i,\textrm{ref}}\mathbf{I} - \boldsymbol\sigma(\mathbf{r}^N_i),
\end{equation}
which is known as the chemical shift tensor, and its isotropic average $\delta$ is called the chemical shift. In the last equation, $\mathbf{I}$ is the unit matrix.

The nuclear magnetic resonance (NMR) shielding tensor is experimentally measurable only at the positions of nuclei with a nonzero spin. However, the above definition shows that the magnetic shielding is a tensor field, which can be computed at any arbitrarily chosen spatial position. These values are known as nucleus-independent chemical shifts and are also available in CP2K.\cite{Sebastiani2006}

The induced current density is therefore the central quantity: once it is known, the shielding tensor follows from an integration over the material and its periodic replicas, as detailed in Section~III~C in the SI.

AIMD configurations and absolutely localized molecular orbital energy decomposition analysis have supported non-CP2K calculations of hydrogen-bond scalar couplings in amorphous ice.\cite{ElgabartyKuehne2026NMR}

Using the approach developed by Sebastiani and Parrinello,\cite{Sebastiani2001} CP2K evaluates the magnetic perturbation using maximally localized Wannier functions (MLWFs) in combination with variational DFPT.\cite{Weber_JCP_2009} The localized nature of the Wannier functions allows the position operator to enter the magnetic perturbation Hamiltonian in a controlled way. The corresponding operator definitions are summarized in Section~III~C in the SI. An underlying core assumption here is that each Wannier function is contained entirely within the simulation cell once it is centered. For insulators, the Wannier functions decay exponentially, so that for sufficiently large simulation cells this assumption is controlled.\cite{Resta2002} In comparison to other approaches developed for condensed-phase systems,\cite{Mauri1996,Sebastiani2001} CP2K differs in its use of local atom-centered Gaussian functions (see Section~\ref{sec:TotalEnergy}), allowing for reduced-complexity algorithms and hence large-scale calculations of magnetic-resonance parameters. With the GAPW method, CP2K/\textsc{Quickstep} is also able to compute all-electron magnetic-resonance parameters. One can also use embedding techniques such as quantum mechanics/molecular mechanics (QM/MM) and further decompose the QM part into GAPW and GPW regions. The reliance on the localization properties of Wannier functions means, however, that this implementation is not suitable for conductors, where the Wannier functions decay only algebraically.

The MLWF formulation turns the magnetic-field response of an extended system into localized response problems. Orbital-specific translations define the magnetic perturbation operators, and the resulting linear-response orbitals yield paramagnetic and diamagnetic current-density contributions for the three applied-field directions. Distributed gauge-origin choices, in particular individual gauges for atoms in molecules (IGAIM) and the continuous set of gauge transformations (CSGT),\cite{Cheeseman1996} control the usual gauge-origin problem of local atomic basis sets. The corresponding vector-potential, operator, current-density, and gauge-origin equations are provided in Section~III~C in the SI.

For periodic systems, CP2K evaluates the shielding integral by decomposing the induced current density in the same spirit as the GAPW electron-density decomposition. The smooth current-density contribution is treated on the PW grid, while local hard and compensation contributions recover the all-electron behavior near nuclei. The \(\mathbf{G}=0\) contribution is handled through a susceptibility correction consistent with the spherical-sample convention, and the local contribution is integrated over atom-centered domains. The explicit decomposition and integration formulas are given in Section~III~C in the SI.
% \begin{verbatim}
% &SUBSYS
%   &KIND H
%     BASIS_SET aug-pcSeg-2
%     POTENTIAL ALL
%     LEBEDEV_GRID 100
%     RADIAL_GRID 200
%   &END KIND
% &END SUBSYS
% \end{verbatim}
% In the \texttt{\&NMR} input section, one can set the value of $R_c$, request a NICS calculation, and specify the spatial points for the latter:
% \begin{verbatim}
% &LOCALIZE
%   &NMR
%     NICS .TRUE.
%     NICS_FILE_NAME nics.xyz
%     SHIFT_GAPW_RADIUS # Cutoff current
%                       # integration
%   &END NMR
% &END LOCALIZE
% \end{verbatim}

%\subsection{Indirect spin-spin couplings}

\subsection{\texorpdfstring{Electron Paramagnetic Resonance \(g\)-Tensor}{Electron Paramagnetic Resonance g-Tensor}}
\label{sec:epr-g-tensor}

CP2K also offers an implementation of the \(g\)-tensor under PBC, which employs the same DFPT approach to compute the induced current densities. This makes it feasible to perform large-scale calculations, including with QM/MM (e.g., paramagnetic active sites in enzymes), to investigate paramagnetic defects in solids under PBC, and to perform large-scale computations of paramagnetic NMR shifts.\cite{Mondal2017}

In electron paramagnetic resonance (EPR) spectroscopy, the \(g\)-tensor plays a role analogous to the shielding tensor in NMR spectroscopy. This is clear in an effective spin-Hamiltonian framework, which gives the coupling between an effective electronic spin $\mathbf{S}$ and an external magnetic field $\mathbf{B}^{\textrm{ext}}$ as the bilinear term
\begin{equation}
  \label{eq:nmr_spinH_g}
  \mathrm{\hat{H}}_{\textrm{eff}}= \frac{1}{2c}\,\mathbf{B}^{\textrm{ext}} \cdot \mathbf{g} \cdot \mathbf{S}
  \ensuremath{,}
\end{equation}
where \(c\) is the speed of light in the atomic-unit convention used for the magnetic interaction.
Like the shielding tensor, the \(g\)-tensor is a second partial derivative of the energy with respect to the external field and a spin magnetic moment, in this case the electronic spin. The corresponding \textit{ab-initio} expression is obtained from the minimal coupling electronic Hamiltonian including relativistic corrections up to $\mathcal{O}(\alpha^3)$, where $\alpha$ is the fine structure constant. Identification of the relevant Hamiltonian terms produces the expression for the \(g\)-tensor as spatially anisotropic relativistic correction terms $\Delta{\mathbf{g}}$ to the free-electron \(g\)-value $g_e$. The implementation in CP2K is based on the Schreckenbach--Ziegler DFT-based approach,\cite{Schreckenbach1997} and its extension by Pickard and Mauri.\cite{Pickard2002,VanYperen-DeDeyne2012} One can identify three contributions to each component of $\Delta\mathbf{g}$ as follows
\begin{equation}
  \label{eq:nmr_g}
  g_{xy}=g_e\delta_{xy} + \Delta{g}_{xy}^{\textrm{ZKE}}  + \Delta{g}_{xy}^{\textrm{SO}}  + \Delta{g}_{xy}^{\textrm{SOO}}
  \ensuremath{,}
\end{equation}
where \(x\) and \(y\) denote Cartesian components and \(\delta_{xy}\) is the Kronecker delta.
The correction terms in Eq.~\eqref{eq:nmr_g} are the Zeeman kinetic energy (ZKE), spin--orbit (SO), and spin-other-orbit (SOO) contributions. ZKE is a kinematic scalar-relativistic correction computed from the spin-polarized KS kinetic energy, while SO and SOO depend on spin-dependent current densities and therefore reuse the Sebastiani--Parrinello MLWF machinery discussed for magnetic shieldings. The effective potential entering the SO term contains the external, Hartree, and XC contributions,\cite{Schreckenbach1997} and the current implementation supports local-density approximation (LDA) and generalized-gradient approximation (GGA) functionals, but not hybrid XC functionals. The detailed ZKE, SO, SOO, corrected-field, and SOO-approximation expressions are summarized in Section~III~D in the SI.

The \(g\)-tensor is generally much less sensitive to the choice of gauge than nuclear magnetic shieldings, and even the computationally convenient CSGT can be used. Tables~1 and 2 in Ref.~\citenum{Mondal2017} offer a comparison of \(g\)-tensors calculated by CP2K with the IGAIM gauge versus those calculated by ORCA,\cite{Neese2005} and MAG-ReSpect,\cite{Malkina2000} using a variety of gauge origins and SOC treatments.

If the simulation cell contains more than one paramagnetic center, the cell response contains additive contributions from the individual centers.\cite{Pigliapochi2017,Mondal2017} The corresponding per-center normalization convention is described in Section~III~D in the SI.
%\TDKtodo{TODO: Check whether the normalization rule should be described as a post-processing convention or as the quantity printed by CP2K, and state this explicitly.}

% To activate the calculation of the g-tensor in the input file, one has to request the spin density from an unrestricted KS calculation (keyword \texttt{\&DFT\%UKS}). One also needs to include the \texttt{\&CURRENT}, \texttt{\&LOCALIZE}, and \texttt{\&EPR} subsections within the \texttt{\&LINRES} section. The various control parameters related to the calculation of the induced current density, in the input section \texttt{\&CURRENT}, were already discussed for the magnetic shielding tensors. The input section
% \begin{verbatim}
% &LINRES%EPR%PRINT%G_TENSOR%XC%XC_FUNCTIONAL
% \end{verbatim}
% sets the functional for the effective XC potential in Eq.~\eqref{eq:nmr_g_so}. %Hybrid functionals are currently not supported.
%\begin{verbatim}
%&PROPERTIES
%  &LINRES
%    &CURRENT
%      ...
%    &END CURRENT
%    &LOCALIZE
%      ...
%    &END LOCALIZE
%    &EPR
%      &PRINT
%        &G_TENSOR
%         &XC
%           &XC_FUNCTIONAL
%             &PBE
%             &END PBE
%           &END XC_FUNCTIONAL
%         &END XC
%        &END G_TENSOR
%      END PRINT
%    &END EPR
%  &END LINRES
%&END PROPERTIES
%\end{verbatim}
% Other XC potentials are activated in a similar fashion, for example \texttt{XALPHA} corresponds to the Dirac-Slater potential, whereas \texttt{BECKE88} together with \texttt{LYP} will use the BLYP XC potential.

\subsection{Hyperfine Couplings}
\label{sec:hyperfine-couplings}
The effective spin-Hamiltonian term for the hyperfine coupling is bilinear in the nuclear and electron spin
\begin{equation}
  \label{eq:nmr_a_Heff}
  \mathrm{\hat{H}}= \mathbf{S} \cdot \mathbf{A} \cdot \mathbf{I}
  \ensuremath{,}
\end{equation}
where \(\mathbf S\) and \(\mathbf I\) are the effective electronic- and nuclear-spin operators, respectively, and \(\mathbf A\) is the hyperfine-coupling tensor.

The \textit{ab-initio} expression for the hyperfine coupling tensor, or $\mathbf{A}$-tensor, is obtained from relativistic quantum mechanics, as in the case of the \(g\)-tensor. For a nucleus \(N\), the dominant contributions are the isotropic Fermi-contact term, which probes the spin density at the nucleus, and the anisotropic dipolar term, which describes the spatially resolved spin--dipole interaction. Their complete integral expressions and prefactors are given in Section~III~E in the SI.
In CP2K, the contact term uses a scalar-relativistic smeared delta function, while the non-relativistic limit recovers a Dirac delta function. The scalar-relativistic smearing and its GAPW integration are also detailed in Section~III~E in the SI.\cite{Declerck2006a} In addition to these two first-order terms, a second-order SOC contribution can become important for heavy atoms. %This term is currently not implemented in CP2K. Yet, for light atoms that are not in the immediate neighborhood of heavy atoms, the first-order terms are usually sufficient.

The isotropic term particularly requires an accurate description of the electron density at the position of the nucleus. Hence, hyperfine couplings are implemented within CP2K's GAPW method.\cite{Declerck2006a} Section~III~E in the SI gives the hard-density, soft-density, and cross-term integration details.

% Unlike the magnetic properties discussed so far, the hyperfine coupling in terms of Eqs.~\ref{eq:nmr_a_iso}-\ref{eq:nmr_delta_smeared} is a first-order property, obtained as an expectation value over the unperturbed spin density. Therefore, the input section for requesting the printing of the hyperfine coupling tensors is found directly under the DFT section via
% \begin{verbatim}
% &DFT%PRINT%HYPERFINE_COUPLING_TENSOR
% \end{verbatim}
% In addition to various print control options, the \texttt{INTERACTION\_RADIUS} can be set inside this input section, which specifies the cutoff radius mentioned above.

%\subsection{Paramagnetic NMR}

%%% Local Variables:
%%% TeX-master: "../main"
%%% End:
% END INLINED FROM DFPT.tex

\section{Density-Functional Theory for Electronic Excitations and Optical Response}
\label{sec:excited-state-dft}
%\section{Time-Dependent Density-Functional Theory}
%%for neutral electronic excitations and fixed nuclei

%Previous version, before TIDFT was added:
%TDDFT describes how the electronic density responds to time-dependent perturbations and is therefore the natural bridge between fixed-nuclei spectroscopy, real-time electron dynamics, and mixed quantum--classical excited-state dynamics. In CP2K, this framework appears in two complementary forms. LR-TDDFT gives excitation energies, oscillator strengths, transition-density analyses, SOC corrections, and excited-state gradients from a response problem around the electronic ground state. Real-time TDDFT (RT-TDDFT), discussed in Section~\ref{sec-rt_tddft}, instead propagates the electronic state explicitly in time and can be coupled to Ehrenfest nuclear motion. This methodological structure links the spectroscopic capabilities discussed below, while the implementation-level working equations are collected in Section~IV in the SI.

Excited electronic states within DFT can be treated via two approaches. The first and most widely used is TDDFT, which describes the response of the ground-state density to time-dependent perturbations. In CP2K, LR-TDDFT provides excitation energies, oscillator strengths, transition-density analyses, SOC corrections, and excited-state gradients, whereas real-time TDDFT (RT-TDDFT) describes explicit electron dynamics coupled to Ehrenfest nuclear motion. The second approach comprises orbital-optimized methods, often called $\Delta$SCF or time-independent DFT (TIDFT), in which excited-state orbitals are variationally optimized using the ground-state KS functional. Although less mature than TDDFT, orbital-optimized approaches can describe charge-transfer, Rydberg, and double-electron excitations that are problematic for standard TDDFT with common XC functionals.\cite{hait2021orbital} This section covers both frameworks, with emphasis on the production-ready TDDFT machinery. The implementation-level working equations are given in Section~IV in the SI.

\subsection{Linear-Response Time-Dependent Density-Functional Theory and Excited-State Properties}\label{sec-lr_tddft}
%{\bf{Anna Hehn}}, Augustin (?)
In the LR-TDDFT implementation, excitation energies and oscillator strengths are obtained from the small-amplitude response around the electronic ground state. CP2K uses the Sternheimer formulation in the Tamm--Dancoff approximation (TDA),\cite{Tamm1945,Dancoff1950} avoiding an explicit construction of the full virtual manifold where possible and connecting naturally to the coupled-perturbed KS (CPKS) or Z-vector machinery used for response properties and gradients.\cite{HandySchaefer1984,Ferrero2008CPKS} MO and AO response formulations, the response-density Poisson equation, exact-exchange and semiempirical kernels, and spin-flip extensions are described in Sections~IV~A and~IV~B in the SI. The discussion below retains only the physical role of these ingredients.
%Restricted excitation window  ?????? %%%%%<<<<<<<

%Excited-state dipole transition moments in the length or velocity gauge or implying a Berry-phase formulation and corresponding oscillator
%strengths $f_{0N}$ are defined as
%\begin{align}
%   f_{0N} = \frac{2}{3} \Omega_N | \langle 0 | \hat{\mu}^{\textnormal{\tiny{L}}} | n \rangle |^2 = \frac{2}{3 \Omega_N}
%   | \langle 0 | \hat{\mu}^{\textnormal{\tiny{V}}} | n \rangle |^2 \, ,
%   f_{0N}^{\textnormal{\tiny{Berry}}} =
%\end{align}
%with $\hat{\mu}^{\textnormal{\tiny{L}}}$, $\hat{\mu}^{\textnormal{\tiny{V}}}$ and
%$\hat{\mu}^{\textnormal{\tiny{Berry}}}$ denoting the dipole operator within the length, velocity, or Berry representation\cite{paperluber}.

SOC effects can be included perturbatively in the TDDFT module, either with an all-electron zeroth-order regular-approximation Hamiltonian or with HGH dual-space GTH-type Gaussian PPs, consistent with the ground-state SOC machinery summarized in Section~\ref{spin_orbit_coupling_section}.\cite{vogt2025}
Applications include Fe L-edge X-ray absorption and defect optical spectra in MgO and HfO$_2$.\cite{StrandWatkins2019,XAS_Nanchen}
For analysis, CP2K provides natural transition orbitals (NTOs), which compress an excitation into dominant particle--hole pairs and rank their contributions through the eigenvalues $\boldsymbol{\lambda}$.\cite{ntos} For the particle states, these orbitals are obtained by solving
\begin{equation}
\left[
\mathbf{T}^N-\boldsymbol{\lambda}\mathbf{S}
\right]\mathbf{U}=0,
\label{nto_eigenvalue_problem}
\end{equation}
with the contracted excitation amplitude matrix
\begin{align}
   T^N_{\mu \nu \sigma} = \sum_{\kappa \lambda i} S_{\mu \kappa} X_{\kappa i \sigma} X_{\lambda i \sigma} S_{\lambda \nu}
   ,
\end{align}
implying the constraint
\begin{align}
   \sum_{\kappa}
\bigl(\Pi_{\mathrm{occ}}\bigr)_{\mu \kappa \sigma}
   U_{\kappa \nu \sigma} = 0.
\end{align}
In these equations, Greek indices label AOs, \(i\) and \(a\) label occupied and virtual MOs, \(\sigma\) labels spin, \(\mathbf X\) contains excitation amplitudes, \(\mathbf S\) is the AO overlap matrix, \(\boldsymbol{\Pi}_{\mathrm{occ}}\) projects onto the occupied subspace, and \(\boldsymbol{\lambda}\) contains the NTO weights.
Eq.~\eqref{nto_eigenvalue_problem} defines the transformation matrix $\mathbf{U}$ for the final NTOs through
\begin{align}
   C^{\mathrm{NTO}}_{\mu a \sigma} = \sum_{\nu} U_{\mu \nu \sigma} C_{\nu a \sigma}
   \ensuremath{.}
\end{align}

%\subsubsection{Core-level time-dependent density functional theory}
%XAS module\cite{Bussy2021}
% END INLINED FROM TDDFT.tex
% BEGIN INLINED FROM TDDFT_gradients.tex

% Harris functional as a specific version
%In case of a perturbative correction such as the Harris functional \cite{Belleflamme2023}, introducing an initial
%''input''
%density $n^{\textnormal{in}} (\mathbf{r})$ next to an improved
%''output'' density $n^{\textnormal{out}} (\mathbf{r})$,
% excited states

Excited-state gradients are the derivative layer of the same LR-TDDFT response hierarchy. They are obtained by differentiating a variational TDA Lagrangian\cite{Hutter_excited_state_forces_2003} and solving the Z-vector response equations. The relaxed and unrelaxed density intermediates and the explicit gradient expression are given in Section~IV~B in the SI. This makes LR-TDDFT useful not only for vertical spectra, but also for excited-state structure optimization, finite-temperature snapshot calculations, and inputs to excited-state dynamics as outlined in Section~\ref{sec:nonadiabatic-md-surface-hopping}.
% END INLINED FROM TDDFT_gradients.tex

Core-level applications such as X-ray absorption spectroscopy (XAS) use the same LR-TDDFT infrastructure with core-specific restrictions and approximations. These aspects are discussed together with transition-potential and real-time X-ray approaches in Sections~\ref{sec:xray-tp}--\ref{sec:xray-cysteine}.\cite{Bussy2021}

\subsection{\texorpdfstring{Toward Time-Dependent Density-Functional Theory for Crystals with $\bk$-Point Sampling: Periodic Transition Dipoles}{Toward Time-Dependent Density-Functional Theory for Crystals with k-Point Sampling: Periodic Transition Dipoles}}\label{TDDFTkpoint}
%Shridhar Shanbhag

For crystals with small primitive cells, a TDDFT description of optical response must be formulated with $\bk$-point sampling and must use optical matrix elements that are compatible with PBC. The key quantity is the transition dipole between Bloch states. Direct numerical differentiation of MO coefficients with respect to $\bk$ is not robust because the phases of neighboring Bloch states are arbitrary. Section~IV~C in the SI therefore gives the gauge-consistent off-diagonal expression in terms of analytic $\bk$ derivatives of the KS and overlap matrices and atom-centered dipole integrals.

\begin{figure}[t!]
    \centering
    \includegraphics[width=\linewidth]{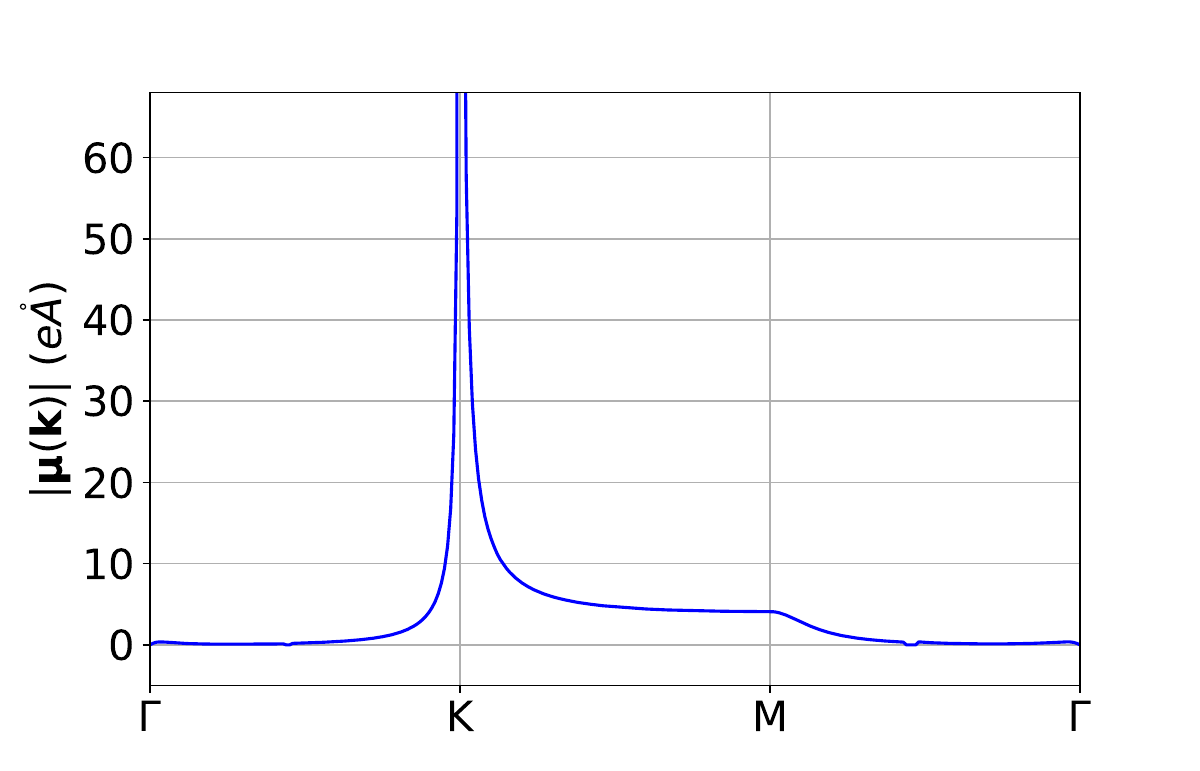}
    \caption{Magnitude of the transition-dipole matrix element for the transition between the highest valence and lowest conduction bands in graphene as a function of $\bk$ in the Brillouin zone, computed with CP2K from the off-diagonal transition-dipole expression detailed in Section~IV~C in the SI.}
    \label{fig-dipoles}
\end{figure}

This expression also explains a useful diagnostic feature: off-diagonal transition dipoles become large near band crossings because they contain the inverse band-energy difference.
Fig.~\ref{fig-dipoles} illustrates this behavior for graphene by plotting the transition-dipole element between the highest occupied and lowest unoccupied bands. The divergence at the \(K\) point reflects the Dirac cone and provides a stringent test of the periodic dipole implementation. The same matrix elements are ingredients for a density-matrix-based RT-TDDFT implementation for crystals that is currently under development.

\subsection{Variable-Metric Time-Independent Density-Functional Theory}
\label{sec:vm-tidft}

Orbital-optimized DFT methods apply the ground-state KS formalism to variationally optimize not only the ground state, but also excited states, computing excitation energies as differences between SCF solutions.\cite{hait2021orbital} The main practical challenge is preventing variational collapse of high-energy states onto lower-energy solutions. Variable-metric time-independent DFT (VM TIDFT) addresses this by allowing electronic states to be nonorthogonal during optimization while gradually enforcing orthogonality through a continuous penalty function.\cite{pham2025tidft}

The loss functional minimized for electronic state $I$ is
\begin{equation}
\Omega = E^{I} + \Omega_{\mathrm{intra}} + \Omega_{\mathrm{inter}} ,
\end{equation}
where $E^{I}$ is the KS energy, $\Omega_{\mathrm{intra}}$ ensures linear independence of the MOs within state $I$,\cite{pham2024scf} and the interstate penalty
\begin{equation}
\Omega_{\mathrm{inter}} = -C_P \sum_{\tau} \ln \det \left(
\boldsymbol{\Phi}_\tau \boldsymbol{\Phi}_{\tau d}^{-1}
\right)
\end{equation}
keeps different electronic states orthogonal. In this expression, \(\boldsymbol{\Phi}_\tau\) is the overlap matrix between Slater determinants for spin $\tau$, with elements \((\boldsymbol{\Phi}_\tau)_{IJ} = \det(\boldsymbol{\sigma}_{IJ\tau})\) constructed from the MO overlap matrices \(\boldsymbol{\sigma}_{IJ\tau}\), and \(\boldsymbol{\Phi}_{\tau d}\) contains the diagonal elements for state normalization. The penalty strength $C_P$ is increased until the orthogonality deviation falls below a prescribed threshold.

A key advantage of this formulation is that MO coefficients serve directly as independent variables, yielding closed-form analytic gradients and enabling the use of standard unconstrained optimization algorithms, specifically preconditioned conjugate gradient in the current CP2K implementation, that guarantee convergence.\cite{pham2024scf,pham2025tidft} This fully variational treatment contrasts with other orbital-optimized DFT methods for excited states that impose orthogonality via projectors or non-aufbau occupation rules and can exhibit convergence difficulties for challenging excited states.

\begin{figure*}[t!]
    \centering
\includegraphics[width=1.0\linewidth]{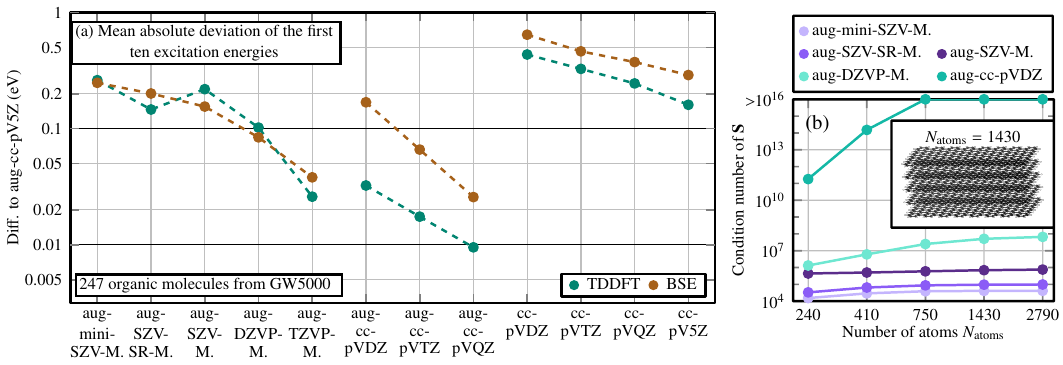}
\caption{Benchmarking the new augmented MOLOPT basis sets. (a) shows the basis-set convergence of the first ten excitation energies from TDDFT (PBE0) and BSE calculations on top of ${G_0W_0}$@PBE0 over 247 molecules from the $GW5000$ database. We report the mean absolute deviation for the augmented MOLOPT, aug-cc-pVXZ, and cc-pVXZ basis sets, with the aug-cc-pV5Z results taken as reference. (b) shows the condition number of the overlap matrix for different basis sets in finite graphite fragments of increasing size. The inset shows the case for 1430 atoms; the number of atoms is varied by changing the horizontal size. Data from Ref.~\citenum{Pasquier2026}.}
    \label{figaugMOLOPT}
\end{figure*}

VM TIDFT correctly reproduces the Coulomb tail of charge-transfer excitation energies at large fragment separations and can describe double-electron excitations that are inaccessible to linear-response TDDFT. Benchmark calculations show MAEs comparable to or better than TDDFT for valence excitations, with marked improvements for charge-transfer states.\cite{pham2025tidft}
Interfaced through the Atomic Simulation Environment, CP2K was used with restricted open-shell KS excited-state MD to resolve competing hydrogen-atom transfer and proton-coupled electron transfer in the photochemical formation of the hydrated electron.\cite{Larsen2017ASE,DiazMiron2026HydratedElectron} The associated spectral signatures were represented by energy-gap distributions rather than a native CP2K spectral implementation.

\subsection{Gaussian Basis Sets for Excited-State Calculations}
% Rémi Pasquier

The computational cost of TDDFT calculations can scale up to the fourth power of the AO basis-set size.\cite{Casida1996} An optimized basis that yields converged results with as few functions as possible is therefore preferable. Most commonly used basis sets have been optimized for the calculation of ground-state properties and therefore converge slowly in excited-state calculations. One way to mitigate this problem is to \textit{augment} an existing basis with additional functions, as in the aug-cc-pVXZ families derived from cc-pVXZ.\cite{Kendall1992} In that case, the additional basis functions were optimized with respect to lowest unoccupied molecular orbital (LUMO) properties, leading to good accuracy for the excited-state properties of smaller molecules but to large condition numbers of the overlap matrix for large molecules and periodic systems because of the presence of very diffuse basis functions (with very small exponents), hindering SCF convergence. Hence, excited-state-optimized basis sets are still needed for larger systems.

Accordingly, Ref.~\citenum{Pasquier2026} introduces a new family of all-electron Gaussian basis sets, augmented MOLOPT, built from the MOLOPT basis sets.
MOLOPT basis sets were specifically designed to balance fast convergence of the DFT ground-state energy and numerical stability by taking into account the overlap condition number during the optimization procedure.\cite{vande2007} The new augmented basis sets were optimized for fast convergence of excitation energies at the ${G_0W_0}$@PBE0 BSE level and for small condition numbers.
The molecular training data comprised the Thiel set\cite{Schreiber2008} and the set used in the original MOLOPT basis optimization.\cite{vande2007}

The new augmented MOLOPT basis sets were then benchmarked on a subset of 247 molecules from the GW5000 database, as shown in Fig.~\ref{figaugMOLOPT}(a).\cite{Stuke2020} We report the basis-set convergence of the first ten excitation energies obtained from BSE calculations on top of ${G_0W_0}$@PBE0 and TDDFT (PBE0) for the augmented MOLOPT, augmented cc-pVXZ, and original cc-pVXZ basis sets, with respect to the aug-cc-pV5Z results. The new augmented MOLOPT basis sets improve substantially upon the original cc-pVXZ basis sets. For BSE calculations, the new basis sets converge faster to the basis-set limit than the aug-cc-pVXZ family, with, for example, a mean absolute deviation (MAD) of 80~meV for the aug-DZVP-MOLOPT basis set, whereas the comparable aug-cc-pVDZ shows a MAD of 170~meV. However, the aug-cc-pVXZ basis sets outperform the aug-MOLOPT family for the TDDFT calculations: for example, the MAD of the aug-DZVP-MOLOPT basis set is 100~meV, whereas the aug-cc-pVDZ basis set achieves a MAD of 32~meV. This difference reflects the fact that the new basis sets were explicitly optimized for BSE excitation energies, whereas the original aug-cc-pVXZ family was optimized for LUMO energies. %Similar conclusions can be drawn from the illustrative example given in Fig.~\ref{figaugMOLOPT} b), where we carry the same calculations for the particular case of 9,10-Dihydroanthracene as a function of the number of basis functions. We show that both augmented MOLOPT and augmented cc-pVXZ basis sets converge much faster to the basis set limit than their non-augmented counterparts, although once again the augmented cc-pVXZ show a better convergence with the number of basis set functions.

The real strength of the new augmented MOLOPT basis sets is shown in Fig.~\ref{figaugMOLOPT}(b), where we study the condition number of the overlap matrix for both augmented basis-set families as a function of the size of a benchmark graphite fragment. The aug-cc-pVDZ condition number is very high (above \(10^{11}\) for a fragment of 240~atoms) and rapidly exceeds the inverse of machine precision (\(10^{16}\)), leading to severe numerical instabilities. By contrast, the condition number of the augmented MOLOPT family never exceeds \(10^8\), even in systems composed of thousands of atoms, and is therefore much more robust for large molecules and crystals.

Overall, the new augmented MOLOPT family shows fast convergence to the basis-set limit for excited-state properties while maintaining a low condition number even for larger systems, thereby enabling numerically stable calculations for systems with up to several thousand atoms. These basis sets are distributed with CP2K in the \texttt{data/} files \mbox{\texttt{BASIS\_AUG\_MOLOPT}} and \mbox{\texttt{BASIS\_RI\_AUG\_MOLOPT}} of the CP2K GitHub repository.

\subsection{Spatial Descriptors for Electronic Excitations}\label{sec-spatial_descriptors}

% Max Graml~\cite{Graml2025}

% Original work: Dreuw, Plasser, Mewes ~\cite{Plasser2012, Plasser2014a,Plasser2014b, Bappler2014, Mewes2015, Mewes2018}
Beyond excitation energies and optical intensities, CP2K provides scalar descriptors of the electron--hole structure of an excitation. These descriptors replace an isosurface-dependent visual analysis with quantitative measures that can be compared systematically across systems and methods. The underlying electron--hole wavefunction, expectation-value convention, and complete descriptor decomposition are given in Section~IV~D in the SI.\cite{Plasser2014a,Plasser2014b,Bappler2014,Mewes2015,Mewes2018,Plasser2015,Graml2025}
The same descriptors are computed by the libwfa library and reported by the standalone TheoDORE toolbox.\cite{Plasser2020,Plasser2022}
The general-purpose Multiwfn program provides related wavefunction-analysis capabilities.\cite{Lu2024Multiwfn}

The central quantity measuring the spatial extent of the \(n\)th excitation is the size of the excitation
\begin{align}
    d_\text{exc} = \sqrt{\langle | \bre - \brh|^2\rangle_\text{exc}}
    \ensuremath{,} \label{eq-exc_size}
\end{align}
where we do not denote the excitation index explicitly to simplify notation.
In this notation, \(\mathbf{r}_e\) and \(\mathbf{r}_h\) denote the electron and hole coordinates, respectively, and \(\langle\cdot\rangle_{\mathrm{exc}}\) denotes an expectation value over the normalized electron--hole wavefunction.

The excitation size combines the separation between the electron and hole centroids, their individual spatial spreads, and their correlation. Positive electron--hole correlation indicates a bound pair, whereas a negative value describes electron and hole densities that avoid one another dynamically.

For anisotropic systems, CP2K also evaluates direction-resolved descriptors. Along the Cartesian \(x\) direction, the excitation size is
\begin{align}
    d_\text{exc}^{(x)} = \sqrt{\langle | x_e - x_h|^2\rangle_\text{exc}}
    \ensuremath{.} \label{eq-longitudinal_excitation_size}
\end{align}
The remaining descriptors can be resolved analogously along any Cartesian direction.
%
% Beyond that, the directional correlation coefficient is normalized with respect to the electron and hole size in the given direction, e.g.,
% %
% \begin{align}
%     R_{eh}^{(x)}
%     = \frac{1}{\sigma_e^{(x)} \sigma_h^{(x)}} \left( \langle x_h \cdot x_e \rangle_\mathrm{exc}
% - \langle x_h \rangle_\mathrm{exc} \cdot \langle x_e \rangle_\mathrm{exc} \right) \, . \label{eq-correlation_coeff_directional}
% \end{align}
%

\begin{figure}
    \centering
    \includegraphics[width=\linewidth]{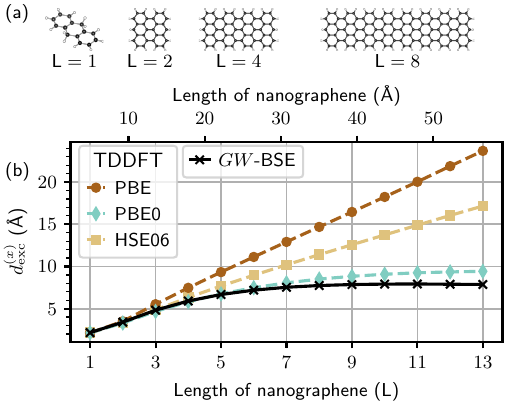}
    \caption{Size of the lowest-energy bright optical excitation in quasi-one-dimensional nanographenes of different length \(\mathsf{L}\) for different excited-state methods.
    (a) Geometries of quasi-one-dimensional nanographenes with increasing length \(\mathsf{L}\), visualized using Ref.~\citenum{Momma2011}.
    (b) Size of the excitation [$d_\text{exc}^{(x)}$, Eq.~\eqref{eq-longitudinal_excitation_size}] along the long nanographene axis for TDDFT (in color, employing the PBE,\cite{perdew1996} PBE0,\cite{Adamo1999} and HSE06,\cite{Heyd2003,Krukau2006} XC functionals) and $GW$ plus the BSE (black curve). $d_\text{exc}^{(x)}$ is strongly dependent on the XC functional in TDDFT, where the hybrid functional PBE0 agrees largely with the $GW$ plus BSE predictions.
    The Tamm--Dancoff approximation is used throughout. Data are taken from Ref.~\citenum{Graml2025}.}
    \label{fig-descriptors}
\end{figure}

%
In Fig.~\ref{fig-descriptors}, we show the size of the lowest-energy bright optical excitation $d_\text{exc}^{(x)}$ in quasi-one-dimensional nanographenes computed from TDDFT and $GW$-BSE (cf. Section~\ref{sec-lr_bse}).
While $GW$-BSE and TDDFT with PBE0 predict a finite size of the excitation for long nanographenes, TDDFT with the PBE and HSE06 functionals predicts roughly a linear growth of $d_\text{exc}^{(x)}$ as the nanographene length increases.
Physically, one expects the size of the excitation to saturate in the solid-state limit due to the Coulomb interaction.
The spurious linear increase of $d_\text{exc}^{(x)}(\mathsf{L})$ is attributed to the absence of long-range electron--hole attraction in PBE and HSE06.
We refer to Ref.~\citenum{Graml2025} for a more detailed discussion of excited-state descriptors and their implementation in CP2K.

\subsection{Real-Time Time-Dependent Density-Functional Theory and Ehrenfest Dynamics}
\label{sec-rt_tddft}

RT-TDDFT is the time-domain counterpart of the linear-response machinery discussed above. Instead of solving for selected excitation vectors, CP2K propagates the occupied KS manifold, or the corresponding density matrix, under an explicitly time-dependent perturbation. This formulation is useful when the perturbation is strong, short, or otherwise not naturally represented by a small set of stationary excited states: a weak impulsive kick gives broadband absorption in the linear regime, whereas shaped fields and pump--probe sequences access nonlinear and nonequilibrium response.

The same propagation framework can be coupled to nuclear motion through Ehrenfest dynamics, where the electrons evolve quantum mechanically and the nuclei move classically on the mean time-dependent electronic force. This mixed quantum--classical approximation is complementary to the trajectory surface-hopping methods discussed in Section~\ref{sec:nonadiabatic-md-surface-hopping}: it preserves a single electronic mean field rather than branching trajectories between adiabatic surfaces. The detailed action formulation, atom-centered-basis Pulay terms, length- and velocity-gauge forms, current response, \(\delta\)-kick spectral extraction, and pump--probe expressions are summarized in Section~IV~E in the SI.

\subsection{Applications of Real-Time Time-Dependent Density-Functional Theory}
\label{sec-rt_tddft_applications}
While LR-TDDFT (Section~\ref{sec-lr_tddft}) provides efficient access to
electronic excitation energies and optical spectra within the frequency domain, it is
inherently limited to the weak-perturbation regime and becomes computationally demanding
for large systems requiring many excited states.  RT-TDDFT offers a
complementary approach in which the KS orbitals are propagated explicitly in time under
the influence of a time-dependent external potential, naturally capturing nonlinear
response, strong-field effects, and the full spectral range within a single
simulation.\cite{Mukamel1995,Yabana1996,Kuehne2020}

\subsubsection{Equations of Motion and Gauge Choice}
\label{sec:rt-tddft-gauge-choice}

For isolated systems, the external field can be coupled in the length gauge through the dipole operator. For periodic systems, where the position operator is incompatible with PBC, CP2K instead uses a velocity-gauge or current-density formulation. A weak $\delta$-kick excites the full linear-response spectrum in one real-time propagation, and the same idea can be applied after a pump pulse to obtain delay-dependent transient absorption. The corresponding equations and spectral-extraction formulas are given in Section~IV~E in the SI.\label{sec:rt-tddft-delta-kick}

\subsubsection{Simulating Pump--Probe Spectroscopy}
\label{sec:rt-tddft-pump-probe}

The explicit time-domain nature of RT-TDDFT makes it uniquely suited to simulate
time-resolved pump--probe experiments.  In such a scheme, a first \emph{pump} pulse
$\mathbf{E}_{\mathrm{pu}}(t)$ is applied to drive the system out of equilibrium. After a
controllable time delay $\tau$, a second \emph{probe} interaction interrogates the
transient electronic structure.

The pump pulse can be chosen to be resonant with a specific transition. For example, one can address a
core-to-valence excitation at an X-ray absorption edge, or deliberately choose a nonresonant pulse,
exciting the system through nonlinear multiphoton processes or generating a nonequilibrium
electron distribution without promoting population into a specific state.  In either case,
the propagation under $\mathbf{E}_{\mathrm{pu}}(t)$ evolves the KS orbitals into a
time-dependent superposition of ground and excited states, which can be characterized
through time-dependent descriptors: the electronic energy $\Delta E(t)$, the induced
dipole $\boldsymbol{\mu}(t)$, the time-dependent electron density $\Delta n(\mathbf{r},t)$,
the spin density, and the projection of the propagated orbitals onto the ground-state
reference manifold to track orbital depopulation, charge transfer, and coherence
decay.

At a selected delay time \(\tau\), a weak \(\delta\)-kick is applied to the nonequilibrium state generated by the pump. Subsequent field-free propagation and Fourier transformation yield the transient absorption difference \(\Delta\sigma(\omega,\tau)=\sigma^{*}(\omega,\tau)-\sigma^{(0)}(\omega)\). The corresponding working equations are given in Section~IV~E in the SI. Negative features represent ground-state bleach or stimulated emission, whereas positive features represent excited-state absorption.\cite{Provorse2016,Repisky2020} Varying \(\tau\) maps charge transfer, exciton dynamics, and core-hole relaxation.

\subsubsection{Illustrative Application: Core Excitation of Aqueous Cysteine}

\begin{figure}[!tbp]
    \centering
    \includegraphics[width=1.0\linewidth]{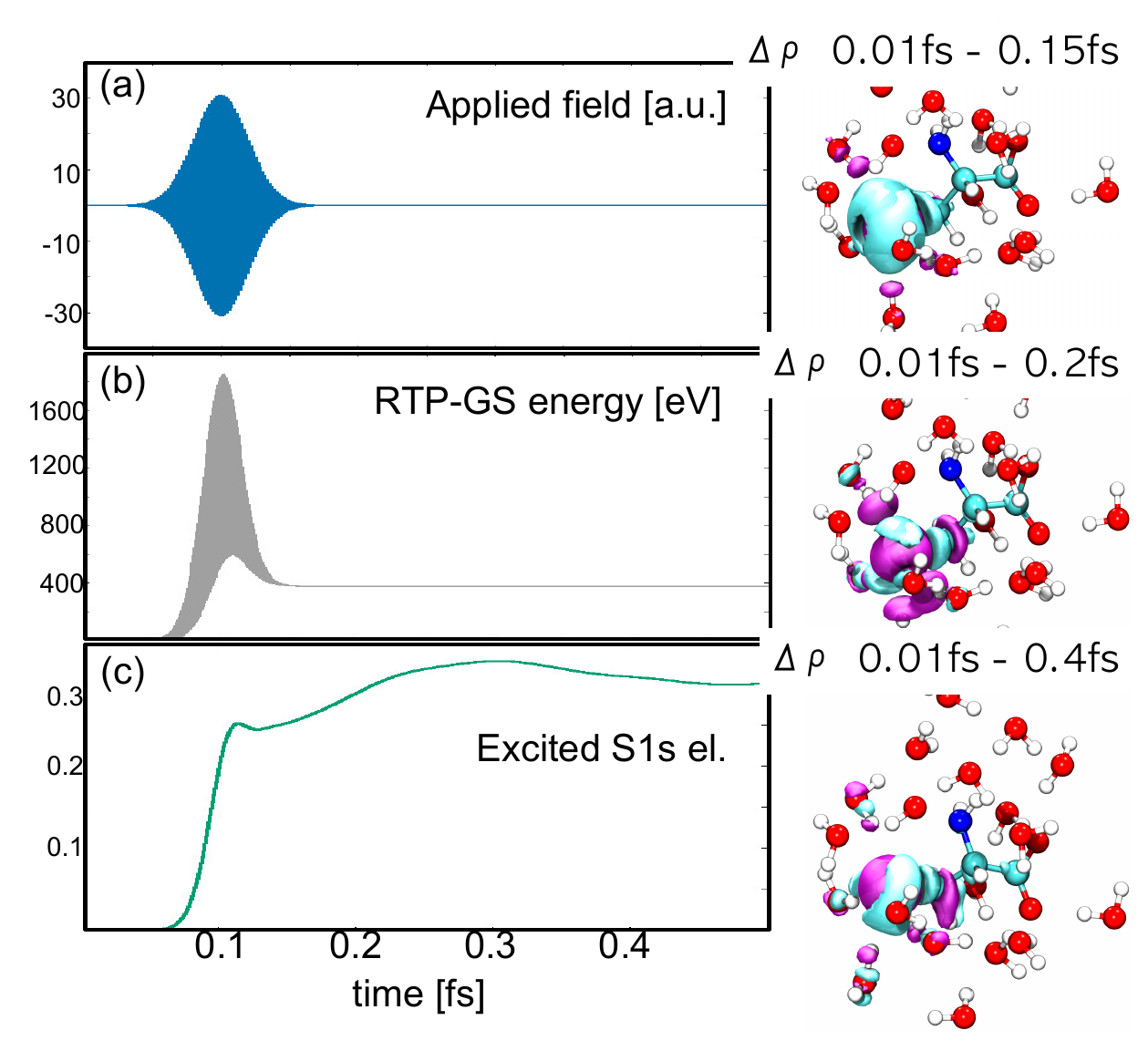}
    \caption{RT-TDDFT simulation of the dianion of cysteine in aqueous solution
    (PBE0, 6-311G** basis set).
    (a) Applied laser pulse with intensity $10^{14}$~W/cm$^2$ and wavelength 0.5073~nm,
    resonant with the S~K-edge.
    (b) Total KS energy change during the pulsed excitation.
    (c) Projection of the time-dependent orbital coefficients onto the ground-state
    reference for all occupied orbitals, quantifying the degree of excitation per orbital.
    Right column: charge-density difference maps at three time steps following the pulse
    onset. Purple indicates charge accumulation, cyan charge depletion.}
    \label{fig:cys_pump}
\end{figure}

As a concrete example, Fig.~\ref{fig:cys_pump} shows an RT-TDDFT simulation of the
dianion of cysteine in aqueous solution at the PBE0/6-311G** level.  A laser pulse resonant
with the S K-edge (wavelength 0.5073~nm, intensity $10^{14}$~W/cm$^2$) promotes the
system into an excited state corresponding to partial depopulation of the S~1s core
orbital.  Panel~(a) shows the time envelope of the applied field. Panel~(b) shows the evolution
of the total KS energy. Panel~(c) shows the projection of the time-dependent orbital
coefficients onto the ground-state reference, which quantifies the degree of excitation.  The charge-density difference maps (right column) illustrate
the spatial redistribution of electron density at successive time steps during and after
the pulse, with purple (cyan) indicating charge accumulation (depletion).  This
nonequilibrium electronic state can subsequently be probed by applying a $\delta$-kick,
as described in Section~\ref{sec:rt-tddft-pump-probe}, to extract the transient X-ray absorption spectrum of the core-excited
intermediate.

% Pump-probe experiments \cite{Siday2024}

% Raman spectroscopy (Dorothea's recent unpublished work)

\section{X-Ray and Core-Level Spectroscopy}
\label{sec:xray-core}

Core-level spectroscopy techniques, such as XAS, X-ray emission spectroscopy (XES), and X-ray photoelectron spectroscopy (XPS), provide element-specific
and site-selective probes of electronic structure, making them indispensable tools for
characterizing the local chemical environment of atoms in molecules, liquids, and solids.
From a theoretical standpoint, their simulation requires an accurate description of
core-electron excitations, including the response of the valence electron density to the
presence of a core hole.  In CP2K, three complementary methods are
available for this purpose: the transition potential (TP) method, LR-TDDFT, and RT-TDDFT, each offering a different
balance between computational cost and physical accuracy.  For all methods, the
all-electron GAPW scheme (Section~\ref{sec:gapw}) is the natural choice, since an accurate
description of the core-region density is essential for reliable core-level energies and
transition moments.

\subsection{The Transition Potential Method}
\label{sec:xray-tp}

Electronic excitations are fundamentally many-electron processes, such that a
single-determinant description is generally insufficient to capture their full complexity.
However, in core-level spectroscopy the created core hole is strongly localized on a
single atomic site, and the problem can often be approximated within a single-particle
framework, in particular for K-edges of light elements.  Building on this simplification,
the one-electron excitation approximation is widely adopted in the theoretical treatment of
XAS, XES, and XPS.\cite{Triguero1998,xray_marcella}

The interaction of the system with the incoming radiation is described by the
electromagnetic vector potential, and the resulting transition probability between
initial state $|i\rangle$ and final state $|f\rangle$ is obtained, to first order
in perturbation theory, from
\begin{subequations}
\begin{align}
    &\mathbf{A}(\mathbf{r},t)
    =
    A_0\,\hat{\mathbf{e}}\cos(\mathbf{k}\cdot\mathbf{r}-\omega t)\;,
    \label{eq:vector_pot_xray}
    \\
    &P_{if}
    =
    \frac{\pi e^2}{2\hbar m^2 c^2}\,A_0^2
    \left|
    \langle f|
    e^{i\mathbf{k}\cdot\mathbf{r}}\,
    \hat{\mathbf{e}}\cdot\mathbf{p}
    |i\rangle
    \right|^2 .
       \label{eq:fermi_golden}
\end{align}
\end{subequations}
The symbols \(A_0\), \(\hat{\mathbf e}\), \(\mathbf k\), and \(\omega\) denote the vector-potential amplitude, polarization unit vector, photon wavevector, and angular frequency. The operator \(\mathbf p\) is the momentum operator, and \(P_{if}\) is the transition probability. The constants \(e\), \(m\), \(c\), and \(\hbar\) denote the elementary charge, electron mass, speed of light, and reduced Planck constant, respectively.
Within the electric-dipole approximation ($e^{i\mathbf{k}\cdot\mathbf{r}} \approx 1$),
Eq.~\eqref{eq:fermi_golden} reduces to the velocity form
$P_{if} \propto |\langle f|\hat{\mathbf{e}}\cdot\mathbf{p}|i\rangle|^2$,
which is equivalent to the length form
$P_{if} \propto (E_f - E_i)^2|\langle f|\hat{\mathbf{e}}\cdot\mathbf{r}|i\rangle|^2$
when the wavefunctions are exact eigenstates of the Hamiltonian.  In the one-electron
excitation approximation, the initial state is identified with a localized core orbital
and the final state with an unoccupied or continuum orbital, so that the many-electron
matrix element reduces to a single-particle overlap evaluated with KS wavefunctions.

Within this framework, Slater's \emph{TP}
method,\cite{slater_self_1972} provides a computationally efficient way to describe core-electron
binding energies, near-edge X-ray absorption fine structure / X-ray absorption near-edge structure spectra, and X-ray emission
spectra using DFT.  The TP approach exploits Slater's transition state theorem, which
treats the total energy as a continuous function of fractional orbital occupation: core
excitation energies are obtained directly from KS eigenvalues computed with a
\emph{transition functional} in which the occupation of the relevant core orbital is set
to a fractional value.  In the original formulation, a \emph{half core-hole} (HCH)
potential is used, corresponding to a core occupation of $1/2$ electron. A \emph{full
core-hole} (FCH) potential (zero core occupation) is also widely employed, and both
variants yield satisfactory agreement with experiment for a broad range of
systems.\cite{Prendergast2006,Iannuzzi2008,Mueller2019XAS}  In the TP picture, the promoted electron is assumed to be
immediately delocalized into the conduction band and its contribution to the spectral
features is considered equivalent regardless of the specific final state. This assumption
means that only a single self-consistent electronic-structure calculation is required to
produce the entire spectrum, rather than one calculation per final state.

In practice, the procedure consists of the following steps: (i) a ground-state KS
calculation is performed. (ii) The target core orbital is identified. In systems with
multiple inequivalent atoms of the same species, localization of the core orbitals
facilitates unambiguous assignment. (iii) The occupation of the selected core orbital is
modified to the desired fractional value and the SCF is restarted with that occupation
fixed. (iv) Excitation energies and transition matrix elements are evaluated from the
converged TP eigenvalues and orbitals.\cite{Iannuzzi2008} The same strategy underlies
the simulation of XPS binding energies, for which the KS eigenvalue of the
half-occupied core orbital is referred to the Fermi energy of the same calculation. The
corresponding TP expressions are
\begin{subequations}
\begin{align}
     \hbar\omega_{if}
    &=
    \varepsilon_f^{\mathrm{TP}}-\varepsilon_i^{\mathrm{TP}},\hspace{1em}
    P_{if}
     =
    \left|
    \langle \psi_f^{\mathrm{TP}}|
    \boldsymbol{\nabla}
    |\psi_i^{\mathrm{TP}}\rangle
    \right|^2 ,
     \label{eq:tp_energies}
     \\[0.5em]
    \mathrm{BE}
    &=
    \varepsilon_{1s}\!\left(\tfrac{1}{2}\right)
    -
    E_F\!\left(\tfrac{1}{2}\right),
    \label{eq:xps_be}
\end{align}
\end{subequations}
respectively. The quantities \(\varepsilon_{i/f}^{\mathrm{TP}}\) and \(\psi_{i/f}^{\mathrm{TP}}\) are the TP orbital energies and orbitals of the initial core and final unoccupied states, \(E_F(1/2)\) is the Fermi energy of the half-core-hole calculation, and \(\mathrm{BE}\) is the core-electron binding energy. The half-occupation ensures a balanced treatment of initial- and final-state
screening effects via Slater's theorem.  An important practical advantage of the TP
approach is that it is applicable with any XC functional, including
hybrid functionals, and is equally robust for metallic, semi-metallic, and zero-gap
systems where methods based on an energy gap (such as LR-TDDFT with the TDA)
may become problematic.

\subsection{Core-Level Spectroscopy from Linear-Response Time-Dependent Density-Functional Theory}
\label{sec:xray-lr-tddft}

An alternative first-principles description of core-level spectra is provided by LR-TDDFT,
which, unlike the TP method, explicitly accounts for electron--hole interactions through
the XC kernel $f_{\mathrm{xc}}$ and therefore captures excitonic
effects that are absent in the independent-particle picture.  The CP2K/\textsc{Quickstep}
implementation of LR-TDDFT for XAS,\cite{Bussy2021} introduces three core-specific
approximations that make the method computationally tractable for large systems.

The first is the \emph{core-valence separation} (CVS) approximation. Because core and
valence states differ greatly in energy and spatial localization, their mutual
coupling is negligible. Valence-origin excitations can therefore be projected out,
reducing the full occupied--virtual response space to the much smaller core--virtual
subspace without loss of accuracy.

The second approximation is the \emph{sudden approximation}, in which the relaxation of
electrons outside the core region upon excitation is neglected.  Combined with the
localized character of core states, this allows excitations to be treated
\emph{independently}, one core state at a time, rather than constructing and diagonalizing
the full response matrix simultaneously. This sequence of small diagonalizations scales
better than one large response calculation, especially with many inequivalent absorbing
sites.

The third approximation is a \emph{core-specific resolution-of-identity} (RI) scheme for
the four-center two-electron repulsion integrals (ERIs).  Since all relevant ERIs involve
at least one core orbital, which is strongly localized on a single atom, the RI
auxiliary basis functions $\{\chi_\mu\}$ can be restricted to those centered on the
excited atom.  For the Coulomb integrals this gives
\begin{equation}
    (pI|Jq)
    \approx
    (pI|\mu)\,(\mu|\nu)^{-1}\,(\nu|Jq),
    \label{eq:ri_coulomb}
\end{equation}
and, for hybrid DFT, the exact-exchange ERIs are approximated analogously as
\begin{equation}
    (pq|IJ)
    \approx
    (pq|\mu)\,(\mu|\nu)^{-1}\,(\nu|IJ).
    \label{eq:ri_exchange}
\end{equation}
The XC-kernel contribution to the response is evaluated as
\begin{equation}
    \begin{aligned}
    (pI|f_{\mathrm{xc}}|Jq)
    &\approx
    (pI|\kappa)\,(\kappa|\lambda)^{-1}\,
    (\lambda|f_{\mathrm{xc}}|\mu)\\
    &\quad\times
    (\mu|\nu)^{-1}\,(\nu|Jq),
    \end{aligned}
    \label{eq:ri_fxc}
\end{equation}
where \(I\) and \(J\) label localized core orbitals, \(p\) and \(q\) label
generic AOs, and Greek
indices label functions in the atom-centered RI auxiliary basis.
The density entering $f_{\mathrm{xc}}[n]$ is projected onto the atom-centered RI
basis via
\begin{subequations}
\label{eq:ri_density_proj}
\begin{align}
    n(\mathbf{r})
    &=
    \sum_{pq} D_{pq}\,\varphi_p(\mathbf{r})\varphi_q(\mathbf{r})
    \approx
    \sum_\nu d_\nu\,\chi_\nu(\mathbf{r}),
    \label{eq:ri_density_proj_density}\\
    d_\nu
    &=
    \sum_{pq}\sum_{\mu} D_{pq}\,(pq|\mu)\,
    \left(\mathbf S^{-1}\right)_{\mu\nu}.
    \label{eq:ri_density_proj_coefficients}
\end{align}
\end{subequations}

where \(D_{pq}\) is an AO density-matrix element, $(pq|\mu)$ denotes the three-center overlap
integrals, and \(\left(\mathbf S^{-1}\right)_{\mu\nu}\) is an element of the inverse overlap matrix of the RI basis functions on the
excited atom.  Together, these three approximations reduce the formal computational cost
of core LR-TDDFT to a level comparable with the TP method while retaining the explicit
treatment of the electron--hole interaction.
For absolute edge positions and core-level binding energies, the same implementation
can be combined with a first-principles GW2X correction that replaces the donor-core
KS eigenvalue by a corrected core-ionization potential, thereby reducing the empirical
energy shifts otherwise required for core-level LR-TDDFT spectra.\cite{Bussy2021Correction}

\begin{figure}[b!]
    \centering
    \includegraphics[width=1.0\linewidth]{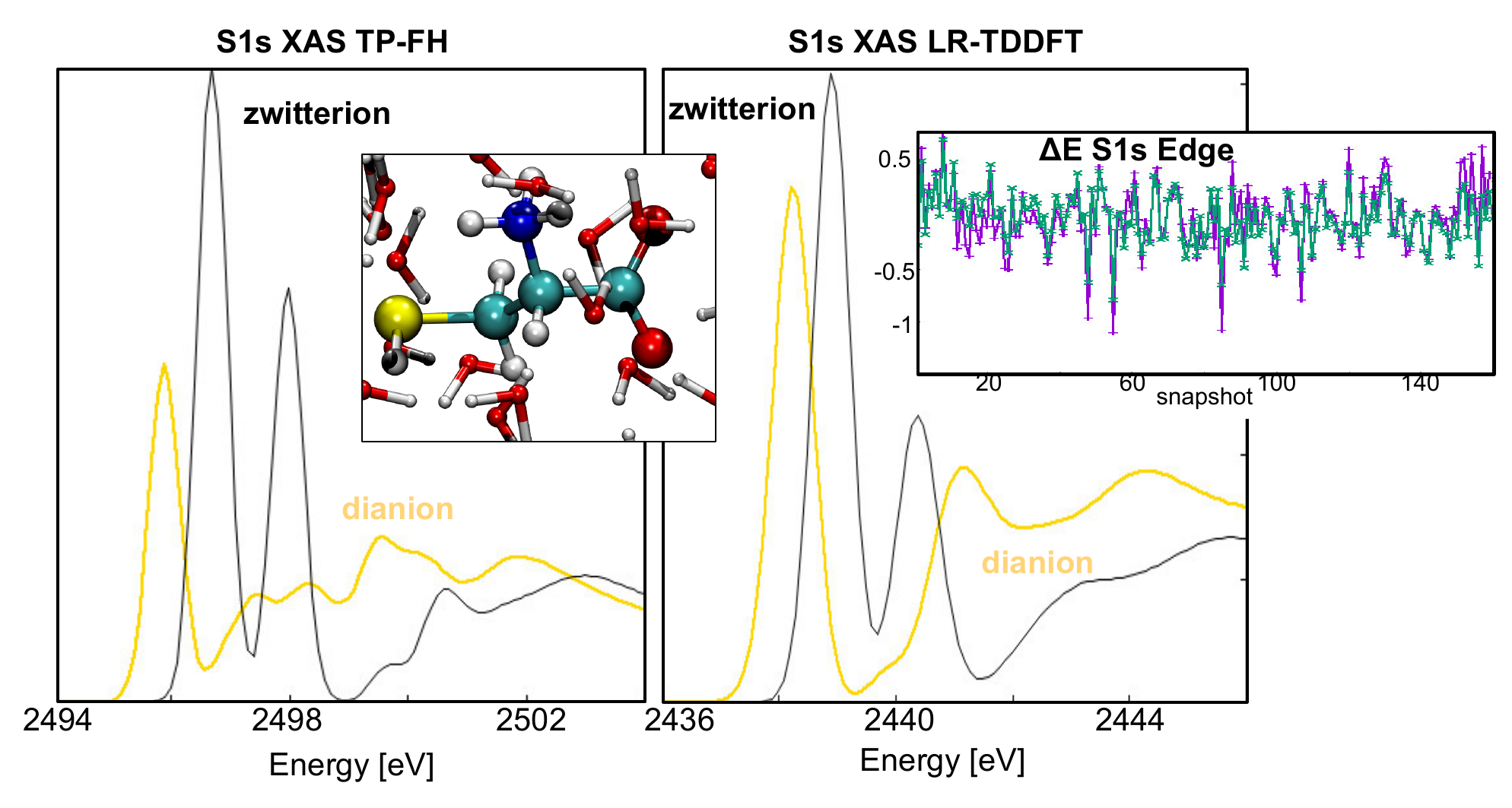}
    \caption{S~1s (K-edge) XAS spectra of aqueous cysteine in the zwitterionic (black)
    and dianionic (yellow) forms, computed with the TP (left panel) and LR-TDDFT (right panel) methods.
    Spectra are averaged over 160 configurations sampled from a 30~ps AIMD trajectory
    at the PBE0/ADMM/PCSEG-2 level within the GAPW all-electron framework.
    The broadened envelope uses a Gaussian linewidth of 0.6~eV.}
    \label{fig:xas}
\end{figure}

\subsection{Illustrative Comparison: S K-Edge Spectra of Aqueous Cysteine}
\label{sec:xray-cysteine}

As an illustrative example, Fig.~\ref{fig:xas} compares the S~1s (K-edge) XAS spectra
of aqueous cysteine in its zwitterionic (black) and dianionic (yellow) forms, computed by
averaging over 160 configurations extracted from a 30~ps \textit{ab-initio} MD trajectory.
Spectra were obtained with both the TP and LR-TDDFT methods, using the
PBE0 hybrid functional in conjunction with
the auxiliary density matrix method (ADMM),\cite{admm} and the PCSEG-2 basis set within the
GAPW all-electron framework.  The spectral differences between the two protonation states
reflect the change in the local electronic structure around the sulfur atom, demonstrating
the sensitivity of S K-edge XAS to the chemical environment and the reliability of
CP2K in resolving these differences by finite-temperature averaging.

\subsection{\texorpdfstring{Resonant Inelastic X-Ray Scattering}{Resonant Inelastic X-Ray Scattering}}\label{sec:rixs}

\begin{figure}[!tbp]
    \centering
    \includegraphics[width=1.0\linewidth]{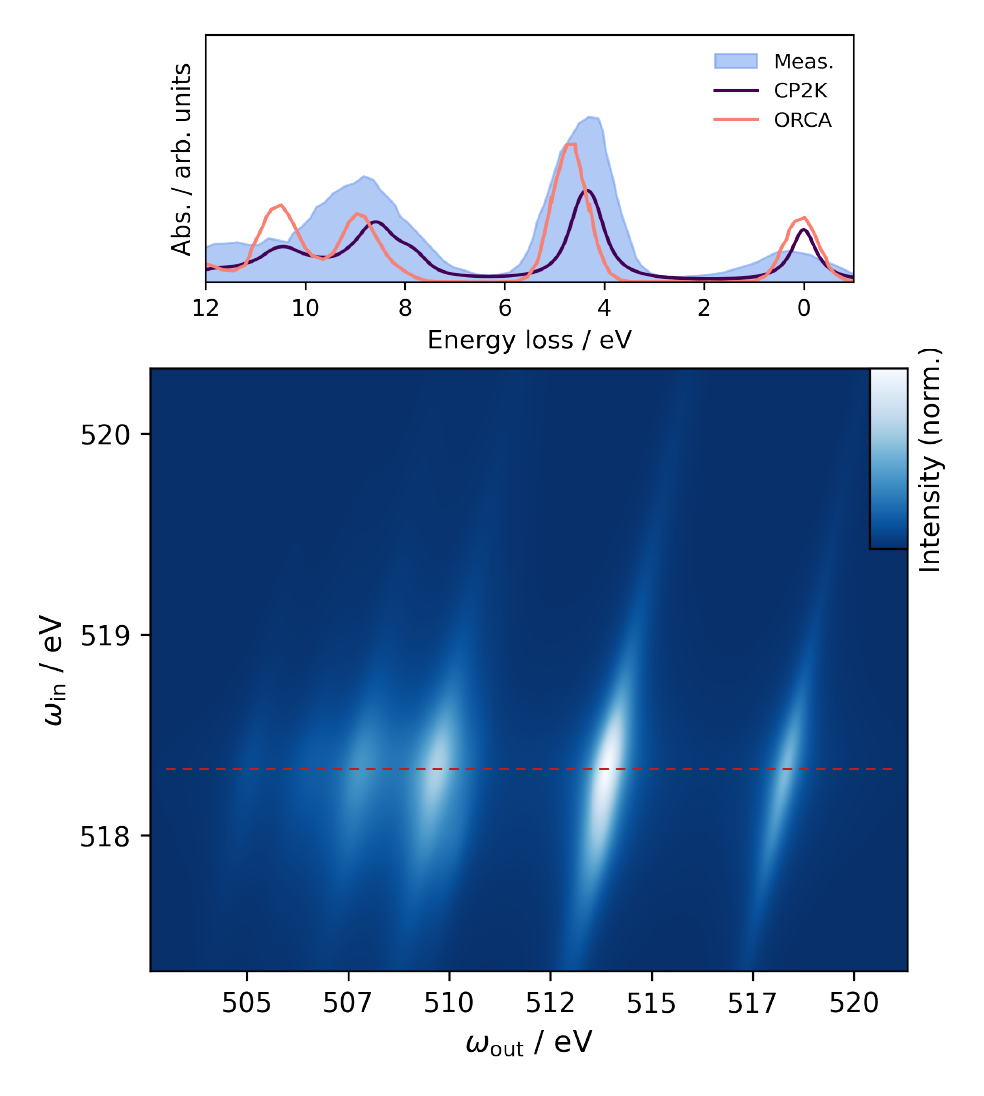}
    \caption{RIXS from the O~1s (K-edge) for one configuration of a liquid-acetone cell. Top: Energy-loss spectrum from the first bright XAS state. CP2K calculations are compared with ORCA results and experimental signals. Bottom: Two-dimensional RIXS map.}
    \label{fig:rixs}
\end{figure}

Resonant inelastic X-ray scattering (RIXS) has emerged as a powerful spectroscopic
probe of electronic structure spanning an exceptionally wide range of energy scales.
The theoretical framework underlying RIXS simulations is the Kramers--Heisenberg (KH) equation.\cite{Ament2011RIXS}
The cross section is expressed as a sum
over intermediate core-excited states $|I\rangle$, weighted by transition dipole moments connecting the ground state $|0\rangle$ to the core-excited manifold and the core-excited manifold to the final valence-excited states $|F\rangle$
\begin{equation}
    F_F = \sum_I \omega_{IF} \omega_{I0}
\frac{\langle 0 | \mathbf{e} \cdot \mathbf{r} | I \rangle
      \langle I | \mathbf{e}' \cdot \mathbf{r} | F \rangle}
     {\omega_{\textrm{in}} - \omega_{I0} + i \Gamma} \,
     \ensuremath{,} \label{eq:kramers}
\end{equation}
where \(F_F\) is the scattering amplitude into final state \(F\), \(\omega_{I0}\) and \(\omega_{IF}\) are transition frequencies, \(\omega_{\mathrm{in}}\) is the incident-photon frequency, \(\mathbf e\) and \(\mathbf e'\) are the incoming and outgoing polarization vectors, and \(\Gamma\) is the core-hole lifetime broadening.
CP2K implements
RIXS for fully periodic systems.\cite{SertcanGoekmen2026RIXS}
The KH cross section is evaluated using transition dipole moments from LR-TDDFT combining
the existing CVS-based X-ray absorption
module,\cite{Bussy2021} with the established valence LR-TDDFT
implementation.\cite{Hehn2022}
The localized character of the core hole
ensures that the dipole matrix elements remain well-defined under periodic boundary
conditions, and the reuse of established infrastructure gives the RIXS
implementation direct access to hybrid functionals, the ADMM, and norm-conserving PPs on non-donor atoms, making hybrid-functional calculations affordable for the system sizes encountered in
liquid and crystalline simulations.
The RIXS module first generates the core-excited intermediate states from an XAS calculation for a selected excited atom and, in a second step, performs a TDA-TDDFT simulation to obtain the final valence-excited states.
For K-edge RIXS we compute the transition dipole moments as
\begin{subequations}
\begin{align}
\mu_{0I}^\alpha& = \sum_{\mu\nu} \tilde{Y}^I_{1s,\nu}\, C_{\mu,\text{1s}}\, R_{\mu\nu}^\alpha\,,
\\
\mu_{IF}^{\alpha}
&=
\sum_k
\sum_{\mu\nu} \tilde{Y}^I_{1s,\mu}\,S_{\mu\nu}\,\tilde{X}^F_{\nu k}
\sum_{\lambda\sigma}
C_{\lambda 1s}
R^\alpha_{\lambda \sigma}
C_{\sigma k}\ensuremath{,}
\end{align}
\end{subequations}
where $\tilde{\mathbf{Y}}$ and $\tilde{\mathbf{X}}$ are covariant AO representations of the transition amplitude vector of
the core-excited intermediate state and of
the excitation amplitude matrix of the valence-
excited final states, respectively.  These are built
by contracting with the matrix of MO coefficients $\mathbf{C}$. The indices \(\mu,\nu,\lambda,\sigma\) label AOs, \(k\) labels an occupied valence MO, and \(\alpha\) is a Cartesian component. The AO overlap-matrix element \(S_{\mu\nu}\) restores orthonormality by reconstructing the
inner product of the MO-space, and $R_{\mu\nu}^\alpha = \braket{\mu | \hat{r}_\alpha | \nu}$ are the dipole integrals
in the AO basis for the Cartesian component $\alpha$. The orientation-averaged RIXS cross section is calculated with a separate post-processing tool distributed with CP2K. A representative liquid-acetone spectrum and two-dimensional RIXS map are shown in Fig.~\ref{fig:rixs}.

\begin{figure*}[t!]
    \centering
    \includegraphics[width=1.0\linewidth]{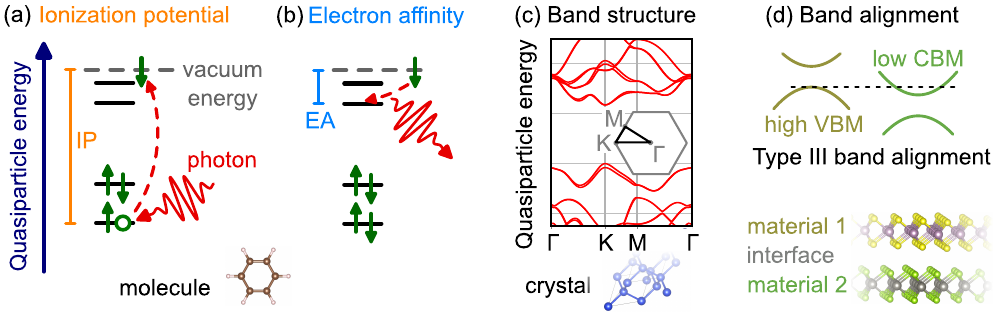}
\caption{Overview of the observables obtained from the $GW$ method.
(a) Ionization potentials correspond to the energy required to remove an electron to the vacuum level.
(b) Electron affinities correspond to the energy gained when adding an electron from the vacuum to the system.
Both quantities arise from the quasiparticle poles of the one-particle Green's function.
(c) In periodic solids, the resulting quasiparticle energies define the electronic band structure.
(d) Aligning quasiparticle levels of two materials enables the prediction of interface band offsets, including type-III (broken-gap) alignment.}
    \label{figGW}
\end{figure*}

Before the native RIXS module, CP2K-based core-excited potentials and dynamical sampling connected XAS and RIXS signatures of water and aqueous ammonia to hydrogen bonding and nuclear motion.\cite{Weinhardt2015AqueousAmmoniaRIXS,VazDaCruz2019WaterRIXS} RT-TDDFT core-hole-clock spectroscopy additionally resolved ultrafast electron delocalization in thiophene-based polymers.\cite{core_hole_clock_spectroscopy_Marcella}

\section{\texorpdfstring{\textit{GW} Methods for Quasiparticle Energies and Electronic Structure}{GW Methods for Quasiparticle Energies and Electronic Structure}}
\label{sec:gw}
%Rémi Pasquier, Leonard Prokisch, Jan Wilhelm \\\\

The $GW$ method is a Green's-function-based approach in which the electron self-energy $\Sigma \eqt iGW$ replaces the DFT exchange-correlation potential and yields single-electron energies~$\varepsilon_n$ that can be interpreted as ionization potentials, electron affinities, and band structures in solids, as discussed in Section~\ref{8A}.
The following subsections summarize the CP2K algorithms used to compute these observables. CP2K employs the space--time formulation of $GW$ (Section~\ref{subsec:GWspacetime}). Sections~\ref{subsec:GWmolecules}, \ref{GWfullkp}, and~\ref{GammaonlyGW} describe low-scaling $GW$ for molecules, for crystals with full $k$-point sampling, and for large supercells in a $\Gamma$-point-only framework. Section~\ref{cohsex} briefly introduces the Coulomb-hole and screened-exchange (COHSEX) approximation to $GW$, which is also implemented in CP2K.

\subsection{\texorpdfstring{Observables Computed with \textit{GW}: Ionization Potentials, Electron Affinities, and Band Structures}{Observables Computed with GW: Ionization Potentials, Electron Affinities, and Band Structures}}\label{8A}

In many areas of materials science one is interested in the energies associated with adding or removing a single electron from a molecule, liquid, solid, or from an interface between different materials.
These quantities are the ionization potentials (IPs) and electron affinities (EAs), which play a central role in many areas of the natural sciences, e.g., in photoelectron spectroscopy and electron transport across interfaces.\cite{Bechstedt2015,martin_reining_ceperley_2016,golze2019}
IPs and EAs determine how easily an electron can be removed from, or added to, a system and therefore govern photochemical reactivity, redox processes, and alignment of molecular levels with electrodes or semiconductors (Fig.~\ref{figGW}).
IPs and EAs are defined in terms of total energy differences between systems that differ by one electron.
For an $N$-electron system with ground-state energy $E_N^0$, the corresponding IPs and EAs are given by the following expressions.\cite{golze2019}
\begin{subequations}
\label{eb7}
\begin{align}
    \varepsilon_n
    &= E_N^0-E_{N-1}^n,
    && \varepsilon_n\leq\varepsilon_{\mathrm F}
    \quad\text{(IPs)},
    \label{eb7_ip}\\
    \varepsilon_n
    &= E_{N+1}^n-E_N^0,
    && \varepsilon_n>\varepsilon_{\mathrm F}
    \quad\text{(EAs)}.
    \label{eb7_ea}
\end{align}
\end{subequations}

The energy $E_{N\pm1}^n$ is that of the $N{\pm}1$ electron system after removing or adding an electron, and the final state can be the $n$th excited state. The quantity $\varepsilon_\text{F}$ is the Fermi energy.
Fig.~\ref{figGW}\,(a) and~(b) illustrate these definitions for a molecule: an IP corresponds to a vertical transition from the neutral ground state to a cation with one electron removed, while an EA corresponds to adding an electron and forming an anion.
In extended systems, the same concepts apply and, in addition, Bloch's theorem holds, so that IPs and EAs become functions of the crystal momentum~$\bk$, written as $\varepsilon_{n\bk}$ [Fig.~\ref{figGW}(c)].

KS DFT,\cite{kohn1965,martin2004} provides the exact electron density and ground-state energy of an interacting system, given that the exact XC functional is used in the calculation.
The orbital eigenvalues~$\eigvalksdft$, however, are auxiliary quantities and do not correspond to the true removal and addition energies defined above, even when the exact XC functional is used.
For the exact functional, the KS eigenvalue of the highest occupied molecular orbital (HOMO) equals minus the vertical IP, $\varepsilon_\text{HOMO}^\text{KS}\eqt{-}\mathrm{IP}$. Janak's theorem, $\partial E/\partial f_i=\varepsilon_i$, provides the related fractional-occupation result but is not identical to the ionization-potential theorem. No analogous exact relation holds for deeper occupied states or for unoccupied states.\cite{Janak1978,martin2004}
In practice, when using LDA or GGA XC functionals, KS band gaps $\varepsilon_\text{LUMO}^\text{KS}\mt\varepsilon_\text{HOMO}^\text{KS}$ systematically underestimate the fundamental gap $\mathrm{IP}\mt\mathrm{EA}$.\cite{martin2004}

A physically meaningful description of electronic levels can be obtained from the one-particle Green's function $G(\br,\br';\omega)$ of many-body perturbation theory,\cite{hedin1965new,Bechstedt2015,martin_reining_ceperley_2016,reining2018,golze2019} which goes beyond ground-state KS DFT.
The density of states~$\rho(\omega)$ of an interacting many-electron system follows from the Green's function.\cite{golze2019}
\begin{align}
  \rho(\omega) = -\frac{1}{\pi}\,\mathrm{Im}\;\mathrm{Tr}\;G(\omega)\,,\label{eb8}
\end{align}
where $\mathrm{Tr}\,G(\omega)\eqt\int d\br\,G(\br,\br;\omega)$.
Peaks of $\rho(\omega)$ occur at the electron removal and addition energies $\varepsilon_n$ defined in Eq.~\eqref{eb7}.
These peak positions are called quasiparticle energies and correspond directly to the IPs and EAs measured in photoemission and inverse-photoemission spectroscopy. In the molecular illustrations of Fig.~\ref{figGW}\,(a) and~(b), each photon-induced transition corresponds to such a quasiparticle excitation.

The central challenge is to obtain the one-particle Green's function $G(\br,\br';\omega)$ of the interacting many-electron system in order to evaluate $\rho(\omega)$ and its peak positions from Eq.~\eqref{eb8}.
In many applications of many-body perturbation theory, $\rho(\omega)$ and $G(\br,\br';\omega)$ are not computed explicitly.
Instead, one introduces the self-energy $\Sigma(\br,\br';\omega)$. Formally, the self-energy plays a role analogous to $v^\text{xc}$, but at the level of a Green's function rather than in an auxiliary single-particle Hamiltonian.
In practice, the self-energy is often approximated by the $G_0W_0$ form.\cite{reining2018,golze2019}
\begin{align}
   \Sigma (\br,\br';\omega)
   ={}& \frac{i}{2\pi}\int d\omega'\,e^{i\omega'\eta}
   G_0(\br,\br';\omega{-}\omega')
   W_0 (\br,\br';\omega'),
   \label{Sigmaconvw}
\end{align}
where $G_0$ is a reference Green's function of noninteracting electrons and $W_0$ is the screened Coulomb interaction of this reference system. The variable \(\omega'\) is the integration frequency, and \(\eta\) is a positive infinitesimal convergence factor.
In the standard $G_0W_0$ scheme, $G_0$ is built from KS orbitals and KS eigenvalues as
\begin{align}
    G_0(\br,\br';\omega)
    =
    \sum_n
    \frac{\psi_n(\br)\,\psi_n^*(\br')}{\omega-\varepsilon_n^\text{KS}
    -i\eta\,\text{sgn}(\varepsilon_\text{F}-\varepsilon_n^\text{KS})}\,,\label{G0}
\end{align}
where \(n\) labels KS states, while \(\psi_n\), \(\varepsilon_n^{\mathrm{KS}}\), and \(\varepsilon_{\mathrm F}\) are the KS orbital, its eigenvalue, and the Fermi energy, respectively.
The screened interaction \(W_0\) is obtained by screening the bare Coulomb interaction $v(\br,\br')\eqt 1/|\br\mt\br'|$ with the inverse dielectric function
\begin{align}
    W_0(\br,\br';\omega)
    =
    \int d\br''\,\epsilon^{-1}(\br,\br'';\omega)\,
    v(\br'',\br')\,.
\end{align}
The dielectric function $\epsilon(\omega)$ is computed from the KS electronic structure, typically within RPA, following standard many-body perturbation theory expressions.\cite{golze2019}
The post-SCF RPA correlation energies and \(\sigma\)-functionals discussed in Section~\ref{sec:TotalEnergy} are distinct from the RPA screening used here to construct \(W_0\).

Within the $G_0W_0$ approximation, one further assumes that the quasiparticle wavefunctions (often called Dyson orbitals in quantum chemistry,\cite{Oana2007,*Mitric2013,*Krylov2020,*Ortiz2020}) are well approximated by the KS orbitals $\psi_n(\br)$.
One then evaluates only the diagonal matrix elements of the self-energy and solves the resulting quasiparticle equation.\cite{golze2019}
\begin{align}
    \varepsilon^{G_0W_0}_n
    =
    \varepsilon_n^\text{KS}
    + \text{Re}\,\Sigma_n(\varepsilon^{G_0W_0}_n)
    - v_n^\text{xc}\,,
    \label{eb9}
\end{align}
where $\Sigma_n(\omega)\eqt\langle \psi_n|\Sigma(\omega)|\psi_n\rangle$ and
$v_n^\text{xc}\eqt \langle \psi_n|v^\text{xc}|\psi_n\rangle$ are the diagonal matrix elements of the self-energy and the KS XC potential, respectively.
Eq.~\eqref{eb9} can be viewed as replacing the erroneous XC contribution contained in the KS eigenvalue $\varepsilon_n^\text{KS}$ by the self-energy $\Sigma(\omega)$, thereby turning the auxiliary KS energy into a physically meaningful quasiparticle energy~$\varepsilon^{G_0W_0}_n$.
These $\varepsilon^{G_0W_0}_n$ values computed from Eq.~\eqref{eb9} are then the peak positions of $\rho(\omega)$ in Eq.~\eqref{eb8} in the $G_0W_0$ approximation.

One often uses the $G_0W_0$ eigenvalues $\varepsilon^{G_0W_0}_n$ to recompute $G_0$ and the self-energy (without updating $W_0$), which gives the sc$GW_0$ scheme and often improves the agreement with experimental IPs and EAs.\cite{klimes2014,golze2019}
Moreover, the XC functional used in the DFT SCF for obtaining $\psi_n$ and $\eigvalksdft$ influences quasiparticle energies. An often adopted procedure is sc$GW_0$@PBE, that is, sc$GW_0$ using PBE KS orbitals.

 For periodic systems, the resulting quasiparticle energies ${\varepsilon_{n\bk}}$ define the band structure shown in Fig.~\ref{figGW}\,(c).
 Because Bloch's theorem applies,\cite{bloch1929,martin2004} each quasiparticle state acquires a crystal momentum $\bk$ as a quantum number, and $G_0W_0$ yields the dispersion $\varepsilon_{n\bk}$ across the Brillouin zone.
 Compared to KS DFT, $G_0W_0$ and sc$GW_0$ typically open the band gap in a way that agrees with photoemission experiments. For sc$GW_0$, the agreement of band gaps is often within 0.1\,--\,0.2~eV compared to measurements.\cite{klimes2014,golze2019}
 %

% Band alignment between different materials can also be obtained from quasiparticle energies, as illustrated in Fig.~\ref{figGW}\,(d).
% %
% Here, screening effects from the surrounding environment play a crucial role: the dielectric function~$\epsilon$ entering $W_0$ determines how strongly electronic levels are renormalized when  two semiconductors form a heterojunction.
% %
% The most extreme and illustrative case is the case of a molecule  adsorbed on a metal surface where  $\epsilon\eqt\infty$ inside the metal reduces the HOMO-LUMO gap of the molecule by serveral eVs~\cite{Neaton2006,golze2019}; this effect is captured by $GW$ and is absent in DFT calculations with GGAs and hybrid functionals.
% %
% Level shifts and gap renormalization at interfaces are a strong and generic feature, essential for describing charge separation in photovoltaic and photocatalytic systems, reactions in electrochemical environments, and charge transport across molecular junctions and solid-state devices.
% %
% $GW$ captures these level shifts and gap renormalizationa and therefore provides a consistent microscopic description of how the electronic structure responds to dielectric screening from its environment and yields reliable band offsets, interface types, and charge-transfer driving forces in complex materials.
% %
% These screening effects are entirely absent from the KS eigenvalues.
% %

Band alignment between different materials can also be obtained from quasiparticle energies, as illustrated in Fig.~\ref{figGW}(d).
Screening by the surrounding environment plays a key role: the dielectric function $\epsilon(\omega)$ entering $W_0$ controls how strongly electron addition and removal energies are renormalized when two materials interact, for example when semiconductors form a heterojunction or when a molecule adsorbs on a surface.
A paradigmatic case is a molecule on a metal surface, where the very large dielectric response of the metal effectively corresponds to $\epsilon(\br,\br') \,{\to}\, \infty$ for $\br,\br'$ inside the metal,  effectively generating an image charge that can reduce the molecular HOMO--LUMO gap by several electronvolts.\cite{Inkson_1973,Neaton2006,golze2019} This gap renormalization is described by $GW$ but is missing in standard DFT with GGA or hybrid functionals.
Such level shifts and gap renormalization at interfaces are generic and are essential for describing charge separation in photovoltaics and photocatalysis, electrochemical reactions at electrodes, and charge transport across molecular junctions and solid-state or spintronic devices.
$GW$ therefore provides a consistent microscopic description of how the electronic structure responds to dielectric screening from its environment and yields reliable band offsets, interfacial band alignment, and charge-transfer driving forces in complex materials.
These environment-induced screening effects are entirely absent in KS eigenvalues.
%

% %
% In summary, KS eigenvalues are not, in general, electron addition or removal energies, and therefore cannot be used to predict fundamental gaps or band alignment.
% %
% The $GW$ method, formulated in terms of the Green’s function and its self-energy, yields quasiparticle energies whose peaks in the density of states correspond to the ionization potentials and electron affinities illustrated in Fig.~\ref{figGW}. This makes $GW$ the natural framework in CP2K for computing IPs, EAs, band structures, and environment-dependent band alignment in a way that can be compared directly to spectroscopic measurements and interface experiments.

\subsection{\texorpdfstring{\textit{GW} Space--Time Formalism}{GW Space-Time Formalism}}\label{subsec:GWspacetime}

For designing efficient $GW$ algorithms, it is crucial to treat the frequency dependence of the self-energy~$\Sigma$, the Green's function~$G$, the dielectric function~$\epsilon$, and the screened interaction~$W$ in a numerically stable and computationally efficient manner.
A direct numerical evaluation of the frequency convolution in Eq.~\eqref{Sigmaconvw} on the real axis is numerically challenging, because the noninteracting Green's function $G_0$ has poles at the KS eigenvalues, see Eq.~\eqref{G0}, which requires very dense real-frequency grids to achieve convergence.
To avoid these difficulties, many $GW$ implementations reformulate the frequency dependence on the imaginary frequency axis, where $\Sigma$, $G$, $\chi$, and $W$ become smooth functions without singularities.
This imaginary-frequency-only formulation already removes the need for real-axis integrations and leads to the standard quartic-scaling ($N^4$) $GW$ algorithms that are widely used in molecular electronic-structure codes.\cite{Ren_2012,BRUNEVAL2016,Holzer2019}
In CP2K, such an imaginary-frequency $GW$ implementation for molecules has been available for many years and was described in Refs.~\citenum{Wilhelm2016,Kuehne2020}.
To treat significantly larger systems and reduce the computational scaling, many modern $GW$ implementations go one step further and reformulate $GW$ in both imaginary time and imaginary frequency.
This combined imaginary-time and imaginary-frequency representation underlies the so-called $GW$ space--time method originally introduced in Ref.~\citenum{rojas_1995} and enables low-scaling $GW$ algorithms, typically with cubic ($N^3$) scaling for nonmetallic systems.
Recent efficient $GW$ implementations based on this space--time approach include Refs.~\citenum{liu2016,wilhelm2018,foerster2020,duchemin2021,wilhelm2021,Graml2024,Pasquier2025,Zhang2026}.
The central practical point is that the space--time formulation moves the numerically delicate real-frequency convolution to smooth imaginary-time and imaginary-frequency representations.
In imaginary time the irreducible polarizability is obtained as a local product of two Green's functions, rather than as an explicit sum over all occupied--empty pairs.
The defining relation is
\begin{equation}
\chi(\br,\br';i\tau)
=
-i\,G_0(\br,\br';i\tau)
G_0(\br,\br';-i\tau),
\label{eq:gw-spacetime-polarizability}
\end{equation}
where \(\chi\) is the irreducible polarizability and \(\tau\) is imaginary time.
This form exposes both the locality of the response and the absence of an explicit occupied--empty double sum.
This enables the transition from conventional quartic-scaling algorithms to reduced-scaling algorithms for nonmetallic systems.
The reduced scaling is only useful when the representation preserves locality.
In a PW representation, the same real-space product becomes a convolution over reciprocal lattice vectors and the practical scaling advantage is lost.
The CP2K implementations therefore combine the space--time idea with localized Gaussian and auxiliary basis functions.
The explicit Green's-function, polarizability, dielectric, screened-interaction, self-energy, analytic-continuation, and minimax-grid working equations are given in Section~V~A in the SI.

% \subsection{Imaginary-time-GW implementation for molecules}

% expansion of KS orbitals in Gaussians, so no space coordinates any more, therefore usually referred to as "imaginary time $GW$" instead of space-time $GW$, original algorithm described in Refs.~\cite{wilhelm2018}, improvement concerning numerical precision via RI with truncated Coulomb metric and larger imaginary time-frequency grids in Ref.~\citenum{wilhelm2021}, and the memory-saving scheme via recomputing three-center integrals in Ref.~\citenum{Pasquier2026}.

% KS orbitals expanded in Gaussian orbitals
% \begin{align}
% \psi_{n}(\br)
%   = \sum_\mu C_{\mu n}
%   \,\phi_\mu(\br)\,,
% \label{kpointsbf}
% \end{align}

\subsection{\texorpdfstring{Imaginary-Time \textit{GW} Implementation for Molecules in a Gaussian Basis}{Imaginary-Time GW Implementation for Molecules in a Gaussian Basis}}
\label{subsec:GWmolecules}

The $GW$ space--time formulation introduced in Section~\ref{subsec:GWspacetime} exploits the locality of the irreducible density response $\chi(\br,\br';i\tau)$ in nonmetallic systems to achieve cubic scaling.
In its original form, this formalism is expressed on real-space grids and relies on FFTs between real-space ($\br,\br'$) and reciprocal space ($\bG,\bG'$).
For atomistic simulations, however, a real-space grid representation is usually prohibitively expensive: achieving sufficient accuracy requires very fine grids near the nuclei and in regions of rapid orbital variation, which leads to a large number of grid points and a substantial computational prefactor.
Recent developments with separable density fitting when using an atom-centered basis enable the creation of compact real-space grids and thus solve this problem,\cite{duchemin2021,delesma2024} but this method still requires an AO basis set.
To retain the favorable scaling while avoiding real-space grids, we reformulate the imaginary-time $GW$ algorithm entirely in a localized Gaussian basis.\cite{wilhelm2018,wilhelm2021,Pasquier2026}

The Gaussian-basis implementation keeps the same space--time logic but changes the representation.
KS orbitals are expanded in atom-centered Gaussian AOs, the density response is assembled in an auxiliary RI basis, and a truncated Coulomb metric is used to make the three-center integrals short-ranged.
This metric is the practical compromise that preserves much of the accuracy of the Coulomb metric while restoring the locality needed for low-scaling calculations.
Because both the AO basis and the auxiliary RI basis are localized, the number of relevant three-center integrals grows only linearly with system size, and the dominant response construction scales quadratically in practice for the molecular systems considered in Ref.~\citenum{wilhelm2021}. The tensor contractions and their scaling analysis are shown in Section~V~B in the SI.
The explicit AO/RI working equations are listed in Section~V~B in the SI.

After the response has been constructed, CP2K follows the same conceptual sequence as in the space--time formulation: transform the response to imaginary frequency, build the dielectric matrix and screened interaction in the auxiliary basis, transform the screened interaction back to imaginary time, evaluate the self-energy, analytically continue it to the real axis, and solve the quasiparticle equation of Section~\ref{8A}.

This low-scaling, imaginary-time $GW$ algorithm for molecules has been described in
Refs.~\citenum{wilhelm2018,wilhelm2021}, and recent developments have significantly reduced its
memory footprint, enabling applications to nanographene molecules containing more than
9000 atoms.\cite{Pasquier2026}
However, the main area of application of the $GW$ method is crystalline materials under
PBC, in particular for computing quasiparticle band structures and
level alignment at interfaces.
The first periodic CP2K $GW$ implementation established the mixed Gaussian/PW framework and its finite-size correction.\cite{wilhelm2017}
Accordingly, the focus of this review is on $GW$ implementations in CP2K  for periodic systems,
which we have only recently realized with high numerical precision,\cite{Graml2024,Pasquier2025} and which we describe in the following.

\subsection{\texorpdfstring{Space--Time \textit{GW} Implementation for Crystals with Full \textit{k}-Point Sampling}{Space-Time GW Implementation for Crystals with Full k-Point Sampling}}
\label{GWfullkp}

We have also adapted the $GW$ space--time method from Sections~\ref{subsec:GWspacetime} and~\ref{subsec:GWmolecules} to periodic crystals with small unit cells and full \textit{k}-point sampling in a Gaussian basis.\cite{Pasquier2025}
In the full-\(\bk\) implementation, the same localized-basis space--time strategy is applied to periodic Bloch states.
The algorithm evaluates the Green's function on the chosen \(\bk\)-point mesh, transforms it to lattice-vector space to exploit the real-space decay of gapped systems, constructs the response and self-energy with truncated-Coulomb three-center integrals, and finally transforms the diagonal self-energy elements back to the Bloch basis.
The practical message is that Gaussian basis functions retain their compactness for two-dimensional crystals, where large PW bases are otherwise needed to describe vacuum.
The full lattice-sum working equations are provided in Section~V~C in the SI.

\begin{figure}[!tbp]
    \centering
\includegraphics[width=0.9\columnwidth]{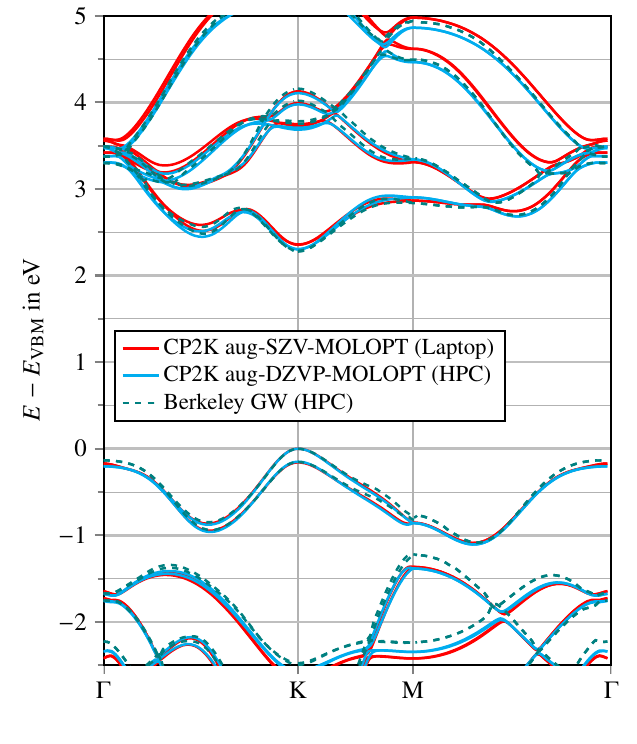}\vspace{-2em}
    \caption{$G_0W_0$@PBE band structure of monolayer MoS$_2$ computed from the algorithm presented in Section~\ref{GWfullkp} and from BerkeleyGW.
    Data from Ref.~\citenum{Pasquier2025}.
    }
    \label{fGWbandstr}
\end{figure}

We have applied this $GW$ algorithm with full \textit{k}-point sampling to monolayer semiconducting transition-metal dichalcogenides (TMDs) MoS$_2$, MoSe$_2$, WS$_2$, and WSe$_2$.
Fig.~\ref{fGWbandstr} compares the $G_0W_0$@PBE band structure of monolayer MoS$_2$ obtained with this algorithm to reference PW calculations from BerkeleyGW.
For monolayer MoS$_2$, MoSe$_2$, WS$_2$, and WSe$_2$, $GW$ band gaps agree on average within $50$\,meV with the PW benchmarks, demonstrating that Gaussian basis sets combined with the space--time formalism can reach the same level of accuracy as high-cutoff, full-frequency PW calculations for two-dimensional semiconductors.\cite{Pasquier2025}
At the same time, the compact Gaussian representation avoids the large PW basis needed to describe vacuum and therefore reduces both memory and computational cost for atomically thin materials:
The corresponding hardware parameters and strong-scaling benchmark are given in Section~V~C in the SI.

\subsection{\texorpdfstring{Space--Time \textit{GW} Implementation for Large Supercells in a $\Gamma$-Point-Only Approach}{Space-Time GW Implementation for Large Supercells in a Gamma-Point-Only Approach}}\label{GammaonlyGW}

An alternative strategy for periodic $GW$ calculations in CP2K is based on a $\Gamma$-point-only treatment for very large supercells.\cite{Graml2024}
The central idea is to compute the irreducible polarizability $\chi$ and the self-energy $\Sigma$ only at the $\Gamma$ point, which reduces the computational cost considerably.
At the same time, the polarizability $\chi_{PQ}(\bk,i\tau)$ is needed on a dense $k$-point mesh because it is multiplied by the bare Coulomb interaction $V(\bk)$ that diverges at the $\Gamma$ point and thus requires fine $k$-point sampling to obtain the screened Coulomb interaction $W(\bk)$, which also diverges at the $\Gamma$ point.
Overall, the $GW$ algorithm still yields finite observable quantities, because $W(\bk)$ is integrated over the whole Brillouin zone to obtain~$W^\bR$ and this integral is convergent.
In practice, we compute the polarizability $\chi^\Gamma$ only at the $\Gamma$ point and reconstruct the $k$-dependence of $\chi$ from $\chi^\Gamma$ using the minimum-image convention (MIC), which becomes exact in the limit of a large unit cell.
For sufficiently large cells, the real-space decay of $\chi$ and $\Sigma$ allows us to reconstruct the full $k$-dependence at negligible additional cost.
This approach has been validated for large two-dimensional (2D) cells with hundreds of atoms in the unit cell.\cite{Graml2024}

The implementation-level reconstruction equations for $\chi^\Gamma$, the MIC mapping of $\chi^\mathbf{R}$, the Brillouin-zone reconstruction of $W^\mathbf{R}$, and the corresponding $\Gamma$-point self-energy are summarized in Section~V~D in the SI.
For sufficiently large nonmetallic supercells, the real-space locality of $G$, $\chi$, and $\Sigma$ allows the $\Gamma$-point response to be unfolded by MIC, transformed back to $k$ space for the screened interaction, and used to evaluate quasiparticle energies, becoming exact in the large-cell limit.

Because MIC is applied to $\chi$, $W$, and $\Sigma$, its validity is crucial.
Tests on monolayer TMDs show that the supercell must be large enough so that $G(\br,\br',i\tau)$ and $\chi(\br,\br',i\tau)$ have decayed before the periodic images overlap.
Our benchmarks indicate that unit cells with a size of $8\times 8$ primitive cells (with lateral dimensions of about \(20\,\text{\AA}\times20\,\text{\AA}\) for typical TMDs) lead to converged $GW$ band gaps, with differences below 20\,meV when increasing the unit cell further.\cite{Graml2024}

A detailed scaling benchmark, including the treatment of spurious negative eigenvalues of $\chi$, is provided in Section~V~D in the SI.

An important motivation for large supercells is the study of moiré systems: 2D van der Waals heterostructures in which a relative twist between layers creates a long-period superlattice.
As a first proof of principle, the algorithm has been applied to twisted MoSe\textsubscript{2}/WS\textsubscript{2} bilayers containing up to $984$ atoms in the unit cell (twist angles $\theta$ between $9.3^{\circ}$ and $26.6^{\circ}$).\cite{Graml2024}
\begin{figure}[!tbp]
    \centering
    \includegraphics[width=0.9\columnwidth]{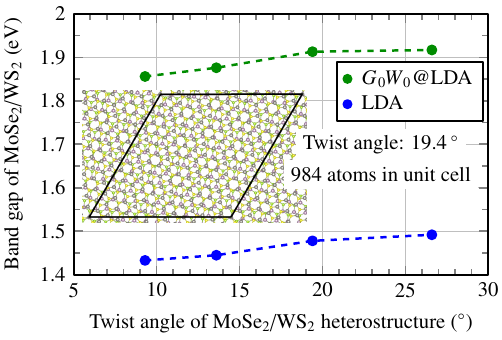}
    \caption{$G_0W_0$@LDA band gap of MoSe$_2$/WS$_2$ twisted bilayers as a function of twist angle.
    The unit cell of the $\theta\eqt19.4^{\circ}$ structure (inset) contains $984$ atoms.
    Reproduced from Ref.~\citenum{Graml2024} under the Creative Commons Attribution 4.0 International License.}
    \label{fig:gw_twisted_bilayers}
\end{figure}
Fig.~\ref{fig:gw_twisted_bilayers} shows the quasiparticle band gap as a function of twist angle.
The band gap changes in a systematic way (from $1.86$\,eV at $\theta\eqt9.3^{\circ}$ to $1.92$\,eV at $\theta\eqt26.6^{\circ}$), which provides microscopic input for understanding twist-dependent exciton physics in van der Waals heterobilayers.
The calculation remains tractable even for the 984-atom moir\'e cell. The corresponding hardware and runtime details are given in Section~V~D in the SI.
Recent code improvements and basis set developments,\cite{Pasquier2026} indicate that supercells with several thousand atoms, and possibly up to $10\,000$ atoms, could become tractable with this algorithm.
%

% In summary, the $\Gamma$-only MIC implementation combines the locality of Gaussian basis functions with real-space reconstruction of $\bk$-dependent quantities.
% %
% This approach enables $GW$ studies of large 2D supercells and twisted bilayers that are beyond the computational reach of PW algorithms, while retaining numerical accuracy at the level of a few $10$\,meV in quasiparticle energies.
% %
% The method is particularly suitable for 2D semiconductors, moiré heterostructures, and nanoscale band-structure engineering in large supercells.

% BEGIN INLINED FROM COHSEX.tex
\subsection{\texorpdfstring{Coulomb-Hole and Screened-Exchange}{Coulomb-Hole and Screened-Exchange}}\label{cohsex}

One of the simplest approximations to the self-energy introduced by Hedin for many-body perturbation theory is the static Coulomb-hole plus screened-exchange (st-COHSEX) approximation.\cite{hedin1965new,hybertsen1986} Using the space--time $GW$ framework in CP2K, we have introduced the st-COHSEX approximation and its scaled variant ($\alpha$-COHSEX). In this approximation, the correlation part of the self-energy simplifies to
\begin{equation}
\Sigma^{\text{c,st-COHSEX}}_{mn} = \frac{1}{2} \sum_{p} W^{\text{C}}_{mppn}(\omega=0)\text{sgn}(\varepsilon_p - \varepsilon_F),
\end{equation}
where \(m\), \(n\), and \(p\) denote one-particle states, \(W^{\mathrm C}_{mppn}\) is a matrix element of the correlation part of the screened interaction, and \(\operatorname{sgn}\) is the sign function. This expression requires only the static limit of the screened Coulomb interaction
\(W^{\mathrm C}(\omega=0)\).
For accurate EAs, the st-COHSEX implementation in CP2K retains off-diagonal self-energy elements.\cite{voora21} The static-response construction and reduced-scaling implementation are detailed in Section~V~E in the SI.

\begin{figure}[tb]
\centering
\includegraphics[width=1.0\linewidth]{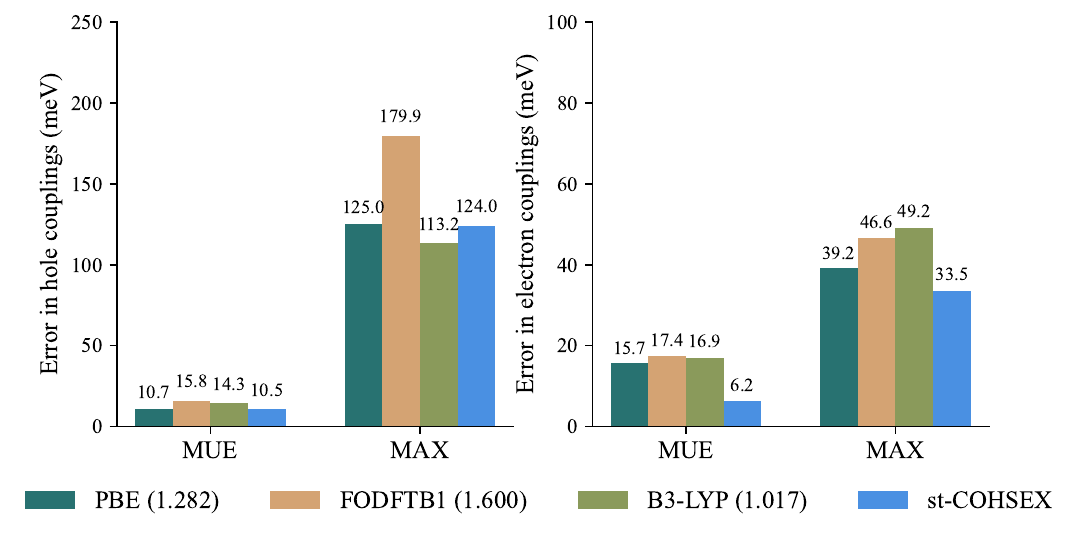}
\caption{Error in hole (left panel) and electron couplings (right panel) for cofacial molecular dimers in the HAB79 dataset.\cite{ziogos21} \(n\)-electron valence state perturbation theory reference values, along with PBE, DFTB, and B3-LYP, are taken from Ref.~\citenum{ziogos21}. Both the reference and st-COHSEX calculations employ the aug-cc-pVDZ basis set (cc-pVDZ for H), whereas PBE and B3-LYP couplings are computed using DZVP-GTH and 6-311G(d) basis sets, respectively. DFTB values are obtained using fragment-orbital first-order DFTB with the MIO parameter set. PBE, DFTB, and B3-LYP couplings are scaled by method-specific factors (given in parentheses), while no scaling is applied to the st-COHSEX results. This figure presents original, unpublished results obtained by the authors.\cite{tyagi_hab}}
\label{fig:hab}
\end{figure}

Despite its simplicity, the st-COHSEX method provides a reliable description of relative electronic-structure quantities. In particular, it accurately reproduces electronic couplings ($H_{ab}$) in charge-transfer systems under the generalized Mulliken--Hush formalism.\cite{CAVE199615,tyagi_hab} For the HAB79 dataset,\cite{ziogos21} (Fig.~\ref{fig:hab}), the st-COHSEX method yields hole and electron couplings in excellent agreement with \(n\)-electron valence state perturbation theory reference values, achieving an MAE of 10.5 meV without using any empirical scaling. In contrast, other KS or generalized KS (GKS) approximations based on semilocal or hybrid functionals or DFTB methods require method-specific scaling factors to reach comparable accuracy. This robustness highlights the suitability of st-COHSEX for studying charge transport in organic semiconductors.

A known limitation of st-COHSEX is the absence of dynamical screening, which leads to systematic overestimation of absolute IPs and EAs.\cite{Tyagi2024} To mitigate this, we employ the $\alpha$-COHSEX scheme,\cite{tyagi_stc_cp2k} in which the screened interaction is scaled by a constant factor according to
\begin{equation}
\Sigma^{\text{c,}\alpha\text{-COHSEX}}_{mn} = \frac{\alpha}{2} \sum_{p} W^{\text{C}}_{mppn}(\omega=0)\text{sgn}(\varepsilon_p - \varepsilon_F),
\end{equation}
where \(\alpha\) is the empirical scaling factor applied to the static screened interaction, thereby partially incorporating dynamical correlation effects.

\begin{figure}[tb]
\centering
\includegraphics[width=1.0\linewidth]{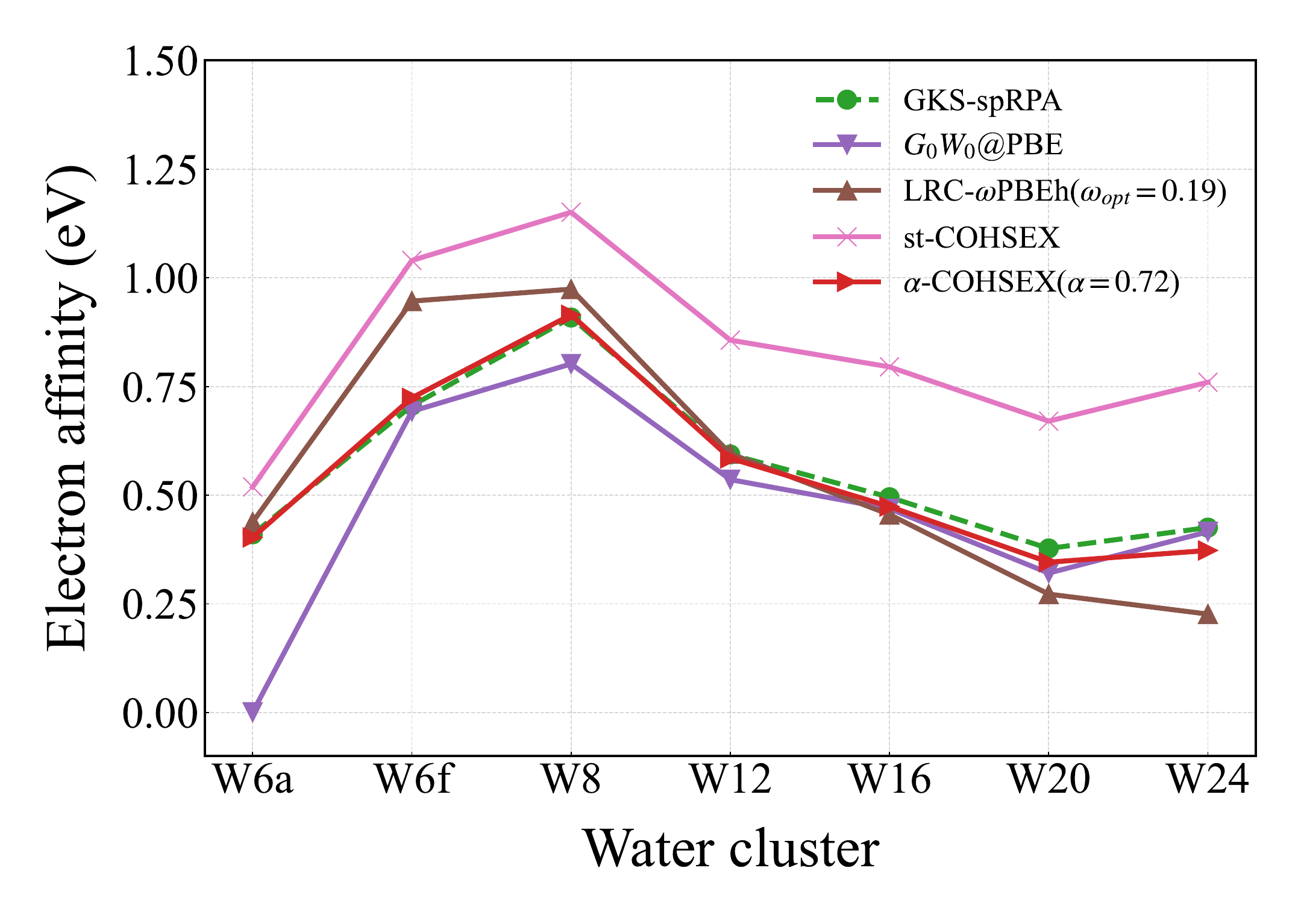}
\caption{Electron affinities of water clusters computed using many-body methods and (G)KS functionals. GKS-spRPA values are taken as reference. W12 is used for parametrization of $\alpha$-COHSEX and LRC-$\omega$PBEh. The aug-cc-pVDZ+7s7p,\cite{dunning1989,Kendall1992,vysotskiy2012} basis set is used for all methods. The nuclear structures are from Ref.~\citenum{vysotskiy2012}. GKS-spRPA and LRC-$\omega$PBEh values are from Ref.~\citenum{Tyagi2024}, while st-COHSEX, $\alpha$-COHSEX, and $G_0W_0$ calculations used the implementation in CP2K. This figure presents original, unpublished results obtained by the authors.\cite{tyagi_stc_cp2k}}
\label{fig:scaled_h2o_clus}
\end{figure}

We assess this approach for electron attachment to water clusters, which are model systems for hydrated electrons. A single scaling parameter fitted to the W12 reference is transferred without refitting across surface- and interior-bound clusters. The cluster set, reference data, and fitted value are specified in Section~V~E in the SI. The $\alpha$-COHSEX method has an MAE of 21 meV, which is much smaller than the st-COHSEX error of 270 meV. Other methods, such as the $G_0W_0$@PBE and tuned long-range-corrected (LRC)-$\omega$PBEh, yield significantly larger errors and fail to reproduce qualitative trends (Fig.~\ref{fig:scaled_h2o_clus}). These results demonstrate that $\alpha$-COHSEX provides an accurate and computationally efficient approach for predicting absolute EAs in extended systems.

Beyond accuracy and efficiency, a key advantage of st-COHSEX for applications in dynamics and spectroscopy lies in the static nature of the self-energy. The static approximation eliminates self-energy poles for finite systems, thereby ensuring smooth and continuous quasiparticle and total energy surfaces with respect to nuclear displacements.\cite{veril2018,berger2021} In contrast, perturbative $GW$ and partially self-consistent $GW$ methods display unphysical discontinuities due to issues with non-unique solutions. As a result, st-COHSEX provides a numerically stable framework for excited-state geometry optimizations and MD,\cite{FaberPRB2015} making it particularly attractive when quasiparticle energies and nuclear motion must be treated consistently.
% END INLINED FROM COHSEX.tex

% BEGIN INLINED FROM GW_BSE.tex
\section{\texorpdfstring{\textit{GW} plus Bethe--Salpeter Equation for Electronic Excitations}{GW plus Bethe-Salpeter Equation for Electronic Excitations}}
Within many-body perturbation theory, the description of electronic excitations requires extending the $GW$ formalism, which predicts accurate quasiparticle energies, by including the interaction of the excited electron--hole pair.
A well-established method to do so is by considering the effect of the $GW$ self-energy $\Sigma$ [Eq.~\eqref{Sigmaconvw}] on the four-point polarizability, which yields the typically employed form of the BSE.\cite{Onida2002,Ljungberg2015,martin_reining_ceperley_2016,Graml2025}
CP2K provides two implementations of the $GW$-BSE formalism. Section~\ref{sec-lr_bse} describes the well-established linear-response BSE (LR-BSE) based on a two-body Schrödinger-like matrix equation.\cite{Graml2025}
Section~\ref{sec-rt_bse} presents the implementation based on the real-time propagation of the single-particle density matrix, denoted real-time BSE (RT-BSE).\cite{Marek2025}
The auxiliary-basis contractions used in LR-BSE and the detailed orbital-basis propagation and kernel expressions used in RT-BSE are given in Sections~VI~A and~VI~B in the SI, respectively.

\subsection{Linear-Response Bethe--Salpeter Equation} \label{sec-lr_bse}
%Max Graml, Jelena Schmitz
%

%
The BSE on top of $GW$ ($GW$-BSE) is a method to compute the response of an electronic system to an optical perturbation,\cite{Strinati1988,Rohlfing2000,Onida2002,Graml2025} which yields both absorption spectra with electronic excitation energies $\Omexc$ and oscillator strengths $f^{(n)}$ as well as wavefunctions of the electronic excitations $\Psiexc(\bre,\brh)$. The explicit construction of the latter is given in Section~IV~D in the SI, and the corresponding spatial descriptors are discussed in Section~\ref{sec-spatial_descriptors}.
In CP2K, the linear-response $GW$-BSE method is currently implemented for finite systems only, where the defining equations of $GW$-BSE are derived from the four-point polarizability.\cite{Onida2002,Ljungberg2015,martin_reining_ceperley_2016,Graml2025}
The resulting generalized eigenvalue equation is reminiscent of a two-body Schrödinger-like equation and looks formally like the Casida equations in linear-response TDDFT (cf. Section~\ref{sec-lr_tddft}).\cite{Ullrich2011}
\begin{align}
\begin{pmatrix}
\mathbf A & \mathbf B\\
\mathbf B & \mathbf A
\end{pmatrix}
\begin{pmatrix}
\bX^{(n)}\\
\bY^{(n)}
\end{pmatrix}
&=
\Omexc
\begin{pmatrix}
\bX^{(n)}\\
-\bY^{(n)}
\end{pmatrix}
\ensuremath{,}
\label{eq-BSE_ABBA}
\end{align}
where $\Omexc$ are the excitation energies, while $\bX^{(n)}$ and $\bY^{(n)}$ are the excitation and deexcitation amplitudes of excitation \(n\). The corresponding left and right generalized eigenvectors are biorthogonal, with their normalization given in Section~VI~A in the SI. For the finite systems considered here, $\mathbf A$ and $\mathbf B$ are real symmetric.
For a closed-shell ground state,\cite{Ljungberg2015,Graml2025} the matrices \(\mathbf A\) and \(\mathbf B\) have the following elements:
\begin{align}
    A_{ia,jb} &= (\varepsilon_a^{GW}-\varepsilon_i^{GW})\delta_{ij}\delta_{ab} + \alpha^\mathrm{(S/T)}
    v_{ia,jb} - W_{ij,ab} \,, \nonumber
    \\[0.5em]
    B_{ia,jb} &= \alpha^\mathrm{(S/T)} v_{ia,bj} - W_{ib,aj} \,,
    \label{eq-BSE_ingredients}
\end{align}
where $\varepsilon_{i/a}^{GW}$ are eigenvalues of a preceding $GW$ calculation (cf. Eq.~\eqref{eb9}) and $\alpha^\mathrm{(S/T)}$ denotes a spin-dependent factor, which is $\alpha^\mathrm{(S)}{=}2$ for singlet excited states and $\alpha^\mathrm{(T)}{=}0$ for triplets. The indices \(i,j\) and \(a,b\) label occupied and virtual states, respectively. The symbol \(\delta\) is the Kronecker delta, \(v\) denotes bare-exchange Coulomb matrix elements, and \(W\) denotes statically screened direct-interaction matrix elements.
Physically, this BSE kernel combines a bare-exchange contribution with an attractive direct term screened in the static limit. The explicit four-index matrix elements, their Coulomb-metric RI factorization, and the connection to the static dielectric matrix are detailed in Section~VI~A in the SI.
%

%
%In CP2K, Eq.\,(\ref{eq-BSE_ABBA}) is reformulated as an hermitian eigenvalue problem under the assumption that $\mathds{A}{-} \mathds{B}$ is positive definite \cite{Ullah1971, Ullrich2011,Graml2025}.
%
% For the Tamm-Dancoff approximation (TDA)~\cite{Benedict1998}, the coupling blocks $\mathds{B}$ are neglected:
% %
% \begin{align}
%     \mathds{A} \bX_\text{TDA}^{(n)} = \Omexc_\text{TDA} \bX_\text{TDA}^{(n)} \,. \label{eq-TDA_of_BSE}
% \end{align}
% %

%
From the solution of Eq.~\eqref{eq-BSE_ABBA}, the optical absorption spectrum can be computed from the imaginary part of the dynamical dipole polarizability tensor $\alpha_{\mu\mu'}$:
\begin{align}
    \alpha_{\mu\mu'}(\omega)
    = \sum_{n>0} \frac{2 \,\Omega^{(n)}\, d^{(n)}_{\mu} d^{(n)}_{\mu'}}
    {\left(\Omega^{(n)}\right)^2-(\omega+i\eta)^2}
    \ensuremath{.}
    \label{eq-BSE_polarizatbility}
\end{align}
The indices \(\mu\) and \(\mu'\) specify Cartesian dipole directions, \(d_\mu^{(n)}\) is a transition-dipole component, \(\omega\) is the probe frequency, and \(\eta>0\) is a spectral broadening parameter.
The LR-BSE eigenvectors determine transition dipoles and oscillator strengths, from which the frequency-dependent polarizability and orientationally averaged absorption spectrum follow. These working expressions are given in Section~VI~A in the SI.
In Fig.~\ref{fig_rtbse}, we showcase the imaginary part of the polarizability in Eq.~\eqref{eq-BSE_polarizatbility} for two small molecules.
Beyond the optical response, the computation of the eigenvectors $\Xian$ and $\Yian$ gives access to the wavefunction of the excited state $\Psiexc$, which can be systematically investigated using the spatial descriptors introduced in Section~\ref{sec-spatial_descriptors}.

\subsection{Real-Time Bethe--Salpeter Equation}\label{sec-rt_bse}

\subsubsection{\texorpdfstring{Theory of the Real-Time Bethe--Salpeter Equation}{Theory of the Real-Time Bethe-Salpeter Equation}}

Nonlinear response can be obtained either by evaluating higher-order polarizabilities in the frequency domain or by propagating the electronic dynamics under an external, oscillating electric field.\cite{Boyd2020,SHGAnalyticRuanLouie,SHGAnalyticRauwolfHolzer,Friedrich2026,Klimmer2026} The real-time approach contains all response orders generated by the applied field and forms the basis of the RT-BSE implementation in CP2K.\cite{Marek2025}

The RT-BSE implementation uses the equation of motion (EOM)
for the single-particle density matrix,\cite{RTBSEAttaccalite} which is derived from
the Kadanoff--Baym equations,\cite{NEGFStefanucciVanLeeuwen} under the
time-local, Hermitian approximation to the self-energy with static screening, namely the static COHSEX approximation. In this implementation, only the screened-exchange (SX) self-energy enters the EOM. It is denoted by $\hat \Sigma^\mathrm{SX}$.
Specifically, the EOM has the form of the standard von Neumann time evolution for the one-electron density operator~$\hat\rho$
\begin{align}
\frac{\partial \hat{\rho}}{\partial t}
&=
-\frac{i}{\hbar}
\left[
\hat{h}^{\mathrm{eff}}(t),\hat{\rho}(t)
\right].
\label{rtbse_eom}
\end{align}

\begin{table*}[t]
\caption{Comparison of real-time propagation methods for neutral electronic
excitations. The common starting point is the RT-BSE equations of motion in
Eqs.~\eqref{rtbse_eom} and~\eqref{rtbse_eom_densmat_1}. The other
methods are obtained by replacing the quasiparticle energies and the
interaction term beyond the Hartree potential. The RT-BSE uses screened exchange
and therefore an explicit screened electron--hole interaction, real-time Hartree--Fock is the
corresponding unscreened limit, and real-time TDDFT represents these effects through
an exchange-correlation potential. Hybrid functionals include nonlocal
exchange, but do not generally reproduce the material- and space-dependent
screened Coulomb interaction.}
		\setlength{\tabcolsep}{0pt}
		\begin{tabular}{llll}
			\hline\hline \\[-0.5em]
		\ReviewTableCell{0.38\textwidth}{\textbf{Method}} &
		\ReviewTableCell{0.09\textwidth}{\textbf{Eigen-\linebreak values}} &
		\ReviewTableCell{0.26\textwidth}{\textbf{Electron--electron interaction beyond \(V^\text{H}\)}} &
		\ReviewTableCell{0.25\textwidth}{\textbf{Linear response}} \\
			\\[-0.5em]
			\hline
			\\[-0.5em]
		\ReviewTableCell{0.38\textwidth}{Real-time BSE} &
		\ReviewTableCell{0.09\textwidth}{\(\varepsilon_n^{GW}\)} &
		\ReviewTableCell{0.26\textwidth}{screened exchange, \(\Sigma^\text{SX}[\hat \rho(t)-\hat \rho_0]\)} &
		\ReviewTableCell{0.25\textwidth}{$GW$ plus BSE (Section~\ref{sec-lr_bse})}\\
			\\[-0.2em]
		\ReviewTableCell{0.38\textwidth}{Real-time Hartree--Fock} &
		\ReviewTableCell{0.09\textwidth}{\(\varepsilon_n^\text{HF}\)} &
		\ReviewTableCell{0.26\textwidth}{bare exchange, \(\Sigma^\text{X}[\hat \rho(t)-\hat \rho_0]\)} &
		\ReviewTableCell{0.25\textwidth}{Linear-response Hartree--Fock} \\
		\\[-0.2em]
		\ReviewTableCell{0.38\textwidth}{Real-time TDDFT (local density approximation / generalized-gradient approximation, Section~\ref{sec-rt_tddft})} &
		\ReviewTableCell{0.09\textwidth}{\(\varepsilon_n^\text{KS}\)} &
		\ReviewTableCell{0.26\textwidth}{\(V^\text{xc}[\hat \rho(t)]-V^\text{xc}[\hat \rho_0]\), typically adiabatic approximation} &
		\ReviewTableCell{0.25\textwidth}{LR-TDDFT (Section~\ref{sec-lr_tddft})} \\
		\\[-0.2em]
		\ReviewTableCell{0.38\textwidth}{Real-time TDDFT (hybrid XC functional)} &
		\ReviewTableCell{0.09\textwidth}{\(\varepsilon_n^\text{GKS}\)} &
		\ReviewTableCell{0.26\textwidth}{\(\Delta\!\left(\alpha V^\mathrm{c}+\beta V^\mathrm{x}+\gamma\Sigma^\mathrm{X}\right)\)} &
		\ReviewTableCell{0.25\textwidth}{LR-TDDFT} \\
		\\[-0.5em]
		\hline\hline
	\end{tabular}
		\label{rtbse_table}
\end{table*}
In the hybrid-functional row of Table~\ref{rtbse_table}, \(\Delta O\equiv O[\hat\rho(t)]-O[\hat\rho_0]\), while \(\alpha\), \(\beta\), and \(\gamma\) weight the semilocal correlation, semilocal exchange, and nonlocal exact-exchange contributions, respectively.

\begin{figure}[!tbp]
	\includegraphics[width=0.9\columnwidth]{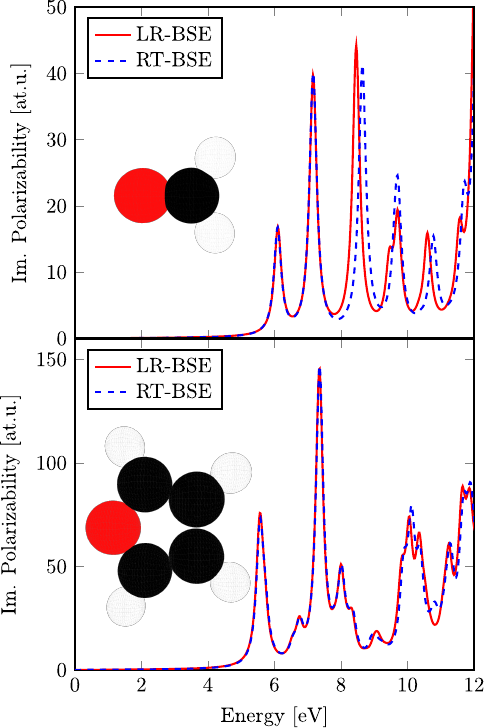}
	\centering
			\caption{Imaginary part of the polarizability
	  of formaldehyde (top) and furan (bottom)
			calculated in the linear-response limit from both RT-BSE propagation (Eq.~\eqref{rtbse_polarizability}) and the linear-response BSE (Eq.~\eqref{eq-BSE_polarizatbility}). For the RT-BSE, the propagation ran for 20~fs using
	the $\delta$-kick excitation scheme, with intensity in the linear regime. The $GW$ calculation used
	the aug-cc-pVTZ basis set with the PBE0 functional. A damping of \(\eta=5~\mathrm{fs}^{-1}\) and a time step of \(1~\mathrm{as}\) were used. Further computational parameters and the plotted data are taken from Ref.~\citenum{Marek2025}. Differences between the spectra arise from the \(G_0W_0\) energies produced by the underlying \(GW\) implementations.}
	\label{fig_rtbse}
\end{figure}

\begin{samepage}

In the dipole approximation, define
\(\Delta\hat\rho(t)=\hat\rho(t)-\hat\rho_0\). The effective one-electron Hamiltonian is then
% \begin{align}
% 	\hat h^\mathrm{eff}(t) =& \,
% 		\hat h ^ {GW} + e \vec E (t) \cdot \vec{\hat \mu}   + \hat V ^ \mathrm{H} [\hat \rho(t)] -
% 		\hat V ^ \mathrm{H} [\hat \rho_0] \nonumber \\[0.3em]
% 		&+ \hat \Sigma ^ \mathrm{COHSEX} [\hat \rho (t)] -
% 		\hat \Sigma ^ \mathrm{COHSEX} [\hat \rho_0]
% \end{align}
\begin{equation}
\hat h^\mathrm{eff}(t)
=\hat h^{GW}+e\bm E(t)\cdot\hat{\bm\mu}
+\bigl(\hat V^\mathrm{H}+\hat\Sigma^\mathrm{SX}\bigr)
[\Delta\hat\rho(t)]\ensuremath{,}
\label{rtbse_eom_densmat_1}
\end{equation}
\end{samepage}
where $\hat h^{GW}$ is the Hamiltonian from a $GW$ calculation,
$\bm E(t)$ is the time-dependent electric field, $\hat{\bm \mu}$ is the dipole-operator vector, $\hat V^\mathrm{H} $ is the Hartree potential operator,
 and
$\hat \Sigma ^ \mathrm{SX}$ is the SX self-energy operator.
The density matrix at the start of the propagation is denoted by $\hat\rho_0$.
Note that the EOM \eqref{rtbse_eom} is equivalent to the effective single-particle
Schrödinger equation
\begin{align}
	i\hbar\frac{\partial}{\partial t}\ket{\psi(t)}=\hat{h}^{\mathrm{eff}}(t)\ket{\psi(t)}\ensuremath{.}
\end{align}

The implementation evaluates the abstract operator equation \eqref{rtbse_eom} in the
basis of AOs, which is extensively covered in the implementation article.\cite{Marek2025}
For clarity, we show the forms of the $\hat V^\text{H}$ and $\hat \Sigma^\text{SX}$ operators in the position basis.\cite{RTBSEAttaccalite}
\begin{subequations}
\label{rtbse_cohsex}
\begin{align}
        \bra{\br_1} \hat V^\text{H} [\hat \rho(t)] \ket{\br_2}
        &= 2\delta (\br_1 {-} \br_2)
        \int \dd^3 r_3\;
        \frac{\rho(\br_3, \br_3, t)}{|\br_1 - \br_3|},
        \label{rtbse_hartree}\\
        \bra{\br_1} \hat{\Sigma}^\text{SX} [\hat \rho(t)] \ket{\br_2}
        &= -W(\br_1,\br_2)\rho(\br_1,\br_2,t).
        \label{rtbse_screened_exchange}
\end{align}
\end{subequations}

The vectors \(\mathbf r_1\), \(\mathbf r_2\), and \(\mathbf r_3\) are spatial coordinates, \(\delta\) is the Dirac delta distribution, \(\rho(\mathbf r_1,\mathbf r_2,t)\) is the one-particle density-matrix kernel, and \(W\) is the statically screened Coulomb interaction. The factor of two accounts for spin degeneracy.
The screened interaction $W$ is obtained in the static limit from the low-scaling $GW$ calculation and held constant during the
time propagation.
The AO density matrix is integrated with an enforced time-reversal symmetry (ETRS) propagator. The corresponding MO-basis equation, Hartree and screened-exchange matrix elements, and propagation details are given in Section~VI~B in the SI.\cite{PropagatorsCastroRubio,Marek2025}

\subsubsection{\texorpdfstring{Comparison of the Real-Time Bethe--Salpeter Equation and Real-Time Time-Dependent Density-Functional Theory}{Comparison of the Real-Time Bethe-Salpeter Equation and Real-Time Time-Dependent Density-Functional Theory}}

The central distinction between RT-BSE and RT-TDDFT is the treatment of the photoexcited electron--hole pair. RT-BSE includes the statically screened interaction \(W\) explicitly through \(\Sigma^{\mathrm{SX}}\), which provides the direct real-time counterpart of the electron--hole kernel in LR-BSE.

Real-time Hartree--Fock (RT-HF) has the same propagation structure but replaces \(W\) by the bare Coulomb interaction \(v\), making it the unscreened reference limit summarized in Table~\ref{rtbse_table}. The explicit exchange kernel is given in Section~VI~B in the SI.

RT-TDDFT instead represents electron--hole effects through an approximate, usually adiabatic XC potential. Local and semilocal functionals do not contain the explicit screened interaction, while hybrids introduce only a global fraction of bare exchange and therefore cannot generally reproduce the material- and position-dependent screening of heterogeneous systems. Section~VI~B in the SI gives the operator-level correspondence.

From this perspective, RT-BSE is the more direct real-time framework for neutral
excitations when screened electron--hole interactions are important. It combines
$GW$ quasiparticle energies with an explicit screened interaction between excited
electrons and holes, and is therefore closely connected to the physical picture
underlying the successful linear-response $GW$-BSE approach.
Relative to linear-response $GW$-BSE, its main limitations are the computational cost of real-time propagation and, in the current implementation, its restriction to molecular systems.

Both RT-BSE and RT-TDDFT can be linearized for small oscillations of the reduced density
matrix \(\hat \rho\). In this limit, RT-BSE reduces to standard $GW$-BSE, RT-TDDFT
reduces to LR-TDDFT, and RT-HF reduces to linear-response Hartree--Fock. The main differences between these
real-time methods are summarized in Table~\ref{rtbse_table}.

\subsubsection{Dynamical Observables}

The central observable obtained from density-matrix propagation is the induced dipole moment
\begin{align}
\Delta\bm{\mu}(t)
&=
\left\langle\bm{\hat{\mu}}\right\rangle(t)-\bm{\mu}_0
\nonumber\\
&=\Tr\!\left[\bm{\hat{\mu}}\left(\hat{\rho}(t)-\hat{\rho}_0\right)\right]
\nonumber\\
&=\sum_{nm}\bm{\mu}_{nm}\left[\rho_{mn}(t)-\rho_{0,mn}\right].
\end{align}

Here, \(\bm{\mu}_0=\Tr[\bm{\hat\mu}\hat\rho_0]\) is the equilibrium dipole moment. The indices \(m\) and \(n\) label one-particle basis states, while \(\bm\mu_{nm}\), \(\rho_{mn}(t)\), and \(\rho_{0,mn}\) are dipole-matrix elements, time-dependent density-matrix elements, and reference density-matrix elements, respectively.
A broadening parameter $\eta$ regularizes its Fourier transform:
\begin{align}
		\Delta\bm \mu(\omega) = \int _ 0 ^ T \dd t \;\mathrm{e}^{i  (\omega + i  \eta) t} \Delta\bm \mu(t)
		\ensuremath{,}
	    \label{rtbse_ft}
\end{align}
where $T$ is the total propagation time.

When the amplitude of the field $\bm E(t)$ restricts the propagation to the linear
regime, the linear polarizability $\alpha _ {j k} (\omega)$ follows from the dipole response.\cite{TDDFTYabanaBertsch}
\begin{align}
		\alpha _ {j k} (\omega) = \frac{\Delta\mu_j (\omega)}{E_k(\omega)}
		\ensuremath{,}
	    \label{rtbse_polarizability}
\end{align}
where $E_k (\omega)$ is the Fourier transform of the applied field component along direction \(k\). CP2K implements all these observables for RT-BSE.

\begin{figure}[!tbp]
	\includegraphics[width=\columnwidth]{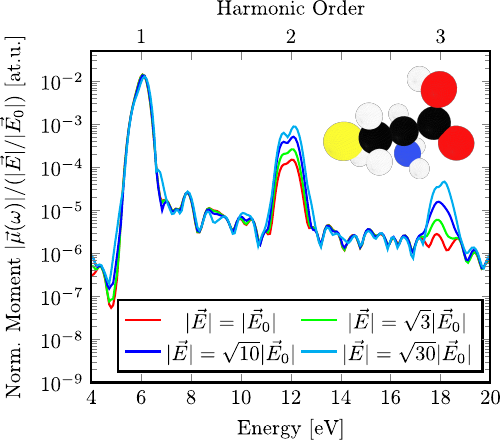}
	\centering
	\caption{Induced-dipole-moment spectrum normalized by the applied-field ratio \(\lvert\bm E\rvert/\lvert\bm E_0\rvert\) of the cysteine
	molecule after excitation by a Gaussian pulse with frequency centered on 6.00~eV and
	spread of 4.2~fs. The propagation time was 52.6~fs. The reference field magnitude \(\lvert\bm E_0\rvert\) corresponds to a peak incident-light intensity of \(10^{10}~\mathrm{W\,cm^{-2}}\). The higher-order peaks (the finite
	spread of the Gaussian pulse induces several oscillations) are observed. Normalized
	moment values below
	$10^{-5}$~a.u. are likely numerical noise.
	The underlying $GW$ calculation used the aug-cc-pVTZ basis set and the PBE0 functional.
	Data taken from Ref.~\citenum{Marek2025}.}
	\label{fig_rtbse_nonlin}
\end{figure}

Fig.~\ref{fig_rtbse} shows the absorption-related imaginary polarizability recovered from both LR-BSE and RT-BSE. In a nonlinear regime, $\Delta\bm \mu(\omega)$ starts to
develop nonlinear components, detected as peaks not centered on the
frequency of the driving field, as illustrated in Fig.~\ref{fig_rtbse_nonlin}.

The present RT-BSE implementation is restricted to isolated molecules. Periodic and substantially larger applications will require distinct formulations and further reductions in propagation cost.
% END INLINED FROM GW_BSE.tex

\section{Active-Space Embedding}
%incl. Quantum Computing
%Stefano Battaglia

% BEGIN INLINED FROM EmbeddingMethods.tex
Strong electronic correlation is often local even when the system of interest is not.
This observation is particularly useful for spectroscopic applications, where a
small number of localized electrons can dominate the measured signal:
examples include defect-center absorption and emission, charge-transfer bands,
open-shell transition-metal centers, and local excitations at surfaces or in
solution. The active space embedding in CP2K provides an orbital-space counterpart
to the real-space and subsystem embedding strategies discussed elsewhere in this
review.  A chemically or physically relevant subset of orbitals and electrons is described by
a correlated active-space solver, while the remaining ones are retained at a
mean-field level through the GPW/GAPW infrastructure.  The same machinery also
provides a compact bridge to quantum-computing workflows, because the embedded
Hamiltonian can be mapped to a qubit Hamiltonian and solved by variational or
equation-of-motion quantum algorithms.\cite{Rossmannek2021,Rossmannek2023,Battaglia2024}
A specific implementation of such an embedding strategy is the periodic range-separated DFT
embedding method, which couples CP2K and Qiskit Nature and has recently
been used to compute the optical absorption and photoluminescence of a neutral oxygen vacancy
in MgO.\cite{Battaglia2024}

\subsection{Active-Space Hamiltonian}

The starting point is a mean-field calculation that defines an orthonormal
one-particle basis.  The MO space is partitioned into inactive
orbitals $\mathcal{I}$, active orbitals $\mathcal{A}$, and external virtual
orbitals $\mathcal{V}$.  In CP2K, the active orbitals may be automatically selected from
canonical ones or by explicit manual selection. The second option allows the use of
any type of localized/projected orbitals, which is essential for point defects and other
local spectroscopic centers in extended systems.  After this partitioning, the correlated
problem is restricted to~$\mathcal{A}$
\begin{align}
   \hat{H}_{\mathcal{A}} &=
   E_{\mathcal{I}} +
   \hat{h}^{\mathrm{emb}}_{\mathcal{A}} +
   \hat{W}_{\mathcal{A}}\,.   \label{eq:as_emb_hamiltonian}
 \end{align}
In this expression, $E_{\mathcal{I}}$ is the scalar energy contribution of the inactive subspace,
$\hat{h}^{\mathrm{emb}}_{\mathcal{A}}$ is the effective one-electron operator in the active subspace, including its mean-field interaction with the inactive environment,
and $\hat{W}_{\mathcal{A}}$ is the electron--electron interaction operator restricted to the active subspace.
The explicit second-quantized forms of the latter two operators are given in Section~VII~A in the SI.
In the range-separated DFT variant, the Coulomb operator is decomposed as
\begin{align}
   \frac{1}{r_{12}} =
   \frac{\operatorname{erf}(\omega r_{12})}{r_{12}}
   +
   \frac{\operatorname{erfc}(\omega r_{12})}{r_{12}} ,
   \label{eq:as_range_separation}
\end{align}
where \(r_{12}=|\mathbf r_1-\mathbf r_2|\) is the interelectronic distance and $\omega$ controls the separation between the long-range and short-range
parts of the interaction.  The long-range part is assigned to the active-space
wavefunction, while short-range Hartree and XC terms remain
in the DFT environment.\cite{Battaglia2024} The corresponding energy contractions of the active-space one-particle density matrix (1-RDM) and two-particle density matrix (2-RDM) are specified in Section~VII~A in the SI. This separation keeps the active-space Hamiltonian compact for exact diagonalization and near-term quantum-computing solvers.

\begin{figure*}
\centering
\includegraphics[width=0.88\linewidth]{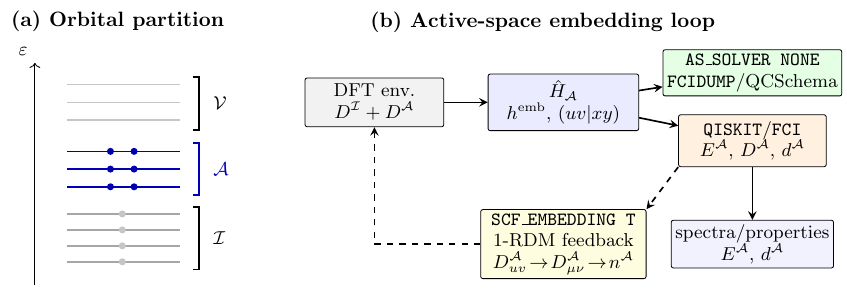}
\caption{Active-space embedding workflow in CP2K, adapted from the
CP2K--Qiskit Nature interface of Ref.~\citenum{Battaglia2024} under the Creative Commons Attribution 4.0 International License and modified to include the
current solver hierarchy.  The correlated solver may use the active-space 2-RDM
$d^{\mathcal{A}}$ for energies and two-particle properties, but the
self-consistent embedding feedback is carried by the active-space 1-RDM
$D^{\mathcal{A}}_{uv}$, which CP2K transforms to the AO density and then to
$n^{\mathcal{A}}$.}
\label{fig:active-space-qiskit-workflow}
\end{figure*}

The active-space solver returns an energy and a 1-RDM. After transformation to the AO and real-space representations, the 1-RDM can be fed back into the embedding potential, whereas the 2-RDM enters the correlated energy and two-particle properties but is not mixed into the mean-field environment. Self-consistent embedding therefore iterates the active-space solution and the surrounding KS response, using linear or quasi-Newton density mixing as needed. The AO back-transformation, density partition, fixed-point map, and mixing equations are detailed in Section~VII~A in the SI.

\subsection{Embedding Hierarchy and Solvers}

The implementation exposes three levels of increasing coupling between the correlated active space and its mean-field environment. Hamiltonian export provides a portable active-space problem for an external solver. One-shot embedding solves the correlated subsystem with Qiskit Nature or full configuration interaction (FCI) in a fixed embedding potential, whereas self-consistent embedding feeds the active-space 1-RDM back into CP2K and relaxes the active region and its polarizable environment together. Their correspondence to mechanical, electrostatic, and polarizable embedding, together with the associated input controls and representative applications, is summarized in Table~S2 in the SI.

\subsection{Implementation in CP2K}

The \texttt{QISKIT} solver follows the CP2K--Qiskit Nature interface introduced in Ref.~\citenum{Battaglia2024} and shown in Fig.~\ref{fig:active-space-qiskit-workflow}. CP2K constructs the embedded Hamiltonian and passes it to Qiskit Nature. The fragment ground state is obtained with the variational quantum eigensolver (VQE),\cite{Peruzzo2014} whereas electronically excited states are accessed through the quantum equation-of-motion algorithm,\cite{Ollitrault2020} making the interface directly relevant to optical spectra of localized centers in extended systems. The current inactive environment is optimized for the ground-state electron density. Its lack of relaxation in response to an excited active-space density remains an approximation for excited-state calculations. The interface and data-flow details are described in Section~VII~B in the SI.

The first application embedded two electrons in five active orbitals for the neutral oxygen vacancy in periodic MgO. It reproduced the photoluminescence energy (2.44~eV versus 2.3~eV experimentally), while the ground-state treatment of the environment was linked to a 0.75~eV overestimate of the singlet absorption.\cite{Battaglia2024}

The internal \texttt{FCI} solver provides a deterministic reference for small active spaces and participates in the same one-shot or self-consistent embedding loop as the Qiskit interface. Excited states are obtained by computing additional roots of the fragment Hamiltonian. Solver-specific data exchange and implementation details are given in Section~VII~B in the SI.

For periodic systems, the active-space infrastructure builds on the same
GPW representation used throughout \textsc{Quickstep}.  The present
implementation supports explicit single-\(\Gamma\)-point calculations with real
wavefunctions.  This is sufficient for the large supercells commonly used for
liquids, surfaces, and localized defects in spectroscopy, while keeping the
active-space Hamiltonian in a real orbital representation. General non-\(\Gamma\)-point
sampling is not currently supported.

Together, these developments make active-space embedding a bridge between
CP2K's established condensed-phase DFT capabilities and correlated electronic
structure methods for local spectra.  The user can stop at a portable
\texttt{FCIDUMP} Hamiltonian, carry out one-shot embedded calculations with a
classical or quantum solver, or iterate the active 1-RDM for a polarizable,
self-consistent treatment of the spectroscopic center and environment.

% END INLINED FROM EmbeddingMethods.tex

\section{Charge Transfer, Open Boundaries, and Transport}
\label{sec:charge-transfer-open-boundaries-transport}

This section brings together electronic-structure methods for charge localization, charge transfer, and transport, ordered by increasing electronic delocalization, environmental openness, and nonequilibrium driving. CDFT first provides localized diabatic charge or spin states and their free-energy surfaces. Absolutely localized molecular orbitals (ALMOs) then impose locality directly on occupied MOs, furnishing both an efficient fragment-localized SCF representation and an energy-decomposition framework for polarization and charge transfer.\cite{shi2018} CDFT and projection operator-based diabatization (POD) are described as complementary approaches for calculating electronic coupling matrix elements, or transfer integrals, for electron-transfer processes. Kubo theory subsequently describes equilibrium linear-response transport without explicit electrodes, whereas Hairy Probes and NEGF introduce electronic reservoirs and, in the latter case, current-carrying steady states under external bias. Parametrized surface hopping closes the section by extending localized-state and coupling concepts to finite-temperature charge, exciton, and exciton-dissociation dynamics. Working equations and implementation details that lie below the common abstraction level of the main text are given in Section~VIII in the SI.

\subsection{Constrained Density-Functional Theory}
\label{sec:cdf-charge-transfer}
%\subsection{Electronic coupled by constrained DFT?}
%+ Gradients incl. DDAPC (ET coupling elements)?

CDFT enables the construction of charge or spin-localized diabatic electronic states.\cite{Wu05,VanVoorhis10} These states suffer less from the electron delocalization error than the adiabatic electronic states obtained from standard DFT. Thus, they tend to give a more reliable description of electron and spin transfer processes and are often used to compute the key quantities of Marcus and other electron transfer theories: reorganization (free) energies, (free) energy differences, and donor--acceptor electronic couplings.\cite{Oberhofer09,Oberhofer10jcp,Holmberg2017} CDFT electronic states also find applications for the parametrization of model Hamiltonians for charge transport calculations,\cite{Giannini20} and for configuration interaction calculations (CDFT-CI).\cite{Kaduk14}

CDFT has been extensively reviewed,\cite{VanVoorhis10} and its static implementation in CP2K was described in detail in previous publications.\cite{Holmberg2017,Kuehne2020} Briefly, charge- or spin-localized states are obtained by minimizing the energy functional \(E[n]\) subject to the constraint
\begin{equation}
  N_{c} = \int w(\mathbf{r})n(\mathbf{r})\,\mathrm{d}\mathbf{r},
  \label{eq:nc}
\end{equation}
where $N_c$ is the target value of the constraint and $w(\mathbf r)$ is a weight function that defines the partitioning of the electron density \(n(\mathbf r)\).
CP2K supports weight functions based on either Becke,\cite{Holmberg2017} or Hirshfeld partitioning.\cite{Ahart22jctc}
The latter yields physically more meaningful partition charges; for water, it assigns a small negative charge to oxygen, whereas Becke partitioning gives a large positive value.\cite{Ahart22jctc}
The constraint is enforced by introducing a Lagrange multiplier $V$ and a new energy functional
\begin{equation}
W[n,V]
= E[n]
+V\left(\int w(\mathbf{r})n(\mathbf{r})\,\mathrm{d}\mathbf{r}-N_{c}\right).
\label{eq:energy}
\end{equation}
\(W[n,V]\) is minimized with respect to \(n\) for a fixed $V$, and $V$ is then adjusted iteratively until the minimized electron density satisfies the
constraint in Eq.~\eqref{eq:nc}.

The CDFT forces resulting from Hirshfeld weight partitioning were recently implemented in CP2K, enabling CDFT geometry optimization and CDFT-based MD simulations (CDFT-MD).\cite{Ahart22jctc}
The total force on atom $i$ in CDFT is given by
\begin{equation}
\mathbf{F}_{\mathrm{tot},i} = \mathbf{F}_{i} + \mathbf{F}_{ci},
\label{eq:total_force}
\end{equation}
where \(\mathbf{F}_{i}\) is the standard DFT force and \(\mathbf{F}_{ci}\) is the additional contribution arising from the constraint:
\begin{equation}
    \mathbf{F}_{ci}= -V \int n(\mathbf{r}) \frac{\partial w(\mathbf{r},\mathbf{R})}{\partial \mathbf{R}_{i}}\,\mathrm{d}\mathbf{r}
    \ensuremath{,}
    \label{eq:force}
\end{equation}
Here, \(\mathbf R\) collects all nuclear coordinates and \(\mathbf R_i\) those of atom \(i\).

The analytic forces enable geometry optimization and energy-conserving CDFT-MD at a computational overhead that remains modest relative to conventional DFT-MD. Their validation, convergence requirements, and representative MgO and aqueous Ru$^{2+}$/Ru$^{3+}$ benchmarks are detailed in Section~VIII~A in the SI.\cite{Ahart22jctc}

\subsection{Absolutely Localized Molecular Orbitals, Energy Decomposition, and Localized-Orbital Molecular Dynamics}
\label{sec:almo-eda-md}

ALMOs are MOs whose expansion coefficients are restricted to the AO basis functions of a chosen molecular fragment.\cite{Khaliullin2007,Khaliullin2008,Khaliullin2013} For a fragment $A$, an occupied ALMO can be written schematically as
\begin{equation}
\psi_i^A(\mathbf{r}) = \sum_{\mu \in A} C_{\mu i}^A \chi_\mu(\mathbf{r}),
\end{equation}
where \(i\) labels an occupied ALMO, \(\mu\) labels an AO \(\chi_\mu\) on fragment \(A\), and \(C_{\mu i}^A\) is the corresponding expansion coefficient.
Consequently, interfragment charge transfer is quenched by construction while each fragment is still allowed to polarize in the electrostatic and XC field of its environment. This locality constraint differs from CDFT: CDFT targets a prescribed density partition and produces charge- or spin-localized diabatic states, whereas ALMO builds a fragment-localized occupied space and then measures how much the energy and density change when the localization constraint is relaxed.

In CP2K, the ALMO implementation is part of the \textsc{Quickstep} DFT machinery and is activated through the \texttt{ALMO\_SCF} infrastructure.\cite{Khaliullin2013} The method exploits the natural molecular block structure of weakly interacting systems: the block-diagonal ALMO reference is optimized with diagonalization, preconditioned conjugate-gradient, or trust-region algorithms, and sparse matrix filtering is used to retain the linear-scaling character for large systems. Atomic or molecular initial guesses and always-stable predictor--corrector (ASPC) extrapolation from previous steps make the approach suitable not only for single-point calculations, but also for Born--Oppenheimer and Car--Parrinello-like AIMD.\cite{Scheiber2018}

The block-diagonal ALMO state provides the localized reference for ALMO energy decomposition analysis (ALMO-EDA).\cite{Khaliullin2007,Khaliullin2008} The periodic condensed-phase formulation and its combination with AIMD were demonstrated and reviewed for bulk liquid water by Khaliullin and K{\"u}hne.\cite{KhaliullinKuehne2013WaterEDA} In this analysis, the interaction energy is separated into a frozen contribution, a polarization contribution, and a charge-transfer or delocalization contribution
\begin{equation}
\Delta E_{\mathrm{INT}} =
\Delta E_{\mathrm{FRZ}} +
\Delta E_{\mathrm{POL}} +
\Delta E_{\mathrm{CT}} .
\end{equation}
The frozen term is obtained from the interaction of unrelaxed fragment densities and contains permanent electrostatics, Pauli repulsion, and dispersion or XC effects at the chosen DFT level. The polarization term is the energy lowering obtained by relaxing each fragment's orbitals while preserving the ALMO locality constraint. The charge-transfer term is the additional energy lowering obtained when occupied ALMOs are allowed to delocalize into virtual orbitals of the other fragments. CP2K can print this charge-transfer energy decomposition and the corresponding charge-transfer analysis in block-resolved form, making ALMO-EDA a practical tool for understanding hydrogen-bonded liquids, molecular crystals, ion pairs, and weakly interacting molecular assemblies.\cite{kuhne2013,zhang2013,kuhne2014,elgabarty2015,shi2018,yun2019,yun2022} The ALMO-EDA framework has also been extended to analyze bonding between molecules and metallic surfaces, enabling the decomposition of adsorbate--metal interactions into physically interpretable contributions.\cite{staub2019}

ALMO-EDA reference data have also been used to train a non-native neural-network surrogate for electron-delocalization energies in large aqueous systems.\cite{Tahmasbi2025MLEDA}

A central extension of the method is the use of extended ALMOs (XALMOs), in which the delocalization space is enlarged from a single fragment to a finite domain of neighboring fragments.\cite{Khaliullin2013,Scheiber2018} CP2K supports non-self-consistent and self-consistent delocalization corrections, as well as fully delocalized reference calculations, allowing one to interpolate between a strictly localized ALMO reference, a compact local charge-transfer description, and the conventional delocalized KS solution. The XALMO radius therefore controls a physically transparent approximation: short-range polarization and charge transfer are included explicitly, while long-range delocalization is suppressed unless it is needed by the system.

This ALMO machinery is particularly useful for dynamics because the localized electronic structure changes smoothly along trajectories of weakly interacting systems. CP2K's ALMO-MD capability uses ASPC-extrapolated ALMO/XALMO orbitals as initial guesses for successive MD steps and can therefore reduce the cost of sampling molecular liquids and molecular crystals when the relevant electronic response remains local.\cite{Scheiber2018} In the present review, ALMO occupies a middle ground: it is not an open-boundary transport method, but it quantifies the local polarization and charge-transfer channels that underlie many condensed-phase electron-transfer and molecular-interaction problems. It therefore naturally precedes the POD and CDFT-coupling discussion, where localized electronic states are converted into explicit electronic couplings.

\subsection{Electronic Couplings from Constrained Density-Functional Theory and Projection-Operator-Based Diabatization}
\label{sec:electronic-couplings-cdft-pod}

Electronic coupling elements $H_{\text{ab}}$ quantify the electronic interaction between two (quasi-)diabatic electronic states $\psi_{\text{a}}$ and $\psi_{\text{b}}$\ensuremath{,}
\begin{equation}
H_{\text{ab}} = \langle \psi_{\text{a}} | \hat{H} | \psi_{\text{b}} \rangle,
\end{equation}
where $\hat{H}$ is the electronic Hamiltonian. Because $\psi_{\text{a}}$ and $\psi_{\text{b}}$ are not eigenstates of the electronic Hamiltonian but rather model states representing a given physical situation, the coupling $H_{\text{ab}}$ is finite. For electron or spin transfer, $\psi_{\text{a}}$ and $\psi_{\text{b}}$ may be
charge- or spin-localized initial (a) and final states (b).
Direct CDFT calculations can provide $H_{\text{ab}}$ from two KS determinants, $\psi_{\text{a}}$ and $\psi_{\text{b}}$, obtained with constraints of the form in Eq.~\eqref{eq:nc} that are chosen to model states a and b.
The CP2K implementation of CDFT electronic-coupling calculations has been described and reviewed previously.\cite{Holmberg2017,Kuehne2020} It has since been applied successfully to modeling charge transport in condensed-phase systems.\cite{Ahart22jacs}

A computationally more efficient and robust way to obtain approximate electronic couplings is via a unitary transformation of the standard KS Hamiltonian, as
in the POD method implemented in CP2K.\cite{Futera17} In this method, however, the diabatic electronic states are no longer KS determinants but simply
one-electron orbitals. Consequently, $H_{\text{ab}}$ is no longer the electronic coupling between $N$-electron wavefunctions as in CDFT, but the electronic coupling between
one-electron orbitals (with indices $\mu$ and $\nu$) in diabatic states a and b, $\bar{\phi}^{(\text{a})}_\mu$ and $\bar{\phi}^{(\text{b})}_\nu$, respectively. Thus, the notation for POD electronic couplings needs
to be supplemented with orbital indices,
$H_{\text{ab}}\rightarrow \bar{H}_{\text{ab},\mu\nu}$. For charge- or spin-transfer problems, $\bar{H}_{\text{ab},\mu\nu}$ closely approximates the electronic coupling from
CDFT, $H_{\text{ab}}$, if $\bar{\phi}^{(\text{a})}_\mu$ and $\bar{\phi}^{(\text{b})}_\nu$ are taken to be the frontier orbitals of the donor and acceptor moieties (\(\mu,\nu=\mathrm{HOMO}\) for hole transfer or \(\mu,\nu=\mathrm{LUMO}\) for electron transfer).\cite{Futera17} The central transformation is summarized below, while its complete derivation and implementation are given in Section~VIII~B in the SI.\cite{Futera17}

Operationally, POD first transforms the KS Hamiltonian to a L\"owdin-orthogonalized AO basis, partitions that basis into fragment subspaces, and diagonalizes each diagonal fragment block. The complete transformation can be summarized as
\begin{equation}
\overline{\mathbf{H}}
=
\mathbf{U}^{\dagger}
\mathbf{S}^{-1/2}\mathbf{H}\mathbf{S}^{-1/2}
\mathbf{U},
\label{eq:pod-compact-transformation}
\end{equation}
where the diagonal blocks of $\overline{\mathbf{H}}$ contain diabatic-state energies and the off-diagonal blocks contain the corresponding interfragment couplings. The matrices \(\mathbf H\) and \(\mathbf S\) are the KS Hamiltonian and AO overlap matrices, while the block-diagonal unitary matrix \(\mathbf U\) diagonalizes the fragment blocks after L\"owdin orthogonalization. The L\"owdin transformation, projector construction, block equations, back-transformation, and post-SCF storage and output strategy are detailed in Section~VIII~B in the SI.\cite{Lowdin70,Kondov07,Futera17}

\begin{figure}[!t]
    \centering
     \includegraphics[width=0.9\linewidth]{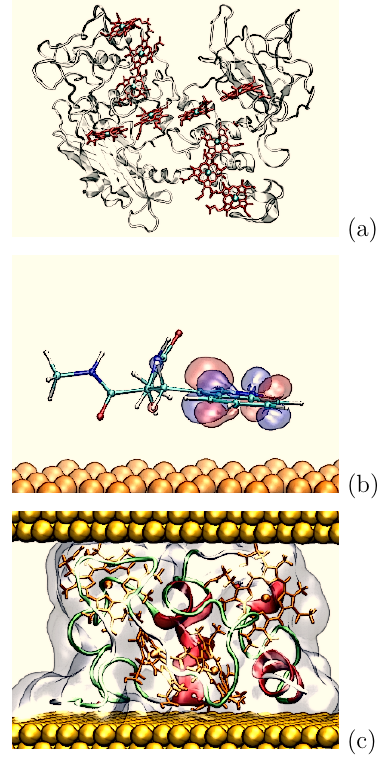}
     \caption{Representative applications of POD electronic couplings. (a) Electron transfer through the heme network of a multi-heme cytochrome. (b) Electronic coupling between a molecular adsorbate and a solid surface. (c) Coherent tunneling across a metal--biomolecule--metal junction.}
    \label{fig-appl}
\end{figure}

POD electronic couplings were benchmarked against high-level \textit{ab-initio} electronic-structure calculations on the HAB11\cite{Futera17} and HAB79 databases,\cite{ziogos21}
of organic dimers and were found to give rather accurate values with errors of typically less than 10\%. The method was then successfully used to compute electronic couplings,
rate constants, and fluxes for electron transfer between heme cofactors in multi-heme cytochromes, such as STC, MtrC, and MtrF [Fig.~\ref{fig-appl}(a)].\cite{Jiang17,vanWonderen19,Jiang20}
A key advantage of POD over other methods for electronic coupling calculations is its ability to obtain a complete set of coupling elements between one-electron
orbital pairs. This feature is particularly beneficial for describing charge transfer at heterogeneous interfaces between molecules and semiconductor or metal surfaces,\cite{Futera17,Futera21}
playing important roles in electrochemistry, molecular electronics, surface catalysis, and dye-sensitized solar cells. It was shown that POD could be used to compute electronic
coupling matrix elements for electron injection from an excited dye molecule to TiO$_2$ and for electron transfer from molecules to metal slabs [Fig.~\ref{fig-appl}(b)].\cite{Futera17}
Finally, the POD method in combination with the DFT+$\Sigma$ method,\cite{Neaton2006} proved very powerful for the calculation of coherent tunneling currents in molecular and
extended biomolecular electronic junctions (Fig.~\ref{fig-appl}c), providing insight into atomistic and electronic structure in molecular electronics and
bioelectronics.\cite{Futera20,Futera23}

\subsection{Linear-Response Transport: Kubo Formalism}
\label{sec:kubo-transport}

The Kubo transport implementation in CP2K provides an equilibrium, linear-response
way to obtain electronic transport coefficients for thermally disordered structures.
It is complementary to the open-boundary approaches of Sections~\ref{sec:hairy-probes} and~\ref{sec:negf-transport}: no external
leads, reservoir self-energies, or finite-bias steady state are introduced.
Instead, the conductivity is obtained from the electronic Hamiltonian of a
finite supercell that represents one thermally accessible configuration of the
material. This is the practical transport step in the finite-temperature
first-principles program developed for disordered crystals: AIMD supplies the
configuration space and Gibbs measure of the atomic degrees of freedom, while a
finite-temperature Kubo formula with dissipation supplies the electronic
transport coefficients for the Hamiltonians associated with those
configurations.\cite{KuhneProdan2018,KuhneHeskeProdan2020,EfremkinHeskeKuehneProdan2026}

For a configuration $\omega$ sampled from the thermodynamic ensemble, the
\textsc{Quickstep} calculation yields the atomic geometry, the AO overlap
matrix \(\mathbf S_\omega\), and the KS Hamiltonian \(\mathbf H_\omega\). The finite-volume
Kubo--Greenwood expression implemented in CP2K evaluates the conductivity tensor
as a function of chemical potential $\mu$, electronic temperature $T$, and a
phenomenological dissipation parameter $\eta$:
\begin{equation}
\begin{aligned}
\sigma_{\alpha\beta}(\mu,T,\eta)
&= \frac{e^2}{\hbar\,|\mathcal{V}_d|}
\sum_s \sum_{mn}
  \Delta X^{(s,\alpha)}_{mn}
  \Delta X^{(s,\beta)}_{nm} \\
&\quad \times
\frac{f_{n s}(\mu,T)-f_{m s}(\mu,T)}
     {\varepsilon_{n s}-\varepsilon_{m s}} \\
&\quad \times
\frac{\eta}{\eta^2+(\varepsilon_{n s}-\varepsilon_{m s})^2} .
\end{aligned}
\label{eq:kubo_transport_sigma}
\end{equation}
The indices \(\alpha\) and \(\beta\) denote Cartesian transport directions, while \(m\) and \(n\) label finite-volume eigenstates.
Here $f_{ns}$ are Fermi occupations, $s$ labels the spin channel,
$\varepsilon_{ns}$ are eigenvalues of the finite-volume Hamiltonian, and
$|\mathcal{V}_d|$ is the normalization measure appropriate to the dimensionality
$d$ of the periodic transport subspace. The matrix elements
$\Delta X^{(s,\alpha)}_{mn}$ are the eigenbasis representation of the
commutator kernel between the position operator and the Hamiltonian. In the
localized Gaussian basis used by CP2K this kernel is constructed from the
minimum-image displacement between the atoms carrying two basis functions
\begin{equation}
  [\mathbf X_\alpha,\mathbf H]_{ij}
  \;\rightarrow\;
  \left[\boldsymbol{\Pi}_d
  \bigl(\mathbf{R}_{a(j)}-\mathbf{R}_{a(i)}\bigr)\right]_\alpha
  H_{ij},
\label{eq:kubo_transport_commutator}
\end{equation}
where $a(i)$ maps AO $i$ to its atom and \(\boldsymbol{\Pi}_d\) projects the displacement onto the
periodic transport subspace. Thus, \(i\) and \(j\) are AO indices, \(H_{ij}\) is a Hamiltonian-matrix element, and \(\mathbf R_{a(i)}\) is the position of the atom carrying AO \(i\). This position--commutator construction is the
finite-supercell analogue of the velocity operator entering the Kubo formula.

Kubo transport is implemented as a native post-SCF property. If the SCF calculation supplies a complete canonical MO set, CP2K reuses its eigenvalues and coefficients. For orbital-transformation (OT) calculations or incomplete canonical spaces, it instead forms the symmetric inverse square root of the AO overlap matrix, diagonalizes the orthonormal Hamiltonian \(\mathbf H^{\perp}=\mathbf S^{-1/2}\mathbf H\mathbf S^{-1/2}\), and transforms the position--Hamiltonian commutator kernels to the resulting eigenbasis. This matrix-based construction is independent of the real-space-grid representation and is therefore shared by GPW and GAPW calculations.

The same $\Gamma$-supercell formulation applies in three, two, and one periodic dimensions because the projector \(\boldsymbol{\Pi}_d\) and normalization measure $|\mathcal{V}_d|$ are constructed from the actual periodic cell vectors. Fully periodic three-dimensional cells and isolated finite boxes use volume normalization and report conductivity in \texttt{S/cm}. Slabs use the periodic area and yield a sheet conductivity in \texttt{S}. Wires use the periodic length and report the corresponding one-dimensional coefficient in \texttt{S*m}. Tilted slabs and non-orthorhombic ribbons are consequently handled without introducing an artificial dependence on vacuum padding. Thermally disordered AIMD snapshots can be analyzed directly without exporting overlap and Hamiltonian matrices. Input controls, carrier-density conventions, neutral-chemical-potential determination, and further normalization details are specified in Section~VIII~C in the SI.

\subsection{Hairy Probes Formalism with Electronic Open Boundaries}
\label{sec:hairy-probes}

%https://pubs.acs.org/doi/full/10.1021/acs.jpclett.3c03615

\begin{figure*}[ht]
  \centering
  \includegraphics[width=0.98\linewidth]{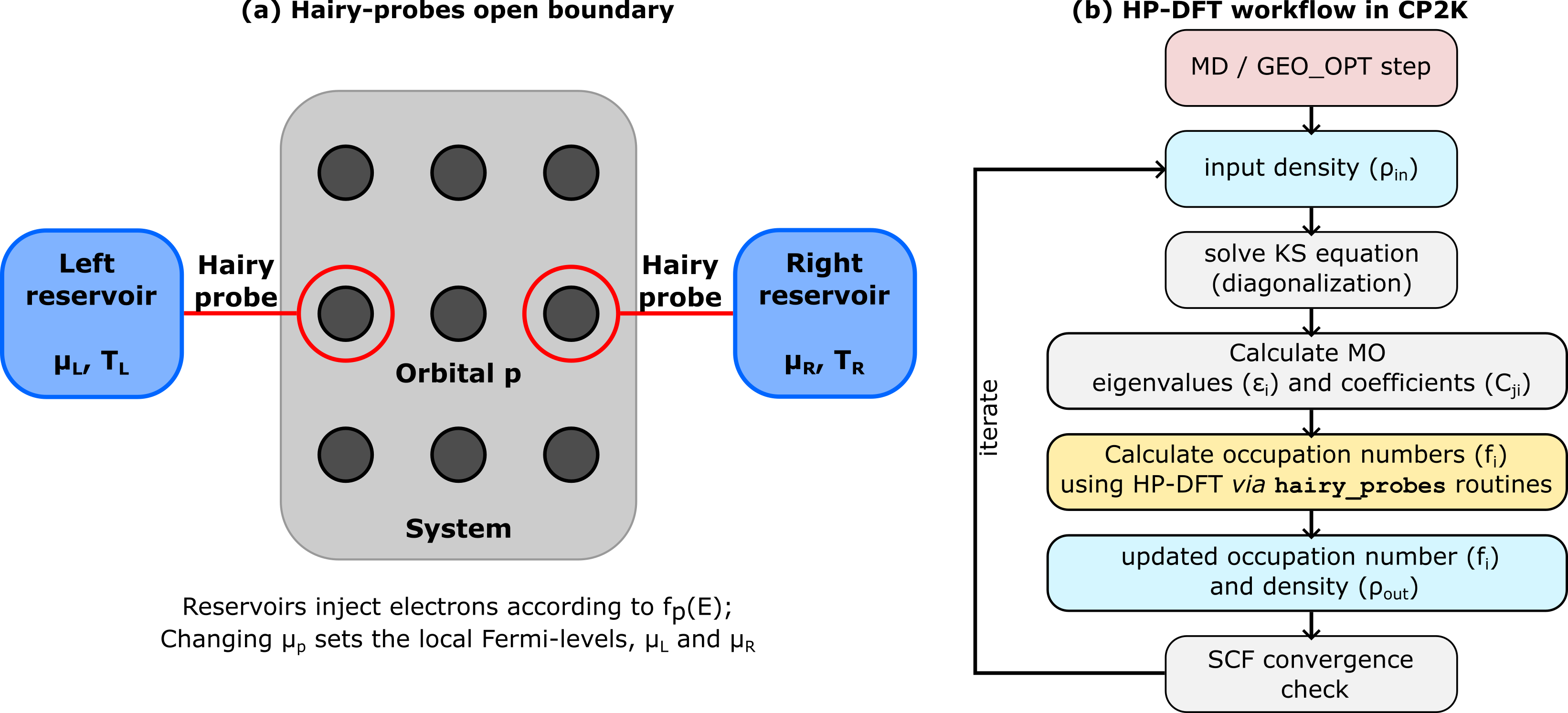}
  \caption{Hairy Probes DFT concept and CP2K implementation workflow.
  (a) Hairy Probes connect selected AOs to reservoirs
  defined by their electrochemical potentials $\mu_{L,R}=
  \bar\mu+\Delta\mu_{L,R}$ and electronic temperatures $T_{L,R}$.
  Changing $\Delta\mu_{L,R}$ imposes local Fermi-level control while
  the reservoir Fermi function controls injection. (b) In \textsc{Quickstep},
  Hairy Probes DFT enters after diagonalization: the \texttt{hairy\_probes}
  routines use the MO eigenvalues and coefficients, the atom-to-AO
  mapping, and the probe input to construct the reservoir-weighted
  occupations $f_i$ and density \(n_{\rm out}\) for the next SCF
  iteration.}
  \label{fig:hpdft_method}
\end{figure*}

Electrified interfaces are electronically open systems: in experiment,
the electrode potential is controlled by an external circuit that can
supply or remove electrons. Conventional KS DFT simulations,
however, are usually performed at fixed electron number and with a
single Fermi level for the whole simulation cell, which makes direct
potential control difficult, especially in explicit-solvent models
where interfacial charge transfer changes the electrode charge and
hence its potential. Hairy Probes DFT (HP-DFT) offers a complementary,
lightweight formulation of electronic open boundaries in CP2K: it makes the
electrons open by connecting selected atoms, and therefore their
associated AO subspaces, to simple electron reservoirs
while retaining the standard \textsc{Quickstep} SCF workflow
(Fig.~\ref{fig:hpdft_method}).\cite{Horsfield2016,Zauchner2018,Buraschi2024}
Because only the construction of the orbital occupations is modified,
the per-step cost is essentially that of an ordinary DFT calculation:
HP-DFT therefore enables AIMD on the
open-boundary stationary electronic state of large electrochemical
cells under direct potential control, at the same cost as standard
AIMD.\cite{Buraschi2024} The dynamics is adiabatic in the
sense of timescale separation: the electrons are assumed to reach
the HP-DFT stationary state instantaneously at each nuclear
configuration, rather than in the thermodynamic sense. We emphasize that the electronic state
propagated here is the open-boundary stationary state of the system
coupled to the reservoirs (in the $\Gamma_p\!\to\!0$ limit
implemented in CP2K, equivalently a constrained-occupation
KS state with local Fermi levels imposed by the probes)
rather than the closed-system Born--Oppenheimer ground state of
ordinary AIMD.

Conceptually, a Hairy Probe is a virtual atomically thin lead attached
to part of the simulated system. Each probe $p$ is connected to a
reservoir defined by an electrochemical potential $\mu_p$ and an
electronic temperature $T_p$. The reservoirs inject electrons
according to their Fermi functions $f_p(E)$. Changing $\mu_p$ is
therefore the practical way of imposing local Fermi-level control on
the electrode region. The formalism originates from the
Lippmann--Schwinger scattering description of electrons coupled to
external reservoirs and, for mean-field electrons at finite probe
coupling, can be cast in the same transport picture as
NEGF.\cite{McEniry2007,Horsfield2016} In the wide-band probe
approximation each probe contributes a local, energy-independent
retarded self-energy on the orbital it is attached to
\begin{equation}
  \Sigma_p^R \;\simeq\; -\frac{i}{2}\Gamma_p,
  \qquad
  \tau_p \;=\; \frac{\hbar}{\Gamma_p},
  \label{eq:hp-selfenergy}
\end{equation}
where the imaginary term allows electronic amplitude to enter or
leave the finite simulation cell. The coupling parameter $\Gamma_p$
sets the reservoir exchange rate, the probed-orbital lifetime $\tau_p$,
and its level broadening.

For electrochemical applications in CP2K, HP-DFT is used in the very
weak-coupling limit, $\Gamma_p\!\to\!0$. In this limit the KS
eigenstates are essentially unchanged by the probes and the steady
electronic current through them vanishes. The open-boundary effect is
retained instead in the occupations, which become a probe-weighted
average of the reservoir Fermi functions. This is the regime
appropriate for electrochemical full cells, where the slow response
is governed by ionic motion, solvent reorganization, and interfacial
chemistry rather than by ballistic electronic transport through the
leads.

In the weak-coupling implementation, the KS eigenstates remain essentially unchanged and the open-boundary condition enters through the orbital occupations. For orbital $i$, the reservoir distributions are weighted by its AO amplitude on the probed atoms:
\begin{equation}
  f_i =
  \frac{
  \alpha\,\bar f(\epsilon_i)\sum_s |C_{j_s i}|^2
  +
  \sum_p f^{(p)}(\epsilon_i)|C_{j_p i}|^2
  }{
  \alpha\sum_s |C_{j_s i}|^2
  +
  \sum_p |C_{j_p i}|^2
  },
  \label{eq:hp-occ}
\end{equation}
where \(\ensuremath{C_{j_p i}}\) and
\(\ensuremath{C_{j_s i}}\) are the coefficients of MO \(i\) on the AOs \(j_p\) and \(j_s\) connected to main and solution probes, respectively, and \(\epsilon_i\) is the MO energy.
The functions
$f^{(p)}(\epsilon_i)$ and $\bar f(\epsilon_i)$ are Fermi-like
distributions for the main and the solution probes, fixed
respectively by the probe parameters $(\mu_p, T_p)$ and by the system
reference $(\bar\mu, \bar T)$. The index $j_p$ labels the AO
components on atoms coupled to a main probe $p$, which imposes the
electrochemical potential of an electrode. The index $j_s$ labels the AO
components coupled to the more weakly weighted \emph{solution
probes}, attached to electrolyte atoms to prevent unphysical
depopulation of states with little overlap with the electrode probes.
The dimensionless parameter $\alpha\!\ll\!1$ controls the weight of
the solution-probe contribution.
The AO density-matrix expansion and explicit reservoir distributions are given in Section~VIII~D in the SI.

The probe offsets are measured relative to a self-consistently floating reference level chosen to preserve the electron count of the neutral, unbiased cell. Potential control therefore maintains overall charge neutrality without a compensating background. The reference-level construction is detailed in Section~VIII~D in the SI.

\subsubsection{Implementation in \textsc{Quickstep}}

In \textsc{Quickstep}, the usual diagonalization-based SCF sequence is
retained. At each iteration CP2K obtains the KS eigenvalues
and MO coefficients in the standard way. A dedicated
\texttt{hairy\_probes} module then replaces the smearing step by
Eq.~\eqref{eq:hp-occ}, and the resulting occupations define the
density matrix used in the next SCF iteration
(Fig.~\ref{fig:hpdft_method}b).
\begin{figure*}[ht]
  \centering
  \includegraphics[width=0.98\linewidth]{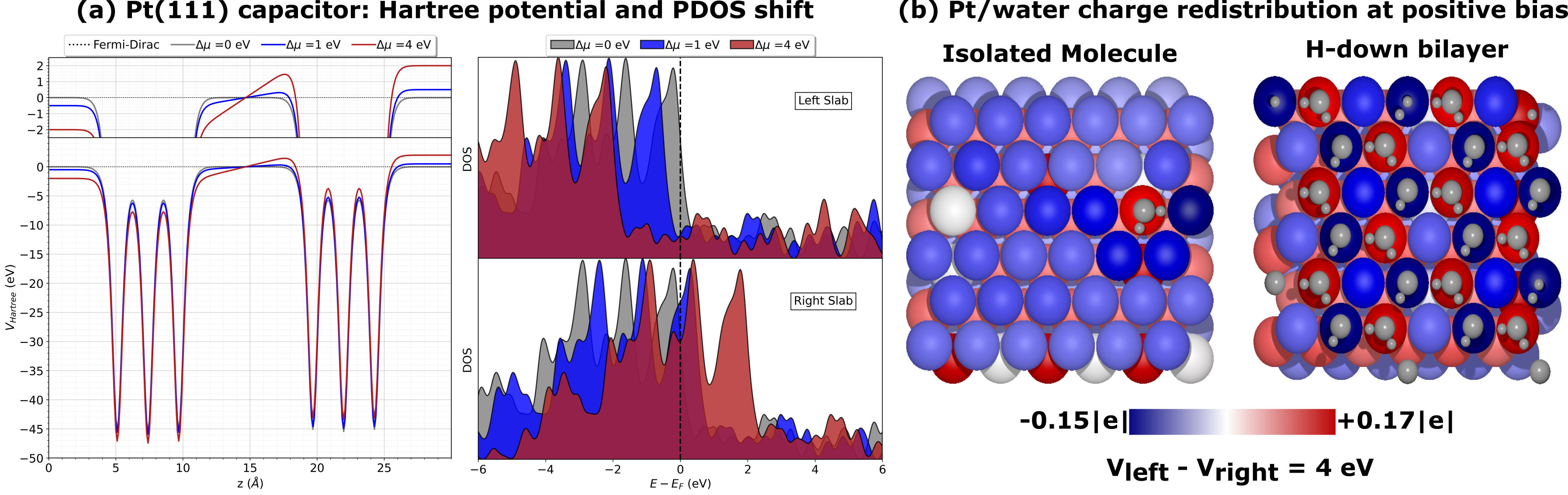}
  \caption{Validation and application of Hairy Probes DFT under potential
  control. (a) Pt(111) parallel-plate capacitor: at finite
  $\Delta\mu$ the planar-averaged Hartree potential develops the
  expected drop and the Pt-slab projected densities of states shift
  rigidly with the imposed probe potentials, validating electrostatic
  and electronic potential control. (b) Pt(111)/water interface at
  positive potential: Hairy Probes DFT reveals the charge redistribution
  associated with water chemisorption: adsorbed water becomes
  positively charged and a local Pt--water polarization center forms,
  with the negative counterpart of the O--Pt bond delocalized over
  the neighboring Pt atoms. Adapted from
  Ref.~\citenum{Buraschi2024} under the Creative Commons Attribution 4.0 International License.}
  \label{fig:hpdft_examples}
\end{figure*}
The implementation supports restricted and unrestricted calculations as well as $\Gamma$-point and $k$-point sampling. Since only the occupation construction is replaced, its overhead is negligible and HP-DFT can be combined directly with the AIMD machinery discussed in Section~\ref{sec:aimd-finite-temperature-spectra}. Probe input, atom-to-AO mapping, solver restrictions, and practical controls are documented in Section~VIII~D in the SI and in the CP2K reference manual.\cite{CP2KHairyProbesManual}

The Pt(111) parallel-plate capacitor is a compact validation system
(Fig.~\ref{fig:hpdft_examples}a). At finite electrochemical-potential
difference $\Delta\mu = \Delta\mu_L - \Delta\mu_R$, the
planar-averaged Hartree potential develops the expected linear drop,
the projected densities of states of the left and right Pt slabs
shift according to the imposed probe potentials, and the capacitance
extracted from the Bader charge accumulated on the internal surfaces,
$C=Q/\Delta\mu$, agrees with the classical parallel-plate
value.\cite{Buraschi2024}

As an application to electrified interfaces, HP-DFT has been used to
follow the charge redistribution at Pt(111)/water under direct
potential control (Fig.~\ref{fig:hpdft_examples}b). Tuning the probe
potentials resolves how small changes in applied bias modify the
charge state of the Pt substrate and of the chemisorbed water layer:
at very positive potential, chemisorbed water donates electron
density to Pt and a locally positive Pt--water center forms, with the
negative counterpart of the O--Pt bond delocalized over neighboring
Pt atoms. This polarization motif rationalizes the potential response
of the electrical double layer: positive potentials stabilize
higher coverages of positively charged chemisorbed water, whereas
negative potentials favor reorientation, flip-flop events, or
desorption. It is also consistent with first-principles analyses
identifying water coverage, water charge, and interfacial
polarization as the key microscopic contributors to surface
potential, adsorbate chemistry, and
capacitance.\cite{DarbyCucinotta2022,Buraschi2024,Raffone2025}

HP-DFT therefore fills an intermediate methodological niche in CP2K. It is more directly connected to electronic open boundaries than the
closed-cell finite-field response approaches of Section~\ref{sec:dens-funct-pert}, and much less expensive
than full DFT+NEGF transport. It is best suited to electrochemical
simulations propagating explicit electrode--electrolyte structures
under potential control without solving for a steady ballistic
current. For current-carrying junctions and transport observables,
the CP2K--SMEAGOL DFT+NEGF interface of Section~\ref{sec:negf-transport} is the appropriate
complementary method.

\begin{figure*}[t]
\centering
\includegraphics[width=0.72\linewidth]{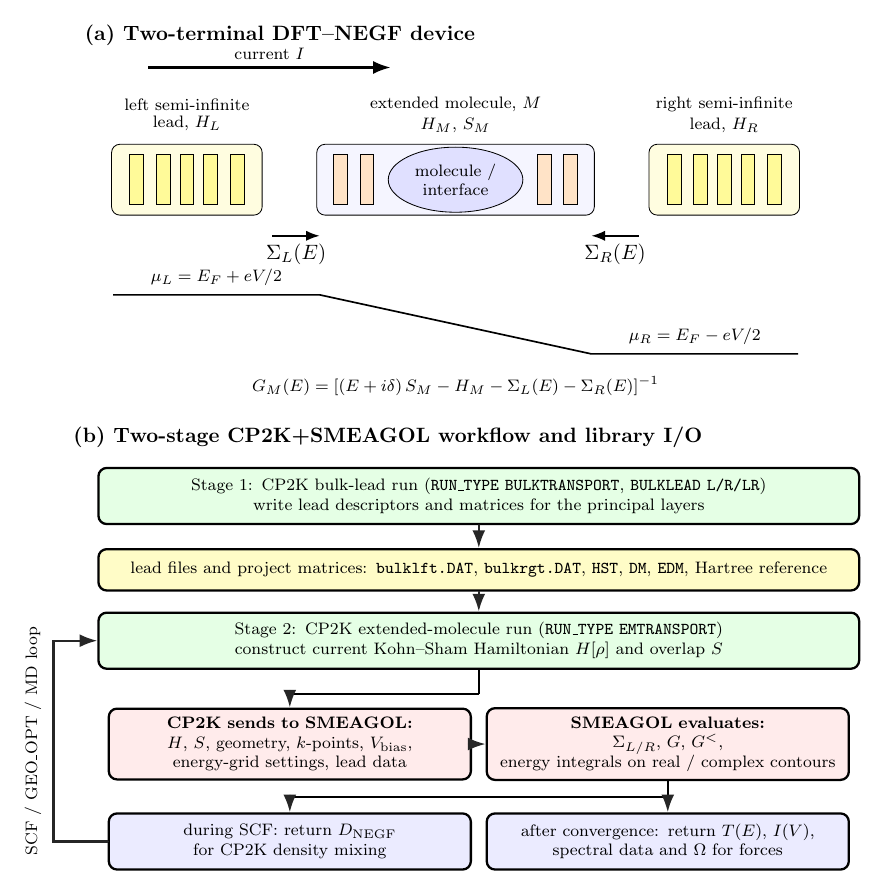}

\caption{Physical and computational structure of CP2K+SMEAGOL.
(a) Two-terminal DFT+NEGF partitioning into left lead, extended molecule, and right lead. The bias-drop line indicates the lead electrochemical potentials $\mu_L=E_F+eV/2$ and $\mu_R=E_F-eV/2$, while the semi-infinite leads enter the finite device through the self-energy matrices \(\boldsymbol{\Sigma}_L(E)\) and \(\boldsymbol{\Sigma}_R(E)\).
(b) Two-stage CP2K+SMEAGOL workflow and library I/O. A bulk-lead CP2K run first writes the principal-layer lead descriptors and matrices. The extended-molecule run then sends \(\mathbf H\), \(\mathbf S\), the geometry, $k$ points, the applied bias and the lead data to SMEAGOL, which constructs \(\boldsymbol{\Sigma}_{L/R}\) and integrates \(\mathbf G\) and \(\mathbf G^<\) on the real and complex contours. During the SCF cycle SMEAGOL returns \(\mathbf D_{\rm NEGF}\) for CP2K density mixing. After convergence it additionally returns $T(E)$, $I(V)$, spectral data and the energy-weighted density matrix \(\boldsymbol{\Omega}\) used in Eq.~\eqref{eq:negf_force}.}
\label{fig:negf_workflow}
\end{figure*}

\subsection{Quantum Transport under External Bias Potentials: Nonequilibrium Green's Functions}
\label{sec:negf-transport}

Many spectroscopic and dynamical observables relevant to operando conditions arise in systems that are not electronically closed. Biased molecular junctions, electrochemical interfaces, atomic contacts, and nanoscale capacitors exchange charge with external reservoirs held at different electrochemical potentials. This situation goes beyond a canonical DFT description with a fixed number of electrons in the simulation cell. It is also distinct from the finite-field response methods of Section~\ref{sec:dens-funct-pert}, which polarize a closed periodic system, and from the Hairy Probes formalism of Section~\ref{sec:hairy-probes}, which imposes local Fermi-level control through weakly coupled probes without sustaining a steady current through the extended molecule. The complementary approach described here is DFT+NEGF, which provides an explicit open-boundary description of current-carrying steady states. The CP2K+SMEAGOL interface couples the GPW/GAPW electronic-structure and MD machinery of CP2K to the SMEAGOL transport engine.\cite{Ahart2024} SMEAGOL was originally developed in conjunction with SIESTA.\cite{Rocha2005,Rocha2006} This makes it possible to compute nonequilibrium densities, electrostatic potential drops, transmission functions, current--voltage characteristics, and current-induced forces within CP2K.

A two-terminal DFT+NEGF device is partitioned into a left lead, an extended molecule $M$, and a right lead, as sketched in Fig.~\ref{fig:negf_workflow}(a). The extended molecule contains the active molecular, interfacial, or nanostructured region together with enough electrode layers to recover bulk-like screening at the boundaries. The semi-infinite leads are held at electrochemical potentials
\begin{equation}
  \mu_L = E_F + \frac{eV}{2}, \qquad
  \mu_R = E_F - \frac{eV}{2},
  \label{eq:negf_mu}
\end{equation}
where $V$ is the applied bias, $E_F$ is the equilibrium Fermi energy, and \(L\) and \(R\) label the left and right leads. In the nonorthogonal AO basis used by CP2K and SMEAGOL, the influence of the semi-infinite leads on the finite extended molecule is encoded in the retarded self-energy matrices \(\boldsymbol{\Sigma}_L(E)\) and \(\boldsymbol{\Sigma}_R(E)\)
\begin{equation}
\begin{gathered}
  \mathbf G_M(E)=
  \left[(E+i\delta)\mathbf S_M-\mathbf H_M
  -\boldsymbol{\Sigma}_L(E)-\boldsymbol{\Sigma}_R(E)\right]^{-1}.
\end{gathered}
  \label{eq:negf_green}
\end{equation}
Here \(\mathbf H_M\) and \(\mathbf S_M\) are the KS Hamiltonian and overlap matrices of the extended molecule. The variable \(E\) is the electron energy, and \(\delta\) is a positive infinitesimal. \(\mathbf G_M(E)\) is the resolvent of the open-boundary effective Hamiltonian \(\mathbf H_M+\boldsymbol{\Sigma}_L+\boldsymbol{\Sigma}_R\): it encodes the response of the extended molecule to a perturbation at energy $E$ and, via the spectral-function matrix \(\mathbf A_M=i(\mathbf G_M-\mathbf G_M^\dagger)\), gives the device's density of states once the leads are coupled in. The self-energies are non-Hermitian matrices: their Hermitian parts shift the bare extended-molecule levels, while their anti-Hermitian parts broaden those levels into resonances of finite lifetime. The corresponding broadening matrices are defined as
\begin{equation}
\begin{gathered}
  \boldsymbol{\Gamma}_\alpha(E)
  =i\left[\boldsymbol{\Sigma}_\alpha(E)
  -\boldsymbol{\Sigma}_\alpha^\dagger(E)\right],
  \qquad \alpha=L,R,
\end{gathered}
  \label{eq:negf_gamma}
\end{equation}
and represent the rate at which electronic amplitude is exchanged between the extended molecule and lead $\alpha$. In this way the infinite open-boundary problem is reduced to a finite, energy-dependent matrix problem on $M$.

The nonequilibrium occupation of the extended molecule is described by the lesser Green's function
\begin{equation}
\begin{aligned}
\mathbf G_M^<(E)
={}&i\mathbf G_M(E)\\[-0.2em]
&\times
\left[\sum_{\alpha=L,R}f_\alpha(E)\boldsymbol{\Gamma}_\alpha(E)\right]
\mathbf G_M^\dagger(E),
\end{aligned}
  \label{eq:negf_lesser}
\end{equation}
where $f_L$ and $f_R$ are the Fermi--Dirac distributions of the two reservoirs. The density matrix is obtained as
\begin{equation}
  \mathbf D_M
  =\frac{1}{2\pi i}\int dE\,\mathbf G_M^<(E),
  \label{eq:negf_density}
\end{equation}
and enters \(\mathbf H_M[\mathbf D_M]\) self-consistently. For numerical integration, CP2K+SMEAGOL separates equilibrium contour contributions from a real-axis nonequilibrium correction and combines left- and right-referenced densities to treat weakly coupled states robustly at high bias. The weighted double-contour expression and integration strategy are given in Section~VIII~E in the SI.\cite{Brandbyge2002}

Once self-consistency is reached, the same Green's functions provide the transport observables. The transmission is
\begin{equation}
\begin{gathered}
  T(E)=\mathrm{Tr}\left[
  \boldsymbol{\Gamma}_L(E)\mathbf G_M(E)
  \boldsymbol{\Gamma}_R(E)\mathbf G_M^\dagger(E)\right],
\end{gathered}
\label{eq:negf_transmission}
\end{equation}
where \(T(E)\) is the energy-resolved transmission probability.
The current follows from the Landauer--B\"uttiker expression
\begin{equation}
  I(V)=\frac{e}{h}\int dE\,T(E)\left[f_L(E)-f_R(E)\right],
\label{eq:negf_current}
\end{equation}
Here, \(I(V)\) is the current and \(h\) is Planck's constant. The spin sum follows the chosen Hamiltonian.

For MD under bias, the central new ingredient is the force on the nuclei in the stationary current-carrying electronic state. In a nonorthogonal localized basis, and with the lead self-energies held fixed for frozen semi-infinite electrodes, the electronic force contribution can be written as
\begin{subequations}
\label{eq:negf_force}
\begin{align}
F_I^{\mathrm{el}}
&=-\mathrm{Tr}\!\left[
\mathbf D_M\frac{\partial \mathbf H_M}{\partial R_I}
\right]
+\mathrm{Tr}\!\left[
\boldsymbol{\Omega}_M\frac{\partial \mathbf S_M}{\partial R_I}
\right],
\label{eq:negf_force_electronic}\\
\boldsymbol{\Omega}_M
&=\frac{1}{2\pi i}\int dE\,E\,\mathbf G_M^<(E).
\label{eq:negf_energy_weighted_density}
\end{align}
\end{subequations}

where \(R_I\) is a nuclear coordinate, \(F_I^{\mathrm{el}}\) is its electronic force component, and \(\boldsymbol{\Omega}_M\) is the energy-weighted density matrix defined in the second line.
The second term is the Pulay contribution associated with the nonorthogonal basis, but the energy-weighted density matrix \(\boldsymbol{\Omega}_M\) cannot be obtained from a closed-system eigenvalue sum: the open device has a continuous spectrum broadened by \(\boldsymbol{\Sigma}_L\) and \(\boldsymbol{\Sigma}_R\). It must instead be evaluated from the same Green-function integration as the density, with an additional factor of $E$ in the integrand. Retaining this term is essential. Omitting it gives qualitatively incorrect zero-bias forces in the Au-wire benchmark. The bias-dependent part of Eq.~\eqref{eq:negf_force} is the electron-wind component of the current-induced force, namely the momentum transfer from the steady electronic flow to the ions. This contribution is, in general, nonconservative and is the microscopic origin of effects such as electromigration, bond elongation, and junction rupture in biased nanoscale conductors,\cite{DiVentra2008,Dundas2009,Zhang2011} and of current-induced phonon renormalization in molecular junctions.\cite{Bai2016}

\subsubsection{CP2K+SMEAGOL Implementation}

The division of labor between CP2K and SMEAGOL preserves the established CP2K infrastructure for Hamiltonian construction, density mixing, forces, geometry optimization, and MD. At each SCF iteration, CP2K constructs the extended-molecule Hamiltonian \(\mathbf H_M[n_{\rm in}]\) and overlap matrix \(\mathbf S_M\) and passes them to SMEAGOL together with the geometry, bias, $k$ points, and precomputed lead data. SMEAGOL constructs the lead self-energies, evaluates the Green functions and energy integrals, and returns the NEGF density matrix \(\mathbf D_M\) for the next CP2K density-mixing step. After convergence, it additionally provides $T(E)$, $I(V)$, optional spectral quantities, and the energy-weighted density matrix \(\boldsymbol{\Omega}_M\) required for forces. The interface translates between CP2K's distributed block-compressed sparse-row (DBCSR) matrices and the sparse SMEAGOL/SIESTA representation, which also enables direct cross-validation against SIESTA+SMEAGOL. Thus, SMEAGOL replaces the closed-system density construction inside the SCF cycle rather than the underlying CP2K electronic-structure machinery. At zero bias and with sufficiently screened boundaries, the resulting density reduces to the closed-cell DFT limit.

As summarized in Fig.~\ref{fig:negf_workflow}(b), the calculation proceeds in two stages. A bulk-lead calculation first generates the principal-layer Hamiltonian and overlap data from which the semi-infinite-lead self-energies are constructed. The subsequent extended-molecule calculation combines these lead data with the active junction or interface and enters the self-consistent transport cycle at the desired bias. The corresponding run modes, integration grids, bias controls, and Hartree-reference alignment are documented in Section~VIII~E in the SI.

The semi-infinite-lead self-energies are held fixed, so the extended molecule must contain enough electrode layers to screen charge rearrangement before the lead boundary. The longer-ranged Gaussian basis functions used by CP2K can consequently require additional screening layers or compact lead bases. Practical basis and boundary considerations are given in Section~VIII~E in the SI.

The independent Green-function evaluations over energy and $k$ points provide the principal level of parallelism. Although biased steps are substantially more expensive than closed-boundary DFT, the implementation remains practical for static transport, optimization, and short biased trajectories in systems exceeding $10\,000$ basis functions. Numerical settings and scaling details are given in Section~VIII~E in the SI.\cite{Ahart2024}

% =====================================================================
% Figure 2: validation panels.
% Top:    Au parallel-plate capacitor (cropped left panel).
% Bottom: solvated Au-wire structure + Bader-charge time evolution.
% =====================================================================
\begin{figure}[!tbp]
\centering
\IfFileExists{Figures/fig_negf_capacitor.png}{%
  \includegraphics[width=0.9\linewidth,trim=0 0 0 6,clip]{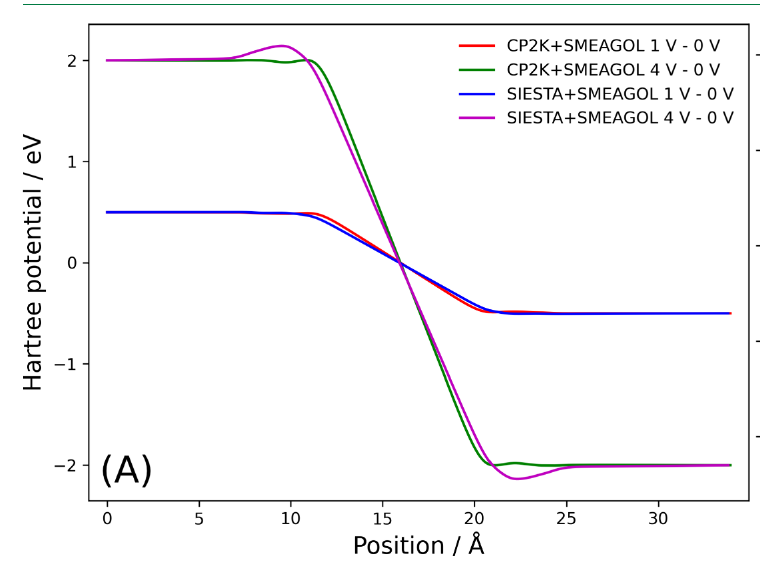}\\[0.5em]
}{%
  \fbox{\parbox{0.62\linewidth}{\centering\itshape Top panel placeholder: Au parallel-plate
    capacitor benchmark. Place \texttt{Figures/fig\_negf\_capacitor.png} for the cropped
    Hartree-potential plot (CP2K+SMEAGOL vs.\ SIESTA+SMEAGOL at 1~V and 4~V).}}%
  \\[0.5em]
}
\IfFileExists{Figures/fig5_solvated_wire.png}{%
  \includegraphics[width=1.0\linewidth,trim=0 0 0 20,clip]{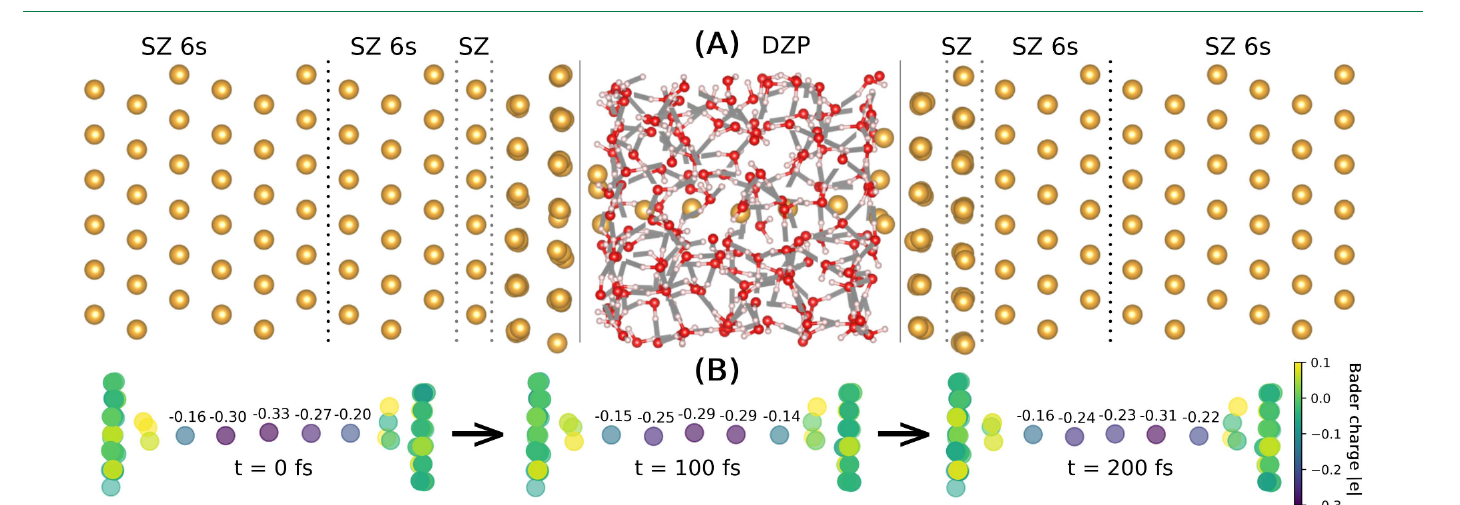}%
}{%
  \fbox{\parbox{1.0\linewidth}{\centering\itshape Bottom panel placeholder: solvated Au-wire
    model and Bader-charge evolution at $t=0,100,200$~fs of biased Born--Oppenheimer MD at
    $V=1$~V. Place \texttt{Figures/fig5\_solvated\_wire.png}.}}%
}
\caption{Representative CP2K+SMEAGOL validation and application results.
(a) Finite-bias Au parallel-plate capacitor benchmark: planar-averaged Hartree-potential drop, with close agreement between CP2K+SMEAGOL and SIESTA+SMEAGOL at 1~V and 4~V applied bias.
(b) Solvated monatomic Au wire used for biased MD at $V=1$~V, together with the evolution of Bader charges along the wire during a 200~fs trajectory.
Together with the Au-wire force and Au--H$_2$--Au relaxation tests discussed in the text, these examples validate electrostatics, forces, transport and biased MD. Panels adapted from Ref.~\citenum{Ahart2024} under the Creative Commons Attribution 4.0 International License.}
\label{fig:negf_results}
\end{figure}

\subsubsection{Validation and Biased Molecular Dynamics}

Au-wire forces, finite-bias capacitor electrostatics, and Au--H$_2$--Au structural response validate the force, potential-drop, and current-induced relaxation components of the interface, with representative results shown in Fig.~\ref{fig:negf_results}(a).\cite{Ahart2024} The complete validation hierarchy is described in Section~VIII~E in the SI.

The most relevant demonstration for condensed-phase operando simulations is biased Born--Oppenheimer MD of a solvated monatomic Au wire connected to Au(111) electrodes, Fig.~\ref{fig:negf_results}(b). The model contains the metallic junction, explicit water, electrode screening layers and semi-infinite leads. The zero-bias transmission remains close to unity near the Fermi level, as expected for an atomic Au contact, while water introduces additional tunneling channels several electronvolts below $E_F$. In a 200~fs trajectory at $V=1$~V, the Bader-charge profile along the wire evolves consistently with migration of electron density to screen the positively biased electrode. These simulations demonstrate the niche of CP2K+SMEAGOL: explicit solvent, realistic electrode models, coherent current flow and current-induced forces can be treated in one DFT+NEGF simulation.

Overall, CP2K+SMEAGOL extends the CP2K spectroscopy and dynamics toolbox from closed-cell polarization response to genuinely open electronic systems under externally imposed bias. It is complementary to finite electric-field, constant-displacement-field, and Hairy Probes approaches. The latter are efficient for closed-cell polarization or weak-coupling potential control, whereas DFT+NEGF is appropriate when observables depend on charge injection, transmission resonances, steady current, or nonconservative current-induced forces.

\subsection{Parametrized Surface Hopping for Charge and Exciton Transport}

The methods discussed in Sections~\ref{sec:cdf-charge-transfer}--\ref{sec:negf-transport} describe either diabatic electronic states and their couplings, equilibrium linear-response transport, or electronically open steady states. A complementary problem arises when electronic charge or excitation energy moves through a molecular material while the nuclei evolve at finite temperature. In such cases the electronic wavefunction may extend over many molecules, and the relevant process is neither a single donor--acceptor electron-transfer event nor a coherent two-terminal current. CP2K therefore also contains parametrized fewest-switches surface-hopping schemes for large-scale nonadiabatic MD: fragment orbital-based surface hopping (FOB-SH) for excess charge transport, Frenkel exciton surface hopping (FE-SH) for excitation-energy transport, and excitonic-state-based surface hopping (X-SH) for exciton dissociation into charge carriers.\cite{Spencer16jcp,Carof17,Carof19,Giannini22nc,Peng22,Ivanovic25,Ivanovic26}

The common strategy is to propagate the nuclei classically while a quantum electronic wavefunction evolves in a compact diabatic basis whose Hamiltonian is parametrized from electronic-structure calculations. This differs from the explicitly electronic-structure-driven trajectory surface-hopping interfaces of Section~\ref{sec:nonadiabatic-md-surface-hopping}: here, expensive excited-state forces and nonadiabatic couplings are replaced by rapidly evaluated site energies, couplings, and their nuclear derivatives. The approach is therefore aimed at nanoscale molecular materials, where charge or exciton dynamics over tens of nanometers and tens of picoseconds would otherwise be inaccessible.

For a basis of $N$ localized electronic states, the electronic Hamiltonian is written in a diabatic state basis as
\begin{equation}
  \hat{H} =
  \sum_{k=1}^{N} H_{kk}(\mathbf{R}) \ket{\phi_k}\bra{\phi_k}
  +
  \sum_{k=1}^{N}\sum_{l\ne k}^{N}
  H_{kl}(\mathbf{R}) \ket{\phi_k}\bra{\phi_l},
  \label{eq:fobsh_hamiltonian}
\end{equation}
where $\phi_k$ denotes a localized diabatic electronic state, $H_{kk}$ is the corresponding site energy, and $H_{kl}$ is the electronic coupling between states $k$ and $l$. In FOB-SH, $\phi_k$ is the state with an excess electron or hole localized on molecule $k$. In FE-SH, $\phi_k$ is a locally excited Frenkel-exciton state. In X-SH, the state space combines locally excited states, charge-transfer states, and the electronic ground state, allowing excitons to dissociate into spatially separated electron--hole pairs. All matrix elements depend on the instantaneous nuclear geometry $\mathbf{R}(t)$ and are updated on the fly during the MD.

The diagonal terms are obtained from a force-field description, for example the General Amber Force Field,\cite{Amber16short} parametrized to reproduce internal reorganization energies or excitation/charge-transfer energetics from DFT, TDDFT, or higher-level \textit{ab-initio} calculations. Electrostatic interactions are evaluated with the damped shifted force (DSF) method.\cite{Giannini25} Off-diagonal charge-transfer and exciton-dissociation couplings are evaluated from the analytic overlap method (AOM), using an approximately linear relation between frontier-orbital overlaps and electronic couplings.\cite{Gajdos14,Ziogos21jcp2} Frenkel-exciton couplings in FE-SH are evaluated in the Coulomb approximation with transition electrostatic potential (TrESP) charges.\cite{Giannini22nc} These ingredients make it possible to evaluate the electronic Hamiltonian, coupling derivatives, and forces at every nuclear time step without repeated full electronic-structure calculations.

The electronic wavefunction associated with each classical trajectory is expanded in the same diabatic basis:
\begin{equation}
  \Psi(t)=\sum_{k=1}^{N} u_k(t)\ket{\phi_k},
\label{eq:fobsh_wavefunction}
\end{equation}
where \(u_k(t)\) is the time-dependent amplitude of diabatic state \(\phi_k\).
The coefficients obey
\begin{equation}
  i\hbar\frac{d u_k(t)}{dt}
  =
  \sum_{l=1}^{N} u_l(t)
  \left[H_{kl}(\mathbf{R})-i\hbar d_{kl}\right],
  \label{eq:fobsh_tdse}
\end{equation}
where $d_{kl}=\langle\phi_k|d/dt|\phi_l\rangle$ is the nonadiabatic coupling element (NACE) between diabatic states. For the localized diabatic bases used in FOB-SH, FE-SH, and X-SH this term is usually small and is commonly neglected after validation.\cite{Carof17,Giannini22nc} The coefficients are propagated in the diabatic representation with a fourth-order Runge--Kutta integrator, while hopping probabilities are evaluated in the adiabatic basis obtained by diagonalizing Eq.~\eqref{eq:fobsh_hamiltonian}.

The active adiabatic state $a$ determines the nuclear force. If \(\mathbf H\) is the matrix representation of Eq.~\eqref{eq:fobsh_hamiltonian} and \(\mathbf U\) diagonalizes it, the force on atom $I$ is obtained from the diabatic Hamiltonian gradients as follows.\cite{Carof17}
\begin{equation}
  \mathbf{F}_{I,a}
  =
  -\left[\mathbf U^{\dagger}
  \left(\boldsymbol{\nabla}_I\mathbf H\right)
  \mathbf U\right]_{aa}.
  \label{eq:fobsh_force}
\end{equation}
The same transformation gives the adiabatic NACE and nonadiabatic coupling vector (NACV) used for Tully hopping, velocity rescaling, and frustrated-hop handling:
\begin{subequations}
\label{eq:fobsh_ad_couplings}
\begin{align}
  d^{\mathrm{ad}}_{ja}
  &=
  \left[\mathbf U^{\dagger}\mathbf D\mathbf U\right]_{ja}
  +
  \left[\mathbf U^{\dagger}\dot{\mathbf U}\right]_{ja},
  \label{eq:fobsh_ad_nace}
  \\
  \mathbf{d}^{\mathrm{ad}}_{I,ja}
  &=
  \frac{
  \left[\mathbf U^{\dagger}
  \left(\boldsymbol{\nabla}_I\mathbf H\right)
  \mathbf U\right]_{ja}}
  {E_j-E_a}
  +
  \left[\mathbf U^{\dagger}\mathbf D_I\mathbf U\right]_{ja}
  \ensuremath{,}
  \label{eq:fobsh_ad_nacv}
\end{align}
\end{subequations}

where \(\mathbf D\) and \(\mathbf D_I\) collect diabatic NACEs and NACVs, respectively. The labels \(a\) and \(j\) denote adiabatic states, \(I\) labels an atom, and \(E_a\) and \(E_j\) are the corresponding adiabatic energies. In typical applications the explicit diabatic-coupling terms are small, so the dominant contributions come from the time dependence and nuclear gradients of the parametrized Hamiltonian. Decoherence corrections, state tracking for trivial crossings, self-consistent hopping probabilities, and a correction for spurious long-range charge-transfer or exciton-transfer events are available. These features are important for maintaining internal consistency and approximate detailed balance at long times and for obtaining converged transport dynamics.\cite{Carof17,Carof19}

\begin{figure*}[p!]
\centering
\includegraphics[width=\textwidth]{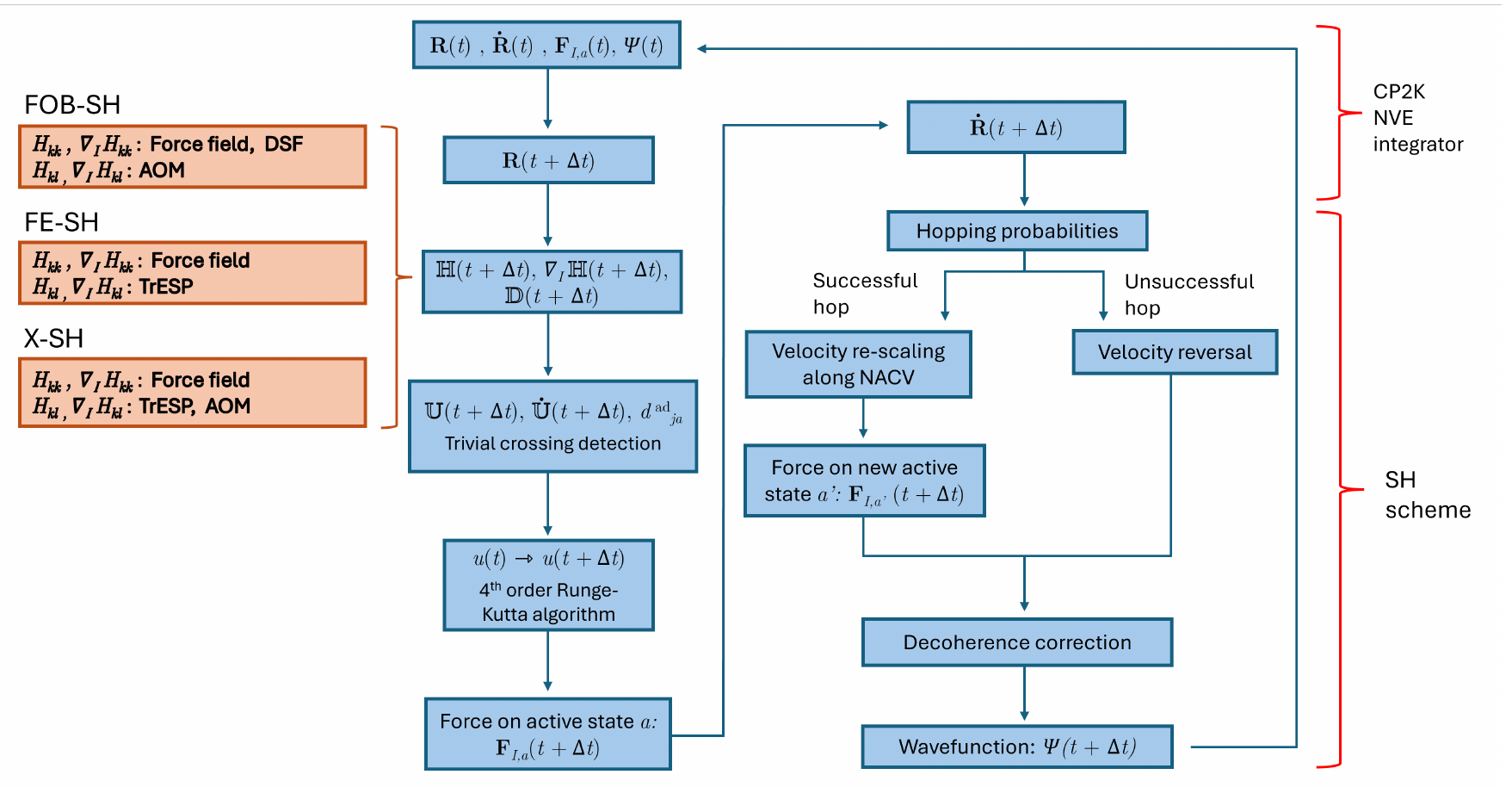}
\caption{Overview of the workflow for FOB-SH, FE-SH, and X-SH.
Methods for calculating the electronic Hamiltonian matrix elements and their nuclear derivatives for the different state spaces are shown in the orange panels. DSF treats electrostatic interactions,\cite{Giannini25} AOM provides electronic coupling matrix elements (transfer integrals),\cite{Gajdos14,Ziogos21jcp2} and TrESP provides excitonic coupling matrix elements.\cite{Giannini22nc} The surface-hopping routines implemented in CP2K propagate the electronic wavefunction $\Psi(t)$ and calculate the nuclear forces $\mathbf{F}_{I,a}(t)$ and $\mathbf{F}_{I,a'}(t)$, which are passed to the built-in CP2K velocity Verlet integrator to propagate the nuclear degrees of freedom (blue panels).  See the main-text discussion for implementation details.}
\label{fig:sh_workflow}
\end{figure*}

The CP2K implementation is summarized in Fig.~\ref{fig:sh_workflow}. The FIST module advances the nuclear positions with the velocity Verlet algorithm and supplies the force-field contribution to the site energies and forces. The FOB-SH, FE-SH, and X-SH routines then assemble the electronic Hamiltonian, compute AOM or TrESP couplings and nuclear derivatives, diagonalize the Hamiltonian, track state order and signs, propagate the electronic wavefunction, evaluate hopping probabilities, and update the active surface. Successful hops trigger velocity rescaling along the NACV, whereas frustrated hops lead to velocity reversal according to the selected scheme. After decoherence correction, the updated wavefunction, active state, forces, positions, and velocities are returned to the next MD step.

\begin{figure*}[p!]
\centering
\includegraphics[width=0.9\linewidth]{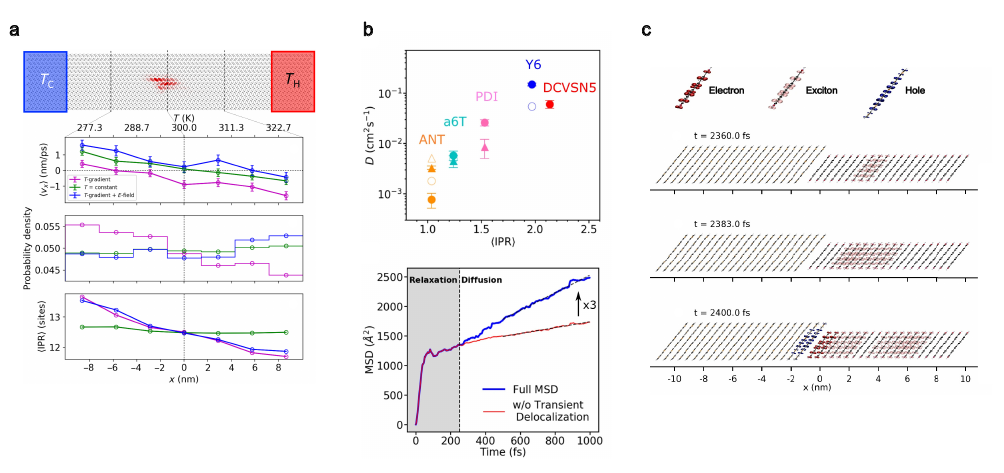}
\caption{Representative applications of FOB-SH, FE-SH, and X-SH. (a) FOB-SH simulation of thermoelectric hole transport in crystalline rubrene, where a hole injected at the hot side drifts toward the cold side as the thermal electronic disorder decreases. (b) FE-SH simulations of exciton transport in molecular organic semiconductors, showing the relation between exciton diffusion and transient wavefunction delocalization. (c) X-SH simulation of exciton dissociation and free-charge generation at a donor--acceptor interface, illustrating hot pathways through delocalized hybrid exciton--charge-transfer states. Panels (a)--(c) are adapted from Refs.~\citenum{Elsner24}, \citenum{Giannini22nc}, and \citenum{Ivanovic25}, respectively, under the Creative Commons Attribution 4.0 International License.}
\label{fig:fobsh_applications}
\end{figure*}

The applications in Fig.~\ref{fig:fobsh_applications} comprise FOB-SH charge transport in organic molecular crystals and thermoelectric transport in rubrene,\cite{GianniniBlumberger2022,Elsner24} FE-SH exciton diffusion linked to transient delocalization,\cite{Giannini22nc,Stojanovic24jctc} and X-SH charge generation through hybrid exciton--charge-transfer states at donor--acceptor interfaces.\cite{Peng22,Ivanovic25,Ivanovic26} In the thermoelectric application, a temperature gradient drives the hole wavefunction from the hot to the cold side of the crystal, while the concomitant decrease in electronic disorder with temperature increases its delocalization. Together, they connect static couplings and steady-state transport to finite-temperature nonadiabatic dynamics and reveal crossovers among localized hopping, band-like transport, and transient delocalization.\cite{GianniniCarofBlumberger2018,Giannini23,Stojanovic24prx}

\section{Nuclear Motion and Finite-Temperature Spectroscopy}
\label{sec:nuclear-motion-finite-temperature-spectroscopy}

Nuclear motion enters spectroscopic observables in two complementary ways: through time-correlation functions, as in infrared, Raman, sum-frequency generation (SFG), and related vibrational spectra, and through configurational or ensemble averages over statistically decorrelated snapshots, as commonly used for electronic, core-level, and magnetic-response spectra. Section~\ref{sec:statistical-spectroscopy-framework} first establishes the sampling and decorrelation framework. Section~\ref{sec:grassmann-extrapolation-aimd} then describes the electronic propagation that makes long AIMD trajectories practical; Section~\ref{sec:aimd-finite-temperature-spectra} connects those trajectories to finite-temperature spectra, and Section~\ref{sec:finite-field-aimd} treats dynamics under imposed electric boundary conditions.

\subsection{Statistical Sampling and Correlation Functions}
\label{sec:statistical-spectroscopy-framework}

MD simulations generate a trajectory in phase space by propagating the equations of motion for a system of interacting particles. For an ergodic stationary process, the long-time average of an integrable observable equals its equilibrium ensemble average; ergodicity does not require a continuous trajectory to visit every individual microstate.\cite{Tuckerman2010,AllenTildesley2017} Formally, for a property $A(\mathbf{r}, \mathbf{p})$ that depends on the positions $\mathbf{r}$ and momenta $\mathbf{p}$ of the particles, this equivalence reads
\begin{equation}
\begin{aligned}
    \langle A \rangle
    &= \lim_{T_s \to \infty} \frac{1}{\ensuremath{T_s}}
    \int_0^{\ensuremath{T_s}} A\bigl(\mathbf{r}(t), \mathbf{p}(t)\bigr)\, \ensuremath{\mathrm{d}t} \\
    &= \int A(\mathbf{r}, \mathbf{p})\,
    \rho(\mathbf{r}, \mathbf{p})\, \ensuremath{\mathrm{d}\mathbf{r}\, \mathrm{d}\mathbf{p}}\ensuremath{,}
\end{aligned}
\end{equation}
where $\rho(\mathbf{r}, \mathbf{p})$ is the stationary equilibrium distribution in the sampled phase space. In practice, the time average is evaluated over a finite trajectory of length $T_s$, and the sampling quality depends on how thoroughly it explores the relevant phase-space regions.

A central quantity in assessing the statistical quality of an MD-generated observable is the time autocorrelation function of its fluctuations, defined by
\begin{equation}
C_A(t) = \langle \delta A(0)\, \delta A(t) \rangle, \quad \delta A(t) = A(t) - \langle A \rangle
\ensuremath{.}
\end{equation}
The normalized integral defines the integrated autocorrelation time
\begin{equation}
\tau_{A,\mathrm{int}}
=
\int_0^\infty
\frac{C_A(t)}{C_A(0)}\,dt .
\label{eq:integrated-autocorrelation-time}
\end{equation}
This quantity measures the loss of statistical information caused by correlations between successive evaluations of \(A\). Configurations separated by \(\Delta t\gg\tau_{A,\mathrm{int}}\) are approximately decorrelated rather than strictly independent. For a sufficiently long stationary trajectory, \(N_{\mathrm{eff}}\simeq T_s/(2\tau_{A,\mathrm{int}})\), and the standard error of the sample mean is approximately \(\operatorname{SE}(\overline A)\simeq\sigma_A/\sqrt{N_{\mathrm{eff}}}\), where \(\sigma_A^2=C_A(0)\) is the variance of the observable. Thus, the required simulation length is tied directly to the target property's relaxation dynamics.

These statistical-mechanical concepts, well established for structural and thermodynamic observables, apply equally to the calculation of spectra from MD trajectories. Spectroscopic properties, including electronic spectra such as ultraviolet--visible (UV--vis) optical absorption and XAS, as well as NMR chemical shifts, vibrational and IR spectra, Raman scattering, and their higher-order derivatives, are all sensitive to the instantaneous atomic configuration and its fluctuations. A rigorous treatment must therefore account for the full distribution of sampled microstates, rather than relying on a single, static representative structure. This is particularly important in condensed-phase systems, where thermal fluctuations, conformational disorder, and explicit environmental effects (solvation, hydrogen bonding, electrostatic screening) can lead to significant broadening and shifting of spectral features that cannot be captured by calculations on a single geometry.

In this framework, the spectrum $S(\omega)$ is obtained as a configurational average over an ensemble of $N$ statistically decorrelated snapshots $\{\mathbf{R}_k\}$ drawn from the MD trajectory:
\begin{equation}
S(\omega) = \frac{1}{N} \sum_{k=1}^{N} S\bigl(\omega; \mathbf{R}_k\bigr),
\end{equation}
where $S(\omega; \mathbf{R}_k)$ is the spectrum calculated for the $k$-th configuration at the level of theory appropriate for the property of interest (e.g., TDDFT for optical spectra, or gauge-including AOs for NMR). The central practical challenge is snapshot selection: the configurations must be sufficiently decorrelated to provide approximately independent samples, yet the spectral calculation is usually too expensive to perform at every MD step.

A pragmatic and recommended strategy is therefore to identify a spectral proxy, i.e., a related property that is computationally inexpensive and can be evaluated on every frame of the trajectory, and to use its autocorrelation function to determine \(\tau_{A,\mathrm{int}}\) and hence the appropriate stride for snapshot selection. For electronic spectra, suitable proxies include low-level estimates of excitation energies, the HOMO--LUMO gap, or simple geometric descriptors known to correlate with the spectral observable of interest. For NMR or vibrational properties, relevant internal coordinates (bond lengths, angles, dihedral angles, or hydrogen bond distances) often serve this purpose. Once \(\tau_{A,\mathrm{int}}\) has been estimated from the proxy, configurations can be selected at intervals of several \(\tau_{A,\mathrm{int}}\) to reduce serial correlation, and the expensive spectral calculation is then performed only on this reduced, decorrelated subset. Because decorrelation is observable-dependent, the proxy should be validated against the target property whenever feasible, for example through block averaging or convergence tests. This procedure provides an operational balance between statistical representativeness and computational cost.
A two-phase thermodynamic analysis of first-principles water trajectories illustrates the same trajectory-level connection by extracting absolute enthalpy, entropy, heat capacity, and free energy from the sampled dynamics.\cite{PascalScharfJungKuehne2012}

\subsection{Grassmann Wavefunction Dynamics}
\label{sec:grassmann-extrapolation-aimd}

Fictitious wavefunction dynamics denotes the numerical propagation of the occupied KS subspace along a nuclear trajectory rather than real-time electronic dynamics. Within CP2K, the same history of occupied subspaces can be used in two conceptually distinct regimes. In Born--Oppenheimer AIMD, the propagated subspace is an initial guess that is subsequently converged to the instantaneous electronic ground state. In second-generation Car--Parrinello AIMD, the Grassmann wavefunction propagation is used within a coupled predictor--corrector dynamics of the electronic and nuclear degrees of freedom.\cite{Kuehne2007,Kuehne2020,HutterIannuzziKuehne2024}

For \(N_{\mathrm{occ}}\) fully occupied states in an AO space of dimension \(N_{\mathrm{AO}}\), the coefficient matrices satisfying \(\mathbf C^\dagger\mathbf S\mathbf C=\mathbf I\) form a generalized Stiefel manifold. Since \(\mathbf C\) and \(\mathbf C\mathbf U\), with \(\mathbf U\) unitary or orthogonal for real \(\Gamma\)-point orbitals, span the same occupied subspace, the physically relevant space is the quotient Grassmann manifold \(\operatorname{Gr}(N_{\mathrm{occ}},N_{\mathrm{AO}})\).\cite{Edelman1998Grassmann,polack_grassmannextrapolation_2021} Equivalently, each subspace can be represented by the \(S\)-idempotent projector \(\boldsymbol{\Pi}=\mathbf P\mathbf S\), where the occupied-space density matrix \(\mathbf P=\mathbf C\mathbf C^\dagger\) satisfies \(\mathbf P\mathbf S\mathbf P=\mathbf P\) and hence \(\boldsymbol{\Pi}^2=\boldsymbol{\Pi}\). For real \(\Gamma\)-point orbitals, \(\mathbf P=\mathbf C\mathbf C^{\mathrm T}\). This gauge-invariant representation motivates the term Grassmann wavefunction dynamics.

\subsubsection{Born--Oppenheimer \textit{ab-initio} Molecular Dynamics}
\label{sec:gext-bomd}

Long AIMD trajectories require an electronic-structure calculation at every nuclear time step, and the iterative solution of the KS equations is often the dominant computational cost. Because consecutive geometries are close in configuration space, previously converged electronic states can provide informed initial guesses for subsequent calculations.\cite{Kuehne2007,niklasson_extendedlagrangian_2009,polack_grassmannextrapolation_2021} Grassmann extrapolation (GExt) represents and extrapolates the occupied subspace in a gauge-invariant form motivated by its nonlinear manifold geometry, while exploiting the temporal coherence between successive configurations, as illustrated in Fig.~\ref{fig:gext-grassmann-dynamics}. The implementation in CP2K combines descriptor-based GExt with the projection used by the ASPC method.\cite{Kuehne2007,polack_approximationstrategy_2020,polack_grassmannextrapolation_2021} It thereby provides the electronic predictor for Born--Oppenheimer MD without changing the nuclear equations of motion or the converged electronic state.
The orbital-transformation method provides an efficient direct minimization of the KS energy under orthogonality constraints.\cite{vandevondele2003}

\begingroup
\begin{figure}[t]
    \centering
    \includegraphics[width=0.96\columnwidth]{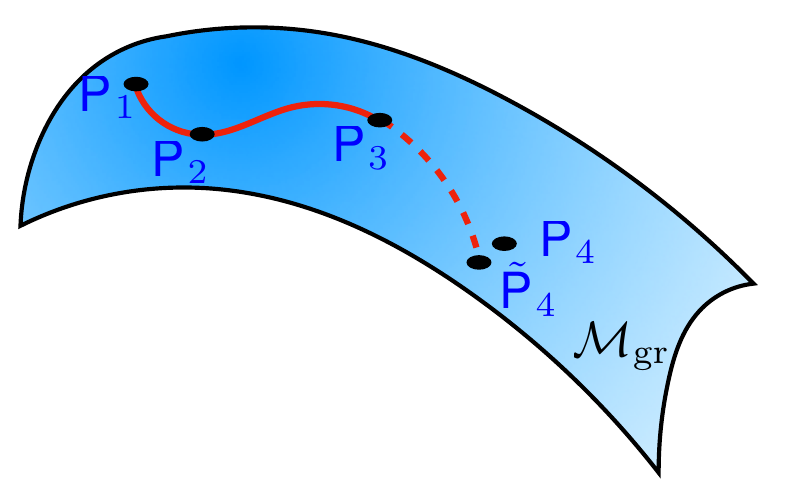}
    \caption{Conceptual representation of Grassmann wavefunction dynamics. Previously converged occupied subspaces, represented by the \(S\)-idempotent projectors associated with $\mathbf{P}_1$, $\mathbf{P}_2$, and $\mathbf{P}_3$, trace a path on the Grassmann manifold $\mathcal{M}_{\mathrm{gr}}$. GExt forms a trial subspace that is reorthonormalized to give \(\widetilde{\mathbf P}_4\) and subsequently initializes the SCF refinement to $\mathbf{P}_4$.}
    \label{fig:gext-grassmann-dynamics}
\end{figure}
\endgroup

Let $\mathbf{C}_j$ and $\mathbf{S}_j$ denote the occupied-orbital coefficient matrix and AO overlap matrix at step $j$, respectively, and let \(\mathbf P_j=\mathbf C_j\mathbf C_j^\dagger\), or \(\mathbf P_j=\mathbf C_j\mathbf C_j^{\mathrm T}\) for real orbitals. At step $n$, the previous coefficient matrix is projected onto an extrapolated occupied subspace according to
\begingroup
\begin{equation}
\begin{aligned}
\widetilde{\mathbf{C}}_n
&=
(\widetilde{\mathbf P\mathbf{S}})_n\mathbf{C}_{n-1}
\\
(\widetilde{\mathbf P\mathbf S})_n
&=
\sum_{i=1}^{q}\alpha_i
(\mathbf P\mathbf{S})_{n-q-1+i}
\ensuremath{,}
\end{aligned}
\label{eq:gext-projected-orbitals}
\end{equation}
\endgroup
where $\widetilde{\mathbf{C}}_n$ is the predicted occupied-orbital coefficient matrix before any SCF or OT correction, \(q\) is the number of stored propagation steps, and the \(\alpha_i\) are extrapolation coefficients obtained from the selected overlap-descriptor fit. The operator \((\widetilde{\mathbf P\mathbf S})_n\) is a linear combination of the historical \(S\)-idempotent projectors \(\mathbf P_j\mathbf S_j\). Because such a linear combination is not generally idempotent, it is an extrapolated projector combination rather than a projector in the strict mathematical sense. CP2K applies it to \(\mathbf C_{n-1}\) and reorthonormalizes the resulting trial orbitals in the current AO metric before the electronic correction. The tilde in the latter expression applies to the product \(\mathbf P\mathbf S\) as a whole. Neither \(\mathbf P_n\) nor $\mathbf{S}_n$ is extrapolated separately.
The overlap matrix serves as an inexpensive molecular descriptor, and the projected form avoids explicit Grassmann exponential and logarithm evaluations. Apart from the sparse overlap matrices, it operates on orbital-shaped matrices and is invariant under rotations within the occupied subspace, making the predictor robust to orbital crossings.

CP2K provides two complementary strategies for determining the extrapolation coefficients. The difference-fit variant represents the current overlap relative to the previous step and enforces a unit sum of the equivalent coefficients, as is advised for geometry optimization.\cite{askarpour_grassmannextrapolation_2025} The quasi-time-reversible variant instead combines the descriptor history in symmetric pairs, which improves the long-time stability of MD trajectories.\cite{pes_quasitimereversible_2023} Table~\ref{tab:gext-benchmarks} illustrates the system- and convergence-dependent balance between initial-guess efficiency and temporal symmetry. The difference-fit variant gives the lowest iteration count in several cases, whereas the quasi-time-reversible form sacrifices some instantaneous efficiency in favor of the temporal symmetry desirable for long MD trajectories. The coefficient construction and CP2K controls are detailed in Section~IX~A in the SI.

\begingroup
\begin{table}[t]
\caption{Average number of OT iterations per AIMD step for 32 water molecules at $300\,\mathrm{K}$ and 216 silicon atoms at $3000\,\mathrm{K}$ using three OT convergence thresholds. The table compares GExt, ASPC, and higher-order PS predictors.}
\label{tab:gext-benchmarks}
\centering
\small
\setlength{\tabcolsep}{3pt}
\renewcommand{\arraystretch}{1.08}
\begin{tabular}{lrrrrrr}
\hline\hline
& \multicolumn{3}{c}{Water} & \multicolumn{3}{c}{Silicon} \\
Method ($q$) & $10^{-2}$ & $10^{-4}$ & $10^{-6}$ & $10^{-2}$ & $10^{-4}$ & $10^{-6}$ \\
\hline
\texttt{GEXT\_PROJ} (5)      & 1.02 & 5.13 &  8.48 & 1.02 &  2.59 & 54.53 \\
\texttt{GEXT\_PROJ} (6)      & 1.02 & 5.13 &  8.48 & 1.02 &  2.80 & 55.90 \\
\texttt{GEXT\_PROJ\_QTR} (5) & 5.79 & 3.81 & 10.10 & 3.24 & 12.81 & 72.86 \\
\texttt{GEXT\_PROJ\_QTR} (6) & 5.96 & 3.80 & 10.10 & 2.97 & 12.94 & 76.46 \\
\texttt{ASPC} (4)            & 5.38 & 3.03 & 10.10 & 3.85 & 15.14 & 81.31 \\
\texttt{PS} (4)              & 7.48 & 6.36 & 10.08 & 5.58 & 22.32 & 97.51 \\
\hline\hline
\end{tabular}
\end{table}
\endgroup

\subsubsection{Second-Generation Car--Parrinello \textit{ab-initio} Molecular Dynamics}
\label{sec:gext-second-generation-cpmd}

In second-generation Car--Parrinello AIMD, a higher-order Gear-type predictor--corrector integrator propagates the electronic wavefunction using just one preconditioned electronic gradient per nuclear time step. It thereby retains a Car--Parrinello-like propagation while remaining close to the Born--Oppenheimer potential energy surface.\cite{Kuehne2007,Kuehne2014SecondGenerationCPMD} A recent CP2K/\textsc{Quickstep} guide complements the formal derivation with practical protocols for pre-equilibration, corrector tuning, and calibration of the modified Langevin friction.\cite{Kuehne2026CP2GBestPractices} The original formulation used fixed-coefficient ASPC propagation.\cite{Kolafa2004ASPC,Kolafa2005GearASPC} In the Grassmann variant described here, that predictor is replaced by the descriptor-adaptive GExt construction of Eq.~\eqref{eq:gext-projected-orbitals}.\cite{polack_grassmannextrapolation_2021,pes_quasitimereversible_2023} In particular, the quasi-time-reversible GExt form combines a geometry-dependent fit with the temporal symmetry needed for stable long trajectories. The explicit symmetric projector combination and the regularized coefficient fit are given in Section~IX~A in the SI.
The resulting orbitals are propagated by the electronic forces acting on the wavefunctions that are calculated by one preconditioned electronic minimization step at the current nuclear geometry. %The predicted orbitals are corrected at the current nuclear geometry by a single preconditioned electronic minimization step.
During the startup phase, the electronic state is fully converged until the history contains the $q$ states required by the predictor. Thereafter, the Born--Oppenheimer and second-generation variants differ in the correction following $\widetilde{\mathbf{C}}_n$: the former continues the SCF cycle to convergence, whereas the latter retains the state after the correction, uses it to evaluate the nuclear forces by the non-self-consistent Harris functional described in Section~\ref{Harris_Functional}, and carries its occupied subspace into the next prediction.

%CP2K provides second-generation Car--Parrinello AIMD as an alternative to conventional Car--Parrinello dynamics. The latter is not implemented in the program.\cite{Kuehne2020,Iannuzzi2025}
The current Grassmann reduced-correction path is available for $\Gamma$-point \textsc{Quickstep} calculations, while the GExt predictor itself also supports complex $\mathbf{k}$-point wavefunctions. Its startup behavior and implementation controls are specified in Section~IX~A in the SI. Unlike conventional Car--Parrinello dynamics, this construction introduces neither a fictitious orbital mass nor a separate, shorter electronic time step.\cite{CarParrinello1985}

Applications span bulk and interfacial water, $h$-BN/Rh(111), aqueous CO/Pt(111), thermally disordered silicon, and photocatalytic interfaces.\cite{Kuehne2009LiquidWater,Kuehne2011WaterVaporInterface,KhaliullinKuehne2013WaterEDA,Kessler2015WaterInterfacePIMD,Musso2018SGCP,Lan2018AqueousCOPt,KuhneProdan2018,Gujt2020WeylWater} Comparative water simulations have validated structural and transport observables across propagation schemes, ensembles, and thermostats.\cite{MunozSantiburcio2022Diffusion} The silicon analysis connects to the Kubo treatment in Section~\ref{sec:kubo-transport}, while DFTB and ALMO studies address on-water catalysis.\cite{Karhan2014OnWater,Henao2026OnWater,SalemKuehne2026OnWater}

Because a single corrector step does not eliminate the entire electronic residual, the resulting non-self-consistent force error can introduce weak systematic dissipation. In the second-generation framework, this effect is represented by an intrinsic friction and compensated through a modified Langevin equation
\begingroup
\begin{subequations}
\label{eq:gext-second-generation-langevin}
\begin{align}
M_I\ddot{\mathbf{R}}_I
={}&
\mathbf{F}^{\mathrm{GExt}}_I
-\gamma_D M_I\dot{\mathbf{R}}_I
+\boldsymbol{\Xi}_I(t),
\label{eq:gext-langevin_motion}\\
\left\langle
\Xi_I^{\alpha}(0)\Xi_J^{\beta}(t)
\right\rangle
={}&
2\gamma_D M_I k_{\mathrm B}T
\delta_{IJ}\delta_{\alpha\beta}\delta(t)
\ensuremath{.}
\label{eq:gext-fluctuation_dissipation}
\end{align}
\end{subequations}

\endgroup
where \(M_I\), \(\mathbf R_I\), and \(\mathbf F_I^{\mathrm{GExt}}\) are the mass, position, and approximate force of nucleus \(I\). The coefficient \(\gamma_D\) accounts for the dissipative component of that force, and \(\boldsymbol{\Xi}_I\) is the corresponding fluctuating force. The indices \(\alpha\) and \(\beta\) label Cartesian components, \(k_{\mathrm B}\) is Boltzmann's constant, \(T\) is temperature, and the three delta symbols are the nuclear-index Kronecker delta, Cartesian Kronecker delta, and Dirac delta in time, respectively. When this weak-friction representation is valid, the fluctuation--dissipation relation restores canonical sampling while preserving the efficiency gained from the single electronic correction.\cite{Kuehne2007,Kuehne2014SecondGenerationCPMD,Kuehne2020} The corresponding CP2K control is specified in Section~IX~A in the SI. The QTR-GExt predictor therefore replaces ASPC at the electronic-propagation level, while the established residual-force treatment remains unchanged.

\subsection{Finite-Temperature Spectroscopy from \textit{ab-initio} Molecular Dynamics}
\label{sec:aimd-finite-temperature-spectra}

The AIMD treatment provides the dynamical core of the finite-temperature spectroscopy strategy summarized in Table~\ref{t4}. Instead of evaluating a response property at a single optimized geometry, one either computes a time-correlation function directly along the trajectory or evaluates an instantaneous spectrum on a decorrelated set of representative configurations. The first strategy is natural for observables that are defined by the spontaneous fluctuations of a dipole, polarization, polarizability, or related response tensor. The second strategy is natural for electronic, core-level, and magnetic-response spectra, where each snapshot is treated as a thermally distorted electronic-structure problem and the final spectrum is obtained by configurational averaging.
For large solvated and biomolecular systems, the fully periodic GROMACS--CP2K interface extends this strategy to QM/MM dynamics and enhanced sampling: GROMACS handles the MM topology and trajectory propagation, whereas CP2K evaluates the QM energies and forces, with long-range electrostatics treated consistently under PBC.\cite{Laino2005QMMM,Laino2006QMMMPBC,Morozov2026GromacsCP2K}

\begin{figure*}
    \centering
    \includegraphics[width=\linewidth]{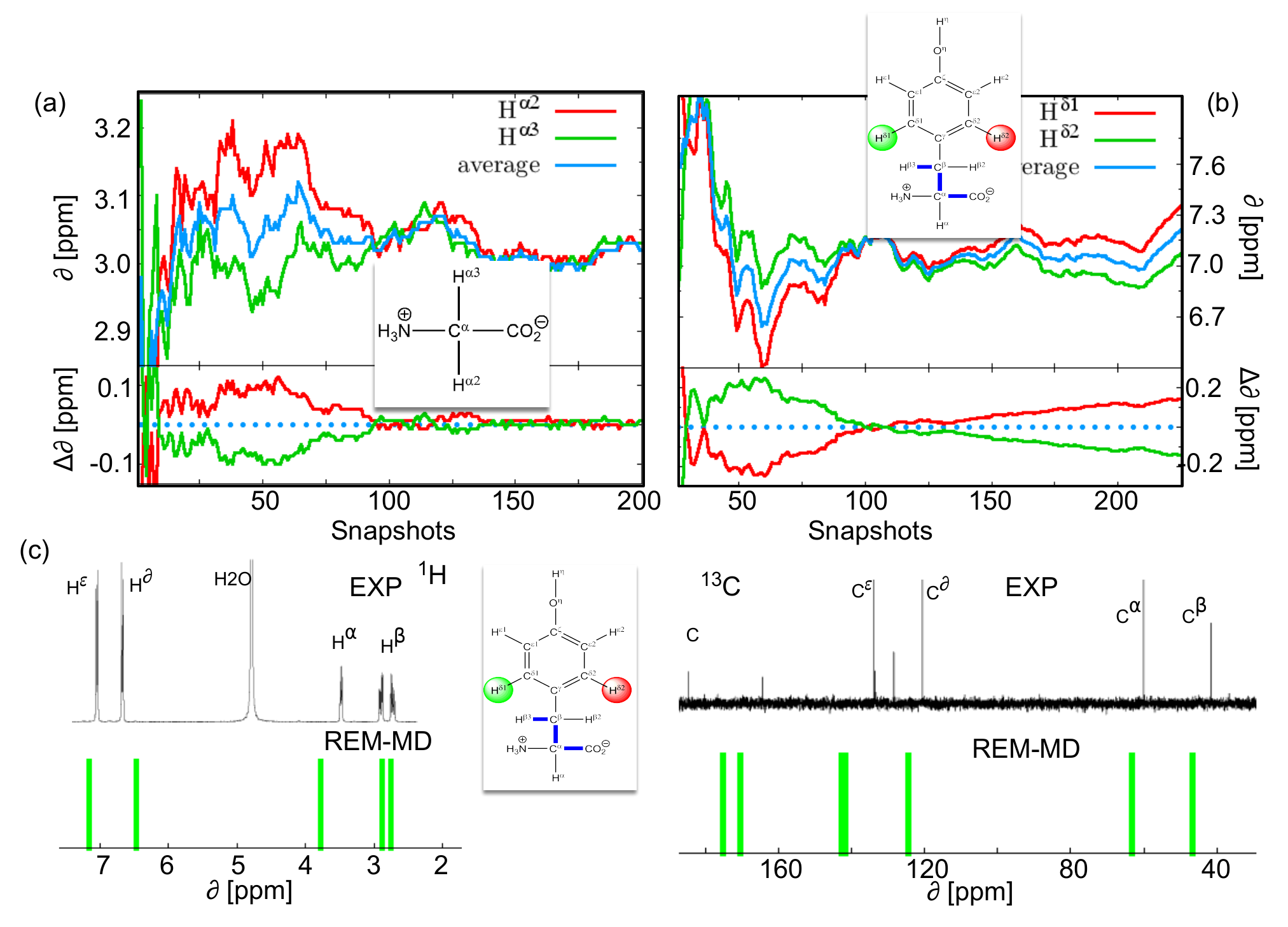}
    \caption{(a) Floating average of glycine(aq) H$^{\alpha 2}$ and H$^{\alpha 3}$ chemical shifts. (b) Incremental average chemical shifts for tyrosine(aq) H$^{\delta 1}$ and H$^{\delta 2}$ over standard MD sampling. (c) Tyrosine $^1$H and $^{13}$C chemical shifts from replica-exchange MD sampling compared with experimental data. Adapted from Ref.~\citenum{Giani2014Thesis}.}
    \label{fig:nmramino}
\end{figure*}

For vibrational spectroscopy, finite-temperature IR absorption follows from dipole or polarization time-correlation functions, while Raman and Raman optical activity require polarizability or optical-activity tensor correlations along the same nuclear trajectory.\cite{vib_martin,raman_luber_marcella,voronoi_vib_martin,roa_martin_2,raman_martin_2019,roa_martin} The same logic extends to interface-sensitive nonlinear spectroscopies such as SFG, where the relevant response is a higher-order polarization correlation and the nuclear trajectory supplies the fluctuating hydrogen-bond, orientation, and local-field environments. In CP2K this connects Born--Oppenheimer or path-integral MD to the finite-field AIMD and response implementations of Section~\ref{sec:finite-field-aimd}: the trajectory supplies the thermodynamic sampling, while dipoles, Berry-phase polarizations, polarizabilities, and related tensor properties supply the spectroscopic signal.
Wannier-based trajectory methods provide finite-temperature Raman spectra and assignments and vibrational SFG, while propagated moments and localized orbitals provide ingredients for chiral spectra.\cite{PartoviAzarKuehne2015Raman,OjhaKuehne2020VSFG,PartoviAzarKuehne2021RamanAssignment,SchrederLuber2024Chiral} Applications include SFG at air--water interfaces, an SFG-based XC benchmark, vibrational and X-ray analyses of liquid water and molten salts, time-resolved terahertz--Raman spectroscopy, and polarization-resolved IR spectroscopy at TiO$_2$(110).\cite{Kimmel2012WaterTiO2IR,OjhaKarhanKuehne2018WaterSpectroscopy,OjhaKaliannanKuehne2019TDVSFG,Ohto2019WaterAirAccuracy,Elgabarty2020EnergyTransfer,OjhaKuehne2021TwoDVSFG,RoyBrehm2021MoltenSalts,Balos2022THzRaman}

For electronic and core-level spectra, the finite-temperature broadening is usually represented by an ensemble of single-snapshot calculations rather than by a real-time correlation function. The relevant configurations should be separated by the decorrelation time of a proxy observable, as discussed in Section~\ref{sec:statistical-spectroscopy-framework}, and each snapshot can then be treated with the method appropriate to the probe, for example LR-TDDFT for optical absorption, $\Delta$SCF or transition-potential approaches for X-ray and photoelectron spectra, or magnetic-response calculations for NMR and EPR. This snapshot-averaging strategy connects naturally to the finite-field machinery used to generate polarization and nonlinear-response observables.
Snapshot ensembles have supported EPR lineshapes, liquid-water X-ray scattering, finite-temperature XAS, and combined Fourier-transform infrared and N K-edge XAS of aqueous ammonia and ammonium.\cite{Krack2002XrayWater,ElgabartySebastiani2013EPR,Ekimova2017AmmoniaSpectroscopy,Mueller2019XAS,Mueller2026FeKedge} For multilevel coupled-cluster water XAS and Na K-edge X-ray absorption features evaluated with external TDDFT, CP2K supplied path-integral or AIMD configurations rather than the spectral engine.\cite{Galib2018NaXANES,Folkestad2024WaterXAS}

Early finite-temperature applications combined DFPT with first-principles trajectories to reproduce NMR shifts of normal and supercritical water.\cite{SebastianiParrinello2002NMRWater} A representative example is the snapshot-averaged calculation of NMR chemical shifts for aqueous glycine and tyrosine.\cite{Giani2014Thesis} Their fluctuating hydrogen-bond networks and conformational states produce broad distributions of instantaneous shieldings. When interconversion between these environments is fast on the NMR timescale, the observed chemical shift is motionally averaged; slow exchange can instead produce distinct resonances. The protocol combined a \(1\,\mathrm{ns}\) classical trajectory in explicit TIP3P water at \(300\,\mathrm{K}\), snapshots separated by \(5\,\mathrm{ps}\), short AIMD refinements, and GAPW shielding calculations with the BLYP functional and cc-pVQZ basis sets. The chemical shifts were referenced to optimized gas-phase tetramethylsilane. Fig.~\ref{fig:nmramino}(a) demonstrates the convergence of equivalent glycine protons, whereas Fig.~\ref{fig:nmramino}(b) shows the slower averaging of tyrosine proton shifts caused by phenyl-ring rotation and solvent reorganization. Replica-exchange sampling improves this conformational coverage and yields the spectra compared with experiment in Fig.~\ref{fig:nmramino}(c).

\subsection{Finite-Field \textit{ab-initio} Molecular Dynamics}
\label{sec:finite-field-aimd}

Finite-field AIMD provides the direct dynamical framework for spectra and response functions that depend on polarization under an externally imposed electric boundary condition. In periodic systems, where the position operator is not well defined, CP2K evaluates the electronic contribution to the polarization through Berry-phase expressions and couples it to functionals at either fixed macroscopic electric field $\mathbf{E}$ or fixed electric displacement field $\mathbf{D}$. This makes it possible to propagate finite-temperature trajectories in which the electronic structure, nuclear forces, and polarization response are mutually consistent.

\subsubsection{Polarization in Periodic Systems}

For a periodic insulating cell of volume $\Omega$, the macroscopic polarization is separated into ionic and electronic contributions according to the modern theory of polarization.\cite{KingSmith:1993hp,resta}
\begin{equation}
\mathbf{P}
=
\mathbf{P}_{\mathrm{ion}}+\mathbf{P}_{\mathrm{el}}
=
\frac{e}{\Omega}\sum_I Z_I\mathbf{R}_I
-\frac{e}{2\pi\Omega}\mathbf{h}\bm{\gamma}^{\mathrm{el}}
\ensuremath{,}
\label{eq:finite-field-polarization}
\end{equation}
where the columns of \(\mathbf h\) are the cell vectors, $Z_I$ is the ionic charge number, and $\bm{\gamma}^{\mathrm{el}}$ is the electronic Berry-phase vector. \(\mathbf R_I\) is the position of ion \(I\), and \(\mathbf P_{\mathrm{ion}}\) and \(\mathbf P_{\mathrm{el}}\) are the ionic and electronic polarization contributions. This expression replaces the ill-defined position operator of an extended system and provides the polarization that enters both the finite-field energy and the nuclear forces.

For a \(\Gamma\)-point supercell, CP2K evaluates each electronic Berry-phase component from the phase of the determinant of an occupied-space twist-operator overlap matrix. The explicit determinant, overlap-matrix, and AO expressions are given in Section~IX~B in the SI. Because the complex logarithm determines $\gamma_k^{\mathrm{el}}$ only modulo $2\pi$, the corresponding polarization component is defined modulo a polarization quantum. A continuous AIMD trajectory must therefore follow a consistent Berry-phase branch so that the polarization and field-coupling energy evolve continuously.

\subsubsection{Coupling to Finite Electric Fields}

At fixed macroscopic electric field, the variational quantity is the electric enthalpy.\cite{Souza:2002fa,Umari:2002eo}
\begin{equation}
\mathcal{F}\!\left(\mathbf{E};\mathbf{C},\{\mathbf{R}_I\}\right)
=
H_{\mathrm{PBC}}\!\left(\mathbf{C},\{\mathbf{R}_I\}\right)
-\Omega\,\mathbf{E}\cdot\mathbf{P}\!\left(\mathbf{C},\{\mathbf{R}_I\}\right) .
\label{eq:finite-field-constant-e}
\end{equation}
By contrast, for selected Cartesian components or for all three directions, an imposed electric displacement field is represented by the internal-energy functional.\cite{Stengel2009}
\begin{equation}
\begin{aligned}
\mathcal{U}\!\left(\mathbf{D};\mathbf{C},\{\mathbf{R}_I\}\right)
={}&
H_{\mathrm{PBC}}\!\left(\mathbf{C},\{\mathbf{R}_I\}\right)
\\
&+\frac{\Omega}{8\pi}
\left|\mathbf{D}-4\pi\mathbf{P}\!\left(\mathbf{C},\{\mathbf{R}_I\}\right)\right|^2 .
\end{aligned}
\label{eq:finite-field-constant-d}
\end{equation}
In Eqs.~\eqref{eq:finite-field-constant-e} and~\eqref{eq:finite-field-constant-d}, \(\mathbf C\) denotes the MO-coefficient matrix, \(H_{\mathrm{PBC}}\) is the field-free periodic KS energy functional, and \(\{\mathbf R_I\}\) are the nuclear coordinates. The vectors \(\mathbf E\) and \(\mathbf D\) are the imposed macroscopic electric and displacement fields, respectively. The factors of \(4\pi\) in Eq.~\eqref{eq:finite-field-constant-d} correspond to Gaussian electrostatic units.
The constant-$\mathbf{E}$ and constant-$\mathbf{D}$ thermodynamic potentials therefore describe distinct electrical boundary conditions: the former fixes the macroscopic field, whereas the latter fixes the displacement field and, consequently, the free charge associated with the selected field components. Because the orbitals and nuclei are propagated from derivatives of the same field-dependent functional, CP2K obtains mutually consistent energies, polarization responses, and forces throughout AIMD. The analytic AO derivatives of the Berry-phase contribution are therefore essential for using these electrical boundary conditions in AIMD rather than only in static response calculations.
Combined constant-\(\mathbf E\) and constant-\(\mathbf D\) simulations provide dielectric correlations and finite-field protocols for electrochemical systems.\cite{ZhangHutterSprik2016,ZhangSayerHutterSprik2020}

In the finite-temperature spectroscopy strategy, finite-field AIMD supplies the thermal ensemble through the trajectory, while Berry-phase polarizations, dipoles, and field-dependent response tensors provide the time-dependent signal used for infrared, Raman, SFG, and related finite-temperature spectra. The AO evaluation of the overlap matrices, their orbital and nuclear derivatives, Pulay-like force contributions, the branch-tracking algorithm, component-selective field coupling, and the CP2K flow chart are detailed in Section~IX~B in the SI.
Applications include pH-dependent TiO$_2$(110) capacitance, field-induced hydrogen-bond asymmetry and spectral diffusion in water, time-resolved interfacial SFG, and electronic response and charge inversion at polarized aqueous gold electrodes.\cite{ElgabartyKaliannanKuehne2019,ZhangHutterSprik2019,OjhaKuehne2023FieldWater,OjhaKuehne2025TRVSFG,AnderssonSprikHutterZhang2025}

\section{Nuclear Quantum Effects}
\label{sec:nuclear-quantum-effects}

CP2K provides two complementary descriptions of NQEs that differ in both representation and approximation. Constrained nuclear-electronic orbital density-functional theory (CNEO-DFT) self-consistently treats some or all nuclei quantum mechanically together with the electrons and incorporates nuclear quantum delocalization into a full-dimensional effective potential energy surface, with the constrained nuclear position expectation values serving as dynamical coordinates. Path-integral methods instead map the finite-temperature quantum Boltzmann distribution of the nuclei onto ring-polymer replicas and converge systematically to the equilibrium quantum statistics on the chosen electronic potential energy surface as the bead number is increased. CNEO-DFT does not require bead replication and provides an efficient approach to incorporating NQEs into geometry optimization and MD, whereas path-integral sampling provides the more general equilibrium treatment and its dynamical variants approximate quantum time-correlation functions. The CNEO formalism and periodic GAPW implementation are summarized below and detailed in Section~X~A in the SI. Path-integral dynamics and its connection to spectroscopy are discussed in Section~\ref{sec:path-integral-dynamics-spectra}. These complementary treatments are particularly relevant at aqueous metal interfaces, where nuclear quantum delocalization influences interfacial structure and dynamics.\cite{Jinggang2022,Jinggang2026}

\subsection{Constrained Nuclear-Electronic Orbitals}
CNEO-DFT
extends KS DFT to incorporate NQEs, especially those of light nuclei, within a
self-consistent electronic-structure framework.
\cite{Xu20084107,Xu20074106,Xu21244110,chen2026constrained}
In this approach, some or all nuclei, typically hydrogen, are treated
quantum mechanically together with all electrons, while any remaining
nuclei are treated classically. The central idea is to assign each
quantum nucleus a classical position by constraining the expectation
value of its position operator. This construction retains a full-dimensional
potential energy surface, analogous to the conventional Born--Oppenheimer
picture. At the same time, nuclear quantum delocalization, zero-point
motion, and shallow tunneling are naturally incorporated into the
CNEO effective potential energy surface, which enables efficient inclusion of
these effects in geometry optimization,\cite{Xu20074106,Zhang23231101}
transition-state searches,\cite{Chen25590}
and MD simulations.\cite{Xu224039}
The CNEO framework was first developed
for molecular systems and implemented in PySCF.
\cite{Sun2026PySCF10Years,pyscfcneo}
More recently, periodic CNEO-DFT has been developed in CP2K,
which provides a practical way to incorporate NQEs,
particularly nuclear quantum delocalization,
in condensed-phase and interfacial simulations at a cost comparable to that of conventional DFT.\cite{Chen2025}

Formally, CNEO-DFT is based on multicomponent DFT\cite{Capitani82568,gidopoulos1998kohn} and constrains the position expectation values of the quantum nuclei. Each quantum nucleus is treated as a distinguishable particle represented by a localized nuclear orbital. Together with the classical nuclear coordinates, the constrained quantum-nuclear position expectation values define a full-dimensional CNEO effective potential energy surface. The detailed energy functional, constrained KS equations, periodic electrostatics, GAPW density partitioning, and analytic gradients are presented in Section~X~A in the SI.

A key conceptual step in extending CNEO-DFT to PBC is reconciling the localized, distinguishable representation of quantum nuclei required to define their position expectation values with the delocalized Bloch-like description natural under PBC. The CP2K implementation represents the quantum nuclei by localized orbitals while constructing the total periodic quantum-nuclear charge density through lattice summation for the electrostatic evaluation.\cite{Chen2025} Within the GAPW framework,\cite{Lippert1999,Krack2000} this periodic quantum-nuclear density is decomposed into smooth grid-based and localized atom-centered parts. The multicomponent Coulomb terms can therefore be evaluated using the same global-plus-local machinery employed for all-electron densities. The implementation augments the electronic SCF procedure with quantum-nuclear KS equations and position-constraint optimization. Because each quantum nucleus is represented by a single localized orbital, the dominant computational effort remains the electronic DFT calculation, and the additional cost increases only slightly with the number of quantum nuclei.\cite{Chen2025}

Hydrogen adsorption on Pt(111) illustrates periodic CNEO-DFT.\cite{Chen2025} Whereas conventional DFT with a classical hydrogen nucleus favors the atop site, nuclear delocalization and zero-point energy alter the relative site stabilities and lower the lateral migration barrier. Analytic gradients thereby determine the quantum-modified adsorption geometry and migration barrier directly on the CNEO effective potential energy surface.

Analytic gradients are the essential bridge from static CNEO-DFT to trajectory-based simulations. Derivatives with respect to both the classical nuclear coordinates and the constrained quantum-nuclear position expectation values provide the forces for Born--Oppenheimer-like CNEO-AIMD. At each geometry, the electronic and quantum-nuclear equations and the position constraints are solved self-consistently.\cite{Xu20074106,Xu224039,Chen2025} Infrared and Raman spectra can then be evaluated from dipole and polarizability autocorrelation functions, respectively,\cite{Xu224039,CNEOPolarizability} along trajectories propagated on the CNEO effective potential energy surface, which incorporates the dominant effects of quantum delocalization and zero-point motion.

This treatment differs from path-integral formalisms because it propagates the classical nuclear coordinates and constrained quantum-nuclear position expectation values on the CNEO effective potential energy surface without ring-polymer replicas. Nevertheless, it is conceptually related to centroid MD (CMD), a path-integral-based method, because both propagate effective nuclear positions as classical dynamical variables: constrained position expectation values in CNEO-MD and ring-polymer centroids in CMD.\cite{Chen23279,chen2026constrained} Molecular CNEO-MD implementations have shown clear improvements for systems with strong hydrogen motion and intermode coupling.\cite{Xu224039,Zhang23231101,Zhang239358,langford2024hidden,liu2025characterizing} The periodic CP2K implementation provides the energies and analytic gradients required for CNEO-AIMD, although published periodic applications have so far focused on geometry optimization and hydrogen migration barriers.\cite{Chen2025} An accelerated CNEO-AIMD scheme based on the Grassmann formulation of second-generation Car--Parrinello AIMD in Section~\ref{sec:gext-second-generation-cpmd} has not yet been implemented. Such a scheme would additionally require consistent predictor--corrector propagation of the electronic and quantum-nuclear orbitals together with the position-constraint variables.

\subsection{Path-Integral Approaches to Spectra from Quantum Dynamics}
%dm \subsection{Path-Integral Approaches to Dynamics and Spectra}
\label{sec:path-integral-dynamics-spectra}

NQEs, including zero-point
%dm energy,
motion,
tunneling, and nuclear
delocalization, strongly affect the structural and dynamical properties of systems
containing light nuclei, most prominently hydrogen-bonded systems and
proton-transfer processes at ambient conditions.
%dm \cite{Marx1999HydratedProton,TODO}
\cite{Marx-2006-ChemPhysChem-7-1848}
Their importance generally increases as the temperature is lowered, and they therefore cannot be neglected in low-temperature regimes.
%dm \cite{TODO}
\cite{brieuc_converged_2020}
In classical MD the nuclei obey Newton's equations of
motion and are therefore treated as classical point particles, thereby neglecting NQEs by construction.
%
%dm -
Path-integral methods overcome this limitation by exploiting the formal isomorphism
between the quantum partition function and the configurational statistics of
classical ring polymers,
%dm a classical ring polymer,
thereby providing a rigorous and
%dm systematically improvable
asymptotically exact
framework for including NQEs within the Born--Oppenheimer approximation
%dm
when computing structural and thermodynamic properties.%
\cite{Feynman1965,ChandlerWolynes1981,
%dm  not review or book:   Parrinello1984,
%dm added:
Ceperley1995,
Marx2009,
%dm added:
Tuckerman2010}
The focus of this section is on the theory of
%dm
numerical path-integral molecular dynamics (PIMD)
%dm PIMD
methods for computing
vibrational and electronic spectra that incorporate NQEs and are implemented in CP2K, with emphasis on condensed-phase aqueous systems.\cite{Perakis2016}
%
%dm this is dynamics, comes later:
%dm For comprehensive recent reviews see
%dm Ref.~\citenum{Althorpe2024/10.1146/annurev-physchem-090722-124705,TODO}.
The connection to spectroscopy relies largely on approximate time-correlation functions (TCFs) generated by quasiclassical dynamics grounded in path-integral representations.\cite{Althorpe2021Path,Althorpe2024/10.1146/annurev-physchem-090722-124705}

% para - PI partition function and unified Hamiltonian
The path-integral formalism relies on the quantum partition function, which is approximated via Trotter
factorization as the configurational integral of a classical ring polymer consisting of
$P$ identical replicas, or so-called beads, connected by harmonic
springs
\cite{Feynman1965,
%dm  not review or book:   Parrinello1984,
%dm added:
Ceperley1995,
Marx2009,
%dm added:
Tuckerman2010}
\begin{equation}
    Z_P(\beta) = \left(\frac{mP}{2\pi\beta\hbar^2}\right)^{\!P/2}
    \int \dd\mathbf{x}\; \exp\!\left[-\beta V_\mathrm{eff}(\mathbf{x})\right] ,
    \label{eq:ZP}
\end{equation}
where \(Z_P\) is the \(P\)-bead approximation to the canonical partition function, \(\beta=(k_{\mathrm B}T)^{-1}\) is the inverse temperature, and \(m\) is the physical nuclear mass.
The vector $\mathbf{x} = (x_1,\ldots,x_P)^T$ denotes the Cartesian bead coordinates and the
effective potential decomposes into harmonic spring and physical potential contributions,
$V_\mathrm{eff} = \tfrac{1}{2}m\omega_P^2\mathbf{x}^T\mathbf{A}\mathbf{x} +
P^{-1}\sum_k\phi(x_k)$, with ring-polymer frequency $\omega_P = \sqrt{P}/(\beta\hbar)$.
This expression is exact in the limit $P \to \infty$ and the generalization to
$N_\mathrm{at}$ interacting particles is straightforward.
Working in the normal-mode representation that diagonalizes the spring matrix $\mathbf{A}$
and introducing fictitious momenta $\mathbf{p}$ with a diagonal mass matrix
$\tilde{\mathbf{M}}$ yields the unified
%dm ring-polymer
path-integral
Hamiltonian below.\cite{Witt2009}
\begin{equation}
    H(\mathbf{q};\mathbf{p}) = \frac{1}{2}\mathbf{p}^T\tilde{\mathbf{M}}^{-1}\mathbf{p}
    + \frac{1}{2}m\omega_P^2\,\mathbf{q}^T\boldsymbol{\Lambda}\mathbf{q} + \Phi(\mathbf{q})
    \ensuremath{.}
    \label{eq:Hunified}
\end{equation}
The vectors \(\mathbf q\) and \(\mathbf p\) contain the ring-polymer normal-mode coordinates and their fictitious conjugate momenta, \(\widetilde{\mathbf M}\) is the diagonal fictitious-mass matrix, and \(\Phi(\mathbf q)\) is the physical potential transformed to normal-mode coordinates.
%dm
The associated equations of motion are
\begin{equation}
    \tilde{\mathbf{M}}\ddot{\mathbf{q}} = -m\omega_P^2\boldsymbol{\Lambda}\mathbf{q}
    - \nabla_{\!\mathbf{q}}\Phi(\mathbf{q}) ,
    \label{eq:EOM}
\end{equation}
where $\boldsymbol{\Lambda} = \mathrm{diag}(\lambda_1,\ldots,\lambda_P)$ is the eigenvalue matrix of $\mathbf{A}$ and the first
normal mode ($\lambda_1 = 0$) corresponds to the centroid of the replicas
$x_{\rm c} \equiv P^{-1}\sum_k x_k$.
%dm $x_c \equiv P^{-1}\sum_k x_k$.
%
Further details of the normal-mode transformation and
eigenvalues are given
%dm
for instance
in Ref.~\citenum{Witt2009}.

In this compact notation, PIMD, CMD, ring-polymer MD (RPMD), thermostatted ring-polymer MD (TRPMD), and Brownian chain MD (BCMD) share the same ring-polymer representation but differ in the fictitious mass matrix $\tilde{\mathbf{M}}$ and the thermostatting scheme. PIMD samples the equilibrium quantum Boltzmann distribution, whereas CMD, RPMD, TRPMD, and BCMD provide distinct approximate dynamics for Kubo-transformed TCFs. The centroid ($j=1$) always carries the physical mass $m$ of the corresponding nucleus, while the fictitious masses of the noncentroid modes ($j \geq 2$) distinguish the methods as follows.\cite{Witt2009}
%
%HF: (as table looks strange)
%dm2 !!! finde ich sinnvoll, aber Christoph, kannst Du die Tabelle noch "aufhuebschen"?
%cs2 done
\begin{equation}
%HF: CMD-> ACMD ??
%dm2  ich wuerde einfach CMD lassen
%cs2  \begin{tabular}{l|lllll}
%cs2  & PIMD & CMD & RPMD & TRPMD & BCMD \\
%cs2  \hline \\
%cs2  $\tilde{m}_j$\, & $\lambda_j m$ & $\frac{\lambda_j m}{\gamma^2}$ &
%cs2  $m$ & $m$ & $\frac{\lambda_j m P \Delta t}{2 \hbar \beta}$ \\
%cs2  \\
%cs2  %HF: all = all beads, yes = non-centroid, special = PILE with infinite friction
%cs2  %HF: maybe use "extreme" instead of special?
%cs2  %dm2 thermostat & all & yes & no & yes & special \\
%cs2  %dm2 finde ich zu kryptisch, daher Vorschlag (Christoph, bitte auf Formatierung achten
%cs2  %dm2 da nicht von mir kompiliert):
%cs2  thermostat & all   & noncentroid & none & noncentroid & see  \\
%cs2             & beads & beads       &      & beads       & text \\
%cs2  %
%cs2  %thermostat & all & non-centroid & no & non-centroid (PILE?) & non-centroid infinite friction PILE \\
%cs2  \end{tabular}
\begin{tabular}{lcc}
Method & $\tilde{m}_j$ ($j \geq 2$) & Thermostat \\
\hline
PIMD  & $\lambda_j m$                              & all beads \\
CMD   & $\lambda_j m/\gamma^2$                     & noncentroid beads \\
RPMD  & $m$                                        & none \\
TRPMD & $m$                                        & noncentroid beads \\
BCMD  & $\lambda_j m P \Delta t / (2\hbar\beta)$   & see text \\
\end{tabular}
\ensuremath{.}
%\end{table*}
%HF: \begin{equation}
%dm !! noch TRPMD und BCMD dazu und wie besproche eine kleine Tabelle mit "masses"
%dm   kannst Du auch KNAPP in STICHWORTEN die Unterschiede
%dm   in "termostatting" einbauen
%HF: schwierig, wird dann schnell zu gross.
%HF:    \tilde{m}_j^{(\mathrm{PIMD})} = \lambda_j m , \quad
%HF:    \tilde{m}_j^{(\mathrm{CMD})}  = \frac{\lambda_j m}{\gamma^2} , \quad
%HF:    \tilde{m}_j^{(\mathrm{RPMD})} = m
%dm .
%cs2 \enspace  ,
    \label{eq:massassignments}
\end{equation}
The corresponding thermostatting schemes are discussed below.
%dm2
$\gamma$ is the CMD adiabaticity parameter,\cite{Martyna_1996_a,Cao_1996_a,Marx_1999_g}
as discussed below. In the BCMD mass, \(\Delta t\) is the integration time step.

% para - PIMD canonical sampling

PIMD was pioneered in the early 1980s and subsequently developed into a powerful \textit{ab-initio} simulation method.\cite{Callaway1982/10.1103/PhysRevLett.49.613,Parrinello1984,Marx_1994_a,Marx1996,Marx2009} It samples the quantum Boltzmann distribution exactly within the Trotter approximation using the mass assignment of Eq.~\eqref{eq:massassignments}.
Direct coupled-cluster PIMD and the quantum ring-polymer contraction (qRPC) method illustrate how correlated and density-functional first-principles descriptions, respectively, can be combined with quantum-nuclear sampling at tractable cost.\cite{Spura2015CCPIMD,John2016QRPC}
Second-generation Car--Parrinello-like propagation has also accelerated coupled-cluster PIMD.\cite{Spura2026AcceleratedCCPIMD}
%
%dm Because the dynamics is entirely fictitious in PIMD and suffers from
For large Trotter numbers, the stiff harmonic internal modes can impair ergodic sampling. Suitable thermostatting restores efficient exploration of the ring-polymer phase space.
%dm thermostats must be attached to every degree of freedom to ensure
%dm ergodic sampling.
%
The standard approach is massive Nos\'e--Hoover chain (NHC)
thermostatting,\cite{Martyna1992,Tuckerman1993,
%dm added for AIPI:
Tuckerman_1996_b}
which couples an independent NHC
thermostat to each normal mode of each atom.
PIMD gives access to exact static quantum averages, including structural properties, free
energies, and isotope-fractionation factors, but its sampling dynamics does not represent real-time quantum dynamics.
First-principles and force-matched PIMD applications have quantified NQEs in protonated water clusters, at the instantaneous water--vapor interface, in dense molecular hydrogen, and in isotope-dependent hydrogen bonding.\cite{Kessler2015WaterInterfacePIMD,Azadi2018HydrogenNQE,Clark2019OpposingNQE}
Equilibrated PIMD configurations nevertheless provide thermalized initial conditions for approximate quantum-dynamics simulations
%dm
in the canonical (\ensuremath{NVT}) ensemble.

% para - PILE thermostat
An efficient alternative to massive NHC thermostatting is the path-integral Langevin
equation (PILE) thermostat,%
%dm of Ceriotti~\textit{et al.},
\cite{Ceriotti2010/10.1063/1.3489925}
which operates directly in normal-mode space with mode-specific Langevin friction.
For the noncentroid modes ($j \geq 2$) the friction coefficient is chosen as
$\gamma_j = 2\omega_{P,j}$, where $\omega_{P,j}$ is the free ring-polymer frequency of
mode $j$. This value corresponds to critical damping of the harmonic ring-polymer motion
and ensures rapid, uniform equilibration of all noncentroid modes without overdamping.
An independent thermostat is applied to the centroid mode separately.
Beyond
%dm
its use in
PIMD sampling, the PILE thermostat is usually the central ingredient for TRPMD, discussed
below.

% para - CMD formalism

Whereas PIMD generates equilibrium statistics, CMD approximates real-time quantum correlations through centroid motion.\cite{Cao1994,Cao1994a,Voth-1996-ACP-93-135} It builds on the observation that quantum thermodynamic properties are encoded in the centroid density $\rho_c(x_c) \propto \int\dd\mathbf{q}'\,e^{-\beta V_\mathrm{eff}}\,\delta(q_1 - x_c)$ and propagates the centroid on the effective centroid potential defined below.\cite{Feynman1965}
\begin{equation}
    V_{\mathrm{eff,c}}(x_{\rm c}) = -\frac{1}{\beta}\ln\rho_{\rm c}(x_{\rm c})
    \ensuremath{.}
    %dm V_{\mathrm{eff},c}(x_c) = -\frac{1}{\beta}\ln\rho_c(x_c)
    \label{eq:ECP}
\end{equation}
The resulting centroid trajectory yields TCFs that approximate the Kubo-transformed quantum TCF.\cite{Ramirez2004,Witt2009}
The centroid carries the physical mass $m$ and moves as a single
%dm
effective or ``dressed''
particle, while the
noncentroid modes are integrated out implicitly through the effective centroid potential.
This centroid dynamics is exact in the high-temperature and harmonic limits.
%
%HF: ?? add note that this is often simply referred to "CMD" since true
%HF: CMD is rarely done/doable.
%dm2   ja, siehe unten
In adiabatic CMD (ACMD),
the effective centroid potential is generated on-the-fly via very small fictitious noncentroid
masses $\lambda_j m/\gamma^2$
%dm2 with
obtained by scaling with the CMD adiabaticity parameter
$\gamma \gg 1$, dictating very small
integration steps.\cite{Martyna_1996_a,Cao_1996_a,Marx_1999_g}
%dm2
In practice, CMD often denotes numerically converged ACMD.
Partially adiabatic CMD
(PA-CMD),\cite{Hone2006}
%dm2 (paCAMD)~\cite{Hone2006}
uses underconverged $\gamma$ to reduce cost but is
known to introduce quantitative errors in computed
spectra.\cite{Witt2009}
PA-CMD has been applied to water absorption, vibrational spectral diffusion, and interfacial SFG analyses.\cite{Spura2015WaterNQE,Ojha2018WaterVibrationsNQE,Ojha2024WaterInterfaceNQE,Kaliannan2026SFG}
%
%dm

At sufficiently low temperatures, the curvature problem in CMD causes characteristic redshifts of high-frequency vibrational bands.\cite{Witt2009,Ivanov-2010-JCP-132-031101} Curvilinear extensions can remove this artifact,\cite{Trenins-2019-JCP-151-054109,Althorpe2024/10.1146/annurev-physchem-090722-124705} but they do not generalize to arbitrary many-body potentials.

% para - RPMD

An alternative to the adiabatic separation used in CMD is provided by RPMD.\cite{Craig2004,Craig2005} It assigns the physical mass $\tilde{m}_j=m$ to every mode [Eq.~\eqref{eq:massassignments}] and propagates the full ring polymer under Newtonian dynamics at the physical temperature, without constructing an effective centroid potential. Comprehensive reviews are given in Refs.~\citenum{rpmd-review-2013,Althorpe2021Path,Althorpe2024/10.1146/annurev-physchem-090722-124705}.
In \textit{ab-initio}-based liquid-water studies, RPMD has been combined with force-matched potentials and qRPC to evaluate diffusion coefficients and velocity autocorrelation functions.\cite{Spura2015WaterNQE,John2016QRPC}
The RPMD approximation to the Kubo-transformed quantum TCF is given by the following expression.\cite{Witt2009}
\begin{align}
    \langle \hat{A}(0)\hat{B}(t) \rangle_{\rm K} \approx
    %dm \langle \hat{A}(0)\hat{B}(t) \rangle_K \approx
    &\frac{1}{(2\pi\hbar)^P Z_P}
    \int\!\dd\mathbf{p}\int\!\dd\mathbf{q}\;
    e^{-\beta H(\mathbf{q};\mathbf{p})}\, \nonumber \\
    &\bar{A}_P\!\left(\mathbf{q}(0)\right)\bar{B}_P\!\left(\mathbf{q}(t)\right) ,
    \label{eq:RPMDtcf}
\end{align}
where the bead-average estimator is $\bar{O}_P(\mathbf{q}) = P^{-1}\sum_{k=1}^P O(q_k)$.
For operators linear in positions or momenta the bead estimator reduces to the centroid,
coinciding formally with the CMD centroid estimator.
However, the centroid dynamics in RPMD (mean bead force $-P^{-1}\sum_k\nabla\phi(x_k)$)
differs from CMD (gradient of the effective centroid potential $-\partial V_{\mathrm{eff,c}}/\partial x_{\rm c}$),
%dm differs from CMD (gradient of the ECP $-\partial V_{\mathrm{eff},c}/\partial x_c$),
so that RPMD and CMD produce different spectra in anharmonic systems even in the dipole-linear approximation.\cite{Witt2009}
A well-known limitation is that the fictitious ring-polymer spring frequencies
%dm
can overlap
with physical vibrational frequencies, generating spurious resonance peaks,
%dm
peak splittings or peak shifts
in computed
spectra, an effect that worsens with decreasing
temperature.\cite{Habershon2008,Witt2009}
%dm temperature.\cite{Witt2009,Habershon2008}
%
Thermostatting the centroid during production would perturb the approximate real-time dynamics.
RPMD production runs are therefore microcanonical, and canonical
%dm
($NVT$)
initial
conditions can, for example, be drawn from a massively thermostatted PIMD trajectory
by appropriate rescaling of the normal-mode velocities.

% para - TRPMD

TRPMD addresses the ring-polymer resonance artifacts of RPMD by attaching a PILE thermostat exclusively to the noncentroid modes ($j \geq 2$), while the centroid evolves freely.\cite{Rossi2014/10.1063/1.4883861}
The PILE friction $\gamma_j = 2\omega_{P,j}$ damps the fictitious ring-polymer
oscillations on their natural timescale, preventing their resonant coupling to physical
vibrational modes without perturbing the centroid trajectory.
The thermostat introduces artificial
%dm
Lorentzian
broadening of spectral peaks
%dm
with width set by $\gamma_j = 2\omega_{P,j}$,
while peak
positions are generally well
reproduced.\cite{Rossi2014/10.1063/1.4883861,Althorpe2024/10.1146/annurev-physchem-090722-124705}

% para - BCMD
%dm Brownian chain molecular dynamics (BCMD), introduced by Shiga\cite{Shiga2022/10.1002/jcc.26989} and implemented in CP2K by H.~Forbert,
BCMD provides a second strategy for removing internal-mode resonances. It combines Newtonian centroid dynamics with overdamped Langevin, or Brownian, dynamics for the noncentroid modes.\cite{Shiga2022/10.1002/jcc.26989} At each step, the
noncentroid velocities are randomized
%dm2 from
according to
the Maxwell--Boltzmann distribution,
which is equivalent to evolving the noncentroid modes under an
overdamped (Smoluchowski-type) Langevin equation
%HF:$\mu_\alpha \eta \dot{q}_\alpha = f_\alpha + \xi_\alpha$ with
%HF: $\eta \equiv 2/\Delta t$.
$\tilde{m}_j \eta \dot{q}_j = f_j + \xi_j$ with
$\eta \equiv 2/\Delta t$,
%HF: \sigma^2_{\xi_j} = \frac{2 \tilde{m}_j \eta}{\beta}
which can also be viewed as thermostatting the noncentroid modes with
a PILE thermostat with infinite friction.
%HF: ($c_1^{(k)}=0$)
%
The absence of an inertia term in the noncentroid equation of motion eliminates
internal-mode resonances by construction, in contrast to TRPMD, which retains the full
inertial noncentroid dynamics and suppresses resonances by friction damping.
%
%dm !! Harald, schau mal wie wir die Massematrix ganz oben dafuer definieren
%dm    also eq 144, 145
Rather than approaching the adiabatic limit, BCMD retains deliberately
nonadiabatic noncentroid masses set from the thermal relaxation time
$\tau_\alpha = \beta\hbar$,
%dm2 .
yielding
%HF: which ever you want... ?
%dm2  this one:
$\tilde{m}_j = \lambda_j m P \Delta t / (2\hbar\beta)
= m \omega^2_P \lambda_j \tau_\alpha / \eta$.
%=....
%
The resulting breaking of adiabaticity is what
alleviates the CMD curvature problem, whereas the limit
%HF: mass here \tilde{m}_j:
%$\mu_\alpha \to 0$ would recover CMD-like adiabatic dynamics.
$\tilde{m}_j \to 0$ would recover CMD-like adiabatic dynamics.
%
%HF: it's not RPMD-like masses, it's CMD like masses but observables
%HF: are RPMD like. (??)
%dm2 BCMD thus inherits the equation of motion and nonadiabatic, RPMD-like masses of RPMD, while removing the chain resonances of RPMD and strongly suppressing the curvature-induced redshift of CMD.
%dm2
Like RPMD, BCMD approximates the Kubo-transformed quantum TCF using the bead-average estimator in Eq.~\eqref{eq:RPMDtcf}.
%dm2 again
\cite{Shiga2022/10.1002/jcc.26989}

% para - Matsubara + QTB/PIGLET/PIQTB -> SI
Matsubara dynamics provides a theoretical framework relating imaginary-time path-integral dynamics to exact quantum dynamics.\cite{Hele2015/10.1063/1.4916311,Althorpe2024/10.1146/annurev-physchem-090722-124705}
Within this framework, CMD is recovered as the $M=1$ (centroid-only)
Matsubara approximation, and RPMD arises from discarding the Matsubara phase, revealing
that both methods incur errors from neglecting real-time quantum coherence.
In practice, Matsubara dynamics is not applicable to condensed-phase systems due to the
intractable sign problem.
Colored-noise
%dm
and thermal bath
approaches,
%dm --
%dm such as the generalised Langevin equation (GLE),\cite{TODO}
such as the generalized Langevin equation,
%dm \cite{TODO}
\cite{Ceriotti2009/10.1103/PhysRevLett.103.030603,Ceriotti2010b/10.1021/ct900563s}
quantum thermal bath,\cite{Dammak2009/10.1103/PhysRevLett.103.190601}
the path-integral generalized Langevin equation thermostat (PIGLET),\cite{Ceriotti2011,
%dm added
Ceriotti2012/10.1103/PhysRevLett.109.100604,
%dm added
Uhl2016/10.1063/1.4959602}
and the path-
%dm -
integral quantum thermal bath
(PIQTB)
%dm \cite{Tsuru2024/10.1002/anie.202416058},
\cite{Brieuc2016/10.1021/acs.jctc.5b01146,Schran2018/10.1021/acs.jctc.8b00705}
%dm  --

are widely used to accelerate equilibrium sampling and to generate initial conditions for subsequent classical or quasiclassical dynamics. Ref.~\citenum{brieuc_converged_2020} provides a concise overview.
%
%dm

PIGLET employs fitted colored-noise correlations to accelerate the convergence of selected equilibrium observables for specified temperatures and bead numbers, whereas PIQTB reduces the bead requirement while retaining systematic convergence with increasing $P$.
%
%dm
These colored-noise and thermal-bath techniques, all available in CP2K, have enabled converged path-integral simulations at temperatures as low as 1~K, thereby opening access to cryochemical regimes.\cite{Uhl2016/10.1063/1.4959602,Schran2018/10.1021/acs.jctc.8b00705,brieuc_converged_2020}

% para - Force engines
All path-integral methods discussed above require only a potential energy surface and its gradient and are
therefore compatible with any force engine in CP2K:
%dm
parametrized force fields,
DFT, hybrid DFT, MP2, and
machine-learning potentials.\cite{Kuehne2020,Iannuzzi2025}
The same CP2K force engines can be connected through the i-PI client-server interface for externally driven path-integral sampling and dynamics.\cite{Kapil2019iPI2}
The MixPI driver also supports mixed-time-slicing PIMD with species-dependent bead numbers.\cite{JohnsonBuMundyAnanth2026MixPI}

%%%%%%%%%%%%%%%%%%%%%%%%%%%%%%%%%%%%%%%%%%%%%%%%%
\subsubsection{Infrared Absorption Spectroscopy}
%%%%%%%%%%%%%%%%%%%%%%%%%%%%%%%%%%%%%%%%%%%%%%%%%

\begin{figure*}[t!]
    \centering
    \includegraphics[width=\textwidth]{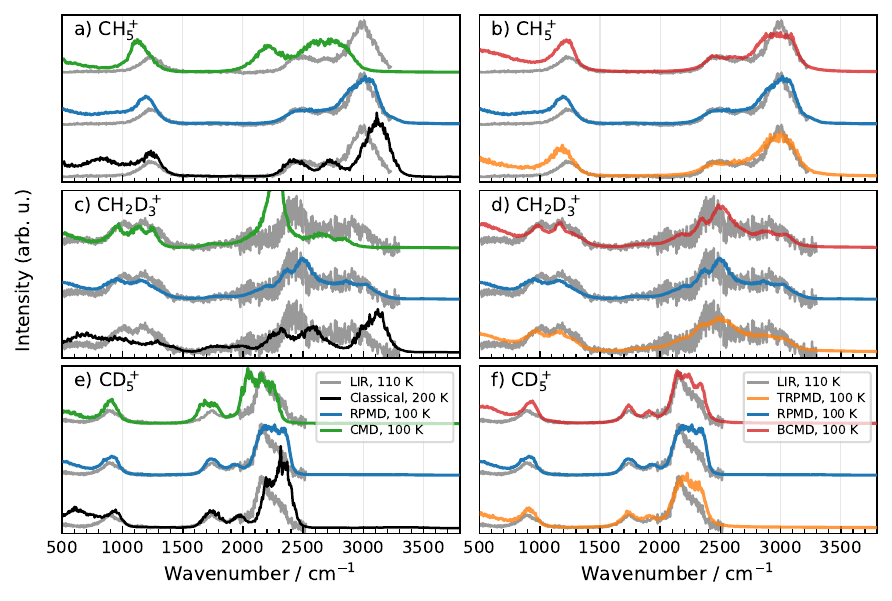}
    \caption{Infrared spectra of \ce{CH5+} (a,b), \ce{CH2D3+} (c,d), and \ce{CD5+}
    (e,f) computed on a
%dm near-
coupled-cluster singles, doubles, and perturbative triples potential-energy surface represented by a high-dimensional neural network and
%dm2
employing a corresponding high-dimensional neural-network
	dipole moment surface using
%HF: ?? together? the NNPs are trained with RubNNet4MD but then CP2K does it alone
%dm2     CP2K together with \textsc{RubNNet4MD}.\cite{RubNNet4MD-code}
%dm2 OK
    \textsc{RubNNet4MD} and CP2K.\cite{ch5p-nnpes,Beckmann2024/10.1039/d4cp02295e,RubNNet4MD-code}
    Left panels: laser-induced-reaction experimental
%HF:    specta
    spectra
    %dm spectrum
%dm ($T=110\,$K)
(at about 110~K)
compared with classical MD ($T=200\,$K),
%dm RPMD ($P=300$, $T=100\,$K), and CMD ($P=300$, $T=200\,$K).
%dm !! check if CMD at 100 (as defined in inset) or 200 K
RPMD ($T=100\,$K), and CMD ($T=100\,$K).\cite{Asvany-2005-Science-309-1219}
    Right panels: laser-induced-reaction experimental spectra
%dm um compared to RPMD, TRPMD and BCMD
compared to TRPMD, RPMD and BCMD
%dm
all
at $T=100\,$K.
    Partially adapted from Ref.~\citenum{Beckmann2024/10.1039/d4cp02295e} under the Creative Commons Attribution 3.0 Unported License.}
    \label{fig:ir_ch5plus}
\end{figure*}

% para - IR absorption and the Kubo time-correlation function
The infrared absorption coefficient of a condensed-phase system is obtained from the quantum dipole autocorrelation function as follows.\cite{Kubo1957/10.1143/JPSJ.12.570,Ramirez2004,
%dm
Ivanov-2013-PCCP-15-10270}
% ,Witt2009}
%
%
\begin{subequations}
\label{eq:IRalpha}
\begin{align}
    \alpha(\omega) &=
%dm \frac{4\pi^2\omega}{3Vc\,n(\omega)}
%dm !! Ich weiss nicht wo dieser Faktor herkommt, hier ist
%dm    das, was ich dem THz review mit Philipp entkommen habe,
%dm    dieser Vorfaktor ist konsistent mit dem, was wir immer
%dm    gemacht haben
%dm Harald, sollte stimmen, aber bitte schau drueber (habe kein PDF erzeugt ...):
\frac{\pi \omega}{3V\, \hbar \, c \, \epsilon_0 \,n(\omega)}
\left(1-e^{-\beta\hbar\omega}\right)I(\omega),
\label{eq:IRalpha_absorption}\\
    I(\omega) &= \frac{1}{2\pi}\int_{-\infty}^{\infty}\!\dd t\; e^{-i\omega t} C(t) ,
\label{eq:IRalpha_spectral_density}
\end{align}
\end{subequations}

where $V$ is the volume, $n(\omega)$ the refractive index (approximately unity for
gas-phase systems), and
$C(t) = \langle\hat{\boldsymbol{\mu}}(0)\cdot\hat{\boldsymbol{\mu}}(t)\rangle$ is the
unsymmetrized (``standard'') quantum dipole autocorrelation function. Here \(\epsilon_0\) is the vacuum permittivity. The variable \(\omega\) is angular frequency and \(I(\omega)\) is the dipole spectral density.
The spectral density $I(\omega)$ must satisfy the quantum detailed-balance condition
$I(-\omega) = e^{-\beta\hbar\omega}I(\omega)$, which is violated by any classical-like
%dm trajectory
TCF.
The Kubo-transformed TCF is defined as follows.\cite{Kubo1957/10.1143/JPSJ.12.570}
\begin{equation}
    C_{\rm K}(t) = \frac{1}{\beta}\int_0^\beta\!\dd\lambda\;
    %dm C_K(t) = \frac{1}{\beta}\int_0^\beta\!\dd\lambda\;
    \langle\hat{\boldsymbol{\mu}}(0)\cdot\hat{\boldsymbol{\mu}}(t+i\hbar\lambda)\rangle
%dm ,
= \langle\hat{\boldsymbol{\mu}}(0)\cdot\hat{\boldsymbol{\mu}}(t)\rangle_{\rm K}
%dm
\ensuremath{.}
    \label{eq:CKubo}
\end{equation}
The integration variable \(\lambda\) spans imaginary time, and the subscript \(\mathrm K\) denotes the Kubo transform.
This function satisfies the classical symmetry condition $C_{\rm K}(t) = C_{\rm K}(-t)$, making it
%dm structurally
%dm satisfies the classical symmetry conditions $C_K(t) = C_K(-t)$, making it structurally
%dm identical
isomorphic
to the classical TCF
%dm and the natural target for quasiclassical approximation.
which provides the proper basis for quasiclassical approximations.
%
%dm siehe meine Ausfuehrung
% The standard and Kubo forms are related in frequency space by\cite{Witt2009}
%
% \begin{equation}
%     \tilde{I}(\omega) = \frac{\beta\hbar\omega}{1-e^{-\beta\hbar\omega}}\,\tilde{C}_K(\omega) .
%     \label{eq:QCF}
% \end{equation}
%
%dm  hier nun
Thus, the Kubo form of Eq.~\eqref{eq:IRalpha} reads
\begin{align}
    \alpha(\omega) =
	&\frac{\pi \omega}{3V\, \hbar \, c \, \epsilon_0 \,n(\omega)} \> (\beta\hbar\omega) \>  \nonumber\\
%dm2 !!! Christoph, koennte man hier einen linebreak haben da die
%dm2     Gleichung zu breit ist. Nun sehr ungerne wuerde ich die
%dm2     Vorfaktoren zusammenziehen da man nur mit dieser Schreibweise den
%dm2     Effekt der Kubotrafo im Vergleich zu oben (149) sieht.
%cs2 Done
	&{}\times\frac{1}{2\pi}\int_{-\infty}^{\infty}\!\dd t\; e^{-i\omega t}
\langle\hat{\boldsymbol{\mu}}(0)\cdot\hat{\boldsymbol{\mu}}(t)\rangle_{\rm K}
\ensuremath{.}
    \label{eq:IRalpha-Kubo}
\end{align}
%
%dm2 it
This expression
readily yields the classical approximation when using
the classical TCF therein,
$\langle {\boldsymbol{\mu}}(0)\cdot {\boldsymbol{\mu}}(t)\rangle$,
and is the basis for various quasiclassical expressions when using
quasiclassical TCFs,
$\langle\hat{\boldsymbol{\mu}}(0)\cdot\hat{\boldsymbol{\mu}}(t)\rangle_{\rm QC}$,
instead of Kubo's exact quantum TCF following Ref.~\citenum{Ramirez2004}.
We note that what is often called the ``harmonic quantum correction factor''
is included without further ado when following this approach.\cite{Ramirez2004}

% para - Quasiclassical approximation: CMD/RPMD/BCMD -> Kubo TCF

For vibrational spectroscopy, CMD, RPMD, TRPMD, and BCMD replace the exact Kubo TCF with a quasiclassical trajectory-based approximation, $\langle\hat{\boldsymbol{\mu}}(0)\cdot\hat{\boldsymbol{\mu}}(t)\rangle_{\rm QC}$.\cite{Ramirez2004,Witt2009,Ivanov-2013-PCCP-15-10270} CMD uses the centroid dipole autocorrelation, whereas RPMD, TRPMD, and BCMD use the bead-average estimator in Eq.~\eqref{eq:RPMDtcf}.
%
%dm   brauchen wir nicht, siehe oben:
%dm When the spectral density from these quasiclassical TCFs is used directly, detailed
%dm balance can be restored \textit{a posteriori} by multiplication with the harmonic quantum
%dm correction factor (QCF) $Q(\omega) = \beta\hbar\omega/(1-e^{-\beta\hbar\omega})$, which
%dm Eq.~\eqref{eq:QCF} shows to be the exact Kubo prefactor in the harmonic
%dm limit.\cite{Ramirez2004,Zwanzig1965/10.1146/annurev.pc.16.100165.000435}

% para - Protocol for computing vibrational spectra
A typical CP2K protocol for path-integral vibrational spectroscopy separates equilibrium sampling from the subsequent approximate dynamics.
%dm \cite{Cao1994,Craig2004,Witt2009} f

First, a massively thermostatted PIMD simulation generates a converged canonical ensemble in the extended ring-polymer phase space. Initial conditions for CMD, RPMD, TRPMD, or BCMD are then drawn from this ensemble, with normal-mode velocities rescaled to preserve the thermal energy of each mode. Multiple independent trajectories are propagated according to the selected method: RPMD production is microcanonical, whereas CMD, TRPMD, and BCMD retain their prescribed treatment of the noncentroid modes. Finally, the dipole TCFs are Fourier transformed and averaged over the ensemble to obtain the canonical vibrational spectrum.
The number of beads required for convergence scales roughly as
$P \sim \omega_{\max}\beta\hbar/2$. For liquid water
at ambient conditions, approximately $P = 32$--$64$
%dm
Trotter replicas are required.

% para - CH5+ as showcase: extreme NQEs on vibrational spectrum
The protonated methane cation \ce{CH5+}
%dm
together with its H/D isotopologues
provides a compelling demonstration of the
%dm
overriding
impact of NQEs on vibrational spectra
%dm
as obtained from approximate quantum dynamics generated by path-integral simulations.
Due to the extremely flat potential energy surface,
%dm along the hydrogen-scrambling coordinates,
all five hydrogen atoms are essentially equivalent on the timescale of
molecular vibrations, rendering \ce{CH5+} a prototypically fluxional molecule whose
spectrum cannot be interpreted in terms of conventional normal modes.
See Refs.~\citenum{Asvany-2005-Science-309-1219,Ivanov-2010-NatChem-2-298,Witt-2011-JPCL-2-1377,Ivanov-2013-PCCP-15-10270} for original literature and a review.
%
%dm
Following the pioneering fully adiabatic \textit{ab-initio} CMD calculation
of IR spectra based on DFT,\cite{Witt-2011-JPCL-2-1377}
a very accurate high-dimensional neural-network (HDNN) potential-energy
surface in conjunction with an HDNN dipole-moment surface, both at the
coupled-cluster singles, doubles, and perturbative triples [CCSD(T)] level,
have been used to compute the spectra for all six isotopologues
using RPMD and CMD.\cite{ch5p-nnpes,Beckmann2024/10.1039/d4cp02295e}
%dm Beckmann~\textit{et al.}\cite{Beckmann2024/10.1039/d4cp02295e} computed RPMD
%dm ($P=300$, $T=100\,$K) and CMD ($T=200\,$K) infrared spectra for all six isotopologues
%dm using a near-CCSD(T) HDNN potential energy
%dm surface~\cite{TODO} and an HDNN dipole moment
%dm surface,\cite{Beckmann2022/10.1021/acs.jctc.2c00511} as implemented in CP2K
%dm together with \textsc{RubNNet4MD}.~\cite{RubNNet4MD-code}
%
We reproduce these spectra together with new results for TRPMD,
BCMD, and CMD at 100~K in Fig.~\ref{fig:ir_ch5plus} for
CH$_{5}^{+}$, CH$_2$D$_{3}^{+}$, and CD$_{5}^{+}$.
%dm CMD ($T=100\,$K), and BCMD in Fig.~\ref{fig:ir_ch5plus} for CH$_{5}^{+}$, CH$_2$D$_{3}^{+}$, and CD$_{5}^{+}$.
%HF: ?? together? the NNPs are trained with RubNNet4MD but then CP2K does it alone
%dm2 CP2K together with \textsc{RubNNet4MD}\cite{RubNNet4MD-code} as interface to HDNNs.
The HDNN potential and dipole-moment surfaces were trained with \textsc{RubNNet4MD},\cite{RubNNet4MD-code} and the path-integral simulations were performed with CP2K.
Classical MD yields
%dm a structured spectrum
artificially overstructured spectra
%dm
with peaks
at too high frequencies, reflecting the
missing zero-point fluctuations, whereas RPMD~-- which captures
%dm
quantum
delocalization over the flat
hydrogen-scrambling surface~-- produces spectra in very good
agreement with the available experimental laser-induced-reaction spectra.%
%dm  of Asvany~\textit{et al.}.
\cite{Asvany-2005-Science-309-1219,
%dm
Ivanov-2010-NatChem-2-298}
CMD exhibits a characteristic redshift in the high-frequency C--H stretching region
%dm
\cite{Witt-2011-JPCL-2-1377}
due to the curvature problem.\cite{Witt2009,Ivanov-2013-PCCP-15-10270}
The right panels of Fig.~\ref{fig:ir_ch5plus} compare RPMD with TRPMD and BCMD,
demonstrating their close similarity.
Overall, these spectra illustrate how
%dm NQEs
NQEs
do not merely shift peak positions but qualitatively alter the
entire spectral
%dm envelope
line shape
through large-amplitude quantum delocalization.

%%%%%%%%%%%%%%%%%%%%%%%%%%%%%%%%%%%%%%%%%%%%%%%%%%
\subsubsection{Ultraviolet--Visible and Electronic Spectroscopy}
%%%%%%%%%%%%%%%%%%%%%%%%%%%%%%%%%%%%%%%%%%%%%%%%%%

% para - UV/vis and electronic spectra from PI trajectory sampling
NQEs influence not only vibrational but also electronic spectra:
%dm the
The
same zero-point nuclear fluctuations that shift and broaden IR bands modify the positions
and line shapes of
UV--vis absorption bands, often significantly even at room
temperature.\cite{%
%dm added
DellaSala_2004_a,
Kaczmarek2009/10.1021/jp8081936,Tsuru2024/10.1002/anie.202416058}
The photoabsorption cross section can be computed via the nuclear ensemble approach (NEA)
as a canonical average of vertical electronic excitation
energies.\cite{Kaczmarek2009/10.1021/jp8081936}
\begin{equation}
    \sigma(\omega) \approx \frac{2\pi^2 e^2}{m_e c\,\epsilon_0}
    \left\langle\sum_m f_{m0}\,\delta(\omega - \omega_{m0})\right\rangle ,
    \label{eq:photoabs}
\end{equation}
where $f_{m0}$ and $\hbar\omega_{m0}$ are the oscillator strength and excitation energy
of the $m$-th excited state, and the average is taken over nuclear configurations from
classical or path-integral trajectories.
In Eq.~\eqref{eq:photoabs}, \(m_e\) is the electron mass, \(\delta\) is the Dirac delta distribution, and \(c\), \(\epsilon_0\), and \(\omega\) have the meanings given above.
%
%dm Kaczmarek, Shiga, and Marx\cite{Kaczmarek2009/10.1021/jp8081936} demonstrated in
This approach underpinned the first on-the-fly \textit{ab-initio} RPMD calculation of IR and ultraviolet spectra, using RI second-order M{\o}ller--Plesset theory and configuration interaction singles with perturbative doubles [CIS(D)] for the floppy hydrazine molecule, and showed that NQEs yield measurable improvements over classical AIMD at room temperature.%
\cite{Kaczmarek2009/10.1021/jp8081936}

\begin{figure}[!tbp]
    \centering
    \includegraphics[width=1.0\linewidth]{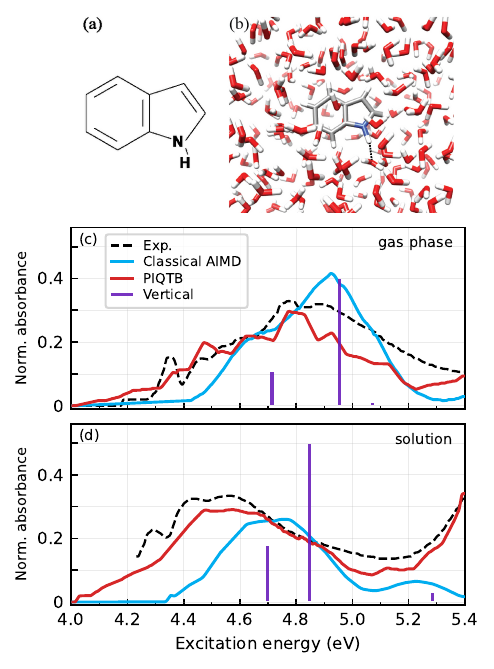}
    \caption{Ultraviolet--visible absorption spectra of indole computed via the nuclear ensemble approach with
    ADC(2)
%dm /aug-cc-pVTZ
vertical excitation energies using \textsc{Turbomole} on
    configurations from classical AIMD and PIQTB ($T=300\,$K)
%dm $P=8$, $T=300\,$K)
    trajectories.\cite{Tsuru2024/10.1002/anie.202416058}
    (a) Chemical structure of indole. (b) Representative snapshot of indole in aqueous
    solution.
    (c) Gas phase and (d) aqueous solution absorption spectra from PIQTB and classical
    AIMD compared with experiment. Vertical bars indicate excitation energies from
    single-point calculations.
%dm     PIQTB reproduces the experimental gas-phase lineshape and solvatochromic redshift in
%dm     solution without artificial shifting, while classical AIMD fails.
    Adapted from Ref.~\citenum{Tsuru2024/10.1002/anie.202416058}, Copyright 2024 Wiley-VCH GmbH.}
    \label{fig:uvvis_indole}
\end{figure}

%dm paragraph
More recently, calculations for indole in aqueous solution showed that NQEs are essential for reproducing the experimental UV--vis absorption spectrum.%
\cite{Tsuru2024/10.1002/anie.202416058}
%
%dm

Within the NEA, PIQTB sampled NQEs in the gas and liquid phases on an equal footing, while ADC(2) described the electronic excitations using a hybrid microsolvation--continuum model.
%
%dm 4000 configurations were extracted from classical AIMD and PIQTB ($P=8$, $T=300\,$K) trajectories; for each snapshot, vertical excitation energies of indole in a cluster of explicitly H-bonded water molecules embedded in a continuum solvation model were computed with correlated wavefunction theory.\cite{Tsuru2024/10.1002/anie.202416058}
%
PIQTB sampling yielded quantitative agreement with the experimental solvatochromic
redshift and line shape of indole in both the gas phase and aqueous solution without any
artificial
%dm
peak
shifting or
%dm
intensity
scaling, while classical AIMD failed to capture these features
%dm (Fig.~\ref{fig:uvvis_indole}).
as shown in Fig.~\ref{fig:uvvis_indole}.

%dm

Together, these studies show that path-integral sampling of the nuclear degrees of freedom is not merely a refinement but an essential ingredient for quantitative electronic spectroscopy of clusters, floppy molecules, and hydrogen-bonded chromophores in solution, where quantum fluctuations strongly influence both peak positions and solvatochromic shifts between the gas and liquid phases.\cite{DellaSala_2004_a,Kaczmarek2009/10.1021/jp8081936,Tsuru2024/10.1002/anie.202416058}

\section{Nonadiabatic Nuclear Dynamics in Electronically Excited States: Surface Hopping and On-the-Fly CP2K Interfaces}
\label{sec:excited-state-nuclear-dynamics}
\label{sec:nonadiabatic-md-surface-hopping}

%TDDFT/TD-DFPT, {\bf{Anna Hehn}}
% BEGIN INLINED FROM Surface_hopping.tex
In contrast to Ehrenfest dynamics, trajectory surface hopping (TSH) propagates each classical trajectory on a single adiabatic (non-averaged) potential-energy surface. An ensemble of independently propagated trajectories with stochastic transitions between surfaces then approximates the branching of the nuclear wave packet.
On the active electronic state $M$, nucleus $\alpha$ follows Newton's equation of motion:
\begin{equation}
M_{\alpha} \frac{\partial^2 \mathbf{R}_{\alpha}(t)}{\partial t^2}
=
-\nabla_{\alpha} E_M\!\left(\mathbf{R}(t)\right).
\label{classical_newtons}
\end{equation}
The symbol $M_{\alpha}$ is the nuclear mass, and $E_M$ is the energy of the active adiabatic state. The electronic wavefunction is expanded in the adiabatic states $|\Psi^M(\mathbf{R}(t))\rangle$ with coefficients $d^M(t)$ as
\begin{equation}
|\Psi(\mathbf{R}(t))\rangle
=
\sum_M d^M(t)|\Psi^M(\mathbf{R}(t))\rangle.
\end{equation}
In atomic units, these coefficients obey the time-dependent Schr\"odinger equation:
\begin{equation}
i\frac{\mathrm{d}d^M(t)}{\mathrm{d}t}
=
\sum_N d^N(t)
\left[
\delta_{MN}E_N(\mathbf{R}(t))
-i\tau_{MN}(t)
\right]
\ensuremath{,}
\label{eom_for_d}
\end{equation}
where \(\delta_{MN}\) is the Kronecker delta, \(E_N\) is the energy of adiabatic state \(N\), and the NACEs are
\begin{equation}
\tau_{MN}(t)
=
\left\langle
\Psi^M(\mathbf{R}(t))
\middle|
\frac{\partial}{\partial t}
\middle|
\Psi^N(\mathbf{R}(t))
\right\rangle .
\end{equation}
Within Tully's fewest-switches surface hopping, the nonadiabatic couplings and excited-state coefficients define the hopping probability $P_{M\rightarrow N}(t)$ during a time step $\Delta t$ as follows.\cite{tully}
\begin{equation}
P_{M\rightarrow N}(t)
=
\max\!\left[
0,
\frac{-2\Delta t}{|d_M(t)|^2}
\tau_{NM}(t)
\operatorname{Re}\!\left(d_M(t)d_N^\ast(t)\right)
\right].
\end{equation}
A hop from state $M$ to state $N$ is accepted when $\sum_{K=1}^{N-1}P_{M\rightarrow K}<r_t\leq\sum_{K=1}^{N}P_{M\rightarrow K}$, where $r_t$ is uniformly distributed on $[0,1]$.
Unlike Ehrenfest dynamics, TSH evaluates the nuclear force from the Lagrangian of the active state $M$:
\begin{equation}
\mathbf{F}^M_{\alpha}(t)
=
-\nabla_{\alpha}L^M(\mathbf{R}(t)).
\label{force_referring_to_lagrangian}
\end{equation}
Within the TDA, the corresponding adiabatic state is represented as
\begin{equation}
\left|\Psi^M\right\rangle
=
\sum_{ia\sigma}
X_{ai\sigma}^M
\left|\Phi_{ai\sigma}\right\rangle .
\label{state_referring_to_eigenvalue}
\end{equation}
In this expansion, \(i\), \(a\), and \(\sigma\) label occupied orbitals, virtual orbitals, and spin channels. The quantity \(X_{ai\sigma}^M\) is a TDA excitation amplitude, and \(\Phi_{ai\sigma}\) is the corresponding singly excited determinant.
Thus, Eqs.~\eqref{force_referring_to_lagrangian} and~\eqref{state_referring_to_eigenvalue} connect the state-specific TDA Lagrangian and eigenvectors used by the excited-state module to the nuclear propagation.

\begin{figure*}[t!]
    \centering
    \includegraphics[width=\textwidth]{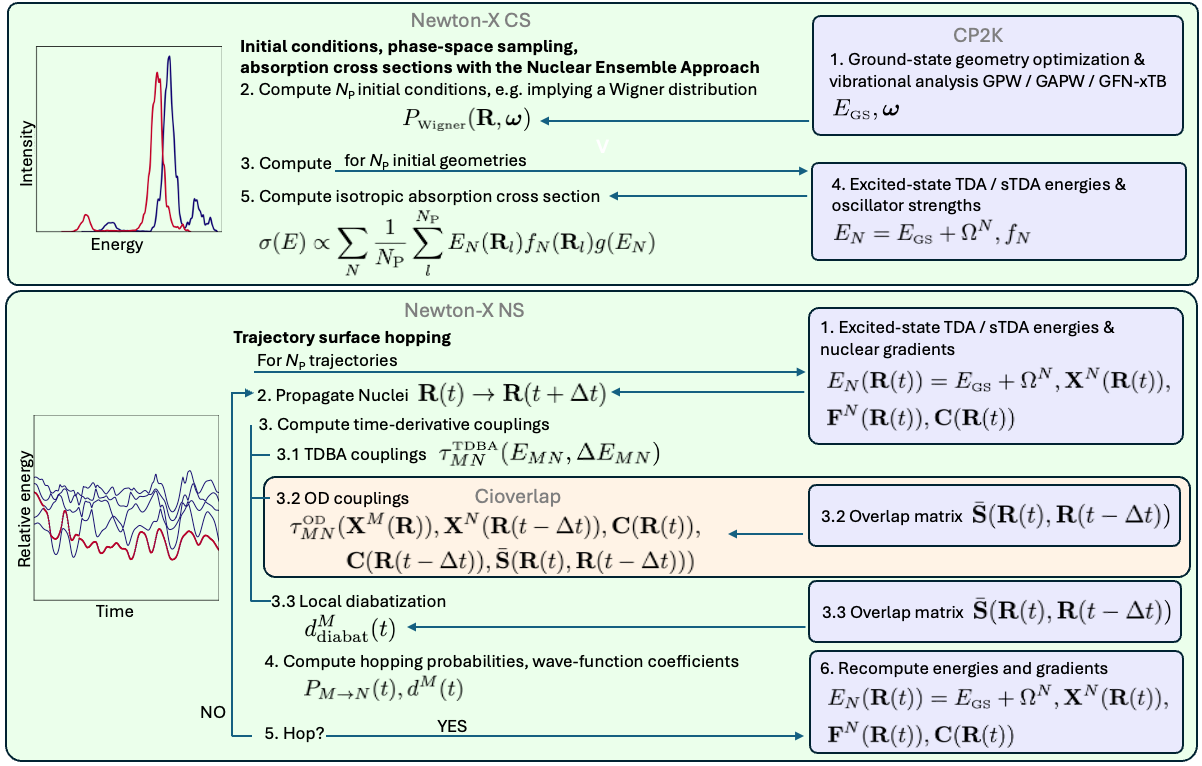}
    \caption{Schematic of the CP2K--Newton-X interface for trajectory surface hopping and nuclear-ensemble spectroscopy.}
    \label{fig:surface_hopping_flow}
\end{figure*}

An important early development toward fully \textit{ab-initio}
nonadiabatic dynamics was the direct combination of Tully-type
surface hopping with on-the-fly DFT within Car--Parrinello MD.
Doltsinis and Marx introduced a scheme that coupled a restricted
open-shell KS excited state to the ground state and evaluated
the nonadiabatic couplings from wavefunction time derivatives, thereby
enabling full-dimensional gas-phase photodynamics at a computational
complexity comparable to conventional Car--Parrinello simulations.
\cite{CarParrinello1985,DoltsinisMarx2002PRL,DoltsinisMarx2002Review}

Subsequent applications addressed aqueous proton--electron transfer and hybrid DFT-QM/MM photoswitching.\cite{LangerDoltsinisMarx2005,BoeckmannDoltsinisMarx2008} Although implemented in CPMD, this work established the condensed-phase lineage of the interfaces discussed here.

% Different interfaces
Because this mixed quantum--classical workflow requires electronic energies, gradients, and couplings at every MD step, interfaces between electronic-structure and nonadiabatic MD (NAMD) programs are common. The CP2K--Newton-X, CP2K--Libra, CP2K--GTSH, and CP2K--Zagreb interfaces link CP2K electronic-structure methods to the propagation and analysis capabilities of the respective external dynamics programs.\cite{cp2k_zagreb,cp2k_libra,cp2k_gtsh,cp2k_newtonx} CP2K--Libra can employ the neglect-of-back-reaction approximation, in which the nuclei follow ground-state forces while excitation energies and couplings are obtained from ground-state KS orbitals.\cite{akimov_nbra,libra_2022} CP2K--Zagreb combines $\Delta$SCF states with Landau--Zener hopping probabilities, whereas CP2K--GTSH implements fewest-switches surface hopping based on orbital-derivative couplings.
The CP2K--Newton-X interface has been assessed using semiempirical and DFT ground-state references and excited-state kernels from \textsc{Quickstep}, while Newton-X handles simulation setup, trajectory propagation, and data analysis.\cite{NewtonX2007,NewtonX2014,Newtonx2022,cp2k_newtonx} The reported benchmarks combine semiempirical, GGA, or ADMM-approximated hybrid references with simplified Tamm--Dancoff approximation (sTDA) or TDA excited states and compare alternative approximations to the nonadiabatic couplings.\cite{admm,stda,xtb_overview,cp2k_newtonx} One option is the time-dependent Baeck--An (TDBA) model, which is defined by
\begin{subequations}
\label{tdba_couplings}
\begin{align}
\mathcal{K}_{MN}
&=
\frac{1}{\Delta E_{MN}}
\frac{\mathrm{d}^2\Delta E_{MN}}{\mathrm{d}t^2},
\label{tdba_curvature}\\
\tau^{\mathrm{TDBA}}_{MN}(t,t-\Delta t)
&=
\begin{cases}
\frac{\operatorname{sgn}(\Delta E_{MN})}{2}
\sqrt{\mathcal{K}_{MN}},
&
\mathcal{K}_{MN}>0,\\[0.6em]
0,
&
\mathcal{K}_{MN}\le 0 .
\end{cases}
\label{tdba_model_coupling}
\end{align}
\end{subequations}

The quantity $\Delta E_{MN}$ is the adiabatic energy gap. Alternatively, numerical time-derivative couplings can be evaluated with the orbital-derivative (OD) ansatz.\cite{od_ansatz,tdba_couplings_2,tdba_couplings,tdba_couplings_mario}
\begin{equation}
\begin{aligned}
\bar{\tau}^{\mathrm{OD}}_{MN}(t,t-\Delta t)
={}&
\sum_{ia\sigma}
\left(X_{ai\sigma}^{M}(t)\right)^{\mathrm{T}}
\Bigg[
X_{ai\sigma}^{N}(t-\Delta t)
\\
&+\sum_b
X_{bi\sigma}^{N}(t-\Delta t)
\bar{S}_{ab\sigma}(t,t-\Delta t)
\\
&-\sum_j
X_{aj\sigma}^{N}(t-\Delta t)
\bar{S}_{ji\sigma}(t,t-\Delta t)
\Bigg].
\end{aligned}
\label{formula_for_overlap_couplings}
\end{equation}
Whereas the TDBA model in Eq.~\eqref{tdba_couplings} requires only adiabatic energy gaps, the OD ansatz thus additionally requires the KS-orbital overlap matrix between consecutive time steps:
\begin{equation}
\bar{S}_{pq\sigma}(t,t-\Delta t)
=
\left\langle
\phi_{p\sigma}(\mathbf{R}(t))
\middle|
\phi_{q\sigma}(\mathbf{R}(t-\Delta t))
\right\rangle \ensuremath{.}
\end{equation}
The indices \(p\) and \(q\) label KS orbitals, \(\sigma\) labels spin, and \(\Delta t\) is the nuclear time step.
CP2K evaluates these overlaps and transfers them, together with the excited-state eigenvectors $\mathbf{X}^M$, to Newton-X to construct $\bar{\tau}^{\mathrm{OD}}_{MN}$ according to Eq.~\eqref{formula_for_overlap_couplings}. The same overlap information can also be used by Newton-X for local diabatization, which provides robust propagation near sharply peaked nonadiabatic couplings.\cite{local_diabatization} Fig.~\ref{fig:surface_hopping_flow} summarizes the resulting CP2K--Newton-X data exchange and explicitly identifies the CP2K response, gradient, and vibrational-analysis modules involved. In addition to trajectory propagation, Newton-X can generate initial geometries, velocities, and electronic states from a Wigner distribution or a ground-state trajectory. Its nuclear-ensemble workflow produces broadened absorption spectra by weighting excitation energies with transition probabilities.\cite{nea_mario}
%\textcolor{red}{Details on how vibrational frequencies are read and transformed etc.}
%%%%%%% Computational results
For crystalline pyrazine, benchmark calculations showed that a single NAMD step with an sTDA kernel, a PBE ground-state reference, a double-zeta basis, and OD couplings can be completed within minutes.\cite{cp2k_newtonx}
%    \centering
%    \resizebox{0.96\linewidth}{!}{%
 %   \begin{tikzpicture}[
 %       font=\tiny,
 %       flow/.style={->, thick, >=stealth},
  %      box/.style={draw, rounded corners=1pt, align=center, inner sep=2pt,
 %           minimum height=0.46cm},
 %       nx/.style={box, fill=orange!10, text width=1.62cm},
 %       cp/.style={box, fill=blue!7, text width=1.86cm},
 %       resultbox/.style={box, fill=green!8, text width=1.65cm}
 %   ]
 %   \node[nx] (setup) at (0.0,1.12) {Newton-X\\setup};
 %   \node[nx] (driver) at (0.0,0.36) {trajectory\\driver};
 %   \node[cp] (scf) at (2.65,1.12) {CP2K\\SCF};
 %   \node[cp] (resp) at (2.65,0.36) {TDA/TDDFT\\response};
 %   \node[cp] (grad) at (2.65,-0.40) {gradients\\OD couplings};
 %   \node[resultbox] (result) at (0.0,-0.40) {hops, %ICs,\\spectra};

  %  \draw[flow] (setup.east) -- (scf.west);
  %  \draw[flow] (scf.south) -- (resp.north);
  %  \draw[flow] (resp.south) -- (grad.north);
  %  \draw[flow] (grad.west) -- (result.east);
  %  \draw[flow] (result.north) -- (driver.south);
  %  \draw[flow] (driver.east) -- (resp.west);
  %  \end{tikzpicture}}
% END INLINED FROM Surface_hopping.tex

%\section{Electron-Phonon Coupling?}
%Frank Ortmann?
%Refs.~\cite{Dorfner2025,Merkel2025}

%\section{Dies und Das}

%Born effective charges (Hossam Elgabarty), NEGF (Matt Watkins, Cuccinotta), RIXS, ROKS

\section{Conclusions}

This review has presented CP2K as an integrated framework in which electronic-structure models, response theory, statistical sampling, and real-time propagation can be combined at a common atomistic level. The resulting hierarchy spans efficient GPW, GAPW, and tight-binding force engines, variational and time-dependent response methods, core-level and many-body spectroscopy, active-space embedding, and methods for charge localization and electronic transport. Their shared value lies not in the number of available methods in isolation, but in connecting a controlled electronic-structure approximation to the perturbation, ensemble, and observable required by experiment.

That connection is especially important beyond a single optimized structure. CP2K can evaluate response properties on thermally sampled configurations, form spectra from trajectory correlation functions, include NQEs through path-integral techniques, and couple electronic excitation to nuclear motion through Ehrenfest or surface-hopping dynamics. The same modular division among force evaluation, nuclear propagation, and property analysis also links closed-system quantities such as polarizabilities and diabatic couplings to Kubo transport and electronically open descriptions under bias. This common infrastructure is a distinctive strength for treating spectroscopy and transport in liquids, solids, interfaces, and heterogeneous environments under realistic conditions.

The developments reviewed here also identify priorities: robust periodic response and excited-state methods, scalable many-body and embedding techniques, consistent finite-temperature and quantum-nuclear sampling, and open-boundary dynamics with reliable forces. Progress along these directions will depend as much on interoperable implementations and validation as on formal method development. CP2K's combination of Gaussian and plane-wave electronic structure with sampling and dynamics provides a strong foundation for that effort and for predictive simulations in which measured response emerges from the coupled electronic, nuclear, and environmental degrees of freedom.

\begin{acknowledgments}
J.W.~acknowledges the German Research Foundation (DFG) for funding via the Emmy Noether Programme (project number 503985532), CRC 1277 (project number 314695032), RTG 2905 (project number 502572516), the German Excellence Strategy--EXC 3112/1--533767171 (Center for Chiral Electronics), and the state of Bavaria for support via the KONWIHR software initiative. T.D.K. would like to thank the European Union's Just Transition Fund, administered by the Sächsische Aufbaubank, under the InfraProNet Research 2021--2027 programme.
Part of the research was funded by the DFG (project numbers 398046241, 417590517/CRC1415 and 519869949).

\end{acknowledgments}

%\nocite{*}

\setlength{\bibsep}{0pt}
%aipnum4-2.bst 2019-01-14 (MD) hand-edited version of apsrev4-1.bst
%aipnum4-2.bst 2019-01-14 (MD) hand-edited version of apsrev4-1.bst

% ============================================================================
% SUPPLEMENTARY INFORMATION
% ============================================================================
\clearpage
\onecolumngrid
\newgeometry{left=2cm,right=2cm,top=1.85cm,bottom=1.85cm}

% Restore the standard REVTeX heading hierarchy used by the standalone SI.
\let\section\SIsection
\let\subsection\SIsubsection
\let\subsubsection\SIsubsubsection

\makeatletter
\newenvironment{fixedfigure}{%
  \par\medskip\noindent\begin{minipage}{\linewidth}%
  \def\@captype{figure}\centering
}{%
  \end{minipage}\par\medskip
}
\makeatother

% Reset visible counters and give hyperref a separate SI anchor namespace.
\setcounter{section}{0}
\setcounter{subsection}{0}
\setcounter{subsubsection}{0}
\setcounter{equation}{0}
\setcounter{figure}{0}
\setcounter{table}{0}
\setcounter{footnote}{0}
\renewcommand{\thetable}{S\arabic{table}}
\providecommand{\theHsection}{}
\providecommand{\theHsubsection}{}
\providecommand{\theHsubsubsection}{}
\providecommand{\theHequation}{}
\providecommand{\theHfigure}{}
\providecommand{\theHtable}{}
\renewcommand{\theHsection}{SI.section.\arabic{section}}
\renewcommand{\theHsubsection}{SI.subsection.\arabic{section}.\arabic{subsection}}
\renewcommand{\theHsubsubsection}{SI.subsubsection.\arabic{section}.\arabic{subsection}.\arabic{subsubsection}}
\renewcommand{\theHequation}{SI.equation.\arabic{equation}}
\renewcommand{\theHfigure}{SI.figure.\arabic{figure}}
\renewcommand{\theHtable}{SI.table.\arabic{table}}

\begin{center}
{\large\bfseries Supplementary Information for\\[0.5em]
\textit{CP2K: An electronic structure and molecular dynamics software package --\\
Dynamics, Transport, and Spectroscopic Response}\par}
\end{center}
\vspace{1em}

\section{Boundary Cases for Spectrum-Like Observables}

\begin{table}[ht!]
  \caption{Boundary cases for magnetic, vibronic, and dielectric-loss observables. These entries identify spectrum-like quantities for which CP2K supplies response tensors, excitation data, trajectories, normal modes, or dielectric-response ingredients, while the complete experimental line shape or loss-function construction requires external spectrum construction.}
\label{tab:si-boundary-cases}
\vspace{0.5em}
  \scriptsize
  \setlength{\tabcolsep}{2pt}
  \renewcommand{\arraystretch}{0.96}

  \begin{tabular}{lllll}

    \hline\hline \\[-0.5em]
    \ReviewTableCell{2.4cm}{\textbf{Observable or experiment}} &
    \ReviewTableCell{3.3cm}{\textbf{CP2K ingredients}} &
    \ReviewTableCell{4.1cm}{\textbf{Missing native capability}} &
    \ReviewTableCell{3.0cm}{\textbf{External completion}} &
    \ReviewTableCell{3.0cm}{\textbf{Boundary reason}} \\
    \\[-0.5em]
    \hline
    \\[-0.5em]
    \ReviewTableCell{2.4cm}{\SpecPartial{Nuclear magnetic resonance $J$-coupling constants and fully simulated multiplet spectra}} &
    \ReviewTableCell{3.3cm}{\SpecPartial{\texttt{SPINSPIN} in \texttt{\&PROPERTIES\%LINRES}, magnetic shieldings, and AIMD or path-integral MD snapshots; main-text NMR and path-integral dynamics sections}} &
    \ReviewTableCell{4.1cm}{\SpecPartial{No built-in spin-Hamiltonian spectrum generator with isotope statistics, scalar-coupling networks, relaxation, pulse sequence, and line-shape model}} &
    \ReviewTableCell{3.0cm}{\SpecPartial{Export tensors and couplings to a nuclear magnetic resonance line-shape or spin-dynamics program. Average over snapshots if needed}} &
    \ReviewTableCell{3.0cm}{\SpecPartial{CP2K computes response constants. The experimental spectrum is assembled from them.}} \\
    \\[-0.5em]
    \ReviewTableCell{2.4cm}{\SpecPartial{Complete electron paramagnetic resonance line shapes and coupling patterns}} &
    \ReviewTableCell{3.3cm}{\SpecPartial{\texttt{EPR} in \texttt{\&PROPERTIES\%LINRES}; \texttt{HYPERFINE\_COUPLING\_TENSOR} in \texttt{\&DFT\%PRINT}; spin densities and ensemble sampling; main-text EPR sections}} &
    \ReviewTableCell{4.1cm}{\SpecPartial{No native powder or orientation averaging, relaxation model, microwave pulse-sequence simulation, or full spin-Hamiltonian fitting procedure}} &
    \ReviewTableCell{3.0cm}{\SpecPartial{Combine CP2K tensors with a dedicated electron paramagnetic resonance simulation package and experimental-broadening model}} &
    \ReviewTableCell{3.0cm}{\SpecPartial{The native output is the magnetic-response input to a spectrum simulation, not the complete spectrum.}} \\
    \\[-0.5em]
    \ReviewTableCell{2.4cm}{\SpecPartial{Extended vibronic line-shape treatments beyond CP2K's bundled VibronicSpec tool}} &
    \ReviewTableCell{3.3cm}{\SpecPartial{\texttt{VIBRATIONAL\_}\linebreak\texttt{ANALYSIS}; \texttt{\&PROPERTIES\%TDDFPT}; excited-state forces, finite-temperature snapshots, and \texttt{tools/}\linebreak\texttt{vibronic\_spec}; main-text TDDFT, finite-temperature spectroscopy, and path-integral dynamics sections}} &
    \ReviewTableCell{4.1cm}{\SpecPartial{The mainline repository includes the VibronicSpec post-processing tool. More general Herzberg--Teller, Duschinsky rotation, solvent/ensemble, and nonadiabatic line-shape variants are not accessible through a single CP2K input procedure}} &
    \ReviewTableCell{3.0cm}{\SpecPartial{Use the bundled post-processing tool where applicable, or transfer geometries, modes, gradients, and excitation data to a specialized vibronic spectrum code}} &
    \ReviewTableCell{3.0cm}{\SpecPartial{CP2K provides native ingredients and a bundled tool, while the most general experimental vibronic spectrum remains model-dependent.}} \\
    \\[-0.5em]
    \ReviewTableCell{2.4cm}{\SpecPartial{True electron-energy-loss spectroscopy and dielectric loss-function spectra}} &
    \ReviewTableCell{3.3cm}{\SpecPartial{\texttt{\&PROPERTIES\%TDDFPT}; \texttt{REAL\_TIME\_}\linebreak\texttt{PROPAGATION} in \texttt{\&DFT}; \texttt{RI\_RPA\%GW\%BSE} in \texttt{\&XC\%WF\_}\linebreak\texttt{CORRELATION}; main-text TDDFT, real-time propagation, and Bethe--Salpeter sections}} &
    \ReviewTableCell{4.1cm}{\SpecPartial{No turnkey momentum-dependent implementation for $\operatorname{Im}[-\epsilon^{-1}(\mathbf q,\omega)]$ with local-field, finite-momentum, and collection-geometry effects}} &
    \ReviewTableCell{3.0cm}{\SpecPartial{Post-process response functions with a dielectric or loss-function implementation}} &
    \ReviewTableCell{3.0cm}{\SpecPartial{CP2K supplies electronic excitations and dielectric-response ingredients. The measured loss spectrum still needs momentum-resolved response and instrument modeling.}} \\
    \hline\hline

  \end{tabular}
\end{table}

\begin{table}[ht!]
  \addtocounter{table}{-1}
  \renewcommand{\theHtable}{\thetable.cont1}
  \caption{Boundary cases for scattering, photoemission, and chiroptical observables. These entries collect cases where CP2K provides electronic-structure, real-time, moment, localized-orbital, molecular-dynamics, or normal-mode ingredients, but not a single native implementation for scattering cross sections, detector-level intensities, rotational or chiral averaging, resonance conditions, or final line-shape modeling.}
\vspace{0.5em}
  \scriptsize
  \setlength{\tabcolsep}{2pt}
  \renewcommand{\arraystretch}{0.96}

  \begin{tabular}{lllll}

    \hline\hline \\[-0.5em]
    \ReviewTableCell{2.4cm}{\textbf{Observable or experiment}} &
    \ReviewTableCell{3.3cm}{\textbf{CP2K ingredients}} &
    \ReviewTableCell{4.1cm}{\textbf{Missing native capability}} &
    \ReviewTableCell{3.0cm}{\textbf{External completion}} &
    \ReviewTableCell{3.0cm}{\textbf{Boundary reason}} \\
    \\[-0.5em]
    \hline
    \\[-0.5em]
    \ReviewTableCell{2.4cm}{\SpecMissing{Dynamic structure factor $S(\mathbf q,\omega)$ for inelastic neutron or X-ray scattering}} &
    \ReviewTableCell{3.3cm}{\SpecMissing{\texttt{\&MOTION\%MD}; \texttt{VIBRATIONAL\_}\linebreak\texttt{ANALYSIS}; path-integral MD, velocities, forces, normal modes, and trajectories; main-text AIMD and path-integral dynamics sections}} &
    \ReviewTableCell{4.1cm}{\SpecMissing{No native coherent or incoherent scattering cross-section driver with neutron scattering lengths, X-ray form factors, resolution convolution, and reciprocal-space averaging}} &
    \ReviewTableCell{3.0cm}{\SpecMissing{Compute intermediate scattering functions from trajectories and Fourier transform externally}} &
    \ReviewTableCell{3.0cm}{\SpecMissing{CP2K is the sampling and force engine. The spectrometer-specific observable is post-processing.}} \\
    \\[-0.5em]
    \ReviewTableCell{2.4cm}{\SpecPartial{Fully simulated angle-resolved photoemission intensities}} &
    \ReviewTableCell{3.3cm}{\SpecPartial{\texttt{GW} in \texttt{\&PROPERTIES\%}\linebreak\texttt{BANDSTRUCTURE}; band structures, DOS, LDOS, PDOS, and orbital projections; main-text band-structure and GW sections}} &
    \ReviewTableCell{4.1cm}{\SpecPartial{No complete one-step photoemission implementation with dipole matrix elements, surface final states, escape depth, polarization geometry, and detector acceptance}} &
    \ReviewTableCell{3.0cm}{\SpecPartial{Combine CP2K electronic structure with dedicated photoemission matrix-element and surface-state modeling}} &
    \ReviewTableCell{3.0cm}{\SpecPartial{The manuscript can discuss quasiparticle energies and dispersions, not a full angle-resolved photoemission intensity map.}} \\
    \\[-0.5em]
    \ReviewTableCell{2.4cm}{\SpecPartial{Vibrational circular dichroism, Raman optical activity, and resonance Raman optical activity}} &
    \ReviewTableCell{3.3cm}{\SpecPartial{\texttt{\&PROPERTIES\%LINRES\%VCD}, \texttt{\&PROPERTIES\%LINRES\%DCDR}, \texttt{\&PROPERTIES\%LINRES\%POLAR}, \texttt{REAL\_TIME\_PROPAGATION}, moment and finite-field outputs; main-text DFPT, real-time propagation, and AIMD spectroscopy sections\cite{VCD_spectra_Luber,Mattiat2019}}} &
    \ReviewTableCell{4.1cm}{\SpecPartial{CP2K provides native static VCD through \texttt{\&VCD}. No single public implementation covers finite-temperature VCD, vibrational or resonance Raman optical activity, rotational/chiral averaging, resonance conditions, tensor-correlation analysis, and experimental line-shape construction}} &
    \ReviewTableCell{3.0cm}{\SpecPartial{Use CP2K response tensors, polarizabilities, optical-activity ingredients, and molecular-dynamics trajectories in an external chiroptical or Raman-optical-activity analysis\cite{roa_martin_2,raman_martin_2019,roa_martin}}} &
    \ReviewTableCell{3.0cm}{\SpecPartial{Static response-based VCD is native. Finite-temperature VCD, Raman optical activity, and complete experimental line shapes require additional trajectory or response analysis.}} \\
    \\[-0.5em]
    \ReviewTableCell{2.4cm}{\SpecPartial{Electronic circular dichroism spectra}} &
    \ReviewTableCell{3.3cm}{\SpecPartial{\texttt{REAL\_TIME\_PROPAGATION} in \texttt{\&DFT}; electric and magnetic delta kicks; \texttt{MOMENTS} in \texttt{\&DFT\%PRINT}; \texttt{MOMENTS\_FT}; gauge ingredients; main-text gauge-choice and delta-kick sections\cite{Mattiat2022}}} &
    \ReviewTableCell{4.1cm}{\SpecPartial{No native electronic-circular-dichroism spectrum driver with the full optical-activity observable, gauge analysis, rotational averaging, and line-shape construction}} &
    \ReviewTableCell{3.0cm}{\SpecPartial{Use CP2K real-time response and moment data as ingredients for an external electronic-circular-dichroism analysis}} &
    \ReviewTableCell{3.0cm}{\SpecPartial{The mainline code contains relevant real-time, moment, and gauge machinery, but not an end-to-end electronic-circular-dichroism spectrum driver.}} \\
    \\[-0.5em]
    \ReviewTableCell{2.4cm}{\SpecPartial{Bulk chiral optical response from local dipoles and propagated localized orbitals}} &
    \ReviewTableCell{3.3cm}{\SpecPartial{\texttt{TOTAL\_DIPOLE} in \texttt{\&DFT\%LOCALIZE\%PRINT}, \texttt{LOCALIZED\_MOMENTS}, Wannier-state and local-dipole machinery, and real-time propagation ingredients}} &
    \ReviewTableCell{4.1cm}{\SpecPartial{No public end-to-end spectroscopy driver that turns these localized-orbital and local-moment ingredients into a complete bulk chiral optical spectrum}} &
    \ReviewTableCell{3.0cm}{\SpecPartial{Post-process localized dipoles, moments, or real-time response data with a dedicated chiral-optical response model}} &
    \ReviewTableCell{3.0cm}{\SpecPartial{The public code contains useful local-response ingredients, but construction of the final chiral optical spectrum requires a dedicated response model and external analysis.}} \\
    \\[-0.5em]
    \hline\hline

  \end{tabular}
\end{table}
\twocolumngrid

\section{\texorpdfstring{Periodic Electronic-Structure Representations and Working Equations}{Periodic Electronic-Structure Representations and Working Equations}}
\label{sec:si-periodic-electronic-structure}

Section~II of the main manuscript introduces electronic-structure approaches that share a common periodic setting but differ in their energy functionals. The working equations presented here therefore distinguish the common ingredients of periodic calculations, including Bloch representations, \(\bk\)-point sampling, Brillouin-zone symmetry, electrostatic boundary conditions, and lattice derivatives, from the method-specific DFTB and GFN-xTB energy models. The common framework applies, with representation-dependent details, to periodic DFT, the Harris functional, and extended tight-binding methods.

\subsection{\texorpdfstring{\ensuremath{k}-Point Sampling and Brillouin-Zone Symmetry}{k-Point Sampling and Brillouin-Zone Symmetry}}
\label{sec:si-kpoint-symmetry-details}

Regular meshes can be generated directly in the \texttt{\&KPOINTS} section, for example, as Monkhorst--Pack or MacDonald-shifted grids.\cite{Monkhorst1976,MacDonald1978} Explicit lists of $\bk$ points are used for user-defined sampling and band-structure paths. General $\bk$ points require complex Bloch wavefunctions, whereas real wavefunctions are restricted to $\Gamma$ and other time-reversal-invariant special points for which the phase factors can be represented without loss of generality by real matrices. Since different $\bk$ points are independent apart from their contribution to the final Brillouin-zone sum, CP2K can distribute them over Message Passing Interface (MPI) subgroups. This distribution is complementary to the real-space grid, sparse-matrix, and diagonalization parallelism already used within each $\bk$ point.

The cost of dense $\bk$-point calculations can be reduced substantially by exploiting symmetry. The full grid can always be evaluated explicitly, but for regular meshes CP2K can instead construct an irreducible set of $\bk$ points and assign each representative its symmetry weight. Conceptually, a full mesh $\mathcal{K}$ is partitioned into symmetry orbits. A space-group operation $g=(\mathcal{R}_g^{(r)},\boldsymbol{\tau}_g)$ acts on real-space positions as $\br'=\mathcal{R}_g^{(r)}\br+\boldsymbol{\tau}_g$. The corresponding reciprocal-space operation is the contragredient one, $\mathcal{R}_g^{(k)}=(\mathcal{R}_g^{(r)})^{-T}$, and maps a point $\bk$ to an equivalent point $\bk'=\mathcal{R}_g^{(k)}\bk+\bG$, with $\bG$ a reciprocal lattice vector that folds the result back into the first Brillouin zone. In an orthonormal Cartesian representation, $\mathcal{R}_g^{(k)}$ and $\mathcal{R}_g^{(r)}$ have the same numerical matrix, but in fractional crystallographic coordinates they generally do not. Time-reversal symmetry adds the antiunitary relation $\bk\rightarrow-\bk$. The representative $\bk_\alpha$ of an orbit contributes with a weight proportional to the number of full-grid points generated from it:
\begin{equation}
\begin{aligned}
    w_\alpha
    &= \frac{|\mathcal{O}(\bk_\alpha)|}{|\mathcal{K}|},~\text{with}\\
    \mathcal{O}(\bk_\alpha)
    &= \{\, \mathcal{R}_g^{(k)}\bk_\alpha+\bG,\,
        -\mathcal{R}_g^{(k)}\bk_\alpha+\bG \,\}.
\end{aligned}
    \label{eq:si-qs-kpoint-orbits}
\end{equation}
The set \(\mathcal O(\bk_\alpha)\) is the symmetry orbit of representative \(\bk_\alpha\), vertical bars denote set cardinality, and \(w_\alpha\) is the normalized orbit weight.
Duplicate points are removed within the mesh tolerance, so that the sum of irreducible weights remains normalized to one.

For a PW representation, this reduction is mostly a reciprocal-space operation. In CP2K's Gaussian representation, it must also be made consistent with the atom-centered basis and with the real-space matrix construction summarized in Section~II~A~2. A symmetry operation is usable only if every atom $A$ is mapped onto an atom $A'$ of the same kind, possibly in a neighboring cell:
\begin{equation}
    \mathcal{R}_g^{(r)} \mathbf{r}_A + \boldsymbol{\tau}_g
    = \mathbf{r}_{A'} + \Delta \bR_{A,g}.
    \label{eq:si-qs-atom-mapping}
\end{equation}
In this relation, \(\mathbf r_A\) and \(\mathbf r_{A'}\) are the positions of the symmetry-related atoms, \(\mathcal R_g^{(r)}\) and \(\boldsymbol{\tau}_g\) are the rotational and translational parts of operation \(g\), and \(\Delta\bR_{A,g}\) is a lattice translation.
The implementation therefore stores, for each accepted operation, the atom permutation, the cell shifts $\Delta\bR_{A,g}$, and the rotation of the local Gaussian basis functions. The latter is trivial for $s$ functions but nontrivial for $p$, $d$, and higher angular momentum shells. When a density, overlap, or Hamiltonian block is transformed between two symmetry-related $\bk$ points, these ingredients produce the required Bloch phases. Schematically, a matrix block transforms as
\begin{equation}
    M_{\mu\nu}^{(g)}(\bk) =
    e^{i\bk\cdot(\Delta\bR_{\nu,g}-\Delta\bR_{\mu,g})}
    \sum_{\mu'\nu'}
    U_{\mu\mu'}^{(g)}
    M_{\mu'\nu'}(\bk')
    U_{\nu\nu'}^{(g)\,*},
    \label{eq:si-qs-kpoint-symm-matrix}
\end{equation}
where \(\mathbf M\) denotes, for example, a density or Hamiltonian matrix and \(\mathbf U^{(g)}\) contains the atom permutation and angular part of the Gaussian-basis rotation. The point \(\bk'\) is the image of \(\bk\) under \(g\), and \(\mu\) and \(\nu\) label Gaussian basis functions. This phase-aware atom mapping is the essential step that makes full atomic $\bk$-point symmetry more delicate than a pure reduction of the list of mesh points.

The default backend follows CP2K's long-standing K290 special-point machinery,\cite{Worlton1972} whereas an optional SPGLIB-based path can be used to obtain and apply space-group operations from an explicit symmetry search. The two tasks are deliberately separated.\cite{spglibv1} In the input, \texttt{SYMMETRY\_REDUCTION\_METHOD} determines how the irreducible mesh is generated, while \texttt{SYMMETRY\_BACKEND} selects the backend used for the actual symmetry transformations of matrices and density information. This separation allows, for example, SPGLIB and K290 to be compared as reduction methods without changing the production transformation path. It is also useful for nonprimitive or nonsymmorphic crystals, where fractional translations and atom-cell shifts are precisely the information that decides whether a formal space-group operation can be applied safely to the Gaussian-basis matrices. The \texttt{FULL\_GRID} keyword disables the reduction, \texttt{INVERSION\_SYMMETRY\_ONLY} restricts the reduction to time-reversal/inversion, and the symmetry tolerance controls how strictly the crystal geometry is matched to the candidate operations. For explicitly supplied \texttt{GENERAL} $\bk$-point lists, the user-provided weights are kept, because there is no guaranteed parent mesh from which symmetry-equivalent weights can be reconstructed unambiguously.

\subsection{\texorpdfstring{Periodic Atom-Centered Representation}{Periodic Atom-Centered Representation}}
\label{sec:si-periodic-xtb-details}

The atom-centered representation below underlies the periodic Gaussian-basis formulation of Section~II~A and the extended tight-binding methods of Section~II~C. The matrix elements and density representations are method dependent, but the Bloch construction, lattice Fourier transformation, and Brillouin-zone summation provide a common framework. The subsequent energy expressions specialize this framework to DFTB,\cite{Seifert2007} SCC-DFTB,\cite{Elstner1998} DFTB3,\cite{Gaus2011} and GFN\(n\)-xTB (\(n=0,1,2\)) methods.\cite{Grimme2017,Bannwarth2019,Pracht2023,Alizadeh2026PeriodicGFN2XTB}

Let the cell matrix be $\mathbf{A}=(\mathbf{a}_1,\mathbf{a}_2,\mathbf{a}_3)$,
with volume $\Omega=\det \mathbf{A}$. Direct and reciprocal lattice vectors are
\begin{equation}
 \mathbf{L}=\mathbf{A}\mathbf{n}, \qquad
 \mathbf{G}=2\pi \mathbf{A}^{-T}\mathbf{m},
 \qquad \mathbf{n},\mathbf{m}\in \mathbb{Z}^3 .
\end{equation}
AOs $\phi_{\mu A}(\mathbf{r}-\mathbf{R}_A)$ are converted
to Bloch sums
\begin{equation}
 \chi_{\mu A}^{\mathbf{k}}(\mathbf{r}) =
 \frac{1}{\sqrt{N_\mathrm{cell}}}\sum_{\mathbf{L}}
 e^{i\mathbf{k}\cdot \mathbf{L}}
 \phi_{\mu A}(\mathbf{r}-\mathbf{R}_A-\mathbf{L}) ,
\end{equation}
where \(\mu\) labels an AO centered on atom \(A\), \(\mathbf R_A\) is its position, \(\mathbf k\) is the crystal momentum, and \(N_{\mathrm{cell}}\) is the number of cells included in the Born--von K\'arm\'an normalization.
This definition leads to lattice-Fourier-transformed matrices
\begin{equation}
\begin{aligned}
 H_{\mu\nu}^{AB}(\mathbf{k})
 &= \sum_{\mathbf{L}} H_{\mu\nu}^{AB}(\mathbf{L})
 e^{i\mathbf{k}\cdot \mathbf{L}},\\
 S_{\mu\nu}^{AB}(\mathbf{k})
 &= \sum_{\mathbf{L}} S_{\mu\nu}^{AB}(\mathbf{L})
 e^{i\mathbf{k}\cdot \mathbf{L}} .
\end{aligned}
 \label{eq:tb-bloch-matrices}
\end{equation}
The band problem is the generalized eigenvalue equation
\begin{equation}
 \mathbf{H}(\mathbf{k})\mathbf{C}_{n\mathbf{k}} =
 \varepsilon_{n\mathbf{k}}\mathbf{S}(\mathbf{k})\mathbf{C}_{n\mathbf{k}}
 \ensuremath{.}
\end{equation}
The matrices \(\mathbf H(\mathbf k)\) and \(\mathbf S(\mathbf k)\) are the Hamiltonian and AO overlap matrices, \(\mathbf C_{n\mathbf k}\) is the coefficient vector of band \(n\), and \(\varepsilon_{n\mathbf k}\) is its energy.
All one-particle quantities are obtained from Brillouin-zone sums. For
example, the real-space density matrix is
\begin{equation}
 D_{\mu\nu}^{AB}(\mathbf{L}) =
 \sum_{\mathbf{k}} w_{\mathbf{k}}
 e^{-i\mathbf{k}\cdot \mathbf{L}}
 \sum_n f_{n\mathbf{k}}
 C_{\mu n}^{A}(\mathbf{k}) C_{\nu n}^{B*}(\mathbf{k}) ,
 \label{eq:tb-density-matrix}
\end{equation}
where $w_{\mathbf{k}}$ are symmetry weights and $f_{n\mathbf{k}}$ are
occupations. The quantity \(D_{\mu\nu}^{AB}(\mathbf L)\) is the real-space AO density-matrix block and \(C_{\mu n}^{A}(\mathbf k)\) is the coefficient of AO \(\mu\) on atom \(A\) in band \(n\). This construction applies generally to periodic atom-centered electronic-structure methods. In DFT, the matrix elements additionally couple the Gaussian representation of the KS orbitals to the PW representation of the density. In DFTB, SCC-DFTB, DFTB3, and GFN-xTB, the differences instead enter through parametrized matrix elements, charge variables, and electrostatic kernels.

\subsection{\texorpdfstring{Density-Functional Tight Binding and GFN-xTB Energy Forms}{Density-Functional Tight Binding and GFN-xTB Energy Forms}}

The DFTB hierarchy follows from a Taylor expansion of the DFT energy around a
reference density \(n_0\):
\begin{equation}
\begin{aligned}
 E[n_0+\Delta n]
 &= E^{(0)}[n_0] + E^{(1)}[n_0,\Delta n]\\
 &\quad + E^{(2)}[n_0,\Delta n^2]
 + E^{(3)}[n_0,\Delta n^3]+\ldots .
\end{aligned}
 \label{eq:periodic-dftb-expansion}
\end{equation}
The band-structure contribution is evaluated from Eq.~\eqref{eq:tb-density-matrix},
while the short-range repulsive terms are lattice sums over atom pairs. In a
compact notation, the second-order SCC-DFTB energy is
\begin{subequations}
\label{eq:dftb-energy-forms}
\begin{equation}
\begin{aligned}
 E_\mathrm{DFTB2} ={}&
 \mathrm{Tr}[\mathbf D\mathbf{H}^0]
 + E_\mathrm{rep} \\
 &+ \frac{1}{2}\sum_{A,B,\mathbf{L}}^{\prime}
 \Delta q_A\,\gamma_{AB}(\mathbf{R}_{AB}^{\mathbf{L}})\,\Delta q_B .
\end{aligned}
\label{eq:dftb2-energy}
\end{equation}
The matrices \(\mathbf D\) and \(\mathbf H^0\) are the periodic AO density matrix and zeroth-order DFTB Hamiltonian, and \(E_{\mathrm{rep}}\) is the short-range repulsive energy.
The quantities $\Delta q_A$ are Mulliken charge fluctuations,
$\mathbf{R}_{AB}^{\mathbf{L}}=\mathbf{R}_B+\mathbf{L}-\mathbf{R}_A$, and the
prime excludes the singular self-interaction. The function $\gamma_{AB}$ is a
damped Coulomb interaction that tends to $1/R$ at long range. In periodic
boundary conditions this asymptotic Coulomb part is treated by Ewald summation,
whereas the remaining short-range difference $\gamma_{AB}(R)-1/R$ is summed in
real space.

DFTB3 adds a third-order charge response, usually written as
\begin{equation}
 E_\mathrm{DFTB3} = E_\mathrm{DFTB2}
 + \frac{1}{3}\sum_{A,B,\mathbf{L}}^{\prime}
 \Gamma_{AB}(\mathbf{R}_{AB}^{\mathbf{L}})
 \Delta q_A^2 \Delta q_B ,
 \label{eq:dftb3-energy}
\end{equation}
\end{subequations}
with element-dependent hardness derivatives entering $\Gamma_{AB}$.\cite{Gaus2011}
For solids the same periodic monopole convention as in
Eq.~\eqref{eq:dftb2-energy} must be used for energy, forces, and stress.

The GFN-xTB family uses the same local-orbital machinery but reorganizes the
energy in terms of method-specific empirical and self-consistent contributions.
For periodic simulations in CP2K the relevant schematic forms are
\begin{subequations}
\label{eq:gfn-periodic-energies}
\begin{align}
E_{\mathrm{GFN0}}
&=
\operatorname{Tr}\!\left[\mathbf D\mathbf{H}^0\right]
+E_{\mathrm{rep}}
+E_{\mathrm{disp}}^{\mathrm{D4},0}\notag\\
&\quad
+E_{\mathrm{EEQ}}
+E_{\mathrm{srb}}
+G_{\mathrm{Fermi}},
\label{eq:gfn0-periodic-energy}\\
E_{\mathrm{GFN1}}
&=
\operatorname{Tr}\!\left[\mathbf D\mathbf{H}^0\right]
+E_{\mathrm{rep}}
+E_{\mathrm{disp}}^{\mathrm{D3}}\notag\\
&\quad
+E_{\mathrm{iso}}^{(2)}(\{\Delta q_l\})
+E_{\mathrm{iso}}^{(3)}(\{\Delta q_l\})\notag\\
&\quad
+E_{\mathrm{XB}}
+G_{\mathrm{Fermi}},
\label{eq:gfn1-periodic-energy}\\
E_{\mathrm{GFN2}}
&=
\operatorname{Tr}\!\left[\mathbf D\mathbf{H}^0\right]
+E_{\mathrm{rep}}
+E_{\mathrm{disp}}^{\mathrm{D4}}(\{q_A\})\notag\\
&\quad
+E_{\mathrm{iso}}^{(2)}
+E_{\mathrm{iso}}^{(3)}
+E_{\mathrm{AES}}^{(2)}
  \!\left(\{q_A,\boldsymbol{\mu}_A,\boldsymbol{\Theta}_A\}\right)\notag\\
&\quad
+E_{\mathrm{AXC}}^{(2)}
+G_{\mathrm{Fermi}} .
\label{eq:gfn2-periodic-energy}
\end{align}
\end{subequations}
In these expressions, \(E_{\mathrm{disp}}\) is the indicated D3 or D4 dispersion energy, \(E_{\mathrm{rep}}\) is the repulsive energy, \(E_{\mathrm{iso}}^{(2/3)}\) are isotropic second- and third-order electrostatic terms, \(E_{\mathrm{AES}}^{(2)}\) and \(E_{\mathrm{AXC}}^{(2)}\) are anisotropic electrostatic and XC terms, and \(G_{\mathrm{Fermi}}\) is the finite-electronic-temperature entropy contribution.
Here $\Delta q_l$ denotes shell or atomic charge fluctuations, $E_\mathrm{EEQ}$
is the electronegativity-equilibration term of GFN0-xTB, $E_\mathrm{XB}$ is the
GFN1-xTB halogen-bond correction, and
$\boldsymbol{\mu}_A$ and $\boldsymbol{\Theta}_A$ are the cumulative atomic
dipoles and quadrupoles used by GFN2-xTB. GFN0-xTB and CP2K-native GFN1-xTB are implemented directly in CP2K. The periodic
GFN2-xTB implementation in CP2K, including multipolar Ewald electrostatics,
\(k\)-point sampling, analytic forces, stress contributions, and molecular-solid
benchmarks, has recently been reported in Ref.~\citenum{Alizadeh2026PeriodicGFN2XTB}.
Additional xTB variants are available through the \texttt{tblite} library interface.\cite{Katbashev2025}
The \texttt{tblite} interface also provides access to the ionization-potential
and electron-affinity targeted IPEA-xTB method,\cite{Asgeirsson2017} which is
intended, for example, for electron-ionization mass spectrometry workflows based
on GFN-xTB geometries.

\subsection{\texorpdfstring{Periodic Electrostatics and Boundary Conditions}{Periodic Electrostatics and Boundary Conditions}}

The periodic Coulomb boundary problem is shared by calculations based on DFT, the Harris functional, and extended tight-binding models, although its numerical representation is method dependent. GPW and GAPW methods treat continuous electronic densities with periodic Poisson solvers, whereas the DFTB and GFN-xTB models below use atom-centered monopolar or multipolar variables. The scalar Ewald decomposition supplies the common long-range kernel for the latter charge models.

The central long-range object is the Ewald decomposition of the Coulomb kernel.\cite{Ewald1921}
For a neutral three-dimensional cell, the decomposition reads
\begin{equation}
\begin{aligned}
 \frac{1}{r}
 &= \frac{\mathrm{erfc}(\eta r)}{r}
 + \frac{4\pi}{\Omega}\sum_{\mathbf{G}\ne\mathbf{0}}
 \frac{\exp[-G^2/(4\eta^2)]}{G^2}
 e^{i\mathbf{G}\cdot\mathbf{r}}\\
 &\quad + v_{\mathbf{G}=\mathbf{0}},
\end{aligned}
\label{eq:ewald-scalar}
\end{equation}
where $\eta$ is the Ewald splitting parameter, \(r=|\mathbf r|\), \(\mathbf G\) is a reciprocal-lattice vector, and \(G=|\mathbf G|\). The last term specifies the
chosen boundary convention for the zero Fourier component. (SCC-)DFTB, DFTB3, and
GFN1-xTB require this kernel for periodic monopole or shell-charge
fluctuations. For example, the periodic monopole energy can be written as
\begin{equation}
\begin{aligned}
 E_{qq}^{\mathrm{per}}
 &= \frac{1}{2}\sum_{A,B,\mathbf{L}}^{\prime}
 \Delta q_A \Delta q_B
 \frac{\mathrm{erfc}(\eta R_{AB}^{\mathbf{L}})}
      {R_{AB}^{\mathbf{L}}} \\
 &\quad + \frac{2\pi}{\Omega}
 \sum_{\mathbf{G}\ne\mathbf{0}}
 \frac{e^{-G^2/(4\eta^2)}}{G^2}
 \left|\sum_A \Delta q_A e^{i\mathbf{G}\cdot\mathbf{R}_A}\right|^2 \\
 &\quad + E_\mathrm{self}+E_{\mathbf{G}=0}.
\end{aligned}
\label{eq:periodic-monopole}
\end{equation}
The terms \(E_{\mathrm{self}}\) and \(E_{\mathbf G=0}\) are the Ewald self-interaction correction and the contribution fixed by the chosen zero-mode boundary convention, respectively.\cite{deLeeuw1980}
The same form applies to the long-range part of the damped DFTB
$\gamma$-interaction and to the EEQ-like monopole terms in GFN0-xTB.

GFN2-xTB is more demanding because its anisotropic electrostatic term contains
charge, dipole, and quadrupole couplings. The compact real-space form can be
written with Cartesian multipoles $M_A^{(l)}$ and interaction tensors
$T^{(n)}=\nabla^{(n)}(1/r)$ as
\begin{equation}
 E_\mathrm{mp} =
 \frac{1}{2}\sum_{A,B,\mathbf{L}}^{\prime}
 \sum_{l,m=0}^{2}
 \frac{(-1)^m}{l!\,m!}
 M_A^{(l)} M_B^{(m)}
 T^{(l+m)}(\mathbf{R}_{AB}^{\mathbf{L}}),
\label{eq:multipole-real-compact}
\end{equation}
where tensor contractions over Cartesian indices are implied. The labels \(A\) and \(B\) denote atoms, \(\mathbf L\) is a lattice vector, \(l\) and \(m\) are multipole ranks, \(\mathbf R_{AB}^{\mathbf L}\) is the periodic atom-pair separation, and the prime excludes the singular self term. The periodic
Ewald representation follows by differentiating Eq.~\eqref{eq:ewald-scalar}.
A useful reciprocal-space form introduces the multipolar structure factor
\begin{equation}
 S(\mathbf{G}) =
 \sum_A e^{i\mathbf{G}\cdot\mathbf{R}_A}
 \left[
 q_A + iG_\alpha \mu_{A\alpha}
 - \frac{1}{2}G_\alpha G_\beta \Theta_{A\alpha\beta}
 \right]
\ensuremath{.}
\label{eq:multipole-structure-factor}
\end{equation}
The quantities \(q_A\), \(\boldsymbol\mu_A\), and \(\boldsymbol\Theta_A\) are the charge, dipole, and quadrupole of atom \(A\). Repeated Cartesian indices \(\alpha\) and \(\beta\) are summed.
This structure factor yields
\begin{equation}
 E_\mathrm{mp}^{\mathrm{rec}} =
 \frac{2\pi}{\Omega}
 \sum_{\mathbf{G}\ne\mathbf{0}}
 \frac{e^{-G^2/(4\eta^2)}}{G^2}
 S(\mathbf{G})S^*(\mathbf{G}) .
 \label{eq:multipole-ewald-rec}
\end{equation}
The corresponding real-space term is obtained by replacing $1/r$ in
Eq.~\eqref{eq:multipole-real-compact} with
$\mathrm{erfc}(\eta r)/r$ and evaluating the differentiated screened tensors.
Self terms, the neutralizing-background convention, and surface terms must be
chosen consistently with the monopole Ewald sum. The corrected multipolar
Ewald formulas of Aguado and Madden and the compact formulation by Laino and
Hutter are the natural reference for this part of the implementation.\cite{Aguado2003,Laino2008}

\subsubsection{\texorpdfstring{Boundary Conditions and Dimensionality}{Boundary Conditions and Dimensionality}}

For molecular calculations, no thermodynamic cell stress is defined, whereas periodic calculations require the electrostatic boundary conditions to match the dimensionality of the cell. GPW and GAPW calculations impose this choice through the Poisson solver and density representation. In three-dimensional extended tight-binding calculations, (SCC-)DFTB, DFTB3, GFN0-xTB, and GFN1-xTB use monopolar Ewald terms, while GFN2-xTB additionally requires the charge--dipole--quadrupole sums of Eqs.~\eqref{eq:multipole-real-compact}--\eqref{eq:multipole-ewald-rec}. For lower-dimensional systems, the Coulomb boundary condition and real-space cell geometry must be chosen consistently. Surfaces and wires are consequently more sensitive to the boundary convention than fully three-dimensional crystals.

\subsection{\texorpdfstring{Analytic Forces, Virials, and Stress}{Analytic Forces, Virials, and Stress}}

The derivative structure is shared by electronic-structure methods formulated in nonorthogonal atom-centered bases: nuclear gradients contain Hellmann--Feynman and Pulay contributions, while the stress additionally differentiates lattice vectors, reciprocal vectors, and volume factors. The decomposition below makes the method-specific DFTB and GFN-xTB contributions explicit.

Analytic gradients require differentiating all explicit lattice sums and all
implicit charge or multipole variables. For a nuclear coordinate
$R_{A\alpha}$
\begin{equation}
\begin{aligned}
 F_{A\alpha}
 &= -\frac{\partial E}{\partial R_{A\alpha}}\\
 &= F_{A\alpha}^{H^0,S}
 + F_{A\alpha}^{\mathrm{rep}}
 + F_{A\alpha}^{\mathrm{disp}}
 + F_{A\alpha}^{\mathrm{SCC}}\\
 &\quad
 + F_{A\alpha}^{\mathrm{mp}}
 + F_{A\alpha}^{\mathrm{CN}},
\end{aligned}
\label{eq:tb-force-decomposition}
\end{equation}
where \(\alpha\) labels a Cartesian component and $F^{H^0,S}$ includes the Hellmann--Feynman and Pulay terms from the
nonorthogonal basis,\cite{Feynman1939Forces,Pulay1969Forces} $F^\mathrm{SCC}$ contains the monopolar charge response,
$F^\mathrm{mp}$ the GFN2 multipolar electrostatics, and $F^\mathrm{CN}$ the
coordination-number chain-rule contributions entering GFN-xTB and dispersion
terms. At self-consistency the response of the variational charge variables
cancels, but non-variational auxiliary quantities such as coordination numbers,
damping functions, and charge-dependent dispersion coefficients still have to
be differentiated explicitly.

The stress follows from a homogeneous strain
$\mathbf{R}_A\rightarrow(\mathbf{1}+\boldsymbol{\epsilon})\mathbf{R}_A$ and
$\mathbf{A}\rightarrow(\mathbf{1}+\boldsymbol{\epsilon})\mathbf{A}$:
\begin{equation}
 \sigma_{\alpha\beta} =
 \frac{1}{\Omega}
 \frac{\partial E}{\partial \epsilon_{\alpha\beta}},
 \qquad
 W_{\alpha\beta}=-\Omega\sigma_{\alpha\beta}.
\label{eq:tb-stress-definition}
\end{equation}
The tensors \(\sigma_{\alpha\beta}\) and \(W_{\alpha\beta}\) are the stress and virial, respectively, while \(\epsilon_{\alpha\beta}\) is the applied homogeneous strain.
For reciprocal-space Ewald terms one also needs
\begin{equation}
 \frac{\partial G_\lambda}{\partial \epsilon_{\alpha\beta}}
 = -\delta_{\lambda\alpha}G_\beta,
 \qquad
 \frac{\partial \Omega}{\partial \epsilon_{\alpha\beta}}
 = \Omega\delta_{\alpha\beta}.
\label{eq:reciprocal-strain}
\end{equation}
The symbol \(\delta_{\alpha\beta}\) denotes the Kronecker delta, and \(\lambda\) labels a Cartesian component of \(\mathbf G\).
For GFN2-xTB, cumulative atomic multipoles are cell-frame tensors and therefore
have their own strain derivatives:
\begin{subequations}
\label{eq:multipole-strain}
\begin{align}
 \frac{\partial \mu_{A\lambda}}{\partial \epsilon_{\alpha\beta}}
 &= \delta_{\lambda\alpha}\mu_{A\beta}, \\
 \frac{\partial \Theta_{A\lambda\tau}}{\partial \epsilon_{\alpha\beta}}
 &= \delta_{\lambda\alpha}\Theta_{A\beta\tau}
  + \delta_{\tau\alpha}\Theta_{A\lambda\beta}.
\end{align}
\end{subequations}
Eqs.~\eqref{eq:multipole-structure-factor}--\eqref{eq:multipole-strain}
make clear why the periodic GFN2-xTB stress is more restrictive than the
corresponding force expression: under strain, real-space distances, reciprocal
vectors, volume prefactors, damping functions, coordination numbers, and the
local multipole tensors all change simultaneously.

\section{Density-Functional Perturbation Theory Working Equations}
\label{sec:si-dfpt-working-equations}

The DFPT discussion in Section~III relies on the working equations assembled here. Section~III~A gives the generic Hellmann--Feynman, density-response, variational, and Sternheimer forms. Section~III~B records the Berry-phase polarization expressions used for periodic electric-field response and trajectory observables. Section~III~C gives the magnetic-shielding current-density machinery. Section~III~D collects the \(g\)-tensor correction terms. Section~III~E gives the hyperfine smearing and GAPW integration details.

\subsection{Generic Density-Functional Perturbation Theory and Sternheimer Response}

The Sternheimer and variational formulations used below follow the standard development of DFPT.\cite{Sternheimer1954,GonzeVigneron1989,Baroni2001}

With
\(
\psi_i^{(1)}=
\left.\partial\psi_i(\lambda)/\partial\lambda\right|_{\lambda=0}
\),
the first-order orbital expansion introduced in Section~III is truncated as
\begin{equation}
  \psi_i(\lambda)
  =
  \psi_i^{(0)}
  +\lambda\psi_i^{(1)}
  +\mathcal{O}(\lambda^2).
  \label{eq:si-dfpt-first-order-truncation}
\end{equation}

For perturbations that enter through a scalar change \(\Delta V(\mathbf{r})\) of the KS effective potential, the Hellmann--Feynman theorem gives
\begin{subequations}
\label{eq:si-dfpt-hellmann-feynman}
\begin{align}
  \frac{\partial E}{\partial \lambda}
  &=
  \int \mathrm{d}\mathbf{r}\,
  \frac{\partial \Delta V(\mathbf{r})}{\partial \lambda}
  n(\mathbf{r}),
  \\
  \frac{\partial^2 E}{\partial \lambda^2}
  &=
  \int \mathrm{d}\mathbf{r}\,
  \frac{\partial^2 \Delta V(\mathbf{r})}{\partial \lambda^2}
  n(\mathbf{r})
  +
  \int \mathrm{d}\mathbf{r}\,
  \frac{\partial \Delta V(\mathbf{r})}{\partial \lambda}
  \frac{\partial n(\mathbf{r})}{\partial \lambda}.
\end{align}
\end{subequations}
The first-order density response is
\begin{equation}
n^{(1)}(\mathbf{r})
=
\sum_i f_i
\left[
\psi_i^{(0)\ast}(\mathbf{r})\,\psi_i^{(1)}(\mathbf{r})
+
\psi_i^{(1)\ast}(\mathbf{r})\,\psi_i^{(0)}(\mathbf{r})
\right].
\label{eq:si-dfpt-rho1}
\end{equation}
The index \(i\) labels reference KS orbitals, \(f_i\) is their occupation, and the asterisk denotes complex conjugation.
The first-order Hamiltonian contains both the explicit derivative of the
external perturbation and the self-consistent response of the Hartree and
XC potentials. The latter is governed by the second-order
energy kernel
\begin{equation}
  \mathcal{K}(\mathbf{r},\mathbf{r}')
  =
  \left.
  \frac{\delta^2 E_{\mathrm{Hxc}}}
  {\delta n(\mathbf{r})\,\delta n(\mathbf{r}')}
  \right|_{n^{(0)}},
  \label{eq:si-dfpt-kernel}
\end{equation}
so that the first-order Hamiltonian can be written as
\begin{subequations}
\label{eq:si-dfpt-first-order-hamiltonian}
\begin{equation}
  H^{(1)}(\mathbf{r})
  =
  \Delta v_{\mathrm{ext}}^{(1)}(\mathbf{r})
  +
  \int
  \mathrm{d}\mathbf{r}'\,
  \mathcal{K}(\mathbf{r},\mathbf{r}')
  n^{(1)}(\mathbf{r}').
  \label{eq:si-dfpt-h1}
\end{equation}
For a nuclear displacement
\(\mathbf{R}_a\rightarrow\mathbf{R}_a+\lambda\hat{\mathbf{e}}\),
the explicit first-order change in the external potential is
\begin{equation}
  \Delta v_{\mathrm{ext}}^{(1)}(\mathbf{r})
  =
  -
  \frac{
  Z_a\hat{\mathbf{e}}\cdot(\mathbf{r}-\mathbf{R}_a)
  }{
  |\mathbf{r}-\mathbf{R}_a|^3
  },
\label{eq:si-dfpt-h1-nuclear-displacement}
\end{equation}
where \(Z_a\) is the nuclear charge. The vector \(\mathbf R_a\) is the position of nucleus \(a\), and \(\hat{\mathbf e}\) is the unit vector along its displacement. For a homogeneous electric field
\(\boldsymbol{\mathcal{E}}\), the field-coupled contribution is
\begin{equation}
  \Delta v_{\mathrm{ext}}^{(1)}
  =
  -\Omega\,
  \boldsymbol{\mathcal{E}}\cdot\mathbf{P}^{\mathrm{elec}},
  \label{eq:si-dfpt-h1-electric-field}
\end{equation}
\end{subequations}
where \(\mathbf{P}^{\mathrm{elec}}\) is the electronic contribution to the
polarization and \(\Omega\) is the simulation-cell volume. The electric-field
amplitude then serves as the perturbation parameter \(\lambda\).
For a noncanonical occupied-orbital representation, linearizing the KS equations gives the coupled Sternheimer problem
\begin{multline}
-\sum_i^{N_{\mathrm{occ}}}
\left(
H^{(0)}\delta_{ij}
-
\left\langle\psi_i^{(0)}\middle|H^{(0)}\middle|\psi_j^{(0)}\right\rangle
\right)
\left|\psi_i^{(1)}\right\rangle
\\
=
H^{(1)}\left|\psi_j^{(0)}\right\rangle .
\label{eq:si-dfpt-sternheimer}
\end{multline}
Expanding the first-order orbitals in atom-centered basis functions
\begin{equation}
  \psi_i^{(1)} = \sum_l c_{li}^{(1)}\, \phi_l,
  \label{eq:si-dfpt-mo-ao}
\end{equation}
and projecting Eq.~\eqref{eq:si-dfpt-sternheimer} onto \(\phi_k\) gives
\begin{multline}
-\sum_i^{N_{\mathrm{occ}}}\sum_l^{N_{\mathrm{basis}}}
\left(
H_{kl}^{(0)}\delta_{ij}
-
S_{kl}
\left\langle\psi_i^{(0)}\middle|H^{(0)}\middle|\psi_j^{(0)}\right\rangle
\right)c_{li}^{(1)}
\\
=
\sum_l H_{kl}^{(1)}c_{lj}^{(0)} .
\label{eq:si-dfpt-sternheimer-ao}
\end{multline}
In this AO representation, \(i\) and \(j\) label occupied orbitals, \(k\) and \(l\) label AOs \(\phi_k\) and \(\phi_l\), \(c_{li}^{(0/1)}\) are zeroth- and first-order orbital coefficients, \(H_{kl}^{(0/1)}\) are Hamiltonian-matrix elements, and \(S_{kl}\) is an AO-overlap-matrix element.
This yields \(N_{\mathrm{basis}}\times N_{\mathrm{occ}}\) simultaneous equations. CP2K solves them self-consistently with a preconditioned conjugate-gradient minimizer.\cite{Gonze1995,Putrino2000} In the parallel-transport gauge, the first-order orbitals are orthogonal to the occupied orbital manifold,\cite{Gonze1995} which is imposed by projecting the right-hand sides onto the unoccupied space. Magnetic perturbations lead to imaginary first-order orbitals, whereas vibrational perturbations in atom-centered bases also generate right-hand-side terms from overlap-matrix derivatives.

The equivalent variational formulation starts from the fact that even perturbation orders obey a stationary principle when the underlying quantity is variational.\cite{GonzeVigneron1989,Gonze1995} Gonze's second-order functional may be written as
\begin{widetext}
\begin{equation}
  \label{eq:si-dfpt-E2-Gonze}
  \begin{aligned}
  E^{(2)}[n(\lambda)]
  =&
  \sum_{i=1}^{N_{\mathrm{occ}}}
  \Bigl[
  \langle \psi_i^{(1)}|
  (H-\epsilon_i)^{(0)}
  |\psi_i^{(1)}\rangle
  +
  \langle \psi_i^{(1)}|
  (T+v)^{(1)}
  |\psi_i^{(0)}\rangle
  \\
  &+
  \langle \psi_i^{(0)}|
  (T+v)^{(1)}
  |\psi_i^{(1)}\rangle
  +
  \langle \psi_i^{(0)}|
  (T+v)^{(2)}
  |\psi_i^{(0)}\rangle
  \Bigr]
  \\
  &+
  \frac{1}{2}
  \int\!\!\int
  \mathrm{d}\mathbf{r}\,
  \mathrm{d}\mathbf{r}'\,
  \mathcal{K}(\mathbf{r},\mathbf{r}')\,
  n^{(1)}(\mathbf{r}')\,
  n^{(1)}(\mathbf{r})
  \\
  &+
  \left.
  \int
  \frac{\mathrm{d}}{\mathrm{d}\lambda}
  \frac{\delta E_{\mathrm{Hxc}}[n^{(0)}]}
  {\delta n(\mathbf{r})}
  \right|_{\lambda=0}
  n^{(1)}(\mathbf{r})\,
  \mathrm{d}\mathbf{r}
  +
  \left.
  \frac{1}{2}
  \frac{\mathrm{d}^2}{\mathrm{d}\lambda^2}
  E_{\mathrm{Hxc}}[n^{(0)}]
  \right|_{\lambda=0}.
  \end{aligned}
\end{equation}
In this functional, \(T\) and \(v\) denote the kinetic-energy and external-potential operators, superscripts \((0)\), \((1)\), and \((2)\) denote perturbation orders, \(\epsilon_i\) are reference orbital energies, and \(E_{\mathrm{Hxc}}\) is the Hartree-plus-XC energy.
Only the first-order density enters the variational variables. Minimization under the parallel-transport constraint \(\langle\psi_i^{(0)}|\psi_j^{(1)}\rangle=0\) leads back to the Sternheimer equation. Putrino et al. recast the same structure in a form that is convenient when the perturbation is supplied directly as an energy functional.\cite{Putrino2000}
\begin{subequations}
\label{eq:si-dfpt-Putrino-functional}
\begin{align}
  E^{(2)}[n(\lambda)]
  =&
  \sum_{i=1}^{N_{\mathrm{occ}}}
  \Biggl[
  \langle \psi_i^{(1)}|
  (H-\epsilon_i)^{(0)}
  |\psi_i^{(1)}\rangle
  +
  \left\langle \psi_i^{(1)}
  \middle|
  \frac{\delta E^{\mathrm{pert}}[n(\lambda)]}
  {\delta \langle \psi_i^{(0)}|}
  \right.
  \nonumber\\
  &\left.
  +
  \frac{\delta E^{\mathrm{pert}}[n(\lambda)]}
  {\delta | \psi_i^{(0)}\rangle}
  \middle|
  \psi_i^{(1)}
  \right\rangle
  \Biggr]
  +
  \frac{1}{2}
  \int\!\!\int
  \mathrm{d}\mathbf{r}\,
  \mathrm{d}\mathbf{r}'\,
  \mathcal{K}(\mathbf{r},\mathbf{r}')\,
  n^{(1)}(\mathbf{r}')\,
  n^{(1)}(\mathbf{r}),
  \label{eq:si-dfpt-E2-Putrino}
  \\
  E[n(\mathbf{r},\lambda)]
  =&
  E^{\mathrm{KS}}[n(\mathbf{r},\lambda)]
  +
  \lambda E^{\mathrm{pert}}[n(\mathbf{r},\lambda)].
  \label{eq:si-dfpt-Epert-Putrino}
\end{align}
\end{subequations}
Terms independent of the first-order density have been dropped because they do not affect the determination of the first-order orbitals. The constrained minimum gives
\begin{equation}
  \begin{aligned}
  \left(
  H^{(0)}
  -
  \epsilon_j^{(0)}
  \right)
  \left|\psi_j^{(1)}\right\rangle
  &=
  -\hat{Q}
  \Biggl[
  \int
  \mathrm{d}\mathbf{r}'\,
  \mathcal{K}(\mathbf{r},\mathbf{r}')\,
  n^{(1)}(\mathbf{r}')
  \left|\psi_j^{(0)}\right\rangle
  \\
  &\qquad\qquad+
  \frac{\delta E^{\mathrm{pert}}(\lambda)}
  {\delta \langle \psi_j^{(0)} |}
  \Biggr]
  \ensuremath{,}
  \end{aligned}
  \label{eq:si-dfpt-sternheimer-Putrino}
\end{equation}
\end{widetext}
which reduces to Eq.~\eqref{eq:si-dfpt-sternheimer} when the perturbation is expressed as an operator.

\subsection{Berry-Phase Polarization and Trajectory Response}

In the modern theory of polarization,\cite{KingSmith:1993hp,resta,Souza:2002fa} the polarization of a periodic system with Bloch states
\(|\psi_{j\mathbf{k}}\rangle=\exp(\mathrm{i}\mathbf{k}\cdot\mathbf{r})|u_{j\mathbf{k}}\rangle\)
is expressed through a \(\mathbf{k}\)-space Berry connection over the lattice-periodic functions
\begin{equation}
  A_{\alpha}(\mathbf{k})
  =
  \mathrm{i}
  \sum_{j=1}^{n}
  \langle u_{j\mathbf{k}}|
  \partial_{k_{\alpha}}u_{j\mathbf{k}}\rangle,
\label{eq:si-berry-connection}
\end{equation}
where the sum is over the \(n\) occupied bands. The quantity \(A_\alpha(\mathbf k)\) is the Berry-connection component along Cartesian direction \(\alpha\), \(u_{j\mathbf k}\) is the lattice-periodic part of Bloch state \(j\), and \(\partial_{k_\alpha}\) differentiates with respect to that component of crystal momentum. The electronic contribution to the macroscopic polarization is then
\begin{equation}
  \mathbf{P}^{(\mathrm{el})}
  =
  -2\mathrm{i}e
  \sum_{j=1}^{n}
  \int_{\mathrm{BZ}}
  \frac{\mathrm{d}\mathbf{k}}{(2\pi)^3}
  \langle u_{j\mathbf{k}}|
  \partial_{\mathbf{k}}u_{j\mathbf{k}}\rangle
  \ensuremath{,}
\label{eq:si-polarization-berry}
\end{equation}
for doubly occupied bands. The integral extends over the Brillouin zone.

CP2K employs the modern theory of polarization in the single-point limit at the \(\Gamma\) point.\cite{Yaschenko1998} In this case, the electronic polarization can be obtained from the phase of a unitary non-Hermitian overlap operator
\begin{subequations}
\label{eq:si-single-point-polarization}
\begin{align}
  \mathbf{P}^{(\mathrm{el})}
  &=
  f_{\mathrm{occ}}
  \frac{e}{2\pi\Omega}\,
  \mathbf{h}\,
  \operatorname{Im}\ln\det\mathbf{S},
  \\
  S_{\alpha,ij}
  &=
  \langle
  \psi_i|
  \exp[-\mathrm{i}2\pi\,\mathbf{h}_{\alpha}^{-1}\cdot\mathbf{r}]
  |\psi_j
  \rangle ,
\end{align}
\end{subequations}
where \(f_{\mathrm{occ}}=2\) for doubly occupied orbitals, \(\mathbf{h}\) is the cell matrix, and \(\Omega\) is the cell volume. The matrix \(\mathbf S_\alpha\) contains twist-operator overlaps between occupied orbitals \(i\) and \(j\), and \(\alpha\) labels a lattice direction. The unitary operator plays the role of the position operator in an extended system,\cite{resta} so that the total electronic and nuclear polarization can be written as
\begin{multline}
  \mathbf{P}
  =
  \frac{e}{2\pi\Omega}
  \mathbf{h}\,
  \operatorname{Im}\ln
  \\
  \left\langle
  \Psi
  \middle|
  \exp\left[
  \mathrm{i}2\pi\,\mathbf{h}_{\alpha}^{-1}\cdot
  \left(
  \sum_l Z_l\mathbf{R}_l-\mathbf{r}
  \right)
  \right]
  \middle|
  \Psi
  \right\rangle .
  \label{eq:si-total-berry-polarization}
\end{multline}
In this expression, \(\Psi\) is the many-electron state, \(l\) labels nuclei of charge \(Z_l e\) at positions \(\mathbf R_l\), and \(\mathbf r\) denotes the collective electronic position entering the many-body twist operator.
For an unbounded periodic system, \(\mathbf{P}\) is ambiguous modulo the polarization quantum \(e\mathbf{R}/\Omega\), where \(\mathbf{R}\) is a lattice vector. CP2K wraps the Berry phase to the \([-\pi,\pi]\) interval. Along an AIMD trajectory, the resulting branch jumps must be unwrapped by adding or subtracting polarization quanta so that the dipole or polarization evolves continuously.

For the electric-field DFPT response used in fixed-geometry polarizabilities, a homogeneous field \(\boldsymbol{\mathcal{E}}\) enters through the field-coupled energy functional
\begin{equation}
  \lambda E^{\mathrm{pert}}[n(\mathbf{r},\lambda)]
  =
  -\Omega\,
  \boldsymbol{\mathcal{E}}\cdot
  \mathbf{P}^{(\mathrm{el})},
  \label{eq:si-dfpt-efield-pert}
\end{equation}
with the Berry-phase polarization from Eq.~\eqref{eq:si-single-point-polarization}. The first-order orbitals for a field along Cartesian direction \(\beta\) are defined by
\begin{equation}
  |\psi_i(\mathcal{E}_{\beta})\rangle
  =
  |\psi_i^{(0)}\rangle
  +
  \mathcal{E}_{\beta}
  |\psi_{i,\beta}^{(1)}\rangle
  +
  \mathcal{O}(\mathcal{E}_{\beta}^{2}),
  \label{eq:si-dfpt-field-orbitals}
\end{equation}
and are obtained from Eq.~\eqref{eq:si-dfpt-sternheimer-Putrino} with the functional derivative of Eq.~\eqref{eq:si-dfpt-efield-pert}. The induced polarization is then
\begin{equation}
  \label{eq:si-dfpt-induced-polarization}
  \begin{aligned}
  \delta P_{\alpha}^{(\mathrm{el})}
  =&
  \sum_{\beta}
  f_{\mathrm{occ}}
  \frac{e}{2\pi\Omega}
  h_{\alpha}\,
  \operatorname{Im}
  \Biggl[
  \sum_{ij}
  \Bigl(
  \langle
  \psi_{i,\beta}^{(1)}
  |
  U_{\alpha}
  |
  \psi_j^{(0)}
  \rangle
  \\
  &+
  \langle
  \psi_i^{(0)}
  |
  U_{\alpha}
  |
  \psi_{j,\beta}^{(1)}
  \rangle
  \Bigr)
  (\mathbf S_{\alpha}^{-1})_{ji}
  \Biggr]
  \mathcal{E}_{\beta},
  \end{aligned}
\end{equation}
where \(U_{\alpha}=\exp[-\mathrm{i}2\pi\,\mathbf{h}_{\alpha}^{-1}\cdot\mathbf{r}]\). The volume factor distinguishes the polarization density from the cell dipole convention used in some fixed-field formulas.

The same polarization enters trajectory-level infrared absorption. In a finite-temperature formulation one may write
\begin{equation}
  \operatorname{Im}\epsilon(\omega)
  =
  \frac{2\pi\Omega\omega}{3k_{\mathrm B}T}
  \int_{-\infty}^{\infty}
  \mathrm{d}t\,e^{-i\omega t}\,
  \left\langle
  \mathbf{P}(t)\cdot\mathbf{P}(0)
  \right\rangle
  \ensuremath{.}
\label{eq:si-dielectric-response}
\end{equation}
The function \(\epsilon(\omega)\) is the frequency-dependent dielectric response, \(\omega\) is angular frequency, and the brackets denote an equilibrium time-correlation average.
This relation connects the Berry-phase polarization machinery to the finite-temperature spectroscopy strategy discussed in Section~X.

\subsection{Magnetic Shieldings from Current-Density Response}

The current density induced by an external magnetic field generates the
induced field through the Biot--Savart relation
\begin{equation}
  \mathbf{B}^{\mathrm{ind}}(\mathbf{r})
  =
  \frac{1}{c}
  \int
  \mathrm{d}\mathbf{r}'\,
  \frac{\mathbf{r}'-\mathbf{r}}
  {|\mathbf{r}'-\mathbf{r}|^3}
  \times
  \mathbf{j}(\mathbf{r}').
  \label{eq:si-nmr-induced-field}
\end{equation}
At a nuclear
position \(\mathbf{r}^N\), the shielding tensor can equivalently be expressed
as the mixed second derivative
\begin{equation}
  \boldsymbol{\sigma}(\mathbf{r}^N)
  =
  \left.
  \frac{\partial^2 E}
  {\partial\mathbf{B}^{\mathrm{ext}}\,\partial\mathbf{m}^N}
  \right|_{|\mathbf{B}^{\mathrm{ext}}|=|\mathbf{m}^N|=0},
  \label{eq:si-nmr-shielding-energy-derivative}
\end{equation}
where \(\mathbf{m}^N\) is the nuclear magnetic moment.

The magnetic-shielding tensor is one branch of a more general Taylor expansion of the induced magnetic field with respect to external fields and nuclear magnetic moments. For a closed-shell reference with nuclear magnetic moments \(\{\mathbf{m}^N\}\), this expansion may be written schematically as
\begin{multline}
  \mathbf{B}^{\mathrm{ind}}
  (\mathbf{r},\mathbf{B}^{\mathrm{ext}},\{\mathbf{m}^N\})
  =
  -
  \boldsymbol{\sigma}(\mathbf{r})\,
  \mathbf{B}^{\mathrm{ext}}
  \\
  -
  \sum_i^{\mathrm{nuclei}}
  \mathbf{K}_i(\mathbf{r})\,
  \mathbf{m}_i^N(\mathbf{r}_i^N)
  +\cdots ,
  \label{eq:si-nmr-Bind-Taylor}
\end{multline}
where \(\mathbf{K}_i(\mathbf{r})\) denotes the reduced indirect spin--spin coupling. Magnetic shielding follows from the first term, whereas the second term identifies the related spin--spin response channel.

For a homogeneous magnetic field, the vector potential can be chosen as
\begin{equation}
  \mathbf{A}(\vecr)
  =
  \frac{1}{2}\mathbf{B}\times(\vecr-\vecr_0),
  \label{eq:si-nmr-vector-potential}
\end{equation}
where \(\vecr_0\) is an arbitrary gauge origin. The explicit position operator is problematic under PBC. In the Sebastiani--Parrinello approach,\cite{Sebastiani2001} each MLWF is assigned its own translated coordinate system, centered at \(\mathbf{d}_i\), so that the localized orbital remains inside the simulation cell and the current is invariant under orbital-specific translations.

The orbital translations generate three perturbation operators:
\begin{subequations}
\label{eq:si-nmr-operators}
\begin{align}
  \hat{H}^{P}
  &=
  \hat{\mathbf{p}},
  \\
  \hat{H}^{Li}
  &=
  (\hat{\mathbf{r}}-\mathbf{d}_i)\times\hat{\mathbf{p}},
  \\
  \hat{H}^{\Delta i}
  &=
  (\mathbf{d}_i-\mathbf{d}_j)\times\hat{\mathbf{p}},
\end{align}
\end{subequations}
where \(\mathbf{d}_i-\mathbf{d}_j\) is evaluated with the MIC. The operator \(\hat{\mathbf p}\) is the electronic momentum, \(\hat{\mathbf r}\) is the position operator, and \(i\) and \(j\) label localized occupied orbitals centered at \(\mathbf d_i\) and \(\mathbf d_j\). The full correction operator is the most expensive term because it requires one response calculation per orbital.

The paramagnetic and diamagnetic contributions to the \(x\)-component of the current-density response induced by a magnetic field along \(y\) are given by the following expressions:\cite{Weber_JCP_2009}
\begin{subequations}
\label{eq:si-nmr-current-response}
\begin{multline}
  j^p_{xy}(\vecr)
  =
  -\frac{1}{2c}
  \sum_{ikl}
  \Bigl[
  C^{(0)}_{ki}
  \bigl(
  C_{li}^{L_y}
  +(\vecr_0-\mathbf{d}_i)_x C_{li}^{P_z}
  \\
  -(\vecr_0-\mathbf{d}_i)_z C_{li}^{P_x}
  -C_{li}^{\Delta i_y}
  \bigr)
  \\
  \times
  \{(\nabla_x\chi_k(\vecr))\chi_l(\vecr)
  -
  \chi_k(\vecr)\nabla_x\chi_l(\vecr)\}
  \Bigr],
\end{multline}
\begin{equation}
  j^d_{xy}(\vecr)
  =
  (\vecr-\vecr_0)_zn(\vecr).
\end{equation}
\end{subequations}
The quantities \(j^p_{xy}\) and \(j^d_{xy}\) are the paramagnetic and diamagnetic \(x\)-current responses to a field along \(y\). The index \(i\) labels occupied orbitals, \(k\) and \(l\) label AOs \(\chi_k\) and \(\chi_l\), \(C^{(0)}\) are ground-state MO coefficients, and \(C^{P}\), \(C^{L}\), and \(C^{\Delta i}\) are the corresponding first-order response coefficients generated by the three operators above.
The two terms are individually gauge-origin dependent, but their sum should be invariant. CP2K can use IGAIM, where the gauge origin is the nearest atom,\cite{Keith1992} or the CSGT, where \(\vecr_0=\vecr\).\cite{Keith1993}

Combining the induced-field relation with the linear response to a field along
the \(x\) direction gives the shielding component
\begin{equation}
  \sigma_{xy}(\mathbf{r})
  =
  \frac{1}{c}
  \int_{\Omega}
  \left[
  \frac{\mathbf{r}'-\mathbf{r}}
  {|\mathbf{r}'-\mathbf{r}|^3}
  \times
  \mathbf{j}_x(\mathbf{r}')
  \right]_y
  \mathrm{d}^3r',
  \label{eq:si-nmr-shielding-component}
\end{equation}
where the integration includes the material and, for an extended system, its
periodic replicas.

For periodic systems, the current density is decomposed analogously to the GAPW electron density
\begin{equation}
\mathbf{j}(\vecr)
=
\tilde{\mathbf{j}}(\vecr)
+
\sum_A^{\mathrm{atoms}}
\left[
\mathbf{j}_A(\vecr)
-
\tilde{\mathbf{j}}_A(\vecr)
\right],
\label{eq:si-nmr-current-gapw}
\end{equation}
where \(\tilde{\mathbf{j}}\) is the smooth contribution, \(\mathbf{j}_A\) is the local hard contribution, and \(\tilde{\mathbf{j}}_A\) prevents double counting. The \(\mathbf{G}=0\) component of the smooth contribution is treated through the susceptibility correction
\begin{equation}
\chi_{xy}
=
\frac{2\pi}{\Omega c}
\int \mathrm{d}\vecr\,
\left[
\vecr\times\tilde{\mathbf{j}}_x(\vecr)
\right]_y
\ensuremath{.}
\label{eq:si-nmr-chi-correction}
\end{equation}
The quantity \(\chi_{xy}\), the indicated Cartesian component of the macroscopic magnetic susceptibility, uses the spherical-sample convention.\cite{Cowan1997} The local contribution to the shielding at nucleus \(\vecr^N\) is
\begin{multline}
  \sigma_{xy}(\vecr^N)
  =
  \frac{1}{c}
  \sum_B
  \int_{\Omega_B}
  \mathrm{d}\vecr\,
  \left[
  \frac{\vecr-\vecr^N}{|\vecr-\vecr^N|^3}
  \right.
  \\
  \left.
  \times
  \left(
  j_{x,B}(\vecr)-\widetilde{\jmath}_{x,B}(\vecr)
  \right)
  \right]_y,
  \label{eq:si-sigma-local}
\end{multline}
where the atom sum is restricted to nuclei within a cutoff radius of \(\vecr^N\).

\subsection{Electron Paramagnetic Resonance \texorpdfstring{\(g\)-Tensor}{g-Tensor} Terms}

The three correction terms entering the \(g\)-tensor in Section~III~D are based on the DFT response formulation and its extension to periodic solids.\cite{Schreckenbach1997,Pickard2002,VanYperen-DeDeyne2012}
They are
\begin{subequations}
\label{eq:si-epr-g-corrections}
\begin{align}
  \Delta g_{xy}^{\mathrm{ZKE}}
  &=
  -\frac{g_e}{c^2}
  (T^{\alpha}-T^{\beta})
  \delta_{xy},
  \\
  \Delta g_{xy}^{\mathrm{SO}}
  &=
  \frac{g_e-1}{c}
  \int_{\Omega_c}
  \mathrm{d}^3\vecr\,
  \left[
  \mathbf{j}_x^{\alpha}(\vecr)
  \times
  \nabla V_{\mathrm{eff}}^{\alpha}(\vecr)
  \right.
  \nonumber\\
  &\qquad\qquad
  \left.
  -
  \mathbf{j}_x^{\beta}(\vecr)
  \times
  \nabla V_{\mathrm{eff}}^{\beta}(\vecr)
  \right]_y,
  \\
  \Delta g_{xy}^{\mathrm{SOO}}
  &=
  \frac{1}{S}
  \int_{\Omega_c}
  \mathrm{d}\vecr\,
  B_{y,B_x}^{\mathrm{corr}}(\vecr)
  \left[
  n^{\alpha}(\vecr)
  -
  n^{\beta}(\vecr)
  \right],
\end{align}
\end{subequations}
where \(\alpha\) and \(\beta\) denote spin channels, \(S\) is the total spin, and \(T\) is the kinetic energy. Moreover, \(\Omega_c\) is the simulation cell, \(\mathbf j_x^\sigma\) is the current response in spin channel \(\sigma\), \(V_{\mathrm{eff}}^\sigma\) is its effective potential, and \(\delta_{xy}\) is the Kronecker delta. The corrected field entering the SOO term is
\begin{multline}
  B_{y,B_x}^{\mathrm{corr}}(\vecr)
  =
  \frac{1}{c}
  \int_{\Omega}
  \mathrm{d}\vecr'\,
  \left[
  \frac{\mathbf{r}'-\mathbf{r}}
  {|\mathbf{r}'-\mathbf{r}|^3}
  \right.
  \\
  \left.
  \times
  \left(
  \mathbf{j}_x(\mathbf{r}')
  -
  \mathbf{j}_x^{\alpha-\beta}(\mathbf{r}')
  \right)
  \right]_y ,
  \label{eq:si-epr-bcorr}
\end{multline}
with \(\mathbf{j}^{\alpha-\beta}=\mathbf{j}^{\alpha}-\mathbf{j}^{\beta}\). Since the SOO term requires the induced field over all space, CP2K uses an approximation in which the atom-centered local current densities are neglected for the \(\mathbf{G}\neq0\) components of the induced field.

If a periodic cell contains \(n\) paramagnetic centers, the individual center contributions add to the simulation-cell response. For reporting a per-center \(g\)-tensor one therefore uses the normalization
\begin{equation}
  \mathbf{g}
  =
  g_e\mathbf{1}
  +
  \frac{1}{n}\Delta\mathbf{g},
  \label{eq:si-epr-g-normalization}
\end{equation}
with \(n\) the number of paramagnetic centers in the cell.

\subsection{\texorpdfstring{Hyperfine Contact Smearing with the Gaussian and Augmented Plane-Wave Method}{Hyperfine Contact Smearing with the Gaussian and Augmented Plane-Wave Method}}

The scalar-relativistic contact term uses the smeared delta function
\begin{equation}
  \delta_T(\vecr)
  =
  \frac{1}{4\pi r^2}
  \frac{2}{Z\alpha^2}
  \frac{1}{\left(1+\frac{2r}{Z\alpha^2}\right)^2},
\label{eq:si-hyperfine-delta-smeared}
\end{equation}
where \(Z\) is the nuclear charge and \(\alpha\) is the fine-structure constant. The variable \(r=|\vecr|\) is the electron--nucleus distance, and \(\delta_T\) is the scalar-relativistically smeared contact distribution. In the nonrelativistic limit, \(\delta_T\) collapses to a Dirac delta function.

For a nucleus \(N\), the isotropic Fermi-contact contribution is
\begin{subequations}
\label{eq:si-hyperfine-components}
\begin{equation}
  A_{\mathrm{iso},N}
  =
  \frac{4\pi}{3}
  \frac{g_e\mu_e g_N\mu_N}{\langle S_z\rangle}
  \int
  \mathrm{d}\vecr\,
  n^{\alpha-\beta}(\vecr)\,
  \delta_T(\vecr),
  \label{eq:si-hyperfine-isotropic}
\end{equation}
whereas the anisotropic spin--dipole contribution is
\begin{equation}
  A_{\mathrm{ani},N}^{ij}
  =
  \frac{1}{2}
  \frac{g_e\mu_e g_N\mu_N}{\langle S_z\rangle}
  \int
  \mathrm{d}\vecr\,
  n^{\alpha-\beta}(\vecr)
  \frac{3r_i r_j-\delta_{ij}r^2}{r^5}.
  \label{eq:si-hyperfine-anisotropic}
\end{equation}
\end{subequations}
Here \(g_N\) is the nuclear \(g\) factor, \(\mu_e\) is the Bohr magneton, \(\mu_N\) is the nuclear
magneton, \(\langle S_z\rangle\) is the expectation value of the
\(z\)-component of the total electronic spin, and \(\vecr\) is measured from
the position of nucleus \(N\).

The Fermi-contact term is a first-order expectation value over the spin density and therefore differs from the response quantities in Sections~III~C and~III~D: no magnetic-field Sternheimer equation is required once the spin-polarized reference density is available. The numerical challenge is instead the near-nuclear behavior of \(n^{\alpha-\beta}\). CP2K therefore evaluates the contact term in the GAPW representation,\cite{Lippert1999,Declerck2006a} where the spin density is decomposed schematically as
\begin{equation}
\begin{aligned}
n^{\alpha-\beta}(\vecr)
&=
\tilde{n}^{\alpha-\beta}(\vecr)
+\sum_A
\Bigl[
n_A^{\alpha-\beta}(\vecr)
\\
&\qquad
-
\tilde{n}_A^{\alpha-\beta}(\vecr)
\Bigr].
\end{aligned}
\label{eq:si-hyperfine-gapw-density}
\end{equation}
The smooth contribution is represented on the PW grid, while the hard local terms are integrated on atom-centered grids. For a nucleus \(N\), the contact integral is dominated by the hard on-site density inside the local atomic domain \(U_N\) and is given by
\begin{subequations}
\label{eq:si-hyperfine-gapw-components}
\begin{equation}
A_{\mathrm{iso},N}
\propto
\int_{U_N}
n_N^{\alpha-\beta}(\vecr)\,
\delta_T(\vecr)\,
\mathrm{d}\vecr ,
\label{eq:si-hyperfine-contact-local}
\end{equation}
with proportionality factor \(4\pi g_e\mu_e g_N\mu_N/(3\langle S_z\rangle)\). The anisotropic term is less singular but still needs the same all-electron reconstruction. With \(T_{ij}(\vecr)=(3r_i r_j-\delta_{ij}r^2)/r^5\), the GAPW partitioning may be written schematically as
\begin{multline}
A_{\mathrm{ani},N}^{ij}
\propto
\int_{\Omega}
\tilde{n}^{\alpha-\beta}(\vecr)\,
T_{ij}(\vecr)\,
\mathrm{d}\vecr
\\
+
\sum_M
\int_{U_M}
\left[
n_M^{\alpha-\beta}(\vecr)
-
\tilde{n}_M^{\alpha-\beta}(\vecr)
\right]
T_{ij}(\vecr)\,
\mathrm{d}\vecr .
\label{eq:si-hyperfine-anisotropic-gapw}
\end{multline}
\end{subequations}
The \(M=N\) term gives the dominant local contribution, while \(M\neq N\) cross terms are small but can be included for neighboring atoms. CP2K therefore restricts these cross terms with a user-controllable interaction radius, whose default value is 10~Bohr. This mirrors the GAPW logic used for magnetic shieldings: the PW part captures the smooth long-range contribution, whereas local hard-density integrations recover the near-nuclear all-electron physics that is essential for hyperfine couplings.

\section{Time-Dependent Density-Functional Theory Working Equations}
\label{sec:si-tddft-working-equations}

The TDDFT discussion in Section~IV connects linear-response excitation energies, excited-state gradients, periodic transition dipoles, spatial excitation descriptors, real-time propagation, and Ehrenfest dynamics. The working equations and implementation definitions used by those subsections are collected here. Section~IV~A gives the CPKS and Z-vector equations used for LR-TDDFT properties and gradients. Section~IV~B collects the
Sternheimer/TDA response problem, spin-flip
kernels, response-density Poisson equation, and explicit gradient
intermediates. Section~IV~C records the finite-\(\bk\) transition-dipole
expression for periodic crystals. Section~IV~D gives the electron--hole quantities used for spatial excitation descriptors. Section~IV~E summarizes the Ehrenfest and RT-TDDFT equations, including gauge choices, current response, and spectral extraction.
Unless stated otherwise in Sections~IV~A and~IV~B, Greek indices label AOs, \(i,j,k,l\) label occupied MOs, \(\sigma,\sigma'\) label spin channels, \(\mathbf C\), \(\mathbf S\), \(\mathbf F\), and \(\mathbf D\) denote the MO-coefficient, AO-overlap, KS, and density matrices, and \(\mathbf Q\) projects onto the virtual AO space.

\subsection{Linear-Response Time-Dependent Density-Functional Theory and Coupled-Perturbed Kohn--Sham Equations}
\label{sec:si-lr-tddft-cpks}

The response and variational-gradient formulations follow the standard LR-TDDFT and analytic-derivative developments.\cite{Casida1996,Furche_Ahlrichs_2002}

For molecular LR-TDDFT properties, in particular excited-state
nuclear gradients, CP2K uses a variational Lagrangian with
CPKS constraints
\begin{align}
L
&=
E_{\mathrm{total}}
+E_{\mathrm{CPKS}}
+\sum_{\alpha}\sum_{\mu\nu}
\bar{W}_{\mu\nu}^{\alpha}g_{\mu\nu}^{\alpha}\nonumber\\
&=
E_{\mathrm{total}}
+\sum_{\mu\nu k\sigma}
\bar{Z}_{\mu k\sigma}
\left(
F_{\mu\nu}C_{\nu k}
-\varepsilon_{k}S_{\mu\nu}C_{\nu k}
\right)\nonumber\\
&\quad
+\sum_{\alpha}\sum_{\mu\nu}
\bar{W}_{\mu\nu}^{\alpha}g_{\mu\nu}^{\alpha}\ensuremath{.}
\label{eq:si-lr-lagrangian}
\end{align}
In this Lagrangian, \(E_{\mathrm{CPKS}}\) imposes the coupled-perturbed KS equations, while \(g_{\mu\nu}^{\alpha}\) collectively denotes the orthogonality constraints associated with the ground-state MO coefficients \(\mathbf C\) and the excited-state eigenvectors \(\mathbf X\). The orbital energy \(\varepsilon_k\) is the eigenvalue of the \(k\)th occupied KS MO. The matrix \(\bar{\mathbf Z}\) serves as the response Lagrange multiplier, whereas \(\bar{\mathbf W}^{\mathrm C}\) and \(\bar{\mathbf W}^{\mathrm X}\) enforce the corresponding orthonormality and orthogonality constraints.
Requiring the Lagrangian to be stationary with respect to the distinct variational variables \(\mathbf C\) and \(\mathbf X\) yields the working equations for these multipliers after projection onto the occupied or virtual MO spaces:
\begin{subequations}
\begin{align}
\frac{\partial L}{\partial \mathbf{C}}\mathbf{C}=0
&\rightarrow
\bar{\mathbf{W}}^{\mathrm{C}}\ensuremath{,}\\
\frac{\partial L}{\partial \mathbf{X}}\mathbf{C}=0
&\rightarrow
\bar{\mathbf{W}}^{\mathrm{X}}\ensuremath{,}\\
\frac{\partial L}{\partial \mathbf{C}}\mathbf{Q}=0
&\rightarrow
\bar{\mathbf{Z}}\ensuremath{.}
\end{align}
\end{subequations}
Introducing Lagrange multipliers to ensure orthonormality of the excited-state eigenvectors and the CPKS orbitals with respect to the occupied ground-state MO manifold is identical to introducing a projector onto the virtual MO space in both the energy expression for the Lagrangian, Eq.~\eqref{eq:si-lr-lagrangian}, and the Z-vector equation, Eq.~\eqref{eq:si-mo-cpks}, with the projection operator defined as
\begin{equation}
   Q_{\mu\nu\sigma}=\delta_{\mu\nu}
   - \sum_{\kappa i}C_{\mu i\sigma}C_{\kappa i\sigma}^{*}
     S_{\kappa\nu} \ensuremath{.}
   \label{eq:si-q-projector}
\end{equation}
Projection onto the virtual subspace, or equivalently the corresponding Lagrange constraints, ensures that only occupied--virtual blocks of the excited-state eigenvectors or CPKS orbitals contribute
\begin{equation}
   \sum_{\mu\nu} C_{\mu i\sigma}^{*}S_{\mu\nu}\bar{Z}_{\nu j\sigma}=0 \ensuremath{,}
   \sum_{\mu \nu} C_{\mu i \sigma}^{*} S_{\mu \nu} X_{\nu j \sigma} = 0 \ensuremath{.}
\end{equation}
The resulting response equations use the Z-vector construction for analytic energy derivatives.\cite{HandySchaefer1984}
The CPKS, or Z-vector, equations are given as
\begin{equation}
\begin{aligned}
&\sum_{\nu j}
\left(F_{\mu\nu\sigma} \delta_{ij}
    -S_{\mu\nu}F_{ij\sigma}\right)\bar{Z}_{\nu j\sigma}\\
&\quad
+ \sum_{\nu\kappa}
  Q_{\nu\mu\sigma}^{*}
  H_{\nu\kappa\sigma}[\mathbf{D}^{\mathrm{Z}}]
  C_{\kappa i\sigma}
= -R_{\mu i\sigma} \ensuremath{.}
\end{aligned}
\label{eq:si-mo-cpks}
\end{equation}
Here, the KS matrix \(\mathbf F\) for hybrid density functionals, the density matrix \(\mathbf D\), the response density \(\mathbf D^{\mathrm Z}\), and the intermediate \(\mathbf H\) are
\begin{subequations}
\begin{align}
   F_{\mu\nu\sigma}
   &= h_{\mu\nu}
      +V_{\mu\nu}^{\mathrm{H}}
      +V_{\mu\nu\sigma}^{\mathrm{XC}}\nonumber\\
   &\quad
      -a_{\mathrm{EX}}
      \sum_{\kappa\lambda}
      D_{\kappa\lambda\sigma}(\mu\kappa|\nu\lambda)\ensuremath{,}
      \label{eq:si-ks-matrix}\\
   D_{\mu\nu\sigma}
   &=
     \sum_i C_{\mu i\sigma}C_{\nu i\sigma}^{*}\ensuremath{,}\\
   D_{\mu\nu\sigma}^{\mathrm{Z}}
   &=\frac{1}{2}\sum_i
     \left(\bar{Z}_{\mu i\sigma}C_{\nu i\sigma}^{*}
          +C_{\mu i\sigma}\bar{Z}_{\nu i\sigma}^{*}\right)\ensuremath{,}\\
   H_{\mu\nu\sigma}[\mathbf{M}]
   &=\sum_{\kappa\lambda\sigma'}M_{\kappa\lambda\sigma'}
     \Bigl[
     2(\kappa\lambda|\mu\nu)\nonumber\\
   &\quad
     -a_{\mathrm{EX}}\bigl((\kappa\mu|\lambda\nu)
     +(\kappa\nu|\lambda\mu)\bigr)\nonumber\\
   &\quad
     +2f_{\kappa\lambda\sigma',\mu\nu\sigma}^{\mathrm{XC}}
   \Bigr]\ensuremath{.}
     \label{eq:si-h-intermediate}
\end{align}
\end{subequations}
\(\mathbf V^{\mathrm H}\) denotes the Hartree-potential matrix, and \(a_{\mathrm{EX}}\) is the exact-exchange fraction of the hybrid density functional.

CP2K also features pure AO formulations using an exponential parametrization of the density matrix
as a localized alternative to the delocalized MO response
\begin{equation}
   \mathbf{D}(\mathbf{X})
   =e^{-\mathbf{X}\mathbf{S}}\mathbf{D}_0 e^{\mathbf{S}\mathbf{X}}\ensuremath{,}
\end{equation}
where \(\mathbf D_0\) is the reference density matrix and \(\mathbf X\) is an anti-Hermitian orbital-rotation matrix. The corresponding energy gradient is
\begin{equation}
   \frac{\partial E^{\mathrm{KS}}}{\partial\mathbf{X}}
   =
    \mathbf{F}\mathbf{D}\mathbf{S}
    -\mathbf{S}\mathbf{D}\mathbf{F}
   \ensuremath{.}
\end{equation}
Redundant blocks are removed with occupied and virtual projectors
\begin{equation}
   \mathcal{P}(\mathbf{X})
   =\boldsymbol{\Pi}_{\mathrm{occ}}
    \mathbf{X}\mathbf{Q}^{\dagger}
    +\mathbf{Q}\mathbf{X}
    \boldsymbol{\Pi}_{\mathrm{occ}}^{\dagger}.
\end{equation}
Here, \(\boldsymbol{\Pi}_{\mathrm{occ}}\) and \(\mathbf Q\) project onto the occupied and virtual AO subspaces, respectively.
The orbital-stationarity constraint can be imposed through the scalar Lagrangian
\begin{equation}
L(\mathbf X,\bar{\mathbf Z})
=
E[\mathbf D(\mathbf X)]
+
\operatorname{Tr}
\left[
\bar{\mathbf Z}^{\dagger}
\left(
\mathbf F\mathbf D\mathbf S
-\mathbf S\mathbf D\mathbf F
\right)
\right]
\ensuremath{.}
\end{equation}
The Z-vector equation follows from stationarity with respect to the exponential parameter \(\mathbf X\)
\begin{equation}
\frac{\partial L}{\partial \mathbf X}
=
\frac{\partial E}{\partial \mathbf X}
+
\frac{\partial}{\partial\mathbf X}
\operatorname{Tr}
\left[
\bar{\mathbf Z}^{\dagger}
\left(
\mathbf F\mathbf D\mathbf S
-\mathbf S\mathbf D\mathbf F
\right)
\right]
=0
\ensuremath{.}
\end{equation}
The trace denotes the Frobenius pairing between the Z-vector and the orbital-stationarity residual. In the real orthogonal AO representation used by the response solver, the Hermitian conjugate reduces to a transpose, and the resulting stationarity condition is solved as the corresponding matrix response equation.
The AO Hessian acting on an anti-Hermitian trial matrix \(\mathbf b\) is
\begin{equation}
   \frac{\partial^2 E}{\partial\mathbf X^2}[\mathbf b]
   =\mathcal{P}^{\dagger}
   \left(
   \mathbf F\mathbf B\mathbf S
   -\mathbf S\mathbf B\mathbf F
   +\mathbf H[\mathbf B]\mathbf D\mathbf S
   -\mathbf S\mathbf D\mathbf H[\mathbf B]
   \right)\ensuremath{,}
\end{equation}
where
\(\mathbf B=\mathcal P(\mathbf b)\mathbf S\mathbf D
-\mathbf D\mathbf S\mathcal P(\mathbf b)\).
CP2K transforms all quantities to an orthogonal AO basis. The corresponding
contravariant and covariant transformations are
\begin{subequations}
\begin{align}
   \tilde{\mathbf{M}}&=\mathbf{S}^{1/2}\mathbf{M}\mathbf{S}^{1/2}\ensuremath{,}&
   \mathbf{M}&=\mathbf{S}^{-1/2}\tilde{\mathbf{M}}\mathbf{S}^{-1/2}\ensuremath{,}\\
   \tilde{\mathbf{N}}&=\mathbf{S}^{-1/2}\mathbf{N}\mathbf{S}^{-1/2}\ensuremath{,}&
   \mathbf{N}&=\mathbf{S}^{1/2}\tilde{\mathbf{N}}\mathbf{S}^{1/2}\ensuremath{.}
\end{align}
\end{subequations}
The energy gradient then simplifies to
\begin{equation}
   \frac{\partial E^{\mathrm{KS}}}{\partial\mathbf{X}}
   =\tilde{\mathbf{F}}\tilde{\mathbf{D}}
    -\tilde{\mathbf{D}}\tilde{\mathbf{F}}
   =[\tilde{\mathbf{F}},\tilde{\mathbf{D}}] \ensuremath{.}
\end{equation}

\subsection{Sternheimer Time-Dependent Density-Functional Theory, Spin-Flip Kernels, and Excited-State Gradients}
\label{sec:si-sternheimer-gradients}

In the TDA, the Sternheimer formulation treats
the virtual space through the projector
\(\mathbf{Q}\), avoiding an explicit virtual-orbital construction. The TDDFT
eigenvalue problem and normalization constraints are
\begin{subequations}
\begin{align}
   \mathbf{A}\mathbf{X}^N
   &=\Omega_N\mathbf{S}\mathbf{X}^N\ensuremath{,}\\
   \left(\mathbf{X}^N\right)^{\dagger}\mathbf{S}\mathbf{X}^M
   &=\delta_{NM}\ensuremath{,}\\
   \mathbf{C}^{\dagger}\mathbf{S}\mathbf{X}^N&=0\ensuremath{.}
\end{align}
\end{subequations}
The Hermitian response matrix \(\mathbf A\) contains the zeroth-order KS matrix \(\mathbf F\) and response-kernel contributions
\begin{equation}
\begin{aligned}
&\sum_{\kappa k}
\left(F_{\mu\kappa\sigma}\delta_{ik}
     -F_{ik\sigma}S_{\mu\kappa}\right)X_{\kappa k\sigma}^N\\
&\quad
+ \sum_{\kappa\lambda}
  Q_{\kappa\mu\sigma}^{*}
  K_{\kappa\lambda\sigma}[\mathbf{D}^{\mathrm{X},N}]
  C_{\lambda i\sigma}\\
&=
\sum_{\kappa}\Omega_N S_{\mu\kappa}X_{\kappa i\sigma}^N \ensuremath{.}
\end{aligned}
\end{equation}
The response kernel is defined as
\begin{align}
 K_{\mu\nu\sigma}[\mathbf{D}^{\mathrm{X},N}]
 &= \sum_{\kappa\lambda\sigma'}
    D_{\kappa\lambda\sigma'}^{\mathrm{X},N}
    \left[
    (\mu\nu|\kappa\lambda)
    +f_{\mu\nu\sigma,\kappa\lambda\sigma'}^{\mathrm{XC}}
    \right] \nonumber\\
 &\quad
 -a_{\mathrm{EX}}\sum_{\kappa\lambda}
  D_{\kappa\lambda\sigma}^{\mathrm{X,EX},N}
  (\mu\kappa|\nu\lambda)\ensuremath{,}
\end{align}
with
\begin{subequations}
\begin{align}
 D_{\mu\nu\sigma}^{\mathrm{X,EX},N}
 &=\sum_k X_{\mu k\sigma}^N C_{\nu k\sigma}^{*}\ensuremath{,}\\
 D_{\mu\nu\sigma}^{\mathrm{X},N}
 &=\frac{1}{2}\sum_k
 \left(X_{\mu k\sigma}^NC_{\nu k\sigma}^{*}
 +C_{\mu k\sigma}(X_{\nu k\sigma}^N)^{*}\right)\ensuremath{.}
\end{align}
\end{subequations}
For spin-flip kernels,\cite{spin_flip_tddft_luber} the kernel becomes
\begin{equation}
\begin{aligned}
&K_{\mu\nu}^{\mathrm{SF}}
[\mathbf{D}^{\mathrm{X}\alpha\rightarrow\beta,N}]\\
&\quad =
\sum_{\tau\eta}
\left[
(\mu\nu|f_{\mathrm{XC}}^{\mathrm{SF}}|\tau\eta)
-a_{\mathrm{EX}}(\mu\tau|\nu\eta)
\right]
D_{\tau\eta}^{\mathrm{X}\alpha\rightarrow\beta,N}\ensuremath{,}
\end{aligned}
\end{equation}
where a thresholded noncollinear XC kernel can be used to
avoid numerical instabilities
\begin{equation}
   f_{\mathrm{XC}}^{\mathrm{SF}}
   \approx
   \frac{V_{\mathrm{XC}}^{\alpha}(\mathbf{r})
        -V_{\mathrm{XC}}^{\beta}(\mathbf{r})}
        {\max\left(
        |n^\alpha(\mathbf{r})
        -n^\beta(\mathbf{r})|,T\right)} \ensuremath{.}
\end{equation}
In the spin-flip XC kernel, \(T\) is a positive density-difference threshold that regularizes the denominator.
The Hartree part of the response kernel is evaluated from the response density
through the Poisson equation
\begin{equation}
 V_{\mathrm{H}}^N(\mathbf{r})
 =
 \int \mathrm{d}\mathbf{r}'
 \frac{n^N(\mathbf{r}')}{|\mathbf{r}-\mathbf{r}'|}
 \rightarrow
 \frac{4\pi}{\Omega}
 \sum_{\mathbf{G}\ne0}\frac{n^N(\mathbf{G})}{\mathbf{G}^2}\ensuremath{,}
\end{equation}
with
\begin{equation}
 n^N(\mathbf{r})
 =\frac{1}{\Omega}\sum_{\mathbf{G}}
 n^N(\mathbf{G})e^{i\mathbf{G}\cdot\mathbf{r}} \ensuremath{.}
\end{equation}

The TDA excited-state Lagrangian and its derivative with respect to a scalar
nuclear coordinate \(\zeta\) are
\begin{subequations}
\begin{align}
   L
   &=E_{\mathrm{KS}}+\Omega_{\mathrm{TDA}}+E_{\mathrm{CPKS}}
     -\sum_{\mu\nu}\Lambda_{\mu\nu}S_{\mu\nu}\ensuremath{,}\\
   \frac{\partial L}{\partial\zeta}
   &=
   \frac{\partial E_{\mathrm{KS}}}{\partial\zeta}
   +\frac{\partial\Omega_{\mathrm{TDA}}}{\partial\zeta}
   +\frac{\partial E_{\mathrm{CPKS}}}{\partial\zeta}
   -\sum_{\mu\nu}\Lambda_{\mu\nu}
   \frac{\partial S_{\mu\nu}}{\partial\zeta}
   \ensuremath{.}
\end{align}
\end{subequations}
The excited-state-force implementation follows the TDA formulation developed for the PW framework.\cite{Hutter_excited_state_forces_2003}
The explicit excited-state gradient is
\begin{align}
   L^\zeta
   &=
   \sum_{\mu\nu\sigma}
   \left[h_{\mu\nu}^{\zeta}
        +V_{\mu\nu\sigma}^{\mathrm{XC}(\zeta)}\right]
        D^{\mathrm{rel}}_{\mu\nu\sigma}
   -\sum_{\mu\nu\sigma}S_{\mu\nu}^{\zeta}\Lambda_{\mu\nu\sigma}
   \nonumber\\
   &\quad
   +\sum_{\mu\nu\kappa\lambda\sigma\sigma'}
   \Bigl[
   (\mu\nu|\kappa\lambda)^\zeta
   \Gamma_{\mu\nu\sigma\kappa\lambda\sigma'}\nonumber\\
   &\qquad
   +f_{\mu\nu\sigma\kappa\lambda\sigma'}^{\mathrm{XC}(\zeta)}
    D_{\mu\nu\sigma}^{\mathrm{X}}
    D_{\kappa\lambda\sigma'}^{\mathrm{X}}
   \Bigr]\ensuremath{.}
   \label{eq:si-tddft-gradient}
\end{align}
The superscript \(\zeta\) denotes a derivative with respect to the nuclear coordinate \(\zeta\), \(\mathbf D^{\mathrm{rel}}\) is the relaxed difference-density matrix, \(\boldsymbol\Lambda\) is the energy-weighted density multiplier, and \(\boldsymbol\Gamma\) is the effective two-particle density entering the differentiated ERIs.
The relaxed and unrelaxed difference-density matrices are
\begin{align}
   \mathbf D^{\mathrm{rel}}
   &=\mathbf{T}+\mathbf{D}^{\mathrm{Z}},\\
   T_{\mu\nu\sigma}
   &=
   \sum_k X_{\mu k\sigma}X_{\nu k\sigma}^{*}\nonumber\\
   &\quad
   -\frac{1}{2}
   \sum_{\kappa\lambda kl}
   \Bigl[
   C_{\mu k\sigma}X_{\kappa k\sigma}^{*}
   S_{\kappa\lambda}X_{\lambda l\sigma}
   C_{\nu l\sigma}^{*}\nonumber\\
   &\qquad
   +C_{\mu l\sigma}X_{\kappa l\sigma}^{*}
   S_{\kappa\lambda}X_{\lambda k\sigma}
   C_{\nu k\sigma}^{*}
   \Bigr]\ensuremath{.}
\end{align}
The resulting computational sequence is to solve the TDA eigenvalue problem,
assemble the terms that depend only on \(\mathbf{D}^{\mathrm{X}}\), solve the
Z-vector equation for \(\mathbf{D}^{\mathrm{Z}}\), and finally evaluate the
gradient terms that contain the relaxed density \(\mathbf D^{\mathrm{rel}}\).

\subsection{\texorpdfstring{$\bk$-Point Transition Dipoles}{k-Point Transition Dipoles}}
\label{sec:si-tddft-kpoint-dipoles}

The finite-\(\bk\) transition-dipole expressions below are implemented in the public CP2K release. Separately, Ref.~\citenum{rttddft_kpoints_luber} reports a CP2K-based RT-TDDFT implementation with explicit \(k\)-point sampling that is formulated in the velocity gauge through the vector potential and current, rather than through these transition-dipole matrix elements. That implementation is not currently available in the public release.

For periodic crystals, the finite-\(\bk\) transition dipole starts from the
length-gauge matrix element
\begin{equation}
 \boldsymbol{\mu}_{nn'}(\bk)
 =
 iq\int_{\mathrm{cell}}\mathrm{d}\mathbf{r}\,
 u_{n\bk}^*(\mathbf{r})
 \frac{\partial u_{n'\bk}(\mathbf{r})}{\partial\bk},
\end{equation}
where $u_{n\bk}(\mathbf{r})=e^{-i\bk\cdot\mathbf{r}}\psi_{n\bk}(\mathbf{r})$. The labels \(n\) and \(n'\) denote bands, \(q\) is the particle charge, and \(\boldsymbol\mu_{nn'}(\bk)\) is the transition-dipole matrix element.
In a periodic Gaussian basis, the lattice-periodic function is expanded as
\begin{equation}
 u_{n\bk}(\mathbf{r})
 =
 e^{-i\bk\cdot\mathbf{r}}
 \sum_\nu C_{\nu n}(\bk)
 \sum_{\mathbf{R}} e^{i\bk\cdot\mathbf{R}}
 \phi_\nu^{\mathbf{R}}(\mathbf{r}),
\end{equation}
with coefficients obtained from
\begin{equation}
 \sum_\nu h_{\mu\nu}(\bk)C_{\nu n}(\bk)
 =
 \sum_\nu S_{\mu\nu}(\bk)C_{\nu n}(\bk)\varepsilon_{n\bk}.
\end{equation}
The phase arbitrariness of \(C_{\nu n}(\bk)\) prevents robust direct numerical
\(\bk\)-derivatives. CP2K therefore evaluates off-diagonal dipoles as
\begin{equation}
\begin{aligned}
 \boldsymbol{\mu}_{nn'}(\bk)
 &=
 q\sum_{\mu\nu}C_{\mu n}^*(\bk)
 \Biggl[
 \frac{-i}{\varepsilon_{n\bk}-\varepsilon_{n'\bk}}
 \frac{\partial h_{\mu\nu}(\bk)}{\partial\bk}\\
 &\quad
 +\frac{i\varepsilon_{n\bk}}{\varepsilon_{n\bk}-\varepsilon_{n'\bk}}
 \frac{\partial S_{\mu\nu}(\bk)}{\partial\bk}\\
 &\quad
 +\mathbf{D}_{\mu\nu}(\bk)
 \Biggr]
 C_{\nu n'}(\bk),
\end{aligned}
\label{eq:si-kpoint-dipole}
\end{equation}
where
\begin{equation}
\begin{aligned}
 \mathbf{D}_{\mu\nu}(\bk)
 =
 \sum_{\mathbf{R}}e^{i\bk\cdot\mathbf{R}}\mathbf{D}_{\mu\nu}^{\mathbf{R}},\\
 \mathbf{D}_{\mu\nu}^{\mathbf{R}}
 =
 \int \mathrm{d}\mathbf{r}\,
 \phi_\mu^{\mathbf{0}}(\mathbf{r})\,\mathbf{r}\,
 \phi_\nu^{\mathbf{R}}(\mathbf{r}).
\end{aligned}
\end{equation}
The $\bk$ derivative of the KS matrix follows analytically
\begin{equation}
 \frac{\partial h_{\mu\nu}(\bk)}{\partial\bk}
 =
 i\sum_{\mathbf{R}}\mathbf{R}e^{i\bk\cdot\mathbf{R}}
 h_{\mu\nu}^{\mathbf{R}},
\end{equation}
and the overlap derivative is obtained analogously. Eq.
\eqref{eq:si-kpoint-dipole} also explains why transition dipoles diverge at
band crossings, where $\varepsilon_{n\bk}-\varepsilon_{n'\bk}$ vanishes.

\subsection{Spatial Descriptors for Electronic Excitations}
\label{sec:si-spatial-descriptors}

The electron--hole descriptors follow the established wavefunction-analysis formalism and its CP2K implementation for TDDFT and BSE excitations, with related implementations in TheoDORE and libwfa.\cite{Plasser2014a,Plasser2014b,Bappler2014,Mewes2015,Plasser2015,Mewes2018,Plasser2020,Plasser2022,Graml2025}
The spatial descriptors introduced in Section~IV~E are constructed from the
electron--hole wavefunction. For an excitation \(n\), this wavefunction is
assembled from the excitation and deexcitation amplitudes as
\begin{align}
\Psi_{\mathrm{exc}}^{(n)}(\mathbf{r}_e,\mathbf{r}_h)
&=
\sum_{ia} X_{ia}^{(n)}
\psi_i(\mathbf{r}_h)\psi_a(\mathbf{r}_e)
\nonumber\\
&\quad+
\sum_{ia} Y_{ia}^{(n)}
\psi_i(\mathbf{r}_e)\psi_a(\mathbf{r}_h),
\label{eq:si-excitation-wavefunction}
\end{align}
where the indices \(i\) and \(a\) label occupied and unoccupied states, respectively.
The orbitals are taken to be real for the finite systems considered here.
For a generic operator \(\hat{O}\), normalized expectation values are evaluated as
\begin{equation}
\langle \hat{O}\rangle_{\mathrm{exc}}
=
\frac{
\langle \Psi_{\mathrm{exc}}^{(n)}|\hat{O}|\Psi_{\mathrm{exc}}^{(n)}\rangle
}{
\langle \Psi_{\mathrm{exc}}^{(n)}|\Psi_{\mathrm{exc}}^{(n)}\rangle
}.
\label{eq:si-excitation-expectation-value}
\end{equation}
The root-mean-square electron--hole separation is
\begin{equation}
d_{\mathrm{exc}}
=
\sqrt{\left\langle
|\mathbf{r}_e-\mathbf{r}_h|^2
\right\rangle_{\mathrm{exc}}},
\label{eq:si-excitation-size}
\end{equation}
which can be decomposed as
\begin{equation}
d_{\mathrm{exc}}
=
\sqrt{
d_{e\rightarrow h}^{\,2}
+\sigma_e^2
+\sigma_h^2
-2\sigma_e\sigma_h R_{eh}
}.
\label{eq:si-excitation-size-decomposition}
\end{equation}
The distance between the electron and hole centroids is
\begin{equation}
d_{e\rightarrow h}
=
\left|
\langle\mathbf{r}_e-\mathbf{r}_h\rangle_{\mathrm{exc}}
\right|,
\label{eq:si-electron-hole-centroid-distance}
\end{equation}
and their individual spatial spreads are
\begin{subequations}
\label{eq:si-electron-hole-spreads}
\begin{align}
\sigma_e
&=
\sqrt{
\left\langle
\left|\mathbf{r}_e-\langle\mathbf{r}_e\rangle_{\mathrm{exc}}\right|^2
\right\rangle_{\mathrm{exc}}
},\\
\sigma_h
&=
\sqrt{
\left\langle
\left|\mathbf{r}_h-\langle\mathbf{r}_h\rangle_{\mathrm{exc}}\right|^2
\right\rangle_{\mathrm{exc}}
}.
\end{align}
\end{subequations}
The dimensionless correlation coefficient entering
Eq.~\eqref{eq:si-excitation-size-decomposition} is
\begin{equation}
R_{eh}
=
\frac{
\langle\mathbf{r}_h\cdot\mathbf{r}_e\rangle_{\mathrm{exc}}
-\langle\mathbf{r}_h\rangle_{\mathrm{exc}}
\cdot\langle\mathbf{r}_e\rangle_{\mathrm{exc}}
}{\sigma_e\sigma_h}.
\label{eq:si-electron-hole-correlation}
\end{equation}
Positive values indicate correlated, bound electron--hole motion, whereas
negative values describe electron and hole densities that avoid one another
dynamically. For anisotropic systems, the same construction is resolved along
individual Cartesian directions. Along \(x\), for example:
\begin{equation}
d_{\mathrm{exc}}^{(x)}
=
\sqrt{
\left\langle|x_e-x_h|^2\right\rangle_{\mathrm{exc}}
}.
\label{eq:si-directional-excitation-size}
\end{equation}
The remaining distance, spread, and correlation descriptors are resolved
analogously by replacing the vector coordinates with their selected Cartesian
components.
\subsection{Ehrenfest and Real-Time Time-Dependent Density-Functional Theory Equations}
\label{sec:si-rt-tddft-ehrenfest}

The mixed quantum--classical equations follow the \textit{ab-initio} Ehrenfest formulation.\cite{LiTullySchlegelFrisch2005}

Ehrenfest dynamics is the mixed quantum--classical limit used in Section~IV~F to connect RT-TDDFT with nuclear motion: the electrons are propagated
quantum mechanically while the nuclei evolve classically on the mean electronic
force. With atom-centered basis functions, the electronic wavefunction is
expanded as
\begin{equation}
 \psi_{i\sigma}(\mathbf{R}(t))
 =
 \sum_\mu C_{\mu i\sigma}(\mathbf{R}(t))\varphi_\mu(\mathbf{R})
 \ensuremath{,}
\end{equation}
where \(C_{\mu i\sigma}\) are the MO coefficients, \(\varphi_\mu\) are the AO basis functions, and \(\psi_{i\sigma}\) are the corresponding time-dependent KS MOs. The indices \(i,j,\ldots\) label occupied orbitals, \(\mu,\nu,\ldots\) label AOs, and \(\sigma\) labels spin.
The corresponding action is
\begin{align}
A
&=
\int \mathrm{d}t
\left[
\sum_\alpha
\frac{M_\alpha}{2}
\left(\frac{\partial\mathbf{R}_\alpha}{\partial t}\right)^2
-V_{\mathrm{nuc}}(\mathbf{R}_\alpha(t))
\right.\nonumber\\
&\quad\left.
+\sum_{j\sigma}
\left\langle
\psi_{j\sigma}
\middle|
i\frac{\partial}{\partial t}-\hat{H}_{\mathrm{el}}
\middle|
\psi_{j\sigma}
\right\rangle
\right]
\ensuremath{.}
\end{align}
The index \(\alpha\) labels nuclei, \(M_\alpha\) and \(\mathbf R_\alpha\) are their masses and positions, \(V_{\mathrm{nuc}}\) is the nuclear repulsion, \(\hat H_{\mathrm{el}}\) is the electronic Hamiltonian, and \(j\) and \(\sigma\) label occupied orbitals and spin.
Variation gives the nuclear equation
\begin{equation}
\begin{aligned}
 M_\alpha\frac{\partial^2\mathbf{R}_\alpha(t)}{\partial t^2}
 &=
 -\frac{\partial}{\partial\mathbf{R}_\alpha}
 V_{\mathrm{nuc}}(\mathbf{R}_\alpha(t))
 \\
 &\quad
 -\sum_{\mu\nu i\sigma}
 C_{\mu i\sigma}(t)C_{\nu i\sigma}(t)
 \Biggl[
 \frac{\partial F_{\mu\nu\sigma}(\mathbf{R}_\alpha(t))}
      {\partial\mathbf{R}_\alpha}\\
 &\qquad
 +D_{\mu\nu}^{\alpha}(\mathbf{R}_\alpha(t))
 \Biggr]
\ensuremath{.}
\end{aligned}
\end{equation}
The quantity \(F_{\mu\nu\sigma}\) is an element of the instantaneous KS matrix, and \(D_{\mu\nu}^{\alpha}\) denotes the basis-motion contribution generated by the nuclear dependence of the atom-centered AOs.
The electronic propagation equation is
\begin{equation}
 -\frac{\partial C_{\mu j\sigma}}{\partial t}
 =
 \sum_{\kappa\lambda}
 \left(\mathbf S^{-1}\right)_{\mu\kappa}
 \left(iF_{\kappa\lambda\sigma}+\tau_{\kappa\lambda\sigma}\right)
 C_{\lambda j\sigma}\ensuremath{.}
\end{equation}
Here
\begin{equation}
\begin{split}
S_{\mu\nu}
&=
\int
\varphi_\mu(\mathbf{r})\varphi_\nu(\mathbf{r})
\,\mathrm{d}\mathbf{r}\ensuremath{,}\\
\tau_{\mu\nu\sigma}
&=
\sum_\alpha
\frac{\partial\mathbf{R}^{\alpha}}{\partial t}
\left\langle
\varphi_\mu
\middle|
\frac{\partial}{\partial\mathbf{R}^{\alpha}}
\middle|
\varphi_\nu
\right\rangle \ensuremath{.}
\end{split}
\end{equation}

The real-time response and impulsive-field construction follow the time-domain TDDFT formulation.\cite{Yabana1996,TDDFTYabanaBertsch}
For RT-TDDFT calculations on isolated systems, a time-dependent electric field
enters naturally in the length gauge
\begin{equation}
\begin{aligned}
i\hbar\frac{\partial}{\partial t}\psi_i(\mathbf{r},t)
&=
\Biggl[
-\frac{\hbar^2\nabla^2}{2m}
+V_{\mathrm{ion}}(\mathbf{r})\\
&\quad
+\int\mathrm{d}\mathbf{r}'\,
\frac{e^2}{|\mathbf{r}-\mathbf{r}'|}
n(\mathbf{r}',t)
\\
&\quad
+V_{\mathrm{xc}}[n(\mathbf{r},t)]
+e\mathbf{E}(t)\cdot\mathbf{r}
\Biggr]\psi_i(\mathbf{r},t)\ensuremath{.}
\end{aligned}
\end{equation}
The quantities \(V_{\mathrm{ion}}\), \(V_{\mathrm{xc}}\), \(n\), and \(\mathbf E(t)\) are the ionic potential, XC potential, time-dependent density, and electric field, respectively, while \(\psi_i(\mathbf{r},t)\) are the time-dependent occupied KS orbitals.
For periodic systems, the length-gauge term is replaced by a velocity-gauge
formulation with
\begin{subequations}
\label{eq:si-rttddft-velocity-gauge}
\begin{align}
\mathbf{A}(t)
&=
-\frac{1}{c}\int^t\mathbf{E}(t')\,\mathrm{d}t',
\label{eq:si-rttddft-vector-potential}\\
\psi_i(\mathbf{r},t)
&=
\exp\!\left[
\frac{ie}{\hbar c}\mathbf{A}(t)\cdot\mathbf{r}
\right]\tilde{\psi}_i(\mathbf{r},t).
\label{eq:si-rttddft-velocity-orbital}
\end{align}
\end{subequations}

The vector \(\mathbf A(t)\) is the electromagnetic vector potential and \(\tilde\psi_i\) is the velocity-gauge orbital.

The \(\delta\)-kick treatment used in Section~IV~F applies an impulsive phase shift
to the ground-state orbitals
\begin{subequations}
\label{eq:si-rttddft-kick-response}
\begin{equation}
 \psi_i(\mathbf{r},0^+)
 =
 e^{i\kappa\hat{\boldsymbol\varepsilon}\cdot\mathbf{r}}
 \psi_i(\mathbf{r},0^-)\ensuremath{,}
\label{eq:si-rttddft-delta-kick}
\end{equation}
equivalent to a field
$\mathbf{E}(t)=\kappa\hat{\boldsymbol\varepsilon}\delta(t)$. The polarizability is then
obtained from the induced dipole
\begin{equation}
 \alpha_{\hat{\boldsymbol\varepsilon}}(\omega)
 =
 -\frac{1}{\kappa}\int_0^\infty
 \Delta\mu_{\hat{\boldsymbol\varepsilon}}(t)
 e^{i\omega t}e^{-\gamma t}\,\mathrm{d}t
 \ensuremath{.}
\label{eq:si-rttddft-polarizability}
\end{equation}
The parameter \(\kappa\) is the kick strength, \(\hat{\boldsymbol\varepsilon}\) is its polarization direction, \(\Delta\mu_{\hat{\boldsymbol\varepsilon}}(t)\) is the induced dipole component, \(\omega\) is angular frequency, and \(\gamma\) is the damping or spectral-broadening parameter.
The photoabsorption cross section follows as
\begin{equation}
 \sigma(\omega)=\frac{4\pi\omega}{c}
 \mathrm{Im}\left[\alpha(\omega)\right]\ensuremath{.}
\label{eq:si-rttddft-photoabsorption}
\end{equation}
\end{subequations}
For periodic systems, the induced dipole is replaced by the macroscopic current
density. The current-density kernel and kinetic momentum operator are
\begin{subequations}
\label{eq:si-rttddft-periodic-response}
\begin{align}
\mathbf{j}(\mathbf{r},t)
&=
\sum_i\frac{e}{2m}
\left[
\tilde{\psi}_i^*(\mathbf{r},t)
\boldsymbol{\pi}
\tilde{\psi}_i(\mathbf{r},t)
+\mathrm{c.c.}
\right],
\label{eq:si-rttddft-current-density}\\
\boldsymbol{\pi}
&=
\mathbf{p}+\frac{e}{c}\mathbf{A}(t).
\label{eq:si-rttddft-kinetic-momentum}
\end{align}

In this expression, \(\boldsymbol\pi\) and \(\mathbf p\) are the kinetic and canonical momentum operators, and \(\mathrm{c.c.}\) denotes the complex-conjugate contribution.
The cell-averaged current is
\begin{equation}
 \mathbf{I}(t)
 =
 -\frac{1}{\Omega}\int \mathrm{d}\mathbf{r}\,\mathbf{j}(\mathbf{r},t)
 \ensuremath{.}
\label{eq:si-rttddft-cell-current}
\end{equation}
The quantity \(\Omega\) is the cell volume and \(\mathbf I(t)\) is the cell-averaged current density.
It gives the optical conductivity
\begin{equation}
 \sigma_{\alpha\beta}(\omega)
 =
 -\frac{1}{\kappa}
 \int_0^\infty
 I_\alpha(t)e^{i\omega t}e^{-\gamma t}\,\mathrm{d}t,
\label{eq:si-rttddft-conductivity}
\end{equation}
and the dielectric tensor
\begin{equation}
 \varepsilon_{\alpha\beta}(\omega)
 =
 \delta_{\alpha\beta}
 +\frac{4\pi i}{\omega}\sigma_{\alpha\beta}(\omega).
\label{eq:si-rttddft-dielectric}
\end{equation}
\end{subequations}
For pump--probe calculations, the same kick can be applied to a nonequilibrium
state at delay time $\tau$
\begin{subequations}
\label{eq:si-rttddft-pump-probe}
\begin{equation}
 \psi_i(\mathbf{r},\tau^+)
 =
 e^{i\kappa\hat{\boldsymbol\varepsilon}\cdot\mathbf{r}}
 \psi_i(\mathbf{r},\tau^-),
\label{eq:si-rttddft-pump-kick}
\end{equation}
and the transient absorption signal is obtained from
\begin{equation}
\Delta\sigma(\omega,\tau)
=
\sigma^*(\omega,\tau)-\sigma^{(0)}(\omega).
\label{eq:si-rttddft-transient-absorption}
\end{equation}
\end{subequations}
The delay is denoted by \(\tau\), while \(\sigma^*(\omega,\tau)\) and \(\sigma^{(0)}(\omega)\) are the pumped-state and unpumped reference spectra, respectively.

\section{\texorpdfstring{Additional \textit{GW} Implementation Details}{Additional GW Implementation Details}}
\label{sec:si-gw-implementation-details}
Throughout Section~V, \(\tau\) and \(\omega\) denote imaginary time and imaginary frequency, \(i\) and \(a\) label occupied and empty KS states, \(n\) labels a general one-particle state, and \(\varepsilon_{\mathrm F}\) is the Fermi energy. The symbols \(G_0\), \(\chi\), \(\epsilon\), \(v\), \(W_0\), and \(\Sigma\) denote the noninteracting Green's function, irreducible polarizability, dielectric operator, bare Coulomb interaction, screened interaction, and self-energy, respectively.

\subsection{\texorpdfstring{Space--Time \textit{GW} Working Equations}{Space-Time GW Working Equations}}
\label{sec:si-gw-spacetime}

The equations combine Hedin's \(GW\) approximation with the space--time representation of the self-energy and dielectric response.\cite{hedin1965new,rojas_1995}

The space--time formulation rewrites the frequency-dependent $GW$ self-energy in
terms of smooth imaginary-time and imaginary-frequency quantities. For a
noninteracting Green's function built from KS orbitals and eigenvalues
\(\Delta_n=\varepsilon_n-\varepsilon_{\mathrm{F}}\)
\begin{equation}
G_0(\mathbf{r},\mathbf{r}';i\tau)
\!=\!
\begin{cases}
i\sum_i^{\mathrm{occ}}
\psi_i(\mathbf{r})\psi_i^*(\mathbf{r}')
e^{-|\Delta_i\tau|},
& \tau<0,\\[0.7em]
-i\sum_a^{\mathrm{empty}}
\psi_a^*(\mathbf{r})\psi_a(\mathbf{r}')
e^{-|\Delta_a\tau|},
& \tau>0,
\end{cases}
\end{equation}
the irreducible polarizability is obtained as a pointwise product
\begin{equation}
\chi(\mathbf{r},\mathbf{r}';i\tau)
=
-i\,G_0(\mathbf{r},\mathbf{r}';i\tau)
G_0(\mathbf{r},\mathbf{r}';-i\tau).
\end{equation}
It is then transformed to imaginary frequency
\begin{equation}
\chi(\mathbf{r},\mathbf{r}';i\omega)
=
i\int_{-\infty}^{\infty}
e^{-i\omega\tau}
\chi(\mathbf{r},\mathbf{r}';i\tau)\,\mathrm{d}\tau.
\end{equation}
The dielectric operator and screened Coulomb interaction are
\begin{align}
\epsilon(\mathbf{r},\mathbf{r}';i\omega)
&=
\delta(\mathbf{r}-\mathbf{r}')
-
\int \mathrm{d}\mathbf{r}''
v(\mathbf{r},\mathbf{r}'')
\chi(\mathbf{r}'',\mathbf{r}';i\omega),\\
W_0(\mathbf{r},\mathbf{r}';i\omega)
&=
\int \mathrm{d}\mathbf{r}''
\epsilon^{-1}(\mathbf{r},\mathbf{r}'';i\omega)
v(\mathbf{r}'',\mathbf{r}').
\end{align}
The inversion of \(\epsilon\) is an operator inversion and becomes a matrix
inversion after discretization in a real-space grid or basis set.
The correlation part of the screened interaction is
\begin{equation}
W_0^{\mathrm{c}}(\mathbf{r},\mathbf{r}';i\omega)
=
W_0(\mathbf{r},\mathbf{r}';i\omega)
-
v(\mathbf{r},\mathbf{r}'),
\end{equation}
and is transformed back to imaginary time
\begin{equation}
W_0^{\mathrm{c}}(\mathbf{r},\mathbf{r}';i\tau)
=
\frac{i}{2\pi}
\int_{-\infty}^{\infty}
e^{i\omega\tau}
W_0^{\mathrm{c}}(\mathbf{r},\mathbf{r}';i\omega)\,
\mathrm{d}\omega .
\end{equation}
The correlation self-energy is then a product in imaginary time
\begin{equation}
\Sigma^{\mathrm{c}}(\mathbf{r},\mathbf{r}';i\tau)
=
i\,G_0(\mathbf{r},\mathbf{r}';i\tau)
W_0^{\mathrm{c}}(\mathbf{r},\mathbf{r}';i\tau),
\end{equation}
followed by the transform to imaginary frequency and projection onto a KS
state
\begin{align}
\Sigma^{\mathrm{c}}(\mathbf{r},\mathbf{r}';i\omega)
&=
i\int_{-\infty}^{\infty}
e^{-i\omega\tau}
\Sigma^{\mathrm{c}}(\mathbf{r},\mathbf{r}';i\tau)\,
\mathrm{d}\tau,\\
\Sigma_n^{\mathrm{c}}(i\omega)
&=
\int \mathrm{d}\mathbf{r}\,\mathrm{d}\mathbf{r}'\,
\psi_n^*(\mathbf{r})
\Sigma^{\mathrm{c}}(\mathbf{r},\mathbf{r}';i\omega)
\psi_n(\mathbf{r}').
\end{align}
The imaginary-frequency self-energy is analytically continued to the real axis
and inserted into the quasiparticle equation of Section~VI. In CP2K, the
time--frequency transforms are evaluated on minimax grids originally developed
for space--time $GW$.\cite{Kaltak2014,liu2016}

\subsection{\texorpdfstring{Localized Gaussian-Basis \textit{GW} Working Equations}{Localized Gaussian-Basis GW Working Equations}}
\label{sec:si-gw-local-gaussian}
In this subsection, Greek indices label Gaussian AOs, capital indices \(P,Q,R,T\) label RI auxiliary functions, \(C_{\mu n}\) are MO coefficients, \(\theta\) is the Heaviside step function, and \(r_{\mathrm c}\) is the Coulomb-truncation radius.

The localized formulation follows the GPW implementation and its reduced-scaling extension.\cite{Wilhelm2016,wilhelm2018}

For finite systems, KS orbitals are expanded in atom-centered Gaussian AO basis
functions
\begin{equation}
\psi_n(\mathbf{r})
=
\sum_\mu C_{\mu n}\phi_\mu(\mathbf{r}),
\end{equation}
and the imaginary-time Green's function is evaluated in the AO basis as
\begin{align}
G_{\mu\nu}(i\tau)
&=
\theta(\tau)
\sum_a^{\mathrm{empty}}
C_{\mu a}C_{\nu a}^*
e^{-(\varepsilon_a-\varepsilon_{\mathrm{F}})\tau}
\nonumber\\
&\quad
-
\theta(-\tau)
\sum_i^{\mathrm{occ}}
C_{\mu i}C_{\nu i}^*
e^{-(\varepsilon_i-\varepsilon_{\mathrm{F}})\tau}.
\end{align}
In an RI basis \(\{\varphi_P\}\), the density response is
\begin{equation}
\chi_{PQ}(i\tau)
=
\sum_{\mu\nu\lambda\sigma}
(\mu\nu|P)
G_{\mu\lambda}(i\tau)
(\lambda\sigma|Q)
G_{\nu\sigma}(-i\tau),
\end{equation}
with three-center integrals
\begin{equation}
(\mu\nu|P)
=
\int \mathrm{d}\mathbf{r}\,\mathrm{d}\mathbf{r}'\,
\phi_\mu(\mathbf{r})\phi_\nu(\mathbf{r})
V_{r_{\mathrm{c}}}(\mathbf{r},\mathbf{r}')
\varphi_P(\mathbf{r}').
\end{equation}
The truncated Coulomb kernel is
\begin{equation}
V_{r_{\mathrm{c}}}(\mathbf{r},\mathbf{r}')
=
\begin{cases}
|\mathbf{r}-\mathbf{r}'|^{-1},
& |\mathbf{r}-\mathbf{r}'|\le r_{\mathrm{c}},\\[0.3em]
0,
& \text{otherwise}.
\end{cases}
\end{equation}
Within the RI approximation, four-center Coulomb integrals are approximated as
\begin{align}
(\mu\nu|\lambda\sigma)_{\mathrm{C}}
\simeq{}&
\sum_{PQRT}
(\mu\nu|P)
\left(\mathbf M^{-1}\right)_{PQ}
 V_{QR}\nonumber\\
&\quad\times
\left(\mathbf M^{-1}\right)_{RT}
(T|\lambda\sigma),
\end{align}
where \(\mathbf M\) is the chosen metric matrix in the auxiliary basis and \(\mathbf V\) is the
Coulomb matrix. The dielectric matrix and screened interaction are then
\begin{align}
\boldsymbol{\epsilon}(i\omega)
&=
\mathbf{1}
-
\mathbf{V}^{1/2}\mathbf{M}^{-1}
\boldsymbol{\chi}(i\omega)
\mathbf{M}^{-1}\mathbf{V}^{1/2},\\
V_{PQ}
&=
\int \mathrm{d}\mathbf{r}\,\mathrm{d}\mathbf{r}'\,
\varphi_P(\mathbf{r})
|\mathbf{r}-\mathbf{r}'|^{-1}
\varphi_Q(\mathbf{r}'),\\
M_{PQ}
&=
\int \mathrm{d}\mathbf{r}\,\mathrm{d}\mathbf{r}'\,
\varphi_P(\mathbf{r})
V_{r_{\mathrm{c}}}(\mathbf{r},\mathbf{r}')
\varphi_Q(\mathbf{r}'),\\
\mathbf{W}(i\omega)
&=
\mathbf{M}^{-1}\mathbf{V}^{1/2}
\boldsymbol{\epsilon}^{-1}(i\omega)
\mathbf{V}^{1/2}\mathbf{M}^{-1}.
\end{align}
After transforming \(W_{PQ}(i\omega)\) to imaginary time, the AO self-energy is
\begin{equation}
\Sigma_{\lambda\sigma}(i\tau)
=
\sum_{\mu\nu PQ}
(\lambda\mu|Q)
G_{\mu\nu}(i\tau)
(\nu\sigma|P)
W_{PQ}(i\tau).
\end{equation}

The contraction order and practical scaling of the molecular response construction are illustrated in Fig.~\ref{fig:si-gw-scaling-chi}. With a local RI metric, sparsity in the AO and auxiliary index pairs reduces the dominant response construction from the nonlocal contraction pattern to approximately quadratic scaling for the molecular systems considered in Ref.~\citenum{wilhelm2021}.

\begin{fixedfigure}
\centering
\includegraphics[width=0.92\linewidth]{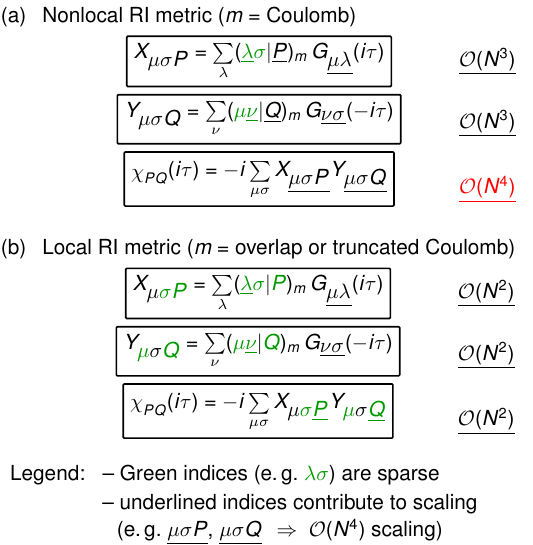}
\caption{Scaling of the imaginary-time density-response construction in a localized basis with (a) a nonlocal and (b) a local RI metric. The tensors $X$ and $Y$ are formed before the final contraction. Green index groups are sparse, and underlined indices determine the formal scaling. Reproduced from Ref.~\citenum{wilhelm2021}, Copyright 2021 The Authors.}
\label{fig:si-gw-scaling-chi}
\end{fixedfigure}

\subsection{\texorpdfstring{Full-\(\mathbf{k}\) Periodic \textit{GW} Working Equations}{Full-k Periodic GW Working Equations}}
\label{sec:si-gw-full-k}
For the periodic equations, \(\mathbf k\) is a crystal momentum, \(\mathbf R,\mathbf R_i,\mathbf S_i,\mathbf T\) are lattice vectors, \(\Omega_{\mathrm{BZ}}\) is the Brillouin-zone volume, and superscript lattice vectors identify real-space matrix blocks.

For periodic crystals, KS orbitals are expanded as Bloch functions in an
atom-centered Gaussian basis:
\begin{equation}
\psi_{n\mathbf{k}}(\mathbf{r})
=
\sum_\mu C_{\mu n}(\mathbf{k})
\sum_{\mathbf{R}}
e^{i\mathbf{k}\cdot\mathbf{R}}
\phi_\mu^{\mathbf{R}}(\mathbf{r}).
\end{equation}
The corresponding AO Green's function is
\begin{align}
G_{\mu\nu}(i\tau,\mathbf{k})
&=
\theta(\tau)
\sum_a^{\mathrm{empty}}
C_{\mu a}(\mathbf{k})C_{\nu a}^*(\mathbf{k})
e^{-(\varepsilon_{a\mathbf{k}}-\varepsilon_{\mathrm{F}})\tau}
\nonumber\\
&\quad
-
\theta(-\tau)
\sum_i^{\mathrm{occ}}
C_{\mu i}(\mathbf{k})C_{\nu i}^*(\mathbf{k})
e^{-(\varepsilon_{i\mathbf{k}}-\varepsilon_{\mathrm{F}})\tau}.
\end{align}
It is transformed to lattice-vector space by
\begin{equation}
G^{\mathbf{R}}_{\mu\nu}(i\tau)
=
\int_{\mathrm{BZ}}
\frac{\mathrm{d}\mathbf{k}}{\Omega_{\mathrm{BZ}}}
e^{-i\mathbf{k}\cdot\mathbf{R}}
G_{\mu\nu}(i\tau,\mathbf{k}).
\end{equation}
The periodic RI response is
\begin{align}
\chi_{PQ}^{\mathbf{R}}(i\tau)
&=
\sum_{\lambda\mathbf{R}_1\nu\mathbf{R}_2}
\left[
\sum_{\mu\mathbf{S}_1}
(\mu\mathbf{R}_1-\mathbf{S}_1\,\nu\mathbf{R}_2\,|\,P\mathbf{0})
G_{\lambda\mu}^{\mathbf{S}_1}(-i\tau)
\right]\nonumber\\
&\quad\times
\left[
\sum_{\sigma\mathbf{S}_2}
(\lambda\mathbf{R}_1\,\sigma\mathbf{R}_2-\mathbf{S}_2\,|\,Q\mathbf{R})
G_{\nu\sigma}^{\mathbf{S}_2}(i\tau)
\right].
\end{align}
The periodic three-center integrals are
\begin{equation}
(\mu\mathbf{R}\,\nu\mathbf{T}\,|\,P\mathbf{P})
=
\int \mathrm{d}\mathbf{r}\,\mathrm{d}\mathbf{r}'\,
\phi_\mu^{\mathbf{R}}(\mathbf{r})
\phi_\nu^{\mathbf{T}}(\mathbf{r})
V_{r_{\mathrm{c}}}(\mathbf{r},\mathbf{r}')
\varphi_P^{\mathbf{P}}(\mathbf{r}').
\end{equation}
The real-space self-energy is then
\begin{align}
\Sigma_{\lambda\sigma}^{\mathbf{R}}(i\tau)
&=
i
\sum_{P\mathbf{R}_1\nu\mathbf{S}_1}
\left[
\sum_{\mu\mathbf{S}_2}
(\lambda\mathbf{0}\,\mu\mathbf{S}_1-\mathbf{S}_2\,|\,P\mathbf{R}_1)
G_{\mu\nu}^{\mathbf{S}_2}(i\tau)
\right]\nonumber\\
&\quad\times
\left[
\sum_{Q\mathbf{R}_2}
(\sigma\mathbf{R}\,\nu\mathbf{S}_1\,|\,Q\mathbf{R}_1-\mathbf{R}_2)
W_{QP}^{\mathbf{R}_2}(i\tau)
\right].
\end{align}
The diagonal Bloch-basis matrix elements enter the usual quasiparticle
equation
\begin{equation}
\varepsilon_{n\mathbf{k}}^{G_0W_0}
=
\varepsilon_{n\mathbf{k}}^{\mathrm{DFT}}
+
\mathrm{Re}\,
\Sigma_{n\mathbf{k}}
\!\left(\varepsilon_{n\mathbf{k}}^{G_0W_0}\right)
-
v_{n\mathbf{k}}^{\mathrm{xc}}.
\end{equation}

The compact Gaussian representation also gives favorable parallel performance for two-dimensional crystals. With an aug-SZV-MOLOPT basis, a $G_0W_0$@PBE band structure of monolayer MoS$_2$ requires approximately one day on a workstation with an Intel i5 processor, 14 cores, and 192\,GB of memory. On a modern high-performance computing system, the same calculation takes substantially less than one hour on several hundred to approximately one thousand cores and exhibits near-ideal strong scaling over more than two orders of magnitude in core count, as shown in Fig.~\ref{fig:si-gw-kpoint-scaling}.\cite{Pasquier2025}

\begin{fixedfigure}
\centering
\includegraphics[width=0.82\linewidth]{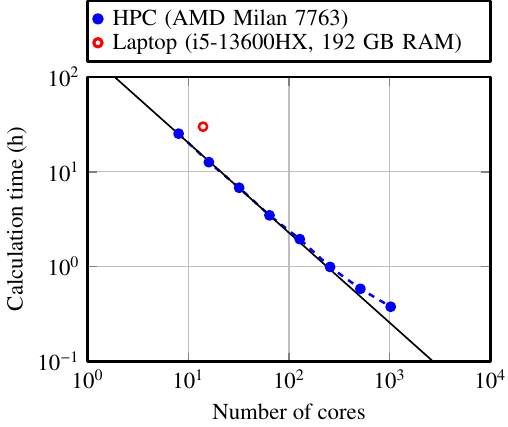}
\caption{Computation time for a $G_0W_0$@PBE band-structure calculation of monolayer MoS$_2$ as a function of the number of processor cores, using an aug-SZV-MOLOPT basis. Reproduced from Ref.~\citenum{Pasquier2025}, Copyright 2025 by The American Physical Society.}
\label{fig:si-gw-kpoint-scaling}
\end{fixedfigure}

An earlier periodic $GW$ formulation introduced in 2017,\cite{wilhelm2017} does not attain the same numerical precision for quasiparticle gaps of crystals with small unit cells. The full-$\mathbf{k}$ implementation described here supersedes that formulation for this regime by retaining the lattice-resolved response and self-energy throughout the space--time construction.

\subsection{\texorpdfstring{$\Gamma$-Point Minimum-Image Reconstruction in Space--Time \textit{GW}}{Gamma-Point Minimum-Image Reconstruction in Space-Time GW}}
\label{sec:si-gw-gamma-mic}

The large-supercell, $\Gamma$-point-only $GW$ implementation in CP2K starts from the $\Gamma$-point polarizability
\begin{align}
\chi_{PQ}^{\Gamma}(i\tau)
&\equiv
\chi_{PQ}(\mathbf{k}=\mathbf{0},i\tau)
=
\sum_{\mathbf{R}}\chi_{PQ}^{\mathbf{R}}(i\tau).
\label{si:gw-chi-gamma}
\end{align}
In a direct periodic implementation, constructing $\chi_{PQ}^{\mathbf{R}}(i\tau)$ involves several lattice sums. For sufficiently large nonmetallic supercells, the Green's function and the polarizability are local in real space and decay exponentially with increasing $|\mathbf{r}-\mathbf{r}'|$.\cite{Kohn1996,martin2004,Prodan2005,Graml2024} In this limit the dominant contribution is contained in a single cell, so the $\Gamma$-point polarizability can be evaluated from lattice-summed three-center integrals as
\begin{align}
\chi_{PQ}^{\Gamma}(i\tau)
&=
\sum_{\lambda\nu\mu\sigma}
(\mu\nu|P)\,
G_{\mu\lambda}^{\Gamma}(-i\tau)\,
(\lambda\sigma|Q)\,
G_{\nu\sigma}^{\Gamma}(i\tau),
\label{si:gw-chi-gamma-ao}\\
(\mu\nu|P)
&=
\sum_{\mathbf{R}_1,\mathbf{R}_2}
(\mu\mathbf{R}_1\,\nu\mathbf{R}_2\,|\,P\mathbf{0}).
\label{si:gw-three-center-sum}
\end{align}
The MIC then assigns each auxiliary-basis pair to the closest periodic image. The real-space polarizability matrix elements are reconstructed according to
\begin{align}
\chi_{PQ}^{\mathbf{R}}(i\tau)
&=
\begin{cases}
\chi_{PQ}^{\Gamma}(i\tau),
& \text{if } \varphi_P^{\mathbf{0}} \text{ and } \varphi_Q^{\mathbf{R}} \text{ are closest,}\\
0,
& \text{otherwise,}
\end{cases}
\ensuremath{,}
\label{si:gw-mic-chi}
\end{align}
which becomes exact in the large-cell limit. The resulting $\chi_{PQ}^{\mathbf{R}}(i\tau)$ is transformed back to the Brillouin zone at negligible additional cost and used to form the screened interaction
\begin{align}
W_{PQ}^{\mathbf{R}}(i\omega)
&=
\int_{\mathrm{BZ}}
\frac{d\mathbf{k}}{\Omega_{\mathrm{BZ}}}\,
\exp(i\mathbf{k}\cdot\mathbf{R})\,
W_{PQ}(\mathbf{k},i\omega).
\label{si:gw-w-real-space}
\end{align}
The Brillouin-zone integral requires special treatment because, for two-dimensional materials and s-type auxiliary functions, $W_{PQ}(\mathbf{k},i\omega)$ diverges at the $\Gamma$ point with $1/k$.\cite{wilhelm2017,ren/etal:2021,zhu2021all,Zhang2026,Gong2026} In the implementation described in Ref.~\citenum{Graml2024}, each matrix element $W_{PQ}^{\mathbf{R}}(i\omega)$ is evaluated on $4\times4$ and $6\times6$ Monkhorst--Pack meshes and extrapolated with $N_k^{-1/2}$, where $N_k$ is the number of $k$ points.\cite{Monkhorst1976,zhu2021all}

The same locality argument is applied to the self-energy. With the MIC-screened interaction $W_{PQ}^{\mathrm{MIC}}(i\tau)$, the $\Gamma$-point self-energy for a large cell is evaluated as
\begin{align}
\Sigma_{\lambda\sigma}^{\Gamma}(i\tau)
&=
i
\sum_{\nu\mu PQ}
(\lambda\mu|Q)\,
G_{\mu\nu}^{\Gamma}(i\tau)\,
(\nu\sigma|P)\,
W_{PQ}^{\mathrm{MIC}}(i\tau).
\label{si:gw-sigma-gamma}
\end{align}
Diagonal matrix elements of this self-energy are then evaluated in the Bloch basis and inserted into the quasiparticle equation. The approximation becomes exact for large nonmetallic supercells, provided that $G$, $\chi$, and $\Sigma$ decay before periodic images overlap.

The scaling benchmark in Fig.~\ref{fig:si-gw-gamma-scaling} covers monolayer MoSe$_2$ supercells from $9\times9$ to $14\times14$, containing up to 588 atoms. A $G_0W_0$ calculation for a $10\times10$ cell with 300 atoms takes approximately 7\,h on 576 processor cores. For the 984-atom twisted MoSe$_2$/WS$_2$ cell discussed in Section~VI, the full calculation requires approximately 42\,h on 1536 cores. Diagonalizing $\chi(\mathbf{k},i\omega)$ removes spurious negative eigenvalues before constructing the screened interaction.

\begin{fixedfigure}
\centering
\includegraphics[width=0.90\linewidth]{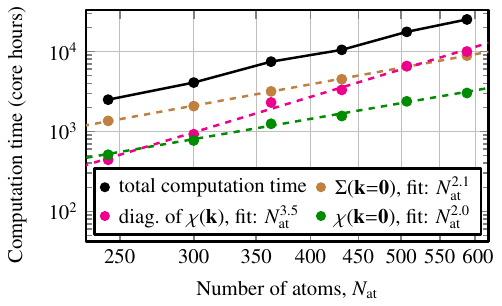}
\caption{Scaling of the $G_0W_0$@PBE computation time for monolayer MoSe$_2$ supercells using the $\Gamma$-point MIC algorithm. The response matrix $\chi(\mathbf{k},i\omega)$ is diagonalized to remove spurious negative eigenvalues before the screened interaction is formed. Reproduced from Ref.~\citenum{Graml2024} under the Creative Commons Attribution 4.0 International License.}
\label{fig:si-gw-gamma-scaling}
\end{fixedfigure}

\subsection{\texorpdfstring{Static Coulomb-Hole and Screened-Exchange Working Equations}{Static Coulomb-Hole and Screened-Exchange Working Equations}}
\label{sec:si-cohsex-working-equations}

The static Coulomb-hole plus screened-exchange (st-COHSEX) approximation requires only the zero-frequency screened interaction.\cite{hedin1965new,hybertsen1986} Within the space--time implementation, its static irreducible polarizability is obtained directly by integrating the imaginary-time response:
\begin{equation}
\chi_{PQ}(\mathbf{k},i\omega=0)
=
\int \mathrm{d}\tau\,
\chi_{PQ}(\mathbf{k},i\tau).
\end{equation}
Constructing $\mathbf{W}^{\mathrm{c}}(i\omega=0)$ formally scales as $\mathcal{O}(N^4)$, whereas the space--time formulation with a truncated Coulomb metric reduces this step to near-$\mathcal{O}(N^3)$ scaling. In contrast to implementations restricted to diagonal quasiparticle corrections, CP2K retains off-diagonal self-energy matrix elements, which are important for accurate EAs.\cite{voora21}

For the water-cluster assessment of $\alpha$-COHSEX, the test set comprises the surface-bound W6a cluster,\cite{hammer05} the internally localized W6f (Kevan) cluster,\cite{feng80} and the interior-bound W8, W12, W16, W20, and W24 clusters.\cite{vysotskiy2012} Generalized KS semi-projected RPA EAs serve as reference values and closely reproduce high-level electron-attachment EOM coupled-cluster singles-and-doubles benchmarks with noniterative triples [EOM-EA-CCSD(T)(a)*].\cite{voora21,Tyagi2024} Fitting the scaling parameter to W12 gives $\alpha=0.72$. This value is then transferred without further adjustment to all remaining clusters.

\section{\texorpdfstring{Bethe--Salpeter Equation Working Equations}{Bethe-Salpeter Equation Working Equations}}
\label{sec:si-bse-working-equations}
Throughout Section~VI, \(p,q,r,s\) label general one-particle states, \(i,j\) occupied and \(a,b\) virtual states, \(P,Q,R,S\) RI auxiliary functions, \(\mu\) a Cartesian direction, and \(n\) an excitation root. The vectors \(\mathbf X^{(n)}\) and \(\mathbf Y^{(n)}\) are the BSE excitation and deexcitation amplitudes, \(\Omega^{(n)}\) is the excitation energy, and \(\eta>0\) is the spectral broadening.

\subsection{\texorpdfstring{Linear-Response Bethe--Salpeter Equation Working Equations}{Linear-Response Bethe-Salpeter Equation Working Equations}}
\label{sec:si-lr-bse-working-equations}

The LR-BSE working equations follow the standard many-body formulation and its CP2K realization for optical excitations.\cite{Strinati1988,Rohlfing2000,Onida2002,Graml2025}

For the finite systems considered here, the orbitals are real, such that \(\mathbf A\) and \(\mathbf B\) are real symmetric and the excitation and deexcitation amplitudes are normalized in the indefinite BSE metric according to
\begin{equation}
\left(\mathbf X^{(n)}\right)^{\mathrm T}\mathbf X^{(m)}
-
\left(\mathbf Y^{(n)}\right)^{\mathrm T}\mathbf Y^{(m)}
=
\pm \delta_{nm},
\label{eq:si-bse-normalization}
\end{equation}
where the sign corresponds to positive and negative excitation energies \(\Omega^{(n)}\), respectively.
This is the biorthonormality condition referred to in Section~VII~A of the main manuscript.

The LR-BSE kernel introduced in Section~VII~A of the main manuscript contains bare-exchange and statically screened direct terms. In the MO basis, their four-index matrix elements are
\begin{subequations}
\label{eq:si-bse-coulomb-elements}
\begin{align}
v_{pq,rs}
&=
\int \mathrm{d}^3r\,\mathrm{d}^3r'\,
\psi_p(\mathbf{r})\psi_q(\mathbf{r})\nonumber\\
&\quad\times
v(\mathbf{r},\mathbf{r}')
\psi_r(\mathbf{r}')\psi_s(\mathbf{r}'),\\
W_{pq,rs}
&=
\int \mathrm{d}^3r\,\mathrm{d}^3r'\,
\psi_p(\mathbf{r})\psi_q(\mathbf{r})\nonumber\\
&\quad\times
W(\mathbf{r},\mathbf{r}';\omega=0)
\psi_r(\mathbf{r}')\psi_s(\mathbf{r}').
\end{align}
\end{subequations}
Dynamic effects in the screening are neglected at this level. CP2K evaluates the matrix elements with RI in the Coulomb metric. Using the three-center integrals introduced in Section~V~B, the bare and screened interactions become
\begin{subequations}
\label{eq:si-bse-ri-elements}
\begin{align}
v_{pq,rs}
&=
\sum_{P,Q}(pq|P)
\left(\mathbf V^{-1}\right)_{PQ}
(Q|rs),\\
W_{pq,rs}
&=
\sum_{P,Q,R,S}
(pq|P)
\left(\mathbf V^{-1/2}\right)_{PQ}
\left(\boldsymbol{\epsilon}^{-1}\right)_{QR}\nonumber\\
&\quad\times
\left(\mathbf V^{-1/2}\right)_{RS}
(S|rs).
\end{align}
\end{subequations}
In these contractions, \(V_{PQ}\) and \(\epsilon_{PQ}(\omega=0)\) are elements of the Coulomb and static dielectric matrices, respectively, both defined in Section~V~B. The Coulomb metric is retained for the BSE contractions rather than replaced by the truncated metric used in the reduced-scaling molecular \(GW\) response.

With the normalization condition in Eq.~\eqref{eq:si-bse-normalization}, the transition dipoles entering the polarizability in Section~VII~A of the main manuscript and their KS-basis matrix elements are
\begin{subequations}
\label{eq:si-bse-transition-dipoles}
\begin{align}
d_\mu^{(n)}
&=
\sqrt{2}\sum_{ia}\mu_{ia,\mu}
\left(X_{ia}^{(n)}+Y_{ia}^{(n)}\right),\\
\bm{\mu}_{ia}
&=
\int \mathrm{d}^3r\,
\psi_i(\mathbf{r})\,\mathbf{r}\,\psi_a(\mathbf{r}).
\end{align}
\end{subequations}
For an isotropic ensemble, orientational averaging gives
\begin{subequations}
\label{eq:si-bse-isotropic-spectrum}
\begin{align}
\bar{\alpha}(\omega)
&=
\frac{1}{3}\sum_{\mu\in\{x,y,z\}}\alpha_{\mu\mu}(\omega)
=
\sum_{n>0}\frac{f^{(n)}}
{\left(\Omega^{(n)}\right)^2-(\omega+i\eta)^2},\\
f^{(n)}
&=
\frac{2}{3}\Omega^{(n)}
\sum_{\mu\in\{x,y,z\}}\left|d_\mu^{(n)}\right|^2.
\end{align}
\end{subequations}

\subsection{\texorpdfstring{Real-Time Bethe--Salpeter Equation Working Equations}{Real-Time Bethe-Salpeter Equation Working Equations}}
\label{sec:si-rt-bse-working-equations}

The position-space Hartree and SX kernels of the effective RT-BSE Hamiltonian are given in Section~VII~B of the main manuscript. To expose their relation to LR-BSE at the implementation level, let \(\hat{h}^{GW}|m\rangle=\varepsilon_m^{GW}|m\rangle\) and expand \(\hat{\rho}(t)=\sum_{mn}\rho_{mn}(t)|m\rangle\langle n|\). The resulting MO-basis equation is
\begin{align}
i\hbar\dot{\rho}_{mn}(t)
&=
\left(\varepsilon_m^{GW}-\varepsilon_n^{GW}\right)\rho_{mn}(t)\nonumber\\
&\quad+
\left[
e\mathbf{E}(t)\cdot\bm{\mu}
+\mathbf V^{\mathrm H}
+\boldsymbol{\Sigma}^{\mathrm{SX}},
\boldsymbol{\rho}(t)
\right]_{mn}.
\label{eq:si-rtbse-mo-eom}
\end{align}
The indices \(m\) and \(n\) label quasiparticle states. The quantities \(\rho_{mn}(t)\) and \(\rho_{0,mn}\) are matrix elements of the time-dependent and reference density matrices \(\boldsymbol\rho(t)\) and \(\boldsymbol\rho_0\), respectively. \(\mathbf E(t)\) is the applied field. The matrix \(\bm\mu\) is the dipole matrix, while \(\mathbf V^{\mathrm H}\) and \(\boldsymbol\Sigma^{\mathrm{SX}}\) are the induced Hartree and screened-exchange matrices.
The interaction matrices entering this equation are
\begin{subequations}
\label{eq:si-rtbse-mo-kernels}
\begin{align}
V_{mn}^{\mathrm{H}}
&=
2\sum_{m'n'}v_{mn,m'n'}
\left(\rho_{m'n'}(t)-\rho_{0,m'n'}\right),\\
\Sigma_{mn}^{\mathrm{SX}}
&=
-\sum_{m'n'}W_{mm',nn'}
\left(\rho_{m'n'}(t)-\rho_{0,m'n'}\right).
\end{align}
\end{subequations}
Their four-index matrix elements are those of Eq.~\eqref{eq:si-bse-coulomb-elements}. For the closed-shell, spin-degenerate density-matrix convention used here, the factor of two in the Hartree term accounts for the two spin channels; the screened-exchange term acts within a single spin channel. CP2K integrates the AO density matrix using ETRS time stepping.\cite{PropagatorsCastroRubio,Marek2025}

RT-HF follows from the same operator structure by replacing the screened interaction with the bare Coulomb interaction:
\begin{equation}
\langle\mathbf{r}_1|
\hat{\Sigma}^{\mathrm{X}}[\Delta\hat{\rho}(t)]
|\mathbf{r}_2\rangle
=
-v(\mathbf{r}_1,\mathbf{r}_2)
\Delta\rho(\mathbf{r}_1,\mathbf{r}_2,t).
\label{eq:si-rthf-exchange}
\end{equation}
Adiabatic RT-TDDFT instead replaces this explicit kernel by the response of the approximate XC potential. A hybrid functional adds a global fraction \(\gamma v\) of bare exchange, which can mimic an average inverse dielectric screening when \(\gamma\approx\epsilon^{-1}\), but it cannot reproduce a spatially varying \(W(\mathbf{r},\mathbf{r}')\) at interfaces or in other heterogeneous environments.

\section{Active-Space Embedding Working Equations and Implementation Details}
\label{sec:si-active-space-details}

The molecular and periodic active-space framework includes self-consistent density feedback and the CP2K--Qiskit Nature interface.\cite{Rossmannek2021,Rossmannek2023,Battaglia2024}

\begin{table*}
\caption{Practical active-space embedding modes in CP2K.}
\label{tab:si-active-space-hierarchy}
\begin{ruledtabular}
\begin{tabular}{llll}
\ReviewTableCell{0.17\textwidth}{Level} &
\ReviewTableCell{0.22\textwidth}{Representative input} &
\ReviewTableCell{0.25\textwidth}{Physical interpretation} &
\ReviewTableCell{0.22\textwidth}{Typical use} \\
\hline
\ReviewTableCell{0.17\textwidth}{Hamiltonian export} &
\ReviewTableCell{0.22\textwidth}{\texttt{AS\_SOLVER NONE}; optional \texttt{FCIDUMP} or QCSchema output} &
\ReviewTableCell{0.25\textwidth}{Mechanical embedding analogue. The active orbital subset is extracted from a frozen reference} &
\ReviewTableCell{0.22\textwidth}{External solvers, reproducible benchmark Hamiltonians, quantum-computing workflows} \\
\ReviewTableCell{0.17\textwidth}{One-shot embedding} &
\ReviewTableCell{0.22\textwidth}{\texttt{AS\_SOLVER QISKIT} or \texttt{FCI}; \texttt{SCF\_EMBEDDING F}} &
\ReviewTableCell{0.25\textwidth}{Electrostatic embedding analogue. Correlated active space in a fixed mean-field potential} &
\ReviewTableCell{0.22\textwidth}{Local excitation energies, solver benchmarking, active-space selection tests} \\
\ReviewTableCell{0.17\textwidth}{Self-consistent embedding} &
\ReviewTableCell{0.22\textwidth}{\texttt{AS\_SOLVER QISKIT} or \texttt{FCI}; \texttt{SCF\_EMBEDDING T}} &
\ReviewTableCell{0.25\textwidth}{Polarizable embedding. Active 1-RDM and environment are converged together} &
\ReviewTableCell{0.22\textwidth}{Strongly coupled defects, charge-transfer states, embedded spectra in polarizable media} \\
\end{tabular}
\end{ruledtabular}
\end{table*}

\subsection{Active-Space Hamiltonian and Density Feedback}
\label{sec:si-active-space-density-feedback}

The active-space embedding formulation in Section~VIII of the main manuscript partitions the molecular-orbital space into inactive, active, and external subspaces. For active spin-orbital indices \(u,v,x,y\in\mathcal{A}\), the one- and two-electron operators entering the embedded Hamiltonian are
\begin{subequations}
\label{eq:si-as-operator-expansion}
\begin{align}
\hat{h}^{\mathrm{emb}}_{\mathcal{A}}
&=
\sum_{uv\in\mathcal{A}}
h^{\mathrm{emb}}_{uv}
\hat{a}^{\dagger}_{u}\hat{a}_{v},\label{eq:si-as-one-electron-operator}\\
\hat{W}_{\mathcal{A}}
&=
\frac{1}{2}\sum_{uvxy\in\mathcal{A}}
(uv|xy)
\hat{a}^{\dagger}_{u}\hat{a}^{\dagger}_{x}
\hat{a}_{y}\hat{a}_{v}.\label{eq:si-as-two-electron-operator}
\end{align}
\end{subequations}
The matrix \(\mathbf h^{\mathrm{emb}}\) combines the one-electron terms with the mean-field interaction generated by the inactive environment, and \((uv|xy)\) denotes a two-electron integral in the active subspace. The operators \(\hat a_u^\dagger\) and \(\hat a_u\) create and annihilate an electron in active spin orbital \(u\), respectively. With the range separation introduced in Section~VIII~A of the main manuscript, the energy partition is
\begin{subequations}
\label{eq:si-as-energy-partition}
\begin{align}
E
&=
E^{\mathcal{I}}+E^{\mathcal{A}},\label{eq:si-as-total-energy}\\
E^{\mathcal{A}}
&=
\sum_{uv\in\mathcal{A}}
h^{\mathrm{emb}}_{uv}D^{\mathcal{A}}_{uv}
+E^{\mathcal{A}}_2,\label{eq:si-as-active-energy}\\
E^{\mathcal{A}}_2
&=
\frac{1}{2}\sum_{uvxy\in\mathcal{A}}
(uv|xy)d^{\mathcal{A}}_{uvxy}.\label{eq:si-as-two-particle-energy}
\end{align}
\end{subequations}
The 1-RDM \(\mathbf D^{\mathcal{A}}\) determines the one-particle contribution and the density returned to the mean-field environment, whereas the 2-RDM \(\boldsymbol d^{\mathcal{A}}\) supplies the correlated two-particle energy and related properties. The active-space 1-RDM is transformed back to the AO representation according to
\begin{align}
D^{\mathcal{A}}_{\mu\nu}
&=
\sum_{uv\in\mathcal{A}}
C_{\mu u}D^{\mathcal{A}}_{uv}C_{\nu v},
\label{eq:si-as-ao-density}
\end{align}
where \(\mu\) and \(\nu\) are AO indices. The coefficient \(C_{\mu u}\) transforms active orbital \(u\) to AO \(\mu\). The density entering the embedding potential is partitioned schematically as
\begin{align}
\mathbf D
&=
\mathbf D^{\mathcal{I}}+\mathbf D^{\mathcal{A}}.
\label{eq:si-as-density-partition}
\end{align}
In self-consistent embedding, CP2K iterates the fixed-point map
\begin{align}
\mathbf D^{\mathcal{A}}_{n+1}
&=
\mathcal{S}\!\left[
\mathbf h^{\mathrm{emb}}\!\left(\mathbf D^{\mathcal{I}},\mathbf D^{\mathcal{A}}_n\right),
(uv|xy)
\right],
\label{eq:si-as-embedding-iteration}
\end{align}
where \(\mathcal{S}\) denotes the selected active-space solver. The returned 1-RDM is converted to the active AO and real-space density before the KS matrix and embedded Hamiltonian are rebuilt. The 2-RDM is not mixed into the environment because the embedding potential depends on the one-particle density. Linear damping updates the input density as
\begin{align}
\mathbf D^{\mathcal{A},\mathrm{in}}_{n+1}
&=
(1-\alpha)\mathbf D^{\mathcal{A},\mathrm{in}}_n
+\alpha\mathbf D^{\mathcal{A},\mathrm{solver}}_{n+1}
\ensuremath{.}
\label{eq:si-as-density-mixing}
\end{align}
The index \(n\) labels embedding iterations, and \(0<\alpha\leq1\) is the linear mixing fraction. Pulay, Broyden, and modified Broyden schemes replace this scalar update by an approximate inverse-Jacobian update in the active-orbital density space. These quasi-Newton variants are particularly useful when the polarizable embedding cycle is tightly coupled to the surrounding KS response.

\subsection{Solver Hierarchy and Interface Details}
\label{sec:si-active-space-solvers}

The implementation exposes three levels that parallel the mechanical, electrostatic, and polarizable limits of multiscale embedding. With \texttt{AS\_SOLVER NONE}, CP2K constructs the active-space Hamiltonian and writes it, for example, as a \texttt{FCIDUMP} or QCSchema file without solving the correlated problem. When the active orbitals form only a subset of the mean-field orbital space, this Hamiltonian-export mode is the mechanical-embedding analogue: the fragment Hamiltonian is extracted from a frozen reference and passed to an external classical or quantum solver.

In one-shot embedding, \texttt{AS\_SOLVER QISKIT} or \texttt{AS\_SOLVER FCI} is combined with \texttt{SCF\_EMBEDDING F}. The active-space solver experiences the potential generated by the mean-field environment, but its correlated 1-RDM is not iterated back to polarize that environment. With \texttt{SCF\_EMBEDDING T}, the returned 1-RDM is instead transformed to the AO density and used to rebuild the real-space density and KS matrix until the range-separated embedding energy is converged. This self-consistent mode is the polarizable-embedding analogue, in which local correlation and the mean-field response of the surrounding material relax together.

For the Qiskit Nature interface, CP2K transfers the active-space one- and two-electron information to a separate solver process through socket-based inter-process communication and receives the active energy and density matrix in return. Ground-state calculations use VQE, whereas the quantum equation-of-motion algorithm provides electronically excited states.\cite{Peruzzo2014,Ollitrault2020} The inactive environment is optimized for its ground-state density and does not presently relax in response to a changed excited-state active density.

The internal \texttt{FCI} backend removes socket communication and external Python dependencies from the solver step and provides a deterministic reference for small active spaces and regression tests. It returns the active-space energy and 1-RDM to the same embedding loop used by the Qiskit interface, so \texttt{SCF\_EMBEDDING F} and \texttt{SCF\_EMBEDDING T} have the same physical meaning for both solvers. Additional eigenvalues of the fragment Hamiltonian provide access to excited states.

\section{Charge-Transfer and Open-Boundary Transport Working Equations and Implementation Details}
\label{sec:si-charge-transport-details}

\subsection{Constrained Density-Functional Theory Dynamics and Validation}
\label{sec:si-cdft-dynamics-validation}

The CDFT force implementation introduced in Section~IX~A of the main manuscript was validated through geometry optimization and MD benchmarks.\cite{Ahart22jctc} For MgO containing neutral and charged oxygen vacancies, PBE geometry optimizations yielded reorganization energies for electron tunneling between the defects.\cite{perdew1996} The results agreed with earlier PW CDFT calculations performed with CPMD.\cite{Oberhofer09,McKenna12prb,Blumberger13}

Because the density constraint is enforced through an additional SCF loop, the inner DFT and outer CDFT iterations must both be converged sufficiently to conserve energy in CDFT-MD. For H$_2^+$ with a target charge difference of $N_c=0.5\,e$, a constraint convergence of $1\times10^{-6}\,e$ gave an energy drift below $1\times10^{-6}$\,H/atom/ps. In practical condensed-phase simulations, a threshold of $1\times10^{-3}\,e$ was sufficient for stable, energy-conserving trajectories of systems including oxygen defects in MgO and aqueous Ru$^{2+}$/Ru$^{3+}$ complexes.\cite{Ahart22jctc}

The computational overhead is typically a factor of 2--3 relative to conventional DFT-MD, which permits condensed-phase sampling with hybrid functionals. CDFT-MD simulations of aqueous Ru$^{2+}$/Ru$^{3+}$ electron self-exchange were performed with BLYP, B3LYP, and the range-separated hybrid $\omega$B97X. The latter yielded reorganization energies closest to experiment.\cite{Becke88,Lee88,Becke93,Chai08,Ahart22jctc}

\subsection{Projection Operator-Based Diabatization Working Equations}
\label{sec:si-pod-working-equations}

POD begins with the KS eigenvalue problem
\begin{equation}
\hat{H}|\psi_i\rangle
=
\epsilon_i|\psi_i\rangle,
\label{eq:si-pod-ks}
\end{equation}
where $|\psi_i\rangle$ and $\epsilon_i$ are the KS orbitals and energies. The AO basis is first orthogonalized by a L\"owdin transformation.\cite{Lowdin70} If $\mathbf{H}$ and $\mathbf{S}$ are the Hamiltonian and overlap matrices in the original nonorthogonal AO basis, the orthogonalized Hamiltonian is
\begin{equation}
\widetilde{\mathbf{H}}
=
\mathbf{S}^{-1/2}\mathbf{H}\mathbf{S}^{-1/2}.
\label{eq:si-pod-lowdin}
\end{equation}

The orthogonalized basis functions are assigned to $N$ spatial or molecular regions, which gives the block structure.\cite{Kondov07}
\begin{equation}
\widetilde{\mathbf{H}}
=
\begin{bmatrix}
\widetilde{\mathbf{H}}_{11} & \widetilde{\mathbf{H}}_{12} & \cdots & \widetilde{\mathbf{H}}_{1N} \\
\widetilde{\mathbf{H}}_{21} & \widetilde{\mathbf{H}}_{22} & \cdots & \widetilde{\mathbf{H}}_{2N} \\
\vdots & \vdots & \ddots & \vdots \\
\widetilde{\mathbf{H}}_{N1} & \widetilde{\mathbf{H}}_{N2} & \cdots & \widetilde{\mathbf{H}}_{NN}
\end{bmatrix}.
\label{eq:si-pod-partition}
\end{equation}
For a donor--acceptor problem, two blocks contain the basis functions of the donor and acceptor. A molecular junction can instead be represented by left-electrode, molecular, and right-electrode blocks. Formally, block $\alpha$ defines the projector
\begin{equation}
\hat{P}_{\alpha}
=
\sum_{\mu\in\alpha}
|\widetilde{\phi}^{(\alpha)}_{\mu}\rangle
\langle\widetilde{\phi}^{(\alpha)}_{\mu}|,
\label{eq:si-pod-projector}
\end{equation}
which gives the method its name.

Each diagonal block is diagonalized with a unitary matrix $\mathbf{U}_{\alpha}$. Defining $\mathbf{U}=\operatorname{diag}(\mathbf{U}_1,\ldots,\mathbf{U}_N)$ gives
\begin{equation}
\overline{\mathbf{H}}
=
\mathbf{U}^{\dagger}\widetilde{\mathbf{H}}\mathbf{U}.
\label{eq:si-pod-block-diagonalization}
\end{equation}
The diagonal blocks $\overline{\mathbf{H}}_{\alpha\alpha}$ contain the diabatic-state energies, whereas the off-diagonal blocks $\overline{\mathbf{H}}_{\alpha\beta}$ contain the electronic couplings between diabatic states on different fragments. These couplings also shift the diabatic energies relative to the original adiabatic KS spectrum, so occupation of the diabatic states can imply a different Fermi level.

POD is implemented as a post-SCF analysis of the $N_{\mathrm{ao}}\times N_{\mathrm{ao}}$ KS matrix.\cite{Futera17} The overlap matrix is diagonalized to construct $\mathbf{S}^{-1/2}$, and the partitioned blocks are stored separately. Only the lower-triangular block set required by subsequent operations is retained or written to binary restart files. After diagonalization of the diagonal blocks, the diabatic coefficients are transformed back to the original AO basis with $\mathbf{S}^{1/2}$ and stored in the internal MO data structures. CP2K can write the resulting diabatic orbitals to CUBE files and the energies and coupling matrices to the log or binary restart files. Restricted and unrestricted KS references are supported.

\subsection{Kubo Transport Implementation and Dimensional Normalization}
\label{sec:si-kubo-implementation}

The Kubo transport property is available through \texttt{\&KUBO\_TRANSPORT} under \texttt{FORCE\_EVAL / PROPERTIES}, with \texttt{METHOD DIAGONALIZATION} providing the production implementation.\cite{Kubo1957/10.1143/JPSJ.12.570,Greenwood1958} The underlying finite-temperature configuration-space sampling follows the finite-volume analysis of disordered crystals.\cite{KuhneProdan2018,EfremkinHeskeKuehneProdan2026} If the SCF calculation already supplies a complete canonical MO set, CP2K reuses its eigenvalues and coefficients. For OT calculations or incomplete canonical spaces, CP2K performs an internal generalized diagonalization. It forms the symmetric inverse square root of the overlap matrix and diagonalizes
\begin{equation}
\mathbf H^{\perp}
=
\mathbf S^{-1/2}\mathbf H\mathbf S^{-1/2}
\ensuremath{.}
\label{eq:si-kubo-orthonormal-hamiltonian}
\end{equation}
The matrices \(\mathbf H\) and \(\mathbf S\) are the Hamiltonian and AO overlap matrices, and \(\mathbf H^\perp\) is the Hamiltonian in the symmetrically orthonormalized basis.
CP2K then transforms the position--Hamiltonian commutator kernels to the orthonormal eigenbasis. This matrix-based path is independent of the GPW real-space-grid representation and is shared by GPW and GAPW calculations.

The periodicity flags and actual cell vectors determine both the projector onto the transport subspace and its normalization measure.\cite{KuhneHeskeProdan2020} Fully periodic cells and isolated finite boxes use volume normalization and report the scalar trace average in \texttt{S/cm}. Slabs with \texttt{PERIODIC XY}, \texttt{XZ}, or \texttt{YZ} use the periodic area and report a sheet conductivity in \texttt{S}, with carrier densities in $\mathrm{m}^{-2}$. Wires with \texttt{PERIODIC X}, \texttt{Y}, or \texttt{Z} use the periodic length and report the transport coefficient in \texttt{S*m}, with carrier densities in $\mathrm{m}^{-1}$. Since the projection is constructed from the actual periodic vectors, tilted slabs and non-orthorhombic ribbons do not acquire an artificial dependence on vacuum padding.

The property input controls the electronic temperature, dissipation broadening, chemical-potential window, number of grid points, and neutral chemical potential. If \texttt{NEUTRAL\_MU} is omitted, CP2K determines the neutral point from the eigenvalue spectrum and total electron count. Electron and hole densities are then reported relative to that point together with the Cartesian conductivity tensor and its trace-averaged scalar value.

\subsection{Hairy Probes Occupations and \textnormal{\textsc{Quickstep}} Implementation}
\label{sec:si-hairy-probes-details}

The weak-coupling Hairy Probes occupations implement electronic open boundaries through atomically attached reservoirs and solution probes.\cite{McEniry2007,Horsfield2016,Zauchner2018}

In the Gaussian AO representation, the density and density matrix are
\begin{subequations}
\label{eq:si-hp-density}
\begin{align}
n(\mathbf{r})
&=
\sum_{nm}D_{nm}\phi_n(\mathbf{r})\phi_m^{*}(\mathbf{r}),
\label{eq:si-hp-real-space-density}\\
D_{nm}
&=
\sum_i f_i C_{ni}C_{mi}^{*}.
\label{eq:si-hp-density-matrix}
\end{align}
\end{subequations}
The quantities $C_{ni}$ are the MO coefficients and $f_i$ are the occupations. The labels \(n\) and \(m\) denote AOs \(\phi_n\) and \(\phi_m\), \(i\) labels MOs, and \(D_{nm}\) is an AO density-matrix element. In weak-coupling HP-DFT, the occupations are constructed from the AO weight of each MO on the main and solution probes:
\begin{equation}
f_i
=
\frac{
\alpha\,\overline{f}(\epsilon_i)\sum_s|\ensuremath{C_{j_s i}}|^2
+\sum_p f^{(p)}(\epsilon_i)|\ensuremath{C_{j_p i}}|^2
}{
\alpha\sum_s|\ensuremath{C_{j_s i}}|^2
+\sum_p|\ensuremath{C_{j_p i}}|^2
}.
\label{eq:si-hp-occupation}
\end{equation}
The reservoir distributions are
\begin{subequations}
\label{eq:si-hp-fermi-functions}
\begin{align}
f^{(p)}(\epsilon_i)
&=
\frac{1}{1+\exp[(\epsilon_i-\mu_p)/(k_{\mathrm B}T_p)]},
\label{eq:si-hp-main-fermi}\\
\overline{f}(\epsilon_i)
&=
\frac{1}{1+\exp[(\epsilon_i-\overline{\mu})/(k_{\mathrm B}\overline{T})]}.
\label{eq:si-hp-solution-fermi}
\end{align}
\end{subequations}
The energy of MO \(i\) is \(\epsilon_i\), \(\mu_p\) and \(T_p\) are the electrochemical potential and temperature of main probe \(p\), and \(\overline\mu\) and \(\overline T\) are their solution-probe counterparts.
The main-probe AO indices $j_p$ impose electrode electrochemical potentials, while $j_s$ labels the more weakly weighted solution probes that prevent unphysical depopulation of electrolyte-localized states. The parameter $\alpha\ll1$ sets their relative weight.

At every SCF iteration, the reference level $\overline{\mu}$ is adjusted so that the total electron count equals that of the neutral, unbiased cell. The probe potential is
\begin{equation}
\mu_p
=
\overline{\mu}+\Delta\mu_p,
\label{eq:si-hp-probe-potential}
\end{equation}
where $\Delta\mu_p$ is supplied in the input. The floating reference preserves overall charge neutrality under applied bias and eliminates the need for a compensating background charge.

In \textsc{Quickstep}, diagonalization first supplies the KS eigenvalues and MO coefficients. The \texttt{hairy\_probes} module then replaces the conventional smearing step by Eq.~\eqref{eq:si-hp-occupation}, after which the resulting density matrix enters the next SCF iteration. Separate \texttt{probe\_occupancy} and \texttt{probe\_occupancy\_kp} routines handle $\Gamma$-point and $k$-point calculations, respectively, and map each selected atom to its AO range.

Repeatable \texttt{\&HAIRY\_PROBES} sections within \texttt{\&DFT} specify atom lists, $\Delta\mu_p$, reservoir temperatures, $\alpha$, and occupation-number tolerances. Restricted and unrestricted references and $k$-point sampling are supported. The present implementation uses the diagonalization solver and is not interfaced with OT. The corresponding input parameters are documented in the CP2K reference manual.\cite{CP2KHairyProbesManual}

\subsection{CP2K+SMEAGOL Density Integration and Interface Details}
\label{sec:si-negf-interface-details}

The transport backend is the established SMEAGOL implementation of DFT+NEGF.\cite{Rocha2005,Rocha2006}

The nonequilibrium density matrix is evaluated by separating an equilibrium contribution, integrated on a complex contour, from a correction on the real axis within the bias window. A single lead can be used as the equilibrium reference, but weakly coupled or bound states may then be populated incorrectly at high bias. CP2K+SMEAGOL therefore uses the weighted double-contour construction.\cite{Brandbyge2002}
\begin{subequations}
\label{eq:si-negf-double-contour}
\begin{align}
\mathbf D_M
&=
\mathbf W^L\left(\mathbf D_{\mathrm{eq}}^L+\boldsymbol{\Delta}_{\mathrm{neq}}^R\right)
\notag\\
&\quad
{}+\mathbf W^R\left(\mathbf D_{\mathrm{eq}}^R+\boldsymbol{\Delta}_{\mathrm{neq}}^L\right),
\label{eq:si-negf-weighted-density}\\
\mathbf W^L+\mathbf W^R
&=\mathbf I
\ensuremath{.}
\label{eq:si-negf-weight-sum}
\end{align}
\end{subequations}
The matrix \(\mathbf D_M\) is the extended-molecule density matrix, \(\mathbf D_{\mathrm{eq}}^{L/R}\) are equilibrium contour contributions referenced to the left or right lead, \(\boldsymbol{\Delta}_{\mathrm{neq}}^{L/R}\) are the complementary real-axis nonequilibrium corrections, \(\mathbf W^{L/R}\) are element-wise mixing-weight matrices, and \(\mathbf I\) is the identity matrix.
The element-wise weights are determined from the left- and right-referenced nonequilibrium contributions.

At each SCF iteration, CP2K constructs the extended-molecule Hamiltonian and overlap matrices and passes them to SMEAGOL together with the geometry, bias, energy grid, $k$ points, and precomputed lead data. SMEAGOL evaluates the self-energies, Green functions, and contour integrals and returns \(\mathbf D_M\) for CP2K density mixing. After convergence, it additionally returns the transmission, current, optional spectral data, and the energy-weighted density matrix required by the force expression. The interface translates between CP2K's DBCSR matrix layout and the sparse SMEAGOL/SIESTA representation, allowing direct cross-validation against SIESTA+SMEAGOL.

A calculation proceeds in two stages. A bulk-lead run with \texttt{RUN\_TYPE BULKTRANSPORT} and \texttt{BULKLEAD L}, \texttt{R}, or \texttt{LR} writes the principal-layer descriptors and matrices. The extended-molecule calculation then uses \texttt{RUN\_TYPE EMTRANSPORT}. \texttt{VBIAS} sets the applied voltage, while \texttt{NENERGIMCIRCLE}, \texttt{NENERGIMLINE}, \texttt{NENERGREAL}, \texttt{DELTA}, and \texttt{TRCOEFFICIENTS} control the density and transmission integrations. \texttt{HARTREELEADSBOTTOM} and related lead-position controls align the bulk-lead and extended-molecule Hartree references and should first be checked at zero bias.

The lead self-energies remain fixed during the SCF cycle. The extended molecule must therefore contain enough electrode layers to screen its density perturbation before the boundary to the semi-infinite leads. Since CP2K Gaussian basis functions are not strictly truncated, larger lead cells, additional screening layers, or compact lead basis sets may be required to avoid couplings between non-neighboring images. A mixed-basis strategy can retain a larger basis in the chemically active region while using compact bases for the lead and screening layers.

The Green-function inversions at different real- and complex-axis energies and $k$ points are independent and distributed over MPI ranks, with OpenMP and threaded LAPACK parallelism within each energy point. The reported solvated Au-wire calculations used 64 real-axis and 32 complex-contour evaluations per SCF step. Biased steps were about one order of magnitude more expensive than closed-boundary CP2K calculations, while systems exceeding $10\,000$ basis functions remained accessible for static transport, geometry optimization, and short biased trajectories.\cite{Ahart2024}

Validation covered forces, electrostatics, transport, and biased MD. At zero bias, the force on a displaced atom in an infinite Au wire agreed with closed-cell CP2K, whereas omission of the energy-weighted overlap-derivative contribution produced qualitatively incorrect forces. An Au parallel-plate capacitor reproduced the expected Hartree-potential drop and agreed closely with SIESTA+SMEAGOL at 1 and 4~V. Optimization of Au--H$_2$--Au showed bias-induced atomic displacements and H--H bond elongation.\cite{Ahart2024}

\begingroup
\section{Nuclear Motion and Finite-Temperature Spectroscopy Working Equations}
\label{sec:si-nuclear-motion-spectroscopy}

\subsection{Grassmann Wavefunction Dynamics}
\label{sec:si-gext-details}

For a fixed number \(N_{\mathrm{occ}}\) of fully occupied states, matrices satisfying \(\mathbf C_j^\dagger\mathbf S_j\mathbf C_j=\mathbf I\) belong to a generalized Stiefel manifold. Identifying coefficient matrices related by occupied-space rotations yields the Grassmann manifold, represented in the nonorthogonal AO metric by \(\boldsymbol{\Pi}_j=\mathbf P_j\mathbf S_j\), with \(\mathbf P_j=\mathbf C_j\mathbf C_j^\dagger\) and \(\boldsymbol{\Pi}_j^2=\boldsymbol{\Pi}_j\).\cite{Edelman1998Grassmann,polack_grassmannextrapolation_2021} For real \(\Gamma\)-point orbitals, \(\mathbf P_j=\mathbf C_j\mathbf C_j^{\mathrm T}\).

Let $\mathbf{C}_j$ and $\mathbf{S}_j$ denote the occupied-orbital coefficient matrix and AO overlap matrix at step $j$, respectively. For a history of length $q$, the projected GExt predictor is
\begin{equation}
\widetilde{\mathbf{C}}_n
=
(\widetilde{\mathbf P\mathbf{S}})_n\mathbf{C}_{n-1}
=
\sum_{i=1}^{q}\alpha_i
(\mathbf P\mathbf{S})_{n-q-1+i}\mathbf{C}_{n-1} .
\label{eq:si-gext-projected-orbitals}
\end{equation}
The integer \(n\) labels the current propagation step, \(i\) labels a history entry, and \(\alpha_i\) are extrapolation coefficients obtained from the selected overlap-matrix fit.
Although each historical \(\mathbf P_j\mathbf S_j\) is idempotent, their linear combination is not generally a projector. CP2K therefore applies this extrapolated projector combination to \(\mathbf C_{n-1}\) and reorthonormalizes the trial orbitals in the current metric \(\mathbf S_n\). The resulting occupied subspace is invariant under rotations among occupied orbitals. The overlap matrices serve as sparse molecular descriptors, while the projected form avoids explicit evaluations of Grassmann exponential and logarithm maps.

In the difference-fit strategy, the current overlap is represented relative to $\mathbf{S}_{n-1}$. The Tikhonov-regularized coefficients are obtained from
\begin{subequations}
\label{eq:si-gext-difference-strategy}
\begin{equation}
\begin{aligned}
\boldsymbol{\beta}^{\star}
={}&
\underset{\boldsymbol{\beta}\in\mathbb{R}^{q-1}}
{\operatorname{arg\,min}}
\Bigg\{
\Bigg\|
(\mathbf{S}_n-\mathbf{S}_{n-1})
\\
&-\sum_{i=1}^{q-1}\beta_i
(\mathbf{S}_{n-q-1+i}-\mathbf{S}_{n-1})
\Bigg\|_{\mathrm F}^{2}
+\varepsilon^2\boldsymbol{\beta}^{\mathrm T}\boldsymbol{\beta}
\Bigg\} ,
\end{aligned}
\label{eq:si-gext-difference-fit}
\end{equation}
with $\varepsilon=10^{-4}$. The corresponding projector is
\begin{equation}
\begin{aligned}
(\widetilde{\mathbf P\mathbf{S}})_n
={}&
\left(1-\sum_{i=1}^{q-1}\beta_i^{\star}\right)
(\mathbf P\mathbf{S})_{n-1}
\\
&+\sum_{i=1}^{q-1}\beta_i^{\star}
(\mathbf P\mathbf{S})_{n-q-1+i} .
\end{aligned}
\label{eq:si-gext-difference-projector}
\end{equation}
\end{subequations}
The equivalent coefficients $\alpha_i$ in Eq.~\eqref{eq:si-gext-projected-orbitals} sum to one, which is advised when the same extrapolation is applied during geometry optimization.\cite{askarpour_grassmannextrapolation_2025}

The quasi-time-reversible strategy instead combines symmetric descriptor pairs. Defining $\widetilde q=q/2$ for even $q$ and $\widetilde q=(q-1)/2$ for odd $q$, its coefficients satisfy
\begin{subequations}
\label{eq:si-gext-qtr-strategy}
\begin{equation}
\begin{aligned}
\boldsymbol{\gamma}^{\star}
={}&
\underset{\boldsymbol{\gamma}\in\mathbb{R}^{\widetilde q}}
{\operatorname{arg\,min}}
\Bigg\{
\Bigg\|
(\mathbf{S}_n+\mathbf{S}_{n-q})
\\
&-\sum_{i=1}^{\widetilde q}\gamma_i
(\mathbf{S}_{n-i}+\mathbf{S}_{n-q+i})
\Bigg\|_{\mathrm F}^{2}
+\varepsilon^2\boldsymbol{\gamma}^{\mathrm T}\boldsymbol{\gamma}
\Bigg\} .
\end{aligned}
\label{eq:si-gext-qtr-fit}
\end{equation}
The extrapolated projector combination is then
\begin{equation}
\begin{aligned}
(\widetilde{\mathbf P\mathbf{S}})_n
={}&
-(\mathbf P\mathbf{S})_{n-q}
\\
&+\sum_{i=1}^{\widetilde q}\gamma_i^{\star}
\left[(\mathbf P\mathbf{S})_{n-i}
+(\mathbf P\mathbf{S})_{n-q+i}\right] .
\end{aligned}
\label{eq:si-gext-qtr-projector}
\end{equation}
\end{subequations}
The symmetric pairing makes the predictor quasi-time-reversible and thereby favors long-time MD stability.\cite{pes_quasitimereversible_2023}

Both strategies are selected in \texttt{\&FORCE\_EVAL\%DFT\%QS} using \texttt{EXTRAPOLATION GEXT\_PROJ} or \texttt{EXTRAPOLATION GEXT\_PROJ\_QTR}. The history length $q$ is controlled by \texttt{EXTRAPOLATION\_ORDER}. Values from 4 to 10 are typically suitable, with Tikhonov regularization becoming increasingly important for larger values of \(q\). In Born--Oppenheimer AIMD, the predicted orbitals initialize the ordinary SCF or OT iterations, which continue to the requested convergence threshold. For the Grassmann realization of second-generation Car--Parrinello AIMD, the $\Gamma$-point \textsc{Quickstep} path combines the quasi-time-reversible predictor with \texttt{MAX\_SCF\_HISTORY 1}.\cite{Kuehne2007,Kuehne2014SecondGenerationCPMD,Kuehne2026CP2GBestPractices} Electronic states are fully converged during startup until the required history has been accumulated. Subsequent steps retain the state after one electronic correction. The compensating friction in the modified Langevin dynamics is controlled by \texttt{\&MOTION\%MD\%LANGEVIN\%NOISY\_GAMMA}. The GExt predictor also supports complex $\mathbf{k}$-point wavefunctions, whereas the reduced-history correction is currently applied directly in the $\Gamma$-point branch.

\subsection{Finite-Field \textit{ab-initio} Molecular Dynamics Working Equations}
\label{sec:si-finite-field-aimd}

\subsubsection{Polarization in Periodic Systems}

For isolated systems, the electronic dipole follows directly from the first moment of the density. Under PBC, however, the position operator is ill-defined, and CP2K therefore evaluates the macroscopic polarization from the Berry phase \(\bm{\gamma}^{\mathrm{el}}\) of the occupied manifold.\cite{KingSmith:1993hp} The polarization $\mathbf{P}$ is the dipole density over the cell volume $\Omega$:
\begin{equation}
    \mathbf{P} = \mathbf{P}_\textrm{ion} + \mathbf{P}_\text{el} = \frac{e}{\Omega}\sum_I
    \mathbf{R}_{I}Z_I - e\frac{\mathbf{h}\bm{\gamma}^{\text{el}}}{2\pi \Omega},
\end{equation}
where $\mathbf{h}=[\mathbf{a},\mathbf{b},\mathbf{c}]$ is the \(3\times3\) matrix composed of cell vectors. Its column $\mathbf{h}_l$ is a direct-lattice vector, $\mathbf{G}_k=2\pi\mathbf{h}^{-1}_k$ is the corresponding primitive reciprocal-lattice vector satisfying $\mathbf{G}_k\cdot\mathbf{h}_l=2\pi\delta_{kl}$, and $Z_I$ denotes the ionic charge number (the valence ionic charge number when a PP is used).

For $\Gamma$-point-only calculations, the formula by Resta,\cite{resta} can be used to obtain the Berry phase $\gamma^\textrm{el}_k$ as the expectation value of a many-body operator
\begin{equation}
    \gamma^\textrm{el}_k = \text{Im} \ \text{ln} \ \langle \Phi | \exp[i2\pi \mathbf{h}^{-1}_k \cdot \sum_i^{N_e}\hat{\mathbf{r}}_i] | \Phi \rangle.
    \label{resta}
\end{equation}
The equality \(N_e=N\) identifies the electron number, and \(k\) labels a primitive lattice direction.

 The overlap of two Slater determinants is given by the determinant of their overlap matrix.\cite{lowdin} Therefore, in terms of the one-electron functions $\phi_i(\br)$, the Berry phase can be evaluated as
\begin{subequations}
\label{eq:si-berry-determinant}
\begin{equation}
    \gamma^{\mathrm{el}}_k
    = \operatorname{Im}\ln\det\mathbf{M}^{k},
    \label{berry_det}
\end{equation}
where the overlap matrix has elements
\begin{equation}
    M_{ij}^{k} = \langle \phi_i | \exp[i2\pi \mathbf{h}^{-1}_k \cdot \hat{\mathbf{r}}] | \phi_j \rangle.
    \label{Eq_Qmatrix}
\end{equation}
\end{subequations}

Within the linear-combination-of-AOs framework, the matrix in Eq.~\eqref{Eq_Qmatrix} is evaluated as
\begin{subequations}
\label{eq:si-berry-ao-representation}
\begin{equation}
    \mathbf{M}^{k}
    = \mathbf{C}^{\top}\mathbf{S}^{k}_{\mathrm{Berry}}\mathbf{C},
\label{eq:si-berry-ao-matrix}
\end{equation}
where $\mathbf{C}_{\mu i}$ = $c_{\mu i}$ and
\begin{equation}
    [{S}^{k}_{\text{Berry}}]_{\mu \nu} = \langle \varphi_\mu |\exp[i2\pi \mathbf{h}^{-1}_k \cdot \hat{\mathbf{r}}]| \varphi_\nu \rangle.
\label{eq:si-berry-ao-integrals}
\end{equation}
\end{subequations}

CP2K evaluates $\mathbf{P}^\textrm{ion}$ analogously to the Berry phase using
\begin{equation}
    \mathbf{P}_\textrm{ion} = \frac{e}{2\pi\Omega}\sum_{k=1}^3 \mathbf{h}_k\, \text{Im}\,\ln \exp\left(i2\pi \mathbf{h}^{-1}_k\cdot\sum_I Z_I  \mathbf{R}_I\right)
    \ensuremath{.}
\end{equation}

\subsubsection{\texorpdfstring{Coupling to the Field Functionals (Constant $\mathbf{E}$ and Constant $\mathbf{D}$)}{Coupling to the Field Functionals (Constant E and Constant D)}}

The electric enthalpy minimized at fixed $\mathbf{E}$ is
\begin{equation}
\mathcal{F} \left(\mathbf{E};\mathbf{C},\{\mathbf{R}_I\}\right)
=
H_{\mathrm{PBC}} \left(\mathbf{C},\{\mathbf{R}_I\}\right)
-\Omega\,\mathbf{E}\cdot \mathbf{P} \left(\mathbf{C},\{\mathbf{R}_I\}\right).
\label{eq:si-constant-E-functional}
\end{equation}

The standard KS or OT machinery supplies
\(\partial H_{\mathrm{PBC}}/\partial\mathbf C\).
Defining the field weights \(w_k\), the complete field-dependent orbital gradient can be written compactly as
\begin{subequations}
\label{eq:si-constant-E-field-gradient}
\begin{equation}
w_k=\frac{e\mathbf{E}\cdot \mathbf{h}_k}{2\pi},
\label{eq:si-constant-E-field-weight}
\end{equation}
\begin{equation}
\begin{aligned}
\left(\frac{\partial \mathcal{F}}{\partial \mathbf{C}}\right)_{\mathbf{E}}
&=
-\Omega\,\mathbf{E}\cdot
\frac{\partial \mathbf{P}}{\partial \mathbf{C}}\\
&=
-\frac{e}{2\pi}\,(\mathbf{h}^{\mathrm T}\mathbf{E})\cdot
\frac{\partial \bm{\gamma}^{\mathrm{el}}}{\partial \mathbf{C}}\\
&=
-\sum_{k=1}^{3} w_k
\frac{\partial \gamma_k^{\mathrm{el}}}{\partial \mathbf{C}}\\
&=
-\sum_{k=1}^{3} w_k
\operatorname{Im}\!\left[
\mathbf{S}^k_{\mathrm{Berry}}\mathbf{C}
\big(\mathbf{M}^k\big)^{-\mathrm T}
\right].
\end{aligned}
\label{eq:kernel_deriv}
\end{equation}
\end{subequations}

Algorithmically, CP2K forms the total orbital gradient by summing the KS contribution and the finite-field (Berry-phase) contribution:
\begin{subequations}
\label{eq:si-constant-E-optimization}
\begin{equation}
\frac{\partial \mathcal{F}}{\partial \mathbf{C}}
=
\frac{\partial H_{\mathrm{PBC}}}{\partial \mathbf{C}}
+
\left(\frac{\partial \mathcal{F}}{\partial \mathbf{C}}\right)_{\mathbf{E}}.
\label{eq:si-constant-E-total-gradient}
\end{equation}

The standard SCF/OT update step then acts on this total gradient in the following schematic form:
\begin{equation}
\mathbf{C}^{(n+1)} = \mathrm{Update}_{\mathrm{OT/SCF}}\!\left(
\mathbf{C}^{(n)},\;
\frac{\partial \mathcal{F}}{\partial \mathbf{C}}
\right).
\label{eq:si-constant-E-update}
\end{equation}
\end{subequations}

At fixed \(\mathbf D\), CP2K minimizes the following functional.\cite{Stengel2009}
\begin{subequations}
\label{eq:si-constant-D-branch}
\begin{equation}
\begin{aligned}
\mathcal{U}\!\left(\mathbf{D};\mathbf{C},\{\mathbf{R}_I\}\right)
&=
H_{\mathrm{PBC}}\!\left(\mathbf{C},\{\mathbf{R}_I\}\right)
\\
&\quad+\frac{\Omega}{8\pi}\sum_{\alpha=x,y,z}
d_\alpha
\\
&\qquad\times
\left(D_\alpha-4\pi P_\alpha(\mathbf{C},\{\mathbf{R}_I\})\right)^2,
\end{aligned}
\label{eq:si-constant-D-functional}
\end{equation}
with $d_\alpha$ (keyword \texttt{D\_FILTER}) optionally masking components.

The field contribution to the orbital gradient is
\begin{equation}
\begin{aligned}
\left(\frac{\partial \mathcal{U}}{\partial \mathbf{C}}\right)_{\mathrm{\mathbf{D}}}
&=
-\Omega\sum_{\alpha=x,y,z}
d_\alpha\left(D_\alpha-4\pi P_\alpha(\mathbf{C},\{\mathbf{R}_I\})\right)\\
&\quad \times
\frac{\partial P_\alpha(\mathbf{C},\{\mathbf{R}_I\})}{\partial \mathbf{C}}.
\end{aligned}
\label{eq:si-constant-D-field-gradient}
\end{equation}
\end{subequations}

Therefore, the constant-$\mathbf{D}$ branch uses the same Berry-phase derivative kernel $\frac{\partial \gamma^\textrm{el}_k(\mathbf{C},\{\mathbf{R}_I\})}{\partial \mathbf{C}}$ as the constant-$\mathbf{E}$ branch, but scaled by $-\Omega(\mathbf{D}-4\pi\mathbf{P})$ (and by $d_\alpha$ per component).

Similarly, CP2K forms the total orbital gradient as
\begin{subequations}
\label{eq:si-constant-D-optimization}
\begin{equation}
\frac{\partial \mathcal{U}}{\partial \mathbf{C}}
=
\frac{\partial H_{\mathrm{PBC}}}{\partial \mathbf{C}}
+
\left(\frac{\partial \mathcal{U}}{\partial \mathbf{C}}\right)_{\mathbf{D}},
\label{eq:si-constant-D-total-gradient}
\end{equation}
and updates the orbitals using the existing SCF/OT optimizer:
\begin{equation}
\mathbf{C}^{(n+1)} = \mathrm{Update}_{\mathrm{OT/SCF}}\!\left(
\mathbf{C}^{(n)},\;
\frac{\partial \mathcal{U}}{\partial \mathbf{C}}
\right).
\label{eq:si-constant-D-update}
\end{equation}
\end{subequations}
Fig.~\ref{fig:finite_field_flow} summarizes the implementation of the finite-field couplings in CP2K.

\begin{fixedfigure}
    \centering
    \includegraphics[width=1.0\linewidth]{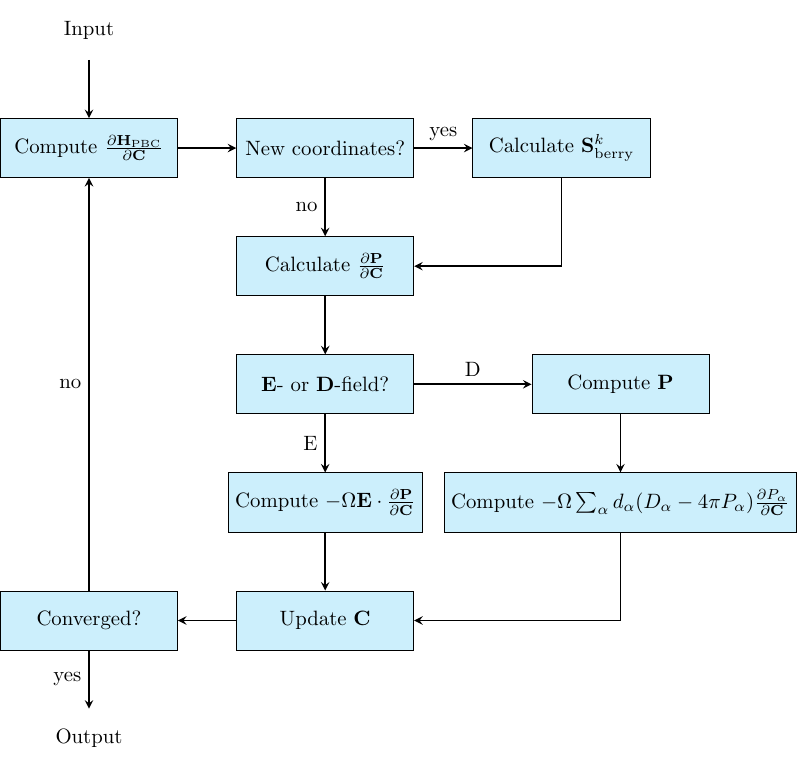}
    \caption{\enspace Flow chart of the finite-field module in CP2K.}
    \label{fig:finite_field_flow}
\end{fixedfigure}

\subsubsection{Forces in Finite-Field Functionals}
\label{sec:forces_finite_field}

At fixed macroscopic electric field \(\mathbf E\), the electric enthalpy is given by Eq.~\eqref{eq:si-constant-E-functional}.
Let $\mathbf{C}_\mathbf{E}(\{\mathbf{R}_I\})$ denote the electronic stationary point at fixed nuclei and fixed $\mathbf{E}$ as defined by
\begin{equation}
\left.\frac{\partial \mathcal{F}}{\partial \mathbf{C}}\right|_{\mathbf{C}_\mathbf{E}}= \mathbf{0}.
\end{equation}
The force on nucleus $I$ is defined from the total derivative of the stationary functional
\begin{equation}
\mathbf{F}_I^{(E)}
=
-\frac{d}{d\mathbf{R}_I}\,
\mathcal{F}\!\left(\mathbf{E};\mathbf{C}_\mathbf{E}(\{\mathbf{R}\}),\{\mathbf{R}\}\right).
\end{equation}
Applying the chain rule yields
\begin{equation}
\frac{d\mathcal{F}}{d\mathbf{R}_I}
=
\left.\frac{\partial \mathcal{F}}{\partial \mathbf{R}_I}\right|_{\mathbf{C}_\mathbf{E}}
+
\left.\frac{\partial \mathcal{F}}{\partial \mathbf{C}}\right|_{\mathbf{C}_\mathbf{E}}
:\frac{d\mathbf{C}_\mathbf{E}}{d\mathbf{R}_I}.
\end{equation}
The stationary condition implies that the second term vanishes. Therefore the constant-$\mathbf{E}$ force can be written as
\begin{equation}
\mathbf{F}_I^{(E)}
=
-\left.\frac{\partial H_{\mathrm{PBC}}}{\partial \mathbf{R}_I}\right|_{\mathbf{C}_\mathbf{E}}
+\Omega\,\mathbf{E}\cdot
\left.\frac{\partial \mathbf{P}(\mathbf{C}_\mathbf{E},\{\mathbf{R}\})}{\partial \mathbf{R}_I}\right|_{\mathbf{C}_\mathbf{E}}.
\label{eq:force_constE_general}
\end{equation}

In a representation where the electronic polarization has no explicit dependence on $\{\mathbf{R}_I\}$ beyond the variational electronic degrees of freedom (e.g., PW basis sets)
\begin{equation}
\mathbf{P}(\mathbf{C}_\mathbf{E},\{\mathbf{R}\}) = \mathbf{P}_{\mathrm{ion}}(\{\mathbf{R}\}) + \mathbf{P}_{\mathrm{el}}(\mathbf{C}_\mathbf{E}),
\label{eq:P_split_no_explicit_R}
\end{equation}
one has
\begin{equation}
\left.\frac{\partial \mathbf{P}_{\mathrm{el}}(\mathbf{C}_\mathbf{E})}{\partial \mathbf{R}_I}\right|_{\mathbf{C}_\mathbf{E}}=\mathbf{0},
\end{equation}
and thus, at the stationary point
\begin{equation}
\left.\frac{\partial \mathbf{P}(\mathbf{C}_\mathbf{E},\{\mathbf{R}\})}{\partial \mathbf{R}_I}\right|_{\mathbf{C}_\mathbf{E}}
=
\frac{\partial \mathbf{P}_{\mathrm{ion}}(\{\mathbf{R}\})}{\partial \mathbf{R}_I}.
\end{equation}
In this case the field contribution to the force reduces to a simple ``ionic'' term, and one may summarize the result as follows.\cite{Souza:2002fa,Umari:2002eo}
\begin{equation}
\mathbf{F}_I^{(E)}=
\mathbf{F}_I^{\mathrm{PBC}}
+ eZ_I\mathbf{E},
\label{eq:force_E_PW}
\end{equation}
where $\mathbf{F}_I^{\mathrm{PBC}}=-\partial H_{\mathrm{PBC}}/\partial \mathbf{R}_I$ is the standard KS force at the finite-field electronic state.

In CP2K's implementation, the electronic polarization is evaluated from overlap-like matrices of the Berry operators in an atom-centered AO basis. Consequently, $\mathbf{P}$ depends explicitly on $\{\mathbf{R}_I\}$ through AO integrals, and Eqs.~\eqref{eq:P_split_no_explicit_R} and~\eqref{eq:force_E_PW} do \emph{not} hold. This introduces additional Pulay-like contributions to the field-dependent forces.

The corresponding nuclear derivative can be written as
\begin{align}
\frac{\partial \gamma^\textrm{el}_k}{\partial \mathbf{R}_I}
={}&
\operatorname{Im}\,\operatorname{Tr}\!\left((\mathbf{M}^k)^{-1}\frac{\partial \mathbf{M}^k}{\partial \mathbf{R}_I}\right)
\nonumber \\
={}&
\operatorname{Im}\,\operatorname{Tr}\!\left(\mathbf{K}^k\,\frac{\partial \mathbf{S}_\textrm{Berry}^k}{\partial \mathbf{R}_I}\right),
\end{align}
where the AO-space kernel $\mathbf{K}^k$ is defined as
\begin{equation}
    \mathbf{K}^k = \mathbf{C}(\mathbf{M}^k)^{-1}\mathbf{C}^\top.
\end{equation}

At fixed displacement field \(\mathbf D\), CP2K employs the functional in Eq.~\eqref{eq:si-constant-D-functional}.
The same stationary argument can be applied,
and the force becomes
\begin{equation}
\begin{aligned}
\mathbf{F}_I^{(D)}
&=
-\left.\frac{\partial H_{\mathrm{PBC}}}{\partial \mathbf{R}_I}\right|_{\mathbf{C}_\mathbf{D}}\\
&\quad +\Omega\sum_{\alpha = x,y,z}
d_\alpha\left(D_\alpha-4\pi P_\alpha\right)
\left.\frac{\partial P_\alpha}{\partial \mathbf{R}_I}\right|_{\mathbf{C}_\mathbf{D}}.
\end{aligned}
\label{eq:force_constD_general}
\end{equation}

Because the Berry-phase polarization under PBC is multivalued modulo a polarization quantum, consistent branch tracking and phase unwrapping of $\bm{\gamma}^{\mathrm{el}}$ are required during geometry optimization or MD. They maintain a continuous field-coupling energy in both electrical ensembles and, because Eq.~\eqref{eq:force_constD_general} depends explicitly on $\mathbf{P}$, continuous constant-$\mathbf{D}$ forces.\cite{KingSmith:1993hp} Finally, it is worth noting that the analytic stress tensor under either $\mathbf{E}$ or $\mathbf{D}$ fields is not yet implemented.

\section{Nuclear Quantum Effects}
\label{sec:si-nuclear-quantum-effects}

\subsection{Periodic Constrained Nuclear-Electronic Orbital Density-Functional Theory}
\label{sec:si-cneo}

CNEO-DFT starts from
multicomponent DFT,\cite{Capitani82568,gidopoulos1998kohn}
in which the total energy is expressed as a functional of the electronic
density $n^{\mathrm{e}}(\bm{r})$ and the quantum-nuclear densities
$\{n_a^{\mathrm{n}}(\bm{r})\}$:
\begin{align}
E_{V_{\mathrm{ext}}}\!\left[
n^{\mathrm{e}},\{n_a^{\mathrm{n}}\}
\right]
&=
T_{\mathrm{s}}^{\mathrm{e}}\!\left[n^{\mathrm{e}}\right]
+\sum_{a\in\mathcal{Q}}T_{\mathrm{s}}^{\mathrm{n},a}\!\left[n_a^{\mathrm{n}}\right]
\nonumber \\
&\quad+ \int\!\dd \bm{r}\,
\!\left[n^{\mathrm{e}}(\bm{r})
- \sum_{a\in\mathcal{Q}} Z_a n_a^{\mathrm{n}}(\bm{r}) \right]
\!V_{\mathrm{ext}}(\bm{r})
\nonumber \\
&\quad+\frac12 \iint \dd \bm{r}\,\dd \bm{r}'\,
\frac{n^{\mathrm{e}}(\bm{r})n^{\mathrm{e}}(\bm{r}')}{|\bm{r}-\bm{r}'|}
\nonumber \\
&\quad-\sum_{a\in\mathcal{Q}} Z_a
\iint \dd \bm{r}\,\dd \bm{r}'\,
\frac{n^{\mathrm{e}}(\bm{r})n_a^{\mathrm{n}}(\bm{r}')}{|\bm{r}-\bm{r}'|}
\nonumber \\
&\quad + \sum_{a<b\in\mathcal{Q}} Z_a Z_b
\iint \dd \bm{r}\,\dd \bm{r}'\,
\frac{n_a^{\mathrm{n}}(\bm{r})n_b^{\mathrm{n}}(\bm{r}')}{|\bm{r}-\bm{r}'|}
\nonumber \\
&\quad+ E_{\mathrm{xc}}^{\mathrm{e}}\!\left[n^{\mathrm{e}}\right]
 + E_{\mathrm{c}}\!\left[n^{\mathrm{e}},\{n_a^{\mathrm{n}}\}\right].
\label{eq:si_mcdft_energy}
\end{align}
The set $\mathcal{Q}$ contains the quantum nuclei, $Z_a$ is the
charge of quantum nucleus $a$, and $V_{\mathrm{ext}}(\bm{r})$ is the
external potential generated by the classical nuclei. The functionals \(T_{\mathrm s}^{\mathrm e}\) and \(T_{\mathrm s}^{\mathrm{n},a}\) are the noninteracting electronic and quantum-nuclear kinetic energies, \(E_{\mathrm{xc}}^{\mathrm e}\) is the electronic XC energy, and \(E_{\mathrm c}\) is the residual electron--nuclear correlation functional. The three Coulomb
terms in Eq.~\eqref{eq:si_mcdft_energy} describe electron--electron
repulsion, electron--quantum-nucleus attraction, and
quantum-nucleus--quantum-nucleus repulsion, respectively. The functional
$E_{\mathrm{c}}[n^{\mathrm{e}},\{n_a^{\mathrm{n}}\}]$ contains the
remaining multicomponent correlation effects beyond the mean-field
description.

CNEO-DFT adopts a distinguishable-particle approximation for the quantum
nuclei, which is justified when neighboring quantum-nuclear densities
have negligible overlap.\cite{chen2026constrained} Each quantum nucleus
is then associated with a prescribed position expectation value:
\begin{equation}
\langle \hat{\bm{r}}_a^{\mathrm{n}} \rangle
\equiv
\int \dd\bm{r}\,
n_a^{\mathrm{n}}(\bm{r})\,\bm{r}
= \bm{R}_a .
\label{eq:si_cneo_constraint}
\end{equation}
Minimization with respect to the electronic and quantum-nuclear orbitals,
subject to orbital orthonormality and the position constraints in
Eq.~\eqref{eq:si_cneo_constraint}, gives coupled KS equations:
\begin{subequations}
\begin{equation}
\left[
-\frac{1}{2}\nabla^2 + V_{\mathrm{eff}}^{\mathrm{e}}(\bm{r})
\right]\psi_i^{\mathrm{e}}(\bm{r})
=
\varepsilon_i^{\mathrm{e}}\,\psi_i^{\mathrm{e}}(\bm{r}),
\label{eq:si_cneo_ks_e}
\end{equation}
\begin{equation}
\begin{aligned}
&\left[
-\frac{1}{2M_a}\nabla^2
+ V_{\mathrm{eff}}^{\mathrm{n},a}(\bm{r})
+ \bm{f}_a \cdot (\bm{r}-\bm{R}_a)
\right] \\
&\quad \times \psi_i^{\mathrm{n},a}(\bm{r})
=
\varepsilon_i^{\mathrm{n},a}\,\psi_i^{\mathrm{n},a}(\bm{r}),
\end{aligned}
\label{eq:si_cneo_ks_n}
\end{equation}
\end{subequations}
where $M_a$ is the mass of quantum nucleus $a$, and $\bm{f}_a$ is the
Lagrange multiplier enforcing the position constraint. The functions \(\psi_i^{\mathrm e}\) and \(\psi_i^{\mathrm{n},a}\) are electronic and quantum-nuclear orbitals, \(\varepsilon_i^{\mathrm e}\) and \(\varepsilon_i^{\mathrm{n},a}\) are their eigenvalues, and \(V_{\mathrm{eff}}^{\mathrm e}\) and \(V_{\mathrm{eff}}^{\mathrm{n},a}\) are the corresponding effective potentials. The effective
electronic potential includes the classical nuclear potential, the
electron--electron Hartree term, attraction to the quantum nuclei, and
XC and multicomponent correlation contributions. The
effective potential for quantum nucleus $a$ contains its
interactions with the classical nuclei, the electronic density, and the
other quantum nuclei. Approximate multicomponent correlation functionals
have been explored,\cite{yang2017development,brorsen2017multicomponent,tao2019multicomponent}
but the dominant CNEO effects are already captured at the constrained
multicomponent mean-field level.\cite{yang2024assessment}

The constrained solution defines a full-dimensional CNEO effective
potential energy surface as a function of the classical nuclear
coordinates and the constrained position expectation values of the quantum
nuclei. This surface embeds the dominant quantum delocalization of the
selected nuclei and can be used for geometry optimization,
transition-state search, and MD.\cite{Xu224039,Chen23279,chen2026constrained}

Under periodic boundary conditions, CNEO-DFT must reconcile localized,
distinguishable quantum nuclei with the translational symmetry of the
electrostatics. In CP2K, the quantum-nuclear orbitals are kept localized
in a reference cell, while the quantum-nuclear charge density entering
the Coulomb terms is made periodic by lattice summation.\cite{Chen2025}
\begin{equation}
n^{\mathrm{n}}(\bm{r})
=
\sum_{a\in \mathrm{cell}}
\sum_{\bm{m}}
n_a^{\mathrm{n}}(\bm{r}+\bm{T}_{\bm{m}}),
\label{eq:si_periodic_nuclear_density}
\end{equation}
where $\bm{T}_{\bm{m}}=m_1\bm{a}_1+m_2\bm{a}_2+m_3\bm{a}_3$ is a
lattice vector. A complementary treatment based on Resta's position
operator has also been discussed.\cite{liu2025constrained}

The periodic CP2K implementation uses the GAPW
framework,\cite{Lippert1999,Krack2000} which naturally separates smooth
grid-based densities from localized atom-centered corrections. For each
quantum nucleus:
\begin{equation}
n_a^{\mathrm{n}}(\bm{r})
=
\tilde{n}_a^{\mathrm{n}}(\bm{r})
+
\left[
n_a^{\mathrm{n}}(\bm{r})
-\tilde{n}_a^{\mathrm{n}}(\bm{r})
\right],
\label{eq:si_gapw_n_density}
\end{equation}
where $\tilde{n}_a^{\mathrm{n}}(\bm{r})$ is the smooth part of the
density. The smooth components contribute to the global periodic charge
density represented on the FFT grid, whereas the localized differences
are handled on atom-centered grids. Atom-centered compensation charges
reproduce the multipole moments of the localized target densities, now
including the quantum-nuclear contributions, so the multicomponent
Coulomb energy can reuse the standard GAPW global-plus-local
partitioning strategy.\cite{Chen2025}

Analytic gradients with respect to both classical nuclear coordinates and constrained quantum-nuclear position expectation values are required for optimization, vibrational analysis, and CNEO-MD.\cite{Xu20074106,Chen2025}
Most force terms are inherited from the standard GAPW machinery, including
contributions from basis functions, projectors, compensation charges, and
the classical external potential.\cite{Chen2025} The new terms arise from
the quantum-nuclear kinetic energy, the quantum-nuclear external
potential, and Coulomb contributions involving the smooth and localized
quantum-nuclear densities. When Gaussian nuclear basis functions are
centered at the constrained positions $\bm{R}_a$, the nuclear overlap and
kinetic matrices are translationally invariant with respect to
$\bm{R}_a$, which removes the corresponding overlap-derivative
contributions. The remaining explicit derivatives are dominated by the
external-potential term and the smooth-density part of the periodic
electrostatics.

The resulting analytic gradients are evaluated through the standard CP2K force interface and can therefore be used by the existing CP2K infrastructure for geometry optimization, vibrational analysis, and MD.

In CNEO-MD, the propagated variables are the classical nuclear
coordinates and the position expectation values of the quantum nuclei.

At each time step, the coupled electronic and quantum-nuclear KS equations are solved together with the position constraints at the instantaneous coordinates, and the corresponding analytic forces are used to advance the trajectory:\cite{Xu224039,Chen23279}
\begin{equation}
M_a \frac{\dd \bm{R}_a}{\dd t}
=
\bm{P}_a,
\qquad
\frac{\dd \bm{P}_a}{\dd t}
=
-\nabla_{\bm{R}_a} V_{\mathrm{CNEO}},
\label{eq:si_cneo_md_eom}
\end{equation}
where $V_{\mathrm{CNEO}}$ is the constrained CNEO potential energy
surface. The vector \(\bm P_a\) is the momentum conjugate to the position expectation value \(\bm R_a\), and \(t\) is time. Although these equations have a classical form, the underlying
CNEO effective potential energy surface already contains the dominant NQEs associated with the quantum nuclei.

This treatment differs from path-integral-based dynamics methods because
CNEO-MD propagates the classical nuclear coordinates and quantum-nuclear
position expectation values on a single CNEO effective potential energy
surface without introducing ring-polymer replicas. It is nevertheless
conceptually related to centroid molecular dynamics because both propagate
effective nuclear positions as classical dynamical variables: constrained
position expectation values in CNEO-MD and ring-polymer centroids in
centroid molecular dynamics.\cite{Chen23279,chen2026constrained}

An accelerated CNEO-AIMD scheme based on the Grassmann formulation of
second-generation Car--Parrinello AIMD described in Section~IX~A has not
yet been implemented. Such a scheme would avoid fully converging the
coupled electronic and quantum-nuclear problem at every time step by
applying a consistent predictor--corrector treatment to the electronic
and quantum-nuclear orbitals together with the position-constraint
variables.
%dm \subsection{Path-Integral Dynamics and Nuclear Quantum Effects for Spectroscopy}
%dm \label{sec:si-pimd-nqe}
%dm % C. Schran, H. Forbert, D. Marx

\endgroup

\end{document}